%% file: ms.tex
\documentclass[acmsmall,screen]{acmart}

\newif\iftechreport
\techreporttrue

\usepackage{array}
\usepackage{nccmath}
\usepackage[inline]{enumitem}
\usepackage{graphicx}
\usepackage{listings}
\usepackage{mathpartir, proof}
\usepackage{thmtools, amsthm, thm-restate}
\usepackage{xcolor}
\usepackage{xspace}
\usepackage{tikz}
\usepackage{balance}
\usepackage{subcaption}
\usepackage[capitalize,noabbrev]{cleveref}

\acmPrice{}
\acmDOI{10.1145/3498701}
\acmYear{2022}
\copyrightyear{2022}
\acmSubmissionID{popl22main-p311-p}
\acmJournal{PACMPL}
\acmVolume{6}
\acmNumber{POPL}
\acmArticle{40}
\acmMonth{1}

\setcopyright{rightsretained}

\definecolor{cornflower}{RGB}{65,105,225}
\definecolor{maroon}{RGB}{178,34,34}
\definecolor{light-gray}{RGB}{245,245,245}
\definecolor{keywordsColor}{RGB}{64,120,242}
\definecolor{keywordsColor2}{RGB}{166,38,164}
\definecolor{keywordsColor3}{RGB}{193,132,1}
\definecolor{keywordsColor4}{RGB}{228,86,73}

\lstdefinestyle{safep4}{
  basicstyle=\ttfamily,
  moredelim=[is][\color{blue}]{@}{@}
}

\lstdefinelanguage{p4}{
  language=C,
  basicstyle=\linespread{0.8}\footnotesize\ttfamily,
  numberstyle=\tiny\ttfamily,
  numbers=left,
  breaklines=true,
  commentstyle=\color{gray},
  frame=none,
  tabsize=2,
  xleftmargin=15pt,
  showstringspaces=false,
  morekeywords={
    extract,
    add,
    emit,
    remit,
    reset,
    if,
    else,
    valid,
    as,
    pkt_out,
    pkt_in,
    length,
    parser,
    transition,
    select,
    control,
    apply,
    isValid,
    state,
    :=
  },
  keywordstyle=\color{keywordsColor},
  otherkeywords={!=,&&,||,:=},
  alsoletter = {>,->,=>,<=>},  
  morekeywords=[2]{>,!,!=,==,=i=,=>,<=>,inout,out,&&,||},
  keywordstyle=[2]{\color{keywordsColor2}},
  morekeywords=[3]{->},
  keywordstyle=[3]{\color{keywordsColor4}},
  moredelim=**[is][\color{keywordsColor2}]{>>}{<<}
}

\lstnewenvironment{p4code}[1][]{
  \lstset{language=p4,mathescape=true,#1}
}{}

\input{macros}

\title{Dependently-Typed Data Plane Programming}
\author{Matthias Eichholz}
\affiliation{
  \institution{Technical University of Darmstadt}
  \country{Germany}
}
\email{eichholz@cs.tu-darmstadt.de}

\author{Eric Hayden Campbell}
\affiliation{
  \institution{Cornell University}
  \country{USA}
}
\email{ehc86@cornell.edu}

\author{Matthias Krebs}
\affiliation{
  \institution{Technical University of Darmstadt}
  \country{Germany}
}
\email{krebs@cs.tu-darmstadt.de}

\author{Nate Foster}
\affiliation{
  \institution{Cornell University}
  \country{USA}
}
\email{jnfoster@cs.cornell.edu}

\author{Mira Mezini}
\affiliation{
  \institution{Technical University of Darmstadt}
  \country{Germany}
}
\email{mezini@cs.tu-darmstadt.de}

\begin{document}

\input{00-abstract}



\maketitle

\input{01-introduction}
\input{02-background}
\input{03-formalization}
\input{04-implementation}
\input{05-experience}
\input{06-case-study}
\input{07-related-work}
\input{08-conclusion}

\begin{acks}
We are grateful to the POPL reviewers for their careful feedback and
many suggestions for improving this paper. Our work has been supported
in part by the German Research Foundation (DFG) as part of the Collaborative
Research Center (CRC) 1053 MAKI, by the National Research Center for Applied 
Cybersecurity ATHENE, by the National Science Foundation under grant FMiTF-1918396
as well as a Graduate Research Fellowship, the Defense Advanced
Research Projects Agency under Contract HR001120C0107, and gifts from
Keysight and InfoSys.
\end{acks}
\balance
\bibliographystyle{ACM-Reference-Format}
\bibliography{ms}
\iftechreport
\clearpage
\appendix
\input{appendix/definitions}
\input{appendix/safety}

\input{appendix/algorithmic_types}
\input{appendix/mtu}
\input{appendix/rewrites.tex}
\fi
\end{document}

%% file: macros.tex
\newcommand{\name}{$\Pi$4\xspace}

\newcommand{\sigmaT}[3]{\ensuremath{\Sigma#1\!:#2.#3}}

\newcommand{\inst}{\ensuremath{\iota}}
\newcommand{\instWeak}{\ensuremath{\inst_\sim}}
\newcommand{\weak}[1]{\ensuremath{{#1}{\texttt{\char`~}}}}
\newcommand{\refT}[3]{\ensuremath{\{#1:#2\mid#3\}}}
\newcommand{\rT}[2]{\ensuremath{\refT{x}{#1}{#2}}}
\newcommand{\cmdType}[3]{\ensuremath{#1\vdash #2\!:\!#3}}
\newcommand{\cmdT}[2]{\ensuremath{\Gamma\vdash #1:#2}}
\newcommand{\bound}[3]{\ensuremath{#1 \vdash #2 \leq #3}}
\newcommand{\tNat}{\ensuremath{\mathbb{N}\xspace}}
\newcommand{\tBool}{\ensuremath{\mathbb{B}\xspace}}
\newcommand{\tBv}{\ensuremath{\mathsf{BV}\xspace}}
\newcommand{\cmdTypeAlgo}[4]{\ensuremath{#1\vdash #2:(#3)\rightsquigarrow#4}}

\newcommand{\typeEquiv}[3]{#1 \vdash #2 \doteq #3}

\newcommand{\eExtract}{\textsc{E-Extract}\xspace}
\newcommand{\eRemit}{\textsc{E-Remit}\xspace}
\newcommand{\eMod}{\textsc{E-Mod}\xspace}
\newcommand{\eModOne}{\textsc{E-Mod1}\xspace}
\newcommand{\eReset}{\textsc{E-Reset}\xspace}
\newcommand{\eAdd}{\textsc{E-Add}\xspace}

\newcommand{\eAscribe}{\textsc{E-Ascribe}\xspace}
\newcommand{\eSeq}{\textsc{E-Seq}\xspace}
\newcommand{\eSeqOne}{\textsc{E-Seq1}\xspace}
\newcommand{\eIf}{\textsc{E-If}\xspace}
\newcommand{\eIfTrue}{\textsc{E-IfTrue}\xspace}
\newcommand{\eIfFalse}{\textsc{E-IfFalse}\xspace}

\newcommand{\tExtract}{\textsc{T-Extract}\xspace}
\newcommand{\tSeq}{\textsc{T-Seq}\xspace}
\newcommand{\tSkip}{\textsc{T-Skip}\xspace}
\newcommand{\tRemit}{\textsc{T-Remit}\xspace}
\newcommand{\tReset}{\textsc{T-Reset}\xspace}
\newcommand{\tIf}{\textsc{T-If}\xspace}
\newcommand{\tMod}{\textsc{T-Mod}\xspace}

\newcommand{\tAdd}{\textsc{T-Add}\xspace}

\newcommand{\tSub}{\textsc{T-Sub}\xspace}
\newcommand{\tAscribe}{\textsc{T-Ascribe}\xspace}
\newcommand{\tExtractAlgo}{\textsc{T-Extract-Algo}\xspace}
\newcommand{\tSeqAlgo}{\textsc{T-Seq-Algo}\xspace}
\newcommand{\tSkipAlgo}{\textsc{T-Skip-Algo}\xspace}
\newcommand{\tRemitAlgo}{\textsc{T-Remit-Algo}\xspace}
\newcommand{\tResetAlgo}{\textsc{T-Reset-Algo}\xspace}
\newcommand{\tIfAlgo}{\textsc{T-If-Algo}\xspace}

\newcommand{\tModAlgo}{\textsc{T-Mod-Algo}\xspace}
\newcommand{\tAddAlgo}{\textsc{T-Add-Algo}\xspace}

\newcommand{\tAscribeAlgo}{\textsc{T-Ascribe-Algo}\xspace}

\newcommand{\entTop}{\textsc{Ent-Top}\xspace}
\newcommand{\entChoiceL}{\textsc{Ent-ChoiceL}\xspace}

\newcommand{\entRefine}{\textsc{Ent-Refine}\xspace}
\newcommand{\entSigma}{\textsc{Ent-Sigma}\xspace}
\newcommand{\entSubst}{\textsc{Ent-Subst}\xspace}

\newcommand{\cExtract}[1]{\ensuremath{\mathit{extract(#1)}}\xspace}
\newcommand{\cIf}[3]{\ensuremath{\mathit{if(#1)\ then\ #2\ else\ #3}}\xspace} %
\newcommand{\cRemit}[1]{\ensuremath{\mathit{remit(#1)}}\xspace}
\newcommand{\cSkip}{\ensuremath{\mathit{skip}}\xspace}
\newcommand{\cReset}{\ensuremath{\mathit{reset}}\xspace}
\newcommand{\cAdd}{\ensuremath{\mathit{add}}\xspace}
\newcommand{\cAscribe}[2]{\ensuremath{#1\ \mathit{as}\ #2}\xspace}
\newcommand{\cMod}[2]{\ensuremath{#1 := #2}\xspace}
\newcommand{\cmdVar}{\ensuremath{\mathsf{heap}}}

\newcommand{\chomp}{\ensuremath{\mathsf{chomp}}}
\newcommand{\chompRef}{\ensuremath{\mathsf{chompRef}}}
\newcommand{\chompRec}{\ensuremath{\mathsf{chompRec}}}
\newcommand{\chompForm}{\ensuremath{\chomp^{\varphi}}}
\newcommand{\chompExpr}{\ensuremath{\chomp^{e}}}
\newcommand{\chompS}{\ensuremath{\mathsf{chomp}^\Downarrow}}
\newcommand{\heapRef}{\ensuremath{\mathsf{heapRef}}}

\newcommand{\true}{\ensuremath{\mathsf{true}}}
\newcommand{\false}{\ensuremath{\mathsf{false}}}
\newcommand{\valid}{\ensuremath{\mathit{valid}}}
\newcommand{\concat}[2]{\ensuremath{#1 +\!\!+\ #2}}
\newcommand{\bvconcat}[2]{\ensuremath{#1 @ #2}}
\newcommand{\bv}[1]{\ensuremath{\langle#1\rangle}}
\newcommand{\bvNil}{\ensuremath{\langle\rangle}}
\newcommand{\length}[1]{\ensuremath{\lvert #1\rvert}}
\newcommand{\includes}[2]{\ensuremath{\mathsf{Includes}~\Gamma~#1~#2}}
\newcommand{\excludes}[2]{\ensuremath{\mathsf{Excludes}~\Gamma~#1~#2}}
\newcommand{\includesCtx}[3]{\ensuremath{\mathsf{Includes}_{#3}~#1~#2}}
\newcommand{\excludesCtx}[3]{\ensuremath{\mathsf{Excludes}_{#3}~#1~#2}}
\newcommand{\sizeof}{\ensuremath{\mathsf{sizeof}}}
\newcommand{\pIn}{{\ensuremath{pkt_{in}}}}
\newcommand{\pOut}{\ensuremath{pkt_{out}}}
\newcommand{\slice}[3]{\ensuremath{#1[#2\!:\!#3]}}
\newcommand{\bvar}[1]{\ensuremath{\mathsf{b_{#1}}}}

\newcommand{\Heaps}{\ensuremath{\mathcal{H}}}
\newcommand{\semantics}[1]{\ensuremath{\llbracket #1 \rrbracket}}
\newcommand{\env}{\ensuremath{\mathcal{E}}}

\newcommand{\dom}{\ensuremath{\mathit{dom}}}

\newcommand{\todo}[1]{\textcolor{red}{\textbf{#1}}}

\newcommand{\HT}{\ensuremath{\mathcal{HT}}}

\newcommand{\PPPP}[1]{\ensuremath{\textrm{P4}_{#1}}}

\newcommand{\kw}[1]{\ensuremath{\mathsf{#1}}}

\newcommand{\subtypeCtx}[3]{#1 \vdash #2 <: #3}
\newcommand{\subtype}[2]{\subtypeCtx{\Gamma}{#1}{#2}}

\newcommand{\bitvar}[1]{\kw{b_{#1}}}

\newcommand{\case}[1]{\item[\textit{Case} #1:]}
\newcommand{\subcase}[1]{\item[\textit{Subcase} #1:]}

\newenvironment{pf}[2][0pt]
{\begin{enumerate}[label=(A\arabic*), leftmargin=#1, series=#2]}
    {\end{enumerate}}
\newenvironment{pf*}[1]
{\begin{enumerate}[label=(A\arabic*), resume*=#1]}
    {\end{enumerate}}

\newcommand{\emit}[1]{\ensuremath{\mathsf{emit}(#1)}}

\newcommand{\appendixref}[1]{
  \iftechreport#1\else{}the companion technical report\fi
}

%% file: 00-abstract.tex
\begin{abstract}
  Programming languages like P4 enable specifying the behavior of
  network data planes in software. However, with increasingly powerful
  and complex applications running in the network, the risk of faults
  also increases. Hence, there is growing recognition of the need for
  methods and tools to statically verify the correctness of P4 code,
  especially as the language lacks basic safety guarantees. Type
  systems are a lightweight and compositional way to establish program
  properties, but there is a significant gap between the kinds of
  properties that can be proved using simple type systems (e.g.,
  SafeP4) and those that can be obtained using
  full-blown verification tools (e.g., \texttt{p4v}).
  In this paper, we close this gap by developing $\Pi$4, a
  dependently-typed version of P4 based on decidable refinements. We
  motivate the design of $\Pi$4, prove the soundness of its type
  system, develop an SMT-based implementation, and present case
  studies that illustrate its applicability to a variety of data plane
  programs.
\end{abstract}

%% file: 01-introduction.tex
\section{Introduction}

Computer networks are becoming increasingly programmable as languages
like P4 \cite{p4} make it possible to specify the behavior of data
planes in software. With the increased availability of programmable
devices, a number of powerful and complex applications having become
viable, ranging from novel network protocols to full-blown in-network
computation (e.g., executing application-level storage queries using
network devices~\cite{netcache}). But as the complexity of these
applications increases, so does the risk of faults, especially as P4's
main abstraction for representing packet data---namely header
types---lacks basic safety guarantees. Experience with a growing
number of programs has shown the risks of the unsafe approach, which
often leads to subtle software bugs~\cite{Eichholz2019,Liu2018}. This
is clearly unacceptable, given the crucial role that networks play in
nearly all modern systems. Hence, we need methods and tools to
statically verify the correctness of data plane programs.

Today, most data plane verification tools \cite{Liu2018,
  Stoenescu2018, Dumitrescu2020} are monolithic in nature. For
example, \texttt{p4v}~\cite{Liu2018}, which is based on software model
checking, operates on whole programs. But while monolithic approaches
have certain advantages---e.g., they minimize the need for programmer
annotations---they also have downsides.
The most fundamental limitation is the inherent tension with modular
design---it is difficult to accommodate an ``open-world'' model, in
which third-party components are plugged into existing programs. For
instance, an equipment vendor might want to provide a ``base'' program
that implements standard packet-processing functionality like Ethernet
switching, and allow customers to add custom functionality of their
own design~\cite{lyra,myP4,daPIPE}. Composable approaches to data
plane programming require compositional reasoning
methods~\cite{Zen-HotNets20}.

Type systems are a lightweight and compositional way to establish
program properties---i.e., the types for individual components
document assumptions about the components they rely upon as well as
the guarantees they offer. However, somewhat surprisingly, types have
rarely been applied in the realm of network programming, and the few
exceptions~\cite{Tussle10, PacLang04, Eichholz2019} are simple type
systems with limited expressive power. For example,
SafeP4~\cite{Eichholz2019} uses regular
types~\cite{xDuce,RegObjTypes,Castagna15} and path-sensitive
occurrence typing~\cite{Tobin-Hochstadt:2010aa} to reason about basic
safety properties, but it cannot capture richer program properties
(e.g., whether the IPv4 and IPv6 headers are only ever accessed on
mutually exclusive execution paths), or track the values of individual
fields (e.g., whether EtherType equals to \texttt{0x0800}, which
indicates an IPv4 packet, or to \texttt{0x86DD}, which indicates an
IPv6 packet). The inability of SafeP4 to reason about the values being
manipulated by the program significantly limits its expressiveness. In
general, there is a significant gap between the kinds of properties
that can be checked using type systems like SafeP4 and full-blown
verification tools like \texttt{p4v}.

Thus, it is natural to ask whether we can design a compositional type
system that has the same expressive power as data plane verification
tools. This paper answers this question in the affirmative, by
presenting \name---a dependently-typed version of P4. \name fits with
the trend of recently proposed dependently-typed languages
\cite{Xi1999,Condit2007,Rondon2008,Vazou2014} that are blurring the
line between type checking and theorem proving. For instance, Liquid
Haskell~\cite{Rondon2008,Vazou2014}
allows programmers to smoothly shift from properties that can be
checked with traditional typing disciplines to more sophisticated ones
that go beyond simple syntactic checks. Under the hood, Liquid Haskell
uses an SMT solver to automatically discharge the logical proof
obligations generated during type checking.

Yet, thus far, the dependently-typed approach has not been explored
for network programming. In this paper, we demonstrate that data plane
programs are a ``killer application'' for dependent types. On the one
hand, they clearly need precise types, as most programs rely on
intricate packet formats (e.g., so-called ``type-length-value''
encodings, where the first few bits determine the type, length, and
structure of the bits that follow). On the other hand, data plane
programs are fundamentally simple (e.g., P4 does not support pointers
or loops) and lack the kinds of complex features that often make
precise type systems complicated to design and implement.

Our main contribution lies in exploring and addressing the subtle
challenges that arise in developing a dependent type system for the P4
programming language, including balancing the tradeoffs between
expressiveness and decidability. \name features a combination of
refinement types, dependent function types, union types, and explicit
substitutions. This combination is key to retain precision during type
checking---e.g., we can compute exact types for conditionals, thereby
having access to an accurate type at any program point. Moreover, our
design enables precise typing in the presence of domain-specific
features that combine packet serialization and deserialization
operations with imperative control-flow. To this end, \name combines a
dependent sum type with a novel ``chomp'' operation that computes the
type that remains after extracting bits from a packet buffer. We
formally prove that \name's type system is sound via standard progress
and preservation theorems.

The chomp operator is reminiscent of regular expression
derivatives~\cite{brzozowski64}. To the best of our knowledge, with a
few notable exceptions (e.g., work by McBride~\cite{McBride01})
derivative-like operations have not been extensively studied the
context of dependent type systems.
Thus, beyond providing an elegant solution to a practical problem, we
believe that \name's innovative combination of dependent types with
regular types and the possibility to compute derivatives of types is
of general theoretical interest and may be useful in other
domains---e.g., one potential direction is verified serializers and
deserializers \cite{EverParse, Delaware2019}.

We have built a prototype implementation of \name in OCaml and the Z3
SMT solver. The type checker determines whether a \name program has a
given type by checking the validity of a series of logical formulae
using an SMT solver. We encode types into the effectively
propositional fragment of first-order logic over fixed-width bit
vectors, which facilitates automating subtyping and equivalence
checks. We prove (cf. Theorem 4.1) that this logical fragment is
sufficient for checking our types under the assumption that the types
written by the programmer denote finite sets. We believe this is a
reasonable assumption, because networks enforce a \emph{maximum
  transmission unit} (MTU) (i.e., a bound on the size of packets
constraining the maximum number of bits that network switches can
receive or transmit\footnote{The MTU is often set to 1500 bytes.})
which bounds the size of bit vectors we need to consider in the
encoding. In the presence of an MTU, our types denote finite sets,
which can be enumerated to decide the key judgments (i.e., subtyping,
size constraints, and inclusion checks).

Using our \name prototype, we develop several case studies,
demonstrating that \name is capable of expressing and (modularly)
reasoning about properties from the literature ranging from basic
safety to intricate invariants: parser round-tripping, protocol
conformance, determined forwarding, etc. We selected properties that
are also studied by recent data plane verification tools like
\texttt{p4v} or Vera, which indicates that \name is capable of
covering properties of interest to the networking community. However,
we leave a careful study of the practical utility of \name (e.g., with
larger examples and user studies) to future work.

\vspace{1em}

\noindent
Overall, the contributions of our work are as follows:
\begin{itemize}
\item \cref{sec:motivation} motivates dependently-typed data plane
  programming.
\item \cref{sec:approach} presents \name, a dependently-typed core of
  P4, featuring a combination expressive types for describing
  structures (regular types as well as decidable refinement and
  dependent function types) combined with a bit-by-bit ``chomp''
  operator.
\item \cref{sec:approach} develops a semantic proof of soundness for
  \name's type system.
\item \cref{sec:impl} defines a decidable algorithmic type system for
  \name.
\item \cref{sec:experience} and \cref{sec:case-study} discusses case
  studies using \name's type system to check realistic program
  properties.
\end{itemize}

%% file: 02-background.tex
\section{Background}
\label{sec:motivation}

Most networks are based on a division of labor between two components:
the control plane and the data plane. The control plane, usually
implemented in software, is responsible for performing tasks such as
learning the topology, computing network-wide forwarding paths,
managing shared resources like bandwidth, and enforcing security
policies. The control plane can either be realized using distributed
routing protocols (e.g., in a traditional network), or as a
logically-centralized program (e.g., in a software-defined network).
The data plane, often implemented in hardware or with highly-optimized
software, is responsible for forwarding packets. It parses packets
into collections of headers, performs lookups in routing tables,
filters traffic using access control lists, applies queueing policies,
and ultimately drops, copies, or forwards the packet to the next
device.

P4 is a domain-specific programming language for specifying the
behavior of network data planes. It is designed to be used with
programmable devices such as PISA
switches~\cite{bosshart2013forwarding},
FPGAs~\cite{p4fpga,p4-netfpga}, or software devices (e.g.,
eBPF~\cite{xdp19}). The language is based on a pipeline abstraction:
given an input packet it executes a sequence of blocks of code, one
per pipeline component, to produce the outputs. Each pipeline
component is either a ``parser,'' which consists of a state machine
that maps binary packets into typed representations, or a ``control,''
which consists of a sequence of imperative commands. To interface with
the control plane, P4 programs may contain ``match-action'' tables,
which contain dynamically reconfigurable entries, each corresponding
to a fixed block of code.

Unfortunately, P4 is an unsafe language that does not prevent
programmers from writing programs such as the one shown in
\cref{fig:unsafe-p4}. The program begins by parsing the Ethernet
header. Then, if the Ethernet header contains the appropriate
EtherType (\texttt{0x0800}), it also parses the IPv4 header. Next, in
the ingress control, if the IPv4 source address matches a specified
address, the packet is marked to be dropped. However, there is no
guarantee that the IPv4 header will be a well-defined value---e.g., if
the EtherType is \texttt{0x08DD}, indicating an IPv6 packet, the value
produced by reading the IPv4 source address (\cref{line:ipv4-src})
will be undefined, making the behavior of the program
non-deterministic and possibly different than what the programmer
intended.

\begin{figure}[t]
\begin{minipage}[t]{0.49\textwidth}
\begin{p4code}
parser P(packet_in p, out hdrs h) {
  state start {
    p.extract(h.ethernet);
    transition select(
      h.ethernet.etherType) {
        0x0800: parse_ipv4;
        default: accept;
    } 
  }
  state parse_ipv4 {
    p.extract(h.ipv4);
    transition accept;
  }
}
\end{p4code}
\end{minipage}
\begin{minipage}[t]{0.49\textwidth}
\begin{p4code}[escapechar=@,firstnumber=15]
control Ingress(inout hdrs h) {
  apply {
    if (h.ipv4.src == 10.10.10.10) { @\label{line:ipv4-src}@
      drop();
    } 
  } 
} 
\end{p4code}
\end{minipage}
\caption{Unsafe P4 program: IPv4 is not guaranteed to be valid in the ingress.}
\label{fig:unsafe-p4}
\end{figure}

SafeP4 addresses the lack of basic safety guarantees for P4 using a simple type system~\cite{Eichholz2019}.
Specifically, its type system 
keeps track of the set of valid header instances at each statement.
For example, starting from the empty heap with no header instances
valid, SafeP4 computes the type of the program above to be $\mathtt{ether.ipv4} +
\mathtt{ether}$ after parsing.
This type reflects that on the first program path
both Ethernet and IPv4 are valid, but on the second program path only Ethernet
is valid. Thus, when type checking the if-condition
in the ingress code, the type checker knows that IPv4 may be
invalid and rejects the program.
A simple fix (shown in \cref{fig:safe-ingress-explicit}) that makes
the program safe is to add an explicit validity check before accessing
the IPv4 header. Because it is aware of the semantics of the
\texttt{isValid} command, the SafeP4 type checker computes the type
before the access to be \texttt{ether.ipv4}---i.e., IPv4 is guaranteed
to be valid.

In practice, however, relying on explicit validity checks is not sufficient. For example, consider the code shown in \cref{fig:safe-ingress-implicit}.
Recall that, given the parser above, the IPv4 header will be present
if the EtherType is \texttt{0x0800}. Hence, the ingress control can be
safely executed. Yet, SafeP4's type checker rejects the program
because the type system is not expressive enough to capture the
dependency between the EtherType value and IPv4's validity.

\begin{figure}[t]
\begin{minipage}[t]{0.49\textwidth}
\begin{p4code}
control Ingress(inout headers h) {
  apply {
    if(h.ipv4.isValid()) {
      if(h.ipv4.src == 10.10.10.10) {
        drop();
}}}}
\end{p4code}
\subcaption{Explicit validity check}
\label{fig:safe-ingress-explicit}
\end{minipage}
\begin{minipage}[t]{0.49\textwidth}
\begin{p4code}
control Ingress(inout headers h) {
  apply {
    if(h.ether.etherType == 0x0800) {
      if(h.ipv4.src == 10.10.10.10) {
        drop();
}}}}
\end{p4code}  
\subcaption{Implicit validity check}
\label{fig:safe-ingress-implicit}
\end{minipage}
\caption{Safe implementation of ingress}
\end{figure}

To address this problem, \name employs a 
dependent type system~\cite{Xi1999},
in which we can compute a precise type for the program after parsing:
\[(x:\refT{y}{\epsilon}{\mathtt{\length{y.pkt_{in}}}>272}) \rightarrow
\left(\begin{array}{l}
   \Sigma y:\mathtt{ether}.\refT{z}{\mathtt{ipv4}}{y.\mathtt{ether.etherType}==0x0800} \\
   +~\refT{z}{\mathtt{ether}}{z.\mathtt{ether.etherType}\neq0x0800})
\end{array}\right)\]
Intuitively, this type says that, starting with the empty heap ($y:
\epsilon$) and a packet buffer that has at least enough bits to
extract both the Ethernet and the IPv4 header ($\length{y.pkt_{in}} >
272$), the parser ends in one of two possible states: 
(1) either both Ethernet and IPv4 are valid
($\sigmaT{y}{\mathtt{ether}}{\refT{z}{\mathtt{ipv4}}{...}}$), if the
EtherType is equal to \texttt{0x0800} (note how $z:\mathtt{ipv4}$ is
conditioned by $y.\mathtt{ether.etherType}==0x0800$), or (2) just
Ethernet is valid, if EtherType is not equal to \texttt{0x0800}. When
checking the ingress control, the type checker uses the predicate
$\mathtt{ether.etherType} == \texttt{0x0800}$ from the conditional to
derive the set of valid header instances, which, in this case,
includes IPv4. Thus, accessing the IPv4 source address is safe and the
program correctly passes the type checker.

While the output type is admittedly notationally heavy---a common
feature in precise type systems---note that the programmer is not 
forced to write down the most precise type! \name only requires the 
annotated type to be sufficiently precise to capture basic safety guarantees 
and other desired invariants. For example, in a program where only the Ethernet 
header needs to be valid at the end of the parser, they could use the type 
$(x:\refT{y}{\epsilon}{\mathtt{\length{y.pkt_{in}}}>272}) \rightarrow \weak{\mathtt{ether}}$,
which indicates that, at the end of the parser, at least Ethernet is valid (and possibly others, too).

This example illustrates how \name's type system
statically checks intricate safety properties with high precision.
Sections \ref{sec:experience} and \ref{sec:case-study} present more case studies
showing how \name's type system can be used to check practical properties
of interest.

%% file: 03-formalization.tex
\section[Dependent Types for P4]{Dependent Types for P4 (\name)}
\label{sec:approach}

This section introduces the design, syntax, and semantics of \name, a core
calculus modeled after P4 and equipped with dependent types.

\subsection[Design of Pi4]{Design of \name}
\name focuses on the aspects of the P4 programming language that
benefit from dependent types, (e.g., parsing, deparsing, validity, and
control flow) and omits features that add clutter (e.g., externs,
registers, checksums, hashing, and packages). Following
\texttt{p4v}~\cite{Liu2018}, we do not explicitly model match-action
tables and instead use ghost state and conditionals to encode them
(see \cref{sec:case-study}). Consequently, \name is a
loop-free\footnote{P4 allows loops within parsers, but because
  programs are restricted to finite state, the language specification
  allows implementations to unroll loops.} imperative language with a
few domain-specific primitive commands: $\cExtract{\inst}$,
$\cRemit{\inst}$, $\cAdd(\inst)$, and $\cReset$.

In P4, the $\texttt{emit}(\inst)$ primitive serializes a header instance
$\inst$ into a bitstring and prepends it to the outgoing packet
payload, \emph{only if $\inst$ is valid}, otherwise it does nothing.
To simplify typing rules and semantics, \name provides the primitive
$\cRemit{\inst}$, which \textit{really} emits $\inst$ if it is valid,
and otherwise gets stuck. Hence, $\texttt{emit}(\inst)$ can be encoded
as syntactic sugar: $\cIf {\inst.\valid} {\cRemit{\inst}} {\cSkip}$.
Another superficial difference is that we model header field accesses
as direct bit-slices into the instance (to avoid another layer of
indirection in our semantics)---i.e., $\texttt{eth.srcAddr}$ is
written $\texttt{eth}[48:96]$.

More substantially, \name diverges from P4 in the way it handles instance
validity. A header instance is valid in two cases: (i) if it has been extracted from
the packet (which automatically populates the instance with the appropriate
bits) or (ii) if its validity bit has manually been set using the
$\texttt{setValid()}$ method (which does nothing to the instance itself). In
P4, reads to uninitialized variables produce undefined values, so a common
programming practice is to follow a call to $\texttt{setValid()}$ with a
sequence of assignments to the header fields---thereby avoiding undefined reads.
In \name, rather than forcing the programmer to manually write default values,
the $\cAdd(\inst)$ command sets $\inst.\valid$ to $\true$, and sets $\inst$ a
pre-determined default value (say $0$).\footnote{The difference here is similar
  to the difference between C's \texttt{malloc} (analogous to P4's semantics) and
  \texttt{calloc} (analogous to \name's semantics).} If required, P4's behavior
could be encoded using an extra 1-bit header to independently track the validity of 
the instance and initialization of its fields. 

We also introduce a new primitive, $\cReset$, which models the behavior of P4
between pipeline stages. In many switch
architectures~\cite{bosshart2013forwarding}, packets are deparsed and then
reparsed between pipelines---e.g., after \texttt{ingress} and before
\texttt{egress}. The $\cReset$ command encodes the behavior of the inner step:
it combines the deparsed bits with the packet's unparsed payload and passes it
along as the input to the next stage. The $\cReset$ command would also be useful
to reason about invariants across multiple switch programs, although we don't
explore that use in this paper.

Finally, 
in designing \name, our primary goal is to enable data plane programmers to make
use of dependent types to verify useful program properties in a compositional
way and without having to write manual proofs. 
To enable modular reasoning,
we need a way to annotate (and modularly check) 
programs with types. We annotate a program $c$ with a type $\sigma$ using an ascription 
operator: $\cAscribe c \sigma$. The ascription has no effect on the runtime behavior of the code
(i.e., $\cAscribe c \sigma$ always just steps to $c$). It does, however, indicate a program 
point where  type checking should occur. Hence, we can independently typecheck $c$ at type $\sigma$ 
and then use $\sigma$ when checking the rest of the program.

\paragraph{Intuition for \name's type system}
Next, we give an intuition for \name's types. 
A command $c$ is always assigned a dependent
function type $(x : \tau_{1}) \to \tau_{2}$, where $x$ may occur in $\tau_{2}$. 
This design allows us to relate the input and output values of commands expressed 
in the \emph{heap types} $\tau_{1}$ and $\tau_{2}$. For example, we may want to 
ensure that the Ethernet header has the same value after being deparsed, reset, 
and then reparsed. To express equations like this, we use refinement types $\refT y \tau \varphi$, 
where $\varphi$ is a formula in the logic of variable-width bit vectors with concatenation 
and length operators. In this example, we could say that the Ethernet header is unchanged 
by using the output type $\refT y {\tau_{2}} {x.\texttt{eth} = y.\texttt{eth}}$.

We also often need to reference intermediate formulae, so we introduce substitution
types, written $\tau_{2}[x \mapsto \tau_{1}]$, where $x$ may occur in $\tau_{2}$
but not in $\tau_{1}$. In such a type, $\tau_{1}$ may represent the type at any
earlier point in the program. \name also supports fine-grained path-dependent
reasoning via union types $(\tau_{1} + \tau_{2})$. It is also convenient to have
trivial ($\top$) and absurd ($\varnothing$) types.

Finally, the design of our type system is also informed by the need to
model parsing operations. Specifically, we must ensure that the
refinements on the input type and on the output type remain consistent
after bits have been shuffled around by a command. For example, given
an input type $\refT x \top {x.\pIn[0:8] == 4 \wedge |x.\pIn| >
  |\texttt{ipv4}|}$, where $x.\pIn$ represents the incoming packet,
and the command $\cExtract{\texttt{ipv4}}$, the output type should
reflect that the \texttt{ipv4} header instance is now valid, that
$\texttt{ipv4}[0:8]$ is $4$, and that $x.\pIn$ may have no more bits
remaining. \name accomplishes this using two key mechanisms: (1) a
dependent sum type $\sigmaT x {\tau_{1}} {\tau_{2}}$ that computes the
disjoint union of the valid instances in $\tau_{1}$ and $\tau_{2}$ and
concatenates the incoming and outgoing packets (\cref{sec:types}), and
(2) a refinement transformer, $\chomp$, that manipulates input
refinements to be consistent with the extraction operation (see
\cref{sec:type-system}).

\subsection{Syntax}

\begin{figure}[t]
  \begin{tabular}{l l r}  
    $\tau$ & $::= \varnothing \mid \top \mid \sigmaT{x}{\tau}{\tau} 
    \mid \tau + \tau
    \mid \rT{\tau}{\varphi} \mid \tau[x\mapsto\tau]$ & (heap types)\\
    $\sigma$ & $::= \tNat \mid \tBool \mid \tBv \mid (x:\tau) \rightarrow \tau $ & (base types)\\
    $\varphi$ & $::= e = e \mid e > e \mid \varphi \land \varphi \mid \neg \varphi \mid x.\inst.valid \mid \true \mid \false$ & (formulae)\\
    $e$ & $::= n \mid bv \mid \length{x.p} \mid e+e \mid \bvconcat{e}{e} \mid x.p \mid \slice{x.p}{l}{r} \mid \slice{x.\inst}{l}{r}$ & (expressions)\\
    $bv$ & $::= \langle\rangle \mid 0::bv \mid 1::bv \mid \bvar{n}::bv$ & (bit vectors)\\
    $p$ & $::= pkt_{in} \mid pkt_{out}$ & (packets)\\
    $c$ & $::= \mathit{extract(\inst)} \mid \mathit{if(\varphi)~c~else~c} \mid \mathit{c ; c} \mid \mathit{\inst.f := e} \mid \mathit{remit(\inst)}$ & (commands)\\
    &$\phantom{{}::=}\mathit{skip} \mid \mathit{reset}
    \mid \cAdd(\inst) \mid \cAscribe{c}{(x:\tau)\rightarrow\tau}$\\
    $d$ & $::= \eta~\{ \overline{f:\tBv} \} \mid \inst \mapsto \eta$ & (declarations)\\
    $P$ & $::= (\overline{d}, c)$ & (programs)
  \end{tabular}\\
  \caption{Syntax of \name}
  \label{fig:syntax}
\end{figure}

\cref{fig:syntax} shows the syntax of \name. 
Boolean formulae $\varphi$ include literals, equality ($=$) and
validity of instances ($x.\inst.\valid$), conjunction ($\land$), and
negation ($\neg$). Expressions $e$ include naturals, bit vectors, packet
length ($\length{x.p}$), addition ($+$), concatenation
($\bvconcat{}{}$), packet access $(x.p)$ and slices of packets
($\slice{x.p}{l}{r}$) and instances ($\slice{x.\inst}{l}{r}$).

To ease the notation, we write $x.\inst[l]$ instead of
$\slice{x.\inst}{l}{l+1}$ for bit-wise access, $x.\inst.f$ instead of
$\slice{x.\inst}{l}{r}$ for ranges matching header instance fields,
$x.\inst$ instead of $\slice{x.\inst}{0}{\sizeof(\inst)}$, 
$\slice{x.\inst}{n}{\!~}$ for the remaining bits of $x.\inst$ starting from bit $n+1$, and
similarly for the corresponding formulae involving packet variables
$x.p$. We use a list-like encoding of bit vectors. A bit vector is
either the empty bit vector ($\bvNil$) or a concatenation of bits. We
assume that bit variables \bvar{n} are not part of the surface syntax and
are only used internally. For singleton bit vectors, we write $\bv{b}$
instead of $b::\bvNil$.

We write
$\epsilon \triangleq \refT{x}{\top}{\bigwedge_{\inst\in\dom(\HT)}\neg x.\inst.valid}$
for the type denoting the empty heap on which no header instances are valid,
$\inst \triangleq \refT{x}{\top}{x.\inst.\valid \land \bigwedge_{\inst'\in\dom(\HT), \inst'\neq\inst} \neg x.\inst'.\valid}$
for the type denoting the heap exclusively containing instance $\inst$,
and $\weak\inst \triangleq \refT{x}{\top}{x.\inst.\valid}$
for the type denoting the heap on which at least instance $\inst$
is guaranteed to be valid.

For formulae, we write $x\equiv y$ (respectively $x\equiv_\inst y$)
as syntactic sugar for the boolean predicates capturing strict
equality (respectively instance equality) between the heaps bound to
$x$ and $y$. Strict equality requires that both the input and output
packets are equivalent as well as all instances contained in the
heap---i.e., $x.\pIn=y.\pIn \land x.\pOut = y.\pOut \land
\bigwedge_{\inst\in\dom(\mathcal{HT})}x.\inst=y.\inst$, while instance
equality only requires that the instances are equivalent in both
heaps. We write $x.\inst.\valid = y.\inst.valid$ as syntactic sugar for 
$(x.\inst.\valid \wedge y.\inst.\valid) \vee 
(\neg x.\inst.\valid \wedge \neg y.\inst.\valid)$.
We use standard encodings using negation and conjunction 
for logical connectives like $\vee$ or $\Rightarrow$.

We write $\overline{x}$ as a shorthand for a possibly empty sequence
$x_1,...,x_n$. A program consists of a sequence of declarations
$\overline{d}$ and a command $c$. Declarations $d$ include header type
declarations $\eta~\{ \overline{f:\tBv} \}$ and header instance
declarations $\inst\mapsto\eta$. Header type declarations specify the
format of network packet headers. They are defined in terms of a name
and a sequence of field declarations, where each field is itself
defined in terms of a field name and a type. We write $f: \tBv$ to
denote that field $f$ has a bit vector type $\tBv$. With $\eta$
ranging over header types, the instance declaration $\inst\mapsto\eta$
assigns the name $\inst$ to header type $\eta$. The global
mapping between header instances and header types is stored in the
so-called header table $\mathcal{HT}$. We assume that names of header
instances and header types are drawn from disjoint sets of names and
that each name is declared only once.

\name~ provides commands for parsing (\textit{extract}), creating
(\textit{add}) and modifying ($\inst.f:=t$) header instances. The
\textit{remit} command serializes a header instance into a bit
sequence. The \textit{reset} command resets the program state---in
particular, the packet buffers. Commands can be sequentially composed
($c_1;c_2$), \textit{skip} is a no-op, and the \textit{if}-command
conditionally executes one out of two commands based on the value of
the boolean formula $\varphi$. We assume that formulae and expressions
used in commands are implicitly prefixed with a variable named
$\cmdVar$, but we often omit it in the surface syntax. For example, we
write $\cIf {\texttt{ether}.\valid} {\cExtract{\texttt{ipv4}}}
{\cSkip}$ instead of $\cIf{\cmdVar.\texttt{ether}.\valid}
{\cExtract{\cmdVar.\texttt{ipv4}}} {\cSkip}$. \name provides modular
reasoning via type ascription
($\cAscribe{c}{(x:\tau_{1})\rightarrow\tau_{2}}$). Finally, we assume
that every header referenced in a program has a corresponding instance
declaration---this could be enforced statically using a simple
analysis.

\subsection{Type System}
\label{sec:types}

As shown in \cref{fig:syntax}, there are two categories of types,
base types $\sigma$ and heap types $\tau$. Base types include natural
numbers, bit vectors, booleans, and dependent function types. Heap
types $\tau$ represent sets of heaps, where each element in the set
describes a different program path. The goal is to capture bit-level
dependencies between header instances and the incoming and outgoing
packet in the type system. A heap $h$ in the set of heaps
$\mathcal{H}$ describes a possible system state, consisting of the
incoming and outgoing packet and the set of valid header instances. We
model heaps as maps from names to bit vectors. A heap contains two
special entries $\pIn$ and $\pOut$ representing the incoming and
outgoing packet buffers, as well as mappings from instance names to
bit vectors for each valid header instance.
\begin{figure}[t]
  \begin{align*}
    \semantics{\tau}_\env &\subseteq \mathcal{H} \\
    \semantics{\varnothing}_\env &= \{\} \\
    \semantics{\top}_\env &= \mathcal{H} \\
    \semantics{\tau_1 + \tau_2}_\env &= \semantics{\tau_1}_\env \cup \semantics{\tau_2}_\env \\
    \semantics{\sigmaT{x}{\tau_1}{\tau_2}}_\env &= \{ \concat{h_1}{h_2} \mid h_1\in\semantics{\tau_1}_\env \land h_2\in\semantics{\tau_2}_{\env[x \mapsto h_1]} \} \\
    \semantics{\tau_1[x\mapsto\tau_2]}_\env &= \{ h \mid h_2 \in \semantics{\tau_2}_\env \land h\in\semantics{\tau_1}_{\env[x\mapsto h_2]} \} \\
    \semantics{\rT{\tau}{\varphi}}_\env &= \{ h \mid h\in\semantics{\tau}_\env \land \semantics{\varphi}_{\env[x \mapsto h]}=\true \}
  \end{align*}
  \caption{Semantics of heap types}
  \label{fig:semantics-of-types}
\end{figure}
The semantics of types is shown in \cref{fig:semantics-of-types}.
Heap types are evaluated in an environment $\env$, which is a mapping
from variable names to heaps. The environment models other heaps
available in the current scope upon which the current header type may
depend.

The type $\varnothing$ denotes the empty set. It is used in situations
where there are unsatisfiable assumptions involving the header
instances or the incoming and outgoing packet buffers.

The top type $\top$ denotes the set of all possible heaps.
The choice type $\tau_1+\tau_2$ denotes the union of the sets of heaps represented by $\tau_1$ and $\tau_2$.
The dependent pair $\sigmaT{x}{\tau_1}{\tau_2}$ denotes the concatenation of heaps from $\tau_1$ and $\tau_2$, where heaps described by $\tau_2$ may depend on heaps from $\tau_1$.
The concatenation $h=\concat{h_1}{h_2}$ of two heaps $h_1$ and $h_2$ requires that header instances contained in $h_1$ and $h_2$ are disjoint.
The resulting heap contains all instances from $h_1$ and from $h_2$, with $\pIn$ and $\pOut$ the concatenation of respective bit vectors in $h_1$ and $h_2$. 
The explicit substitution $\tau_1[x\mapsto\tau_2]$ denotes the set of heaps obtained by evaluating $\tau_1$ for every heap described by $\tau_2$.
Finally, the refinement type $\rT{\tau}{e}$ denotes the set of heaps described by $\tau$ for which the predicate $e$ holds.

The refinement predicate $\varphi$ is evaluated
in the same type of environment as heap types.
Formulae evaluate to a boolean value, i.e.,
$\semantics{\varphi}_\env\in\tBool$. %
The semantics of expression equality ($e_1=e_2$) is defined
as the semantic equality between expressions $e_1$ and $e_2$. 
If the semantics of one of the expressions is undefined,
expression equality always evaluates to false.
The semantics of expression comparison ($e_1 > e_2$) is defined 
analogously. Instance validity $x.\inst.\valid$ evaluates to true
if header instance $\inst$ is contained in the heap
bound to $x$ in $\env$, otherwise it evaluates to false.
The remaining operations have standard semantics.

\begin{figure}[t]
  \begin{align*}
    \semantics{e}_\env &\in \tBv \cup \tNat\\
    \semantics{\length{x.p}}_\env &= \begin{cases}
      0 & \text{if } \env(x)(p) = \langle\rangle \\
      n & \text{if } \env(x)(p) = \langle b_1,...,b_n\rangle\\
      \bot & \text{otherwise}
    \end{cases}\\
    \semantics{x.p}_\env &= \begin{cases}
      \langle b_1,...,b_n\rangle & \text{if }\env(x)(p)=\langle b_1,...,b_n\rangle \\
      \bot & \text{otherwise}
    \end{cases}\\
    \semantics{x.p[n:m]}_\env &= \begin{cases}
    \langle b_n, ..., b_{m-1} \rangle &\text{if } \semantics{x.p}_\env=\langle b_0, ..., b_k\rangle~\land \\
    & 0 \leq n < m \leq k+1 \\
    \bot & \text{otherwise}
    \end{cases}\\
    \semantics{x.\inst[n:m]}_\env &= \begin{cases}
      \langle b_n, ..., b_{m-1} \rangle &\text{if } \semantics{x.\inst}_\env=\langle b_0, ..., b_k\rangle~\land \\
      & 0 \leq n < m \leq k+1 \\
      \bot & \text{otherwise}
      \end{cases}
  \end{align*}
  \caption{Selected cases of the semantics of expressions}
  \label{fig:semantics-of-terms}
\end{figure}

The semantics of expressions is defined in \cref{fig:semantics-of-terms}.
For brevity, we omit standard cases.
Expressions evaluate to either a bit vector or a natural number.
For the semantic addition ($+$) and bit vector concatenation operator ($@$),
we assume that as soon as evaluating one operand results in an error ($\bot$)
the whole expression also evaluates to $\bot$.
To evaluate the length of a packet $\length{x.p}$,
we compute the length of the bit vector of $\pIn$ or $\pOut$ respectively
in the heap bound to $x$ in the environment.
If no heap is bound to variable name $x$, the expression evaluates to $\bot$.
The semantics of bit vectors is as expected, except for variables,
which we look up from a designated location in the environment. 
If no binding for the bit variable exists, it evaluates to $\bot$.
The semantics of bit vector concatenation is standard.
A packet access $x.p$ looks up the respective entry
from the heap bound to variable $x$ in $\env$.
A packet slice $\slice{x.p}{l}{r}$ is evaluated in the same way,
but additionally the designated slice is obtained from the bit vector. 
Again, if the variable is not bound or an index is greater than
the length of the bit vector, the expression evaluates to $\bot$.
The semantics of instance slices $\slice{x.\inst}{l}{r}$ is defined similarly
but the lookup occurs on header instance $\inst$.
We interpret slices as half-open intervals,
where the left bound is included and the right bound is excluded. 
For example, given a bit vector $bv=1010$ we have $\slice{bv}{1}{4}=010$. 

We define two semantic operations on heap types: \textit{inclusion} and \textit{exclusion} of instances.
The first, $\includes{\tau}{\inst}$, traverses $\tau$ and checks that instance $\inst$ is valid in every path.
Semantically this says that $\inst$ is a member of every element of $\semantics{\tau}_\env$---i.e., if $\env \models \Gamma$, then $\forall h\in\semantics{\tau}_\env.\inst\in dom(h)$.
The second, $\excludes{\tau}{\inst}$, traverses $\tau$ and checks that instance $\inst$ is invalid in every path.
Semantically this says that $\inst$ is no member of every element of $\semantics{\tau}_\env$---i.e., if $\env \models \Gamma$, then $\forall h\in\semantics{\tau}_\env.\inst\not\in dom(h)$.

\subsection{Operational Semantics}

The small-step operational semantics of \name, is shown in
\cref{fig:operational-semantics}. It is defined in terms of a
four-tuple $\langle I,O,H,c\rangle$, where $I$ and $O$ are the
bitstrings for the incoming and outgoing packets respectively, $H$ is
a map that relates instance names to records containing the field
values, and $c$ is a command.

\begin{figure}[t]
  \begin{mathpar}
    \inferrule[\eExtract]{
      \mathcal{HT}(\inst) = \eta \\
      \mathit{deserialize_\eta}(I) = (v, I')\\
    }{
      \langle I,O,H,\cExtract{\inst}\rangle \rightarrow \langle I',O,H[\inst \mapsto v], \cSkip\rangle
    }
    \and
    \inferrule[\eRemit]{
      \inst\in dom(H)\\
      \mathcal{HT}(\inst) = \eta \\\\
      \mathit{serialize_\eta}(H(\inst)) = bv
    }{
      \langle I,O,H,\cRemit{\inst}\rangle \rightarrow \langle I,O::bv,H,\cSkip\rangle
    }
    \and
    \inferrule[\eMod]{
      H(\inst) = r \\
      r' \triangleq \{r\ with\ f = v\}
    }{
      \langle I,O,H,\inst.f:=v\rangle \rightarrow \langle I,O,H[\inst \mapsto r'],\cSkip\rangle
    }
    \and
    \inferrule[\eReset]{
      I'=\bvconcat{O}{I}
    }{
      \langle I,O,H,\cReset\rangle \rightarrow \langle I',\langle\rangle,[],\cSkip\rangle
    }
    \and
    \inferrule[\eAdd]{
      \inst\not\in\dom(H)\\
      \mathcal{HT}(\inst)=\eta\\
      \mathit{init_\eta}=v
    }{
      \langle I,O,H,\cAdd(\inst)\rangle \rightarrow \langle I,O,H[\inst\mapsto v],\cSkip\rangle
    }
    \and
    \inferrule[\eAscribe]{\ }{
      \langle I,O,H,\cAscribe{c}{(x:\tau_1)\rightarrow\tau_2} \rangle
      \rightarrow \langle I,O,H,c \rangle
    }
  \end{mathpar}
  \caption{Small-step operational semantics of \name}
  \label{fig:operational-semantics}
\end{figure}

The $\cExtract{\inst}$ command (\eExtract) first looks up the header 
type from the header table ($\HT$), and uses a
deserialization function to copy the appropriate number of bits from
the input packet into the deserialized representation of the instance
$v$. This value is added to the map of valid header instances $H$. We
assume there exists a deserialization function for every header
instance. For example, assuming $I=110011B$, where $B$ is the rest of
the bitstring, and $\eta=\{ f:4; g:2 \}$, then
$\mathit{deserialize}_\eta(I)=(\{f=1100;g=11\}, B)$.

The $\cRemit{\inst}$ command (rule \eRemit) requires that the header 
instance is valid---i.e., it is contained in $H$. Similar to \eExtract, we 
assume there is a serialization function for every heap type, which turns 
a record representing the instance back into a bit sequence. For example,
$\mathit{serialize}_\eta(\{f=1100;g=11\})=11011$. The serialized bit
sequence is appended to the end of the outgoing packet. Both the input
packet and the set of valid headers remain unchanged.

The rule \eMod defines the semantics of assigning a value to a header
field. Assuming $r$ is the record storing the values of the fields, an
updated record $r'$ with the modified field value is stored in $H$.
The input and output packets remain unchanged. If the assigned expression is
not a value, it is reduced first.

The rules for sequencing (\eSeq, \eSeqOne) are standard. Sequences of
commands evaluate from left to right---i.e., the left-hand command is
reduced to skip before the right-hand command is evaluated. The
evaluation rules for conditionals (\eIf, \eIfTrue, \eIfFalse) are also
standard. All standard rules are omitted.

The rule \eReset defines the semantics of the command $\cReset$. It
would be invoked between the ingress and egress pipelines, when the
packet emitted by the ingress becomes the input packet for the egress.
Operationally, the bits contained in the output packet are prepended
to the bits of the input packet. This concatenated bit sequence serves
as the new input packet. The output packet is emptied and all valid
header instances are discarded.

The rule \eAdd initializes a header instance if it is not already
valid. The evaluation is similar to rule \eExtract, except that no
bits are taken from the input bitstring. Instead we assume that there
exists an initialization function $\mathit{init}_\eta$ for every heap
type $\eta$ ---similar to \textit{deserialize}---that initializes all
fields of an instance to a fixed value.
If an instance is already valid, this operation is a no-op.
An ascribed command $\cAscribe{c}{\sigma}$ (rule \eAscribe) evaluates 
to $c$ trivially, without modifying the heap.

\subsection{Typing Judgement}
\label{sec:type-system}

The typing judgement has the form $\cmdT{c}{(x:\tau_1)\rightarrow\tau_2}$.
Intuitively, type $\tau_1$ describes the input heap and $\tau_2$ describes the 
output heap obtained after the execution of $c$.
$\Gamma$ is a variable context that maps variable names to heap types 
and is used to capture additional dependencies of the input type. 
If a command typechecks in a context where $x$ maps to $\tau$ (i.e., 
$\Gamma,x:\tau \vdash c: (x:\tau_1) \rightarrow \tau_2$) it means that given 
some heap described by type $\tau$ on which the input  heap might depend, 
executing $c$ on the input heap described by $\tau_1$ will 
result in a heap described by $\tau_2$.

\begin{figure}[t]
  \begin{mathpar}
    \inferrule[\tExtract]{
      \Gamma \vdash \sizeof_{\pIn}(\tau_1) \geq \sizeof(\inst) \\
      \varphi_1 \triangleq z.\pIn = z.\pOut = \bvNil\\\\
      \varphi_2 \triangleq \bvconcat{y.\inst}{z.\pIn} = x.\pIn \land z.\pOut = x.\pOut \land z \equiv_\inst x
    }{
      \cmdT{\mathit{extract(\inst)}}{(x:\tau_1) \rightarrow \sigmaT{y}{\refT{z}{\inst}{\varphi_1}}{\refT{z}{\chomp(\tau_1, \inst, y)}{\varphi_2}}}
    }
    \and
    \inferrule[\tSeq]{
      \cmdType{\Gamma}{c_1}{(x:\tau_1)\rightarrow\tau_{12}}\\\\
      \cmdType{\Gamma, x:\tau_1}{c_2}{(y:\tau_{12})\rightarrow\tau_{22}}\\
    }{
      \cmdType{\Gamma}{c_1;c_2}{(x:\tau_1)\rightarrow\tau_{22}[y\mapsto \tau_{12}]}
    }
    \and
    \inferrule[\tSkip]{
      \tau_2 \triangleq \refT{y}{\tau_1}{y \equiv x}
    }{
      \cmdT{\mathit{skip}}{(x:\tau_1) \rightarrow \tau_2}
    }
    \and
    \inferrule[\tRemit]{
      \includes{\tau_1}{\inst} \\
      \varphi \triangleq z.\pIn = \bvNil \land z.\pOut=x.\inst %
    }
    {
      \cmdT{\mathit{remit(\inst)}}{(x:\tau_1) \rightarrow \sigmaT{y}{\refT{z}{\tau_1}{z\equiv x}}{\refT{z}{\epsilon}{\varphi}}}
    }
    \and
    \inferrule[\tReset]{
      \varphi_1 \triangleq z.\pOut = \bvNil \land z.\pIn = x.\pOut\\\\
      \varphi_2 \triangleq z.\pOut = \bvNil \land z.\pIn = x.\pIn
    }{
      \cmdT{\mathit{reset}}{(x\!: \tau_1) \rightarrow 
      \sigmaT{y}
             {\refT{z}{\epsilon}{\varphi_1}}
             {\refT{z}{\epsilon}{\varphi_2}}}
    }
    \and
    \inferrule[\tAscribe]{
      \cmdT{c}{\sigma}
    }{
      \cmdT{\cAscribe{c}{\sigma}}{\sigma}
    }
    \and
    \inferrule[\tIf]{
      \Gamma;\tau_1 \vdash \varphi : \tBool \\
      \cmdType{\Gamma}{c_1}{(x\!:\refT{y}{\tau_1}{\varphi[y/\cmdVar]}) \rightarrow \tau_{12}} \\
      \cmdType{\Gamma}{c_2}{(x\!:\refT{y}{\tau_1}{\neg \varphi[y/\cmdVar]}) \rightarrow \tau_{22}}
    }{
      \cmdType{\Gamma}{\mathit{if(\varphi)\ c_1\ else\ c_2}}{(x\!:\tau_1) \rightarrow  \refT{y}{\tau_{12}}{\varphi[x/\cmdVar]} + \refT{y}{\tau_{22}}{\neg \varphi[x/\cmdVar]}}
    }
    \and
    \inferrule[\tMod]{
      \includes{\tau_1}{\inst} \\
      \mathcal{F}(\inst, f) = \tBv \\
      \Gamma;\tau_1 \vdash e : \tBv \\
      \varphi_{pkt} \triangleq y.\pIn = x.\pIn \land y.\pOut = x.\pOut~ \\
      \varphi_\inst \triangleq \forall \kappa \in \dom(\HT).~\inst\neq\kappa \rightarrow y.\kappa = x.\kappa\\
      \varphi_{f}\triangleq\forall g \in \dom(\HT(\inst)).~f \neq g \rightarrow y.\inst.g = x.\inst.g
    }{
      \cmdT{\inst.f := e}{(x\!:\tau_1) \rightarrow \refT{y}{\top}{\varphi_{pkt} \land \varphi_\inst \land \varphi_f \land y.\inst.f = e[x/\cmdVar]
      }}
    }
    \and
    \inferrule[\tAdd]{
      \excludes{\tau_1}{\inst} \\
      \mathit{init}_{\HT(\inst)} = v \\\\
      \varphi \triangleq z.\pIn=z.\pOut=\bvNil \wedge z.\inst = v
    }{
      \cmdT{\cAdd(\inst)}{(x:\tau_1) \rightarrow \sigmaT{y}{\refT{z}{\tau_1}{z \equiv x}}{\refT{z}{\inst}{\varphi}}}
    }
    \and
    \inferrule[\tSub]{
      \subtype{\tau_1}{\tau_3} \\\\
      \subtypeCtx{\Gamma,x:\tau_1}{\tau_4}{\tau_2} \\\\
      \cmdT{c}{(x:\tau_3) \rightarrow \tau_4}
    }{
      \cmdT{c}{(x:\tau_1) \rightarrow \tau_2}
    }
  \end{mathpar}
  \caption{Command typing rules for \name.}
  \label{fig:command-typing-rules}
\end{figure}

The typing rules are presented in \cref{fig:command-typing-rules}.
The typing rule \tExtract captures that $\inst$ must be valid after an \textit{extract} command is executed
and the input packet of type $\tau_1$ provides enough bits for the instance 
($\sizeof_\pIn(\tau) \ge n \text{ iff } \forall \env,h\in\semantics{\tau}_\env, \length{h(\pIn)} \ge n$).
Intuitively, the $\chomp$ operator ensures that the output type reflects that the first $n$ bits, where $n$ is 
the number of bits contained in header instance $\inst$, are removed from the input packet and copied 
into instance $\inst$.

The typing rule for sequencing \tSeq is mostly standard, with one peculiarity:
because our typing judgement assigns dependent function types to commands,
the result type $\tau_{22}$ of command $c_2$ might depend on its input type $\tau_{12}$---i.e., variable $y$ might appear free in $\tau_{22}$.
Hence, we must also capture the type $\tau_{12}$ in the result type. 
The typing rule \tSkip is standard, except that it strictly enforces that the
heaps described by the output type are equivalent to the heaps described by the
input type. 
To typecheck the command \textit{remit}, we check whether the instance to be emitted is guaranteed to be valid in the input type.
The assigned output type ensures that emitting a header instance
appends the value of the instance to the end of the outgoing packet (second projection of the assigned $\Sigma$-type)
but leaves the input packet and all other validity information unchanged (first projection of the assigned $\Sigma$-type). 
The rule \tReset resets all assumptions about header validity,
empties $\pOut$ and refines $\pIn$ to be the concatenation of $\pOut$ and $\pIn$ of the input type.
In the output type, we use a $\Sigma$-type to model the concatenation.

The rule \tIf typechecks each branch of the conditional with the additional
assumption that the condition $\varphi$ holds respectively does not hold.
The resulting type is a path-sensitive union type, which includes the types
of both paths. By default, all variables in formula $\varphi$ in the command
are bound to $\cmdVar$. To turn $\varphi$ into a refinement on a type,
we substitute every occurrence of $\cmdVar$ with the respective 
binder of the type we want to refine.
We write $\varphi[x/\cmdVar]$ to denote the formula obtained from $\varphi$
in which $\cmdVar$ is replaced with $x$.
For example, if the command is $\cIf {\texttt{ether}.\texttt{etherType} = \texttt{0x0800}} {\cExtract{\texttt{ipv4}}} {\cSkip}$, we typecheck the then-branch with
type $(x:\refT{y}{\tau_1}{y.\texttt{ether}.\texttt{etherType} = \texttt{0x0800}}) \rightarrow \tau_{12}$. The full command is checked
with type $(x:\tau_1) \rightarrow \refT{y}{\tau_{12}}{x.\texttt{ether}.\texttt{etherType} = \texttt{0x0800}} + \refT{y}{\tau_{22}}{\neg x.\texttt{ether}.\texttt{etherType} = \texttt{0x0800}}$.

To typecheck a modification of an instance field, the typing rule \tMod first
checks if the instance to be modified is guaranteed to be valid in the input
type. The output type is similar to the strongest-postcondition of the input
type: everything in the output type is the same as in $x$, except for the
modified instance field $y.\inst.f$, which must be equal to $e[x/\cmdVar]$.

Rule \tAdd first checks that the instance is not yet included in the type and assigns an output type that reflects that all information from the input type $\tau_1$ are retained and just instance $\inst$ is added.
The typing rule for ascription \tAscribe is standard.
The typing rule for subsumption \tSub is also standard.
We write $\subtype{\tau_1}{\tau_2}$ to denote the subtyping check between $\tau_1$ and $\tau_2$. The contexts $\Gamma_1$ and $\Gamma_2$ capture external dependencies of $\tau_1$ and $\tau_2$ respectively.

\begin{figure}[t]
  \begin{minipage}{.45\textwidth}
    \begin{equation*}
        \subtype{\tau_1}{\tau_2}\\
        \overset{\Delta}{\Leftrightarrow}\\
        \forall \env \models \Gamma.
  \semantics{\tau_{1}}_{\env} \subseteq \semantics{\tau_{2}}_{\env}
    \end{equation*}
  \end{minipage}%
  \begin{minipage}{.55\textwidth}
    \begin{equation*}
      \env \models \Gamma \overset{\Delta}{\Leftrightarrow}\forall x_i\in\dom(\Gamma).\env(x_i)=h_i \land h_i\models_\env\Gamma(x_i)
    \end{equation*}
  \end{minipage}
  \caption{Left: Subtyping. Right: Entailment between environments and subtyping contexts.}
  \label{fig:subtyping}
\end{figure}

We take a semantic approach for defining subtyping as shown in the left of \cref{fig:subtyping}.
Type $\tau_1$ with context $\Gamma_1$ is a subtype of type $\tau_2$ with context $\Gamma_2$, if and only if for all environments $\env_1$ and $\env_2$ such that $\env_1$ entails the context $\Gamma_1$ for subtype $\tau_1$ and $\env_2$ entails the context $\Gamma_2$ for supertype $\tau_2$, the set of heaps described by $\tau_1$ evaluated in environment $\env_1$ is a subset of the set of heaps described by $\tau_2$ evaluated in environment $\env_2$.

The entailment between environments and typing contexts is defined in the right of \cref{fig:subtyping}. 
An environment $\env$ entails a context $\Gamma$, iff for every mapping from a variable name $x_i$ to some heap type $\tau_i$ in $\Gamma$ there exists a mapping from variable $x_i$ to some heap $h_i$ in environment $\env$ and that heap $h_i$ entails type $\tau_i$.
The entailment between a heap and a type is defined in \cref{fig:entailment}.
A heap $H[\pIn\mapsto I,\pOut\mapsto O]$, in
short $(I,O,H)$ entails a type $\tau$, if it is contained in the type.

\begin{figure}[t]
  \begin{mathpar}
    \inferrule[Ent-Top]{\ }{
    	(I,O,H)\models_\env\top
    }
    \and
    \inferrule[Ent-ChoiceL]{
      (I,O,H)\models_\env\tau_1
    }{
      (I,O,H)\models_\env\tau_1+\tau_2
    }
    \and
    \inferrule[Ent-ChoiceR]{
      (I,O,H)\models_\env\tau_2
    }{
      (I,O,H)\models_\env\tau_1+\tau_2
    }
    \and
    \inferrule[Ent-Refine]{
      (I,O,H)\models_\env\tau \\\\
      \semantics{\varphi}_{\env[x\mapsto (I,O,H) %
      ]} = \true
    }{
      (I,O,H)\models_\env\rT{\tau}{\varphi}
    }
    \and
    \inferrule[Ent-Sigma]{
      (I_1,O_1,H_1)\models_\env\tau_1\\\\
      (I_2,O_2,H_2)\models_{\env[x\mapsto (I_1,O_1,H_1)%
      ]}\tau_2
    }{
      (\bvconcat{I_1}{I_2},\bvconcat{O_1}{O_2},H_1\cup H_2)\models_\env\sigmaT{x}{\tau_1}{\tau_2}
    }
    \and
    \inferrule[Ent-Subst]{
      (I_2,O_2,H_2) \models_\env \tau_2\\\\
      (I,O,H) \models_{\env[x\mapsto (I_2,O_2,H_2)]%
      } \tau_1
    }{
      (I,O,H) \models_\env \tau_1[x\mapsto\tau_2]
    }
  \end{mathpar}
  \caption{Entailment between heaps and heap types.}
  \label{fig:entailment}
\end{figure}

\subsection{Chomp}
When an instance $\inst$ is extracted, $\sizeof(\inst)$ bits are moved from the
input bitstring to the instance---we call this
\emph{chomping}. To reflect it in the type that we assign to an extract
command, we define a syntactic operation $\chomp$ that transforms a heap type
into the heap type that would result from extracting an instance.

We first specify a semantic chomp operation on a single heap
($\chompS(h,n)$) in \cref{def:semantic-chomp}---it removes the first
$n$ bits from the input packet in heap $h$.

\begin{definition}[Semantic Chomp]
  \label{def:semantic-chomp}
  $\chompS(h,n) = h[\pIn\mapsto h(\pIn)[n\!:]]$
\end{definition}

Intuitively, syntactic $\chomp$ lifts $\chompS$ to heap types (written formally in
\cref{lem:semantic-chomp-body}). For example, given a header instance A of
type \texttt{A\_t \{ f: 2 \}},
$\chomp(\rT{\epsilon}{\slice{x.\pIn}{0}{2}=11}, A, y) = \refT{x}{\epsilon}{\slice{y.A}{0}{2}=11}$,
i.e., because header instance A contains two bits, the first two bits are moved
from $\pIn$ to instance A, bound by $y$.
It turns out that we can define $\chomp$ via a simple syntactic
transformation. To do this, we first define a single-bit operation,
$\chomp_1$, that processes only a single bit. Then $\chomp$
recursively lifts $\chomp_{1}$ to the appropriate length.

\subsubsection{Chomp\textsubscript{1}}

To chomp a single bit from a heap type, we will need to update all references to
the length, as well as to the first bit of $\pIn$. This computation resembles
Brzozowski derivatives~\cite{brzozowski64}.
For the complete definition of $\chomp_1$, we refer the reader
to\appendixref{the appendix}. Here we provide some intuition for how
it works. Semantically, $\chomp_1(\tau, \bvar{0})$ transforms for each
heap $h$ denoted by a heap type $\tau$, into the heap $h[\pIn \mapsto
  \slice{h(\pIn)}{1}{~}]$. The variable $\bvar{0}$ is a placeholder
corresponding to the missing bit. Then, a helper function $\heapRef_1$
replaces the placeholder bits introduced by $\chomp_1$ with references
to the extracted bit. In particular, the $i$-th call to $\heapRef_1$,
replaces variable $\bvar{i}$ with $\slice{x.\inst}{i-1}{i}$.

Syntactically, when chomping a heap type $\tau$ we update
each occurrence of $\pIn$ in a refinement, if that occurrence describes
the first bit of $\pIn$ of a heap in the semantics of $\tau$.
Types $\varnothing$ and $\top$ are not affected by chomping.
For a choice type $\tau_1+\tau_2$, $\chomp_1$ is applied
to both $\tau_1$ and $\tau_2$ individually,
as each branch of the choice type describes isolated heaps of $\tau$.
In the substitution type, $\tau_1[x\mapsto\tau_2]$, only $\tau_1$ is chomped,
as $\tau_2$ only captures information relevant for evaluating refinements.

For the refinement type $\refT{x}{\tau_1}{\varphi}$, chomp is applied
recursively to $\tau_1$ and all references to the first bit of $\pIn$
as well as the length of $\pIn$ are updated accordingly. We increment
numeric expressions referencing $x.\pIn$ (e.g. the refinement
$\length{x.\pIn}$ becomes $\length{x.\pIn}+1$ and we prepend the
placeholder bit $\bvar{n}$ for bit-vector expressions referencing
$x.\pIn$).

When we apply $\chomp_1$ to type $\sigmaT{x}{\tau_1}{\tau_2}$, we have
to distinguish two cases, either the input packet described by
$\tau_1$ contains at least one bit or the input packet described by
$\tau_1$ is empty. In the first case, $\chomp_1$ removes the first bit
of $\pIn$ in $\tau_1$ and in the second case it removes the first bit
of $\pIn$ in $\tau_2$. If we chomp in $\tau_1$ we need to update all
refinements to $x.\pIn$ in $\tau_2$, as $\tau_1$ is bound to $x$ in
$\tau_2$; otherwise chomping could cause contradictions between
refinements referencing the same component. Similar to the computation
of a Brzozowski derivative of a product, the result is the union of
the type obtained by chomping $\tau_1$ and $\tau_2$ respectively,
where we additionally assert in the second case that $\pIn$ of
$\tau_1$ must be empty.

\paragraph{Example}
Given type $\tau = \sigmaT{x}{\refT{y}{\epsilon}{\length{y.\pIn}=1}}{\refT{z}{\epsilon}{\length{x.\pIn}=1}}$
\begin{align*}
  \chomp_1(\tau, \bvar{0})
  &= \sigmaT{x}{\refT{y}{\epsilon}{\length{y.\pIn}+1=1}}{\refT{z}{\epsilon}{\length{x.\pIn}+1=1}} ~+ \\
  &\phantom{{}={}}\sigmaT{x}{\refT{y}{\epsilon}{\length{y.\pIn}=1 \wedge \length{y.\pIn} = 0}}{\refT{z}{\epsilon}{\length{x.\pIn}=1}}
\end{align*}

\paragraph{Example}
Given a type $\tau = \refT{x}{ \refT{y}{\inst}{\length{y.\pIn}=8} }{\slice{x.\pIn}{0}{8}=x.\inst[4:12]}$
\begin{align*}
\chomp_1(\tau, b_0)
  &= \refT{x}{\refT{y}{\inst}{\length{y.\pIn}+1=8}}{b_0::\slice{x.\pIn}{0}{7}=x.\inst[4:12]}
\end{align*}

\paragraph{Example}
Given a header instance $A$ and a heap type $\tau = \refT{x}{\epsilon}{\bvconcat{\bvar{0} :: \bvNil}{x.\pIn[0]} = 10}$.
The first call to $\heapRef_1$ returns type $\refT{x}{\epsilon}{\bvconcat{(\bvconcat{y.\inst[0]}{\bvNil})}{x.\pIn[0]} = 10}$.

\subsubsection{Correctness of Chomp}

We prove that $\chomp$ is correct with respect to $\chompS$.
Specifically, \cref{lem:semantic-chomp-body} states that---given some heap $h\in\semantics{\tau}_{\env}$---there exists a corresponding heap $h'$ in the semantics of the chomped type
that is equivalent to the heap obtained after applying $\chompS$ to $h$.
Since $\chomp$ adds a refinement on $x.\inst$, we evaluate the chomped type
in an environment, where $x$ maps to the heap in which $\inst$ contains the first
$\sizeof(\inst)$ bits from $h(\pIn)$. This reflects the intuition that
$\chomp$ populates the header instance $\inst$ with the first $\sizeof(\inst)$ bits
from the input packet.

\begin{lemma}[Semantic Chomp]
  \label{lem:semantic-chomp-body}
  If $x$ does not appear free in $\tau$, then for all heaps $h\in\semantics{\tau}_\env$ where $\length{h(\pIn)}\geq \sizeof(\inst)$,
  there exists $h'\in\semantics{\chomp(\tau,\inst,x)}_{\env'}$ such that $h'=\chompS(h,\sizeof(\inst))$ where $\env'=\env[x\mapsto(\bvNil, \bvNil, [\inst\mapsto h(\pIn)[0:\sizeof(\inst)]])]$.
\end{lemma}

\subsection[Safety of Pi4]{Safety of \name}

We prove safety of \name in terms of standard progress and
preservation theorems. That is, well-typed programs do not get stuck
and when well-typed programs are evaluated, they remain well typed.
Both theorems make use of the entailment relation defined in
\cref{fig:entailment}.

\begin{theorem}[Progress]
If $\cmdType{\cdot}{c}{(x:\tau_1)\rightarrow\tau_2}$ and $(I,O,H)\models\tau_1$, then either $c=\cSkip$ or \\
$\exists\langle I',O',H',c'\rangle.\langle I,O,H,c\rangle\rightarrow\langle I',O',H',c'\rangle$.
\end{theorem}
\begin{proof}
  By induction on the typing derivation. For details, see\appendixref{\cref{sec:safety}, \cref{thm:progress}}.
\end{proof}

As usual, progress says that if a command is well-typed, it is either
\textit{skip} or can take a step.

\begin{theorem}[Preservation]
If $\Gamma \vdash c:(x:\tau_1) \rightarrow \tau_2$,
  $\langle I,O,H,c\rangle \rightarrow \langle I',O',H',c'\rangle$, and
  $\env \models \Gamma$ and $(I,O,H)\models_\env \tau_1$, then there exists
  $\Gamma',\env',x',\tau_1',\tau_2'$, such that $\Gamma' \vdash c':(x':\tau_1') \rightarrow \tau_2'$ and
  $\env' \models \Gamma'$ and 
  $(I',O',H')\models_{\env'}\tau_1'$ and
  $\semantics{\tau_2'}_{\env'[x' \mapsto (I',O',H')]} \subseteq \semantics{\tau_2}_{\env[x \mapsto (I,O,H)]} $
\end{theorem}
\begin{proof}
  By induction on the typing derivation. For details, see\appendixref{\cref{sec:safety}, \cref{thm:preservation}}.
\end{proof}

The preservation theorem says that if a command is a well-typed
command $c$ that can step to $c'$, and a heap entails the input type
$\tau_1$, then $c'$ is well-typed from $\tau_1'$ to $\tau_2'$ for some
$\tau_1'$ and $\tau_2'$ such that the final heap entails $\tau_1'$ and
the set of heaps described by $\tau_2'$ is a subset of the heaps
denoted by $\tau_2$ with their respective input heaps bound to
variable $x$.

%% file: 04-implementation.tex
\section{Implementation}
\label{sec:impl}

We have built a prototype implementation of \name's type system in OCaml and Z3.
Under the hood, it uses an encoding of \name's types
into a decidable theory of first-order logic, facilitating use of an SMT solver
to automatically discharge the various side conditions that arise during type
checking. We describe the algorithmic type system and its decidability, some
optimizations we use to simplify our types, and our $\PPPP{16}$ frontend.

\subsection{Algorithmic Type System and Decidability}
For our implementation, we define an algorithmic version of our type 
system whose rules are mostly identical to the rules from our 
declarative type system. 
\cref{fig:algo-typing-selected-rules} shows two selected algorithmic 
typing rules that demonstrate the key differences from our declarative 
system.   
Many of the typing rules have semantic conditions that must be checked
during type checking. 
In the algorithmic type system, we encode these constraints as subtype constraints.
For example, when we type-check the command \cAdd(\inst), we must check
that the newly added instance is not already valid in the input type $\tau_1$, 
i.e., $\excludes{\tau_1}{\inst}$.
As shown by rule \tAddAlgo in \cref{fig:algo-typing-selected-rules}, 
$\excludes{\tau_1}{\inst}$ becomes the subtype check 
$\Gamma \vdash \tau_1 <: \refT{x}{\top}{\neg x.\inst.\valid}$.
Similarly, rule \tMod and \tRemit require $\includes{\tau_1}{\inst}$, which becomes
$\Gamma \vdash \tau_1 <: \refT{x}{\top}{x.\inst.\valid}$ in \tModAlgo and \tRemitAlgo 
respectively.
The check $\sizeof_\pIn(\tau) \geq \sizeof(\inst)$ required by \tExtract becomes 
$\Gamma \vdash \tau_1 <: \refT{x}{\top}{\length{x.\pIn} \ge \sizeof(\inst)}$ in rule \tExtractAlgo.

The second major difference is the rule for type ascription
\tAscribeAlgo. In our implementation we check if the input type
$\tau_1$ is a subtype of the ascribed input type $\hat\tau_1$. We then
use the ascribed input type to compute an output type $\tau_c$.
Finally, we check if the computed output type is a subtype of the
ascribed output type. Note, that our type checking algorithm can be
used to obtain a weak form of type inference. Given an input type that
describes the state before the execution, our algorithm computes an
output type, which describes the state after the execution of the
program. However, a full-blown treatment of type reconstruction (such
as the one used by Liquid Haskell~\cite{vazou18}) is left for future
work.

\begin{figure}[t]
  \begin{mathpar}
    \inferrule[\tAddAlgo]{
      \subtype{\tau_1}{\refT{x}{\top}{\neg x.\inst.\valid}} \\
      \mathit{init}_{\HT(\inst)} = v
    }{
      \cmdTypeAlgo{\Gamma}{\cAdd{\inst}}{x:\tau_1}{\sigmaT{y}{\refT{z}{\tau_1}{z \equiv x}}{\refT{z}{\inst}{z.\pIn=z.\pOut=\bvNil \wedge z.\inst = v}}}
    }
    \and
    \inferrule[\tAscribeAlgo]{
      \cmdTypeAlgo{\Gamma}{c}{x:\hat{\tau_1}}{\tau_c}\\
      \subtype{\tau_1}{\hat{\tau_1}} \\
      \subtypeCtx{\Gamma,x:\hat{\tau_1}}{\tau_c}{\hat{\tau_2}} \\
    }{
      \cmdTypeAlgo{\Gamma}{\cAscribe{c}{(x:\hat{\tau_1}) \to \hat{\tau_2}}}{x:\tau_1}{\hat{\tau_2}}
    }
  \end{mathpar}
  \caption{Selected rules of the algorithmic type system}
  \label{fig:algo-typing-selected-rules}
\end{figure} 

We convert every check
$\subtype{\tau_1}{\tau_2}$
into a formula in the theory of fixed-width
bit vectors. This is largely straightforward, except for the encoding
of $\pIn$ and $\pOut$, which may be arbitrarily long. However, 
network switches have a maximum number of bits that they can receive
or transmit, called the maximum transmission unit (MTU). So when
compiling a $\PPPP{}$ program to a given switch, we know that the
transmitted packets must be smaller than MTU. We exploit this fact to
prove a bound on the size of the bit vectors that must be considered.

More formally, we say that a type $\tau$ is bounded by $N$ in a
context $\Gamma$, written $\bound \Gamma \tau N$, iff for
every $\env \models \Gamma$, and $h \in \semantics{\tau}_{\env}$,
$|h(\pIn)| + |h(\pOut)| \leq N$. We need to bound both $h(\pIn)$ and
$h(\pOut)$ by $N$ because (as seen in the \cReset command), the
emitted packet is $h(\pOut)@h(\pIn)$. 
\cref{thm:fwd-mtu-bound} (\emph{MTU-Bound}) says that given
an algorithmic typing judgement on a program $c$, for which the input type and
all ascribed types in $c$ respect the MTU $N$, the output type will require no
more than $N + \emit c$ bits, where $\emit c \in \mathbb N$ is the number of
bits that could possibly be emitted in $c$. The details are shown in\appendixref{\cref{fig:max-emit}}. Note that even though the real input type is
constrained by the same MTU $N$, intermediate states may require more than just
$N$ bits. \emph{MTU-Bound} shows that $N + \emit c$ suffices as the
maximum combined width of $\pIn$ and $\pOut$.

\begin{theorem}[MTU-Bound]
  \label{thm:fwd-mtu-bound}
  For every $\Gamma$, $c$, $x$, $\tau_{1}$, $\tau_{2}$, and $N$, 
  if $\Gamma \vdash \tau_{1} \leq N$ and 
  $\cmdTypeAlgo{\Gamma}{c}{(x:\tau_1)}{\tau_2}$ and 
  every ascribed type in $c$ is also bounded by $N$, 
  then $\bound {\Gamma, (x : \tau_{1})} {\tau_{2}} {N + \emit c}$.
 \end{theorem}

 \begin{proof}
   By induction on the typing derivation. For details, see\appendixref{\cref{sec:mtu}, \cref{thm:forwards-mtu-bound}}.
 \end{proof}

\cref{thm:algo-typ-correct} establishes the correctness of the algorithmic typing relation.
It states that a program $c$ typechecks in the declarative system with type $(x:\tau_1) \to \tau_2$ if and only if it 
also typechecks in the algorithmic system with type $(x:\tau_1) \rightsquigarrow \tau_2'$ and the output type of the 
algorithmic system $\tau_2'$ is a subtype of the output type $\tau_2$ of the declarative system.

\begin{theorem}[Algorithmic Typing Correctness]
  \label{thm:algo-typ-correct}
  For all $\Gamma$, $c$, $x$, $\tau_{1}$, and $\tau_{2}$,
  where $x$ is not free in $\tau_{1}$,
  $\cmdType{\Gamma}{c}{(x:\tau_{1}) \to \tau_{2}}$
  if and only if
  there is some $\tau_{2}'$ such that
  $\cmdTypeAlgo{\Gamma}{c}{x:\tau_{1}}{\tau_{2}'}$,
  and
  $\subtypeCtx{\Gamma,(x: \tau_{1})}{\tau_{2}'}{\tau_{2}}$.
\end{theorem}
\begin{proof}
  By induction on the typing derivation. 
  For details, see\appendixref{\cref{sec:mtu}, \cref{thm:algorithmic-typing-correctness}}.
\end{proof}

With Theorems~\ref{thm:algo-typ-correct} and~\ref{thm:fwd-mtu-bound} in hand, it
is straightforward to show the decidability of the declarative type system,
i.e., that $\cmdType{\Gamma}{c}{(x:\tau_1) \to \tau_2}$ is decidable (cf.
\cref{thm:typing-decidable}). \cref{thm:algo-typ-correct} allows us to
equivalently show that typechecking the command in the algorithmic type
system---i.e., $\cmdTypeAlgo{\Gamma}{c}{(x:\tau_1)}{\tau_2'}$---terminates and
that checking $\Gamma,(x:\tau_{1}) \vdash \tau_2' <: \tau_2$ terminates, which
follows by finite enumeration using the bounds guaranteed by
\cref{thm:fwd-mtu-bound}.

\begin{theorem}[Decidability]
\label{thm:typing-decidable}
If $\Gamma$, $\tau_1$, $\tau_2$ and every ascribed type in $c$ are bounded by the MTU $N$, 
then $\cmdType{\Gamma}{c}{(x:\tau_1) \to \tau_2}$ is decidable.
\end{theorem}
\begin{proof}
  Proof by \emph{Algorithmic Typing Correctness}, \emph{MTU-Bound} and by induction on the typing derivation.
  For details, see\appendixref{\cref{sec:mtu}, \cref{thm:decidability}}.
\end{proof}


\subsection{Rewriting Optimizations}
The final major difference between the declarative type system and our
implementation is that we exploit two type equivalences to eliminate
$\Sigma$-types and $\chomp$. We exploit the fact that $\Sigma$-types
can be written using refinement and substitution types. In other
words, in any context $\Gamma$, the type
$\sigmaT{x}{\tau_{1}}{\tau_{2}}$ is equivalent to

\[\left\{x\!:\!\top\,\middle|\,
    \left(\begin{array}{l}
            x.\pIn = \bvconcat{l.\pIn}{r.\pIn}\,\wedge \\
            x.\pOut = \bvconcat{l.\pOut}{r.\pOut}
          \end{array}\right)
    \wedge\!\!\!\!\!\!\!\!
    \bigwedge_{\inst\in\dom(\HT)}\!\!
    \left(\begin{array}{l}
            x.\inst.\valid = l.\inst.\valid \oplus r.\inst.\valid\,\wedge \\
            l.\inst.\valid  \implies x.\inst = l.\inst\,\wedge \\
            r.\inst.\valid \implies x.\inst = r.\inst
          \end{array}\right)
      \right\}\!\!\!\!
  \begin{array}{l}
    {[r \mapsto \tau_2]}\\
    {[l\mapsto \tau_1]}
  \end{array}\]

We also can eliminate occurrences of $\chomp$ produced by extractions. Observe
that in the context where $(x: \tau)$, the following two types are
equivalent
\[\begin{array}{c}
    \text{\Large $\Sigma$} y : \left\{z : \inst\, \middle|\,
    \begin{array}[center]{l}
      z.\pIn = \bv{}\,\wedge \\
      z.\pOut = \bv{}
    \end{array}\right\} \mathrel{.}
    {\left\{ z : \chomp(\tau, \inst, y)\, \middle|\,
    \left(\begin{array}[center]{l}
            y.\inst@z.\pIn = x.\pIn\,\wedge \\
            \,z.\pOut = x.\pOut\\
    \end{array}\right)\wedge z \equiv_{\inst} x \right\}}\\
    \doteq \\
    {\refT y \top {y.\inst.\valid \wedge 
                    x.\pIn = \bvconcat{y.\inst}{y.\pIn} \wedge 
                    x.\pOut = y.\pOut \wedge 
                    \bigwedge_{\kappa\in\dom(\HT) \wedge \kappa\neq\inst} y.\kappa = x.\kappa }}
  \end{array}\]

Because these types are \emph{equivalent} we can give $\cExtract \inst$ commands the following type:
\[(x: \tau) \to {\refT y \top {y.\inst.\valid \wedge 
x.\pIn = \bvconcat{y.\inst}{y.\pIn} \wedge 
x.\pOut = y.\pOut \wedge\!\!\!\!\!\!
\bigwedge_{\kappa\in\dom(\HT) \wedge \kappa\neq\inst}\!\!\!\!\!\ y.\kappa = x.\kappa }}\]
This optimization, along with similar changes to the types for $\cAdd(\inst)$ and
$\cRemit \inst$ greatly reduce the size of the generated Z3 formulae, making
typechecking tractable.

\subsection{Beyond the Core Calculus}

Our prototype is equipped with a $\PPPP{16}$ frontend that uses Petr4's
parser~\cite{Doenges2021} to translate a subset of type-annotated $\PPPP{16}$
programs into \name programs. We leverage P4's builtin annotation mechanism to
allow users to annotate control and parser blocks with types using the custom
$\texttt{@pi4}(\sigma)$ annotation, where $\sigma$ is the desired type. We also
provide convenience notation such as $\texttt{@pi4\_roundtrip}(\tau)$, which
ensures, as elaborated in \cref{sec:roundtrip} that the composition of
deparser, $\cReset$, and parser has type
$(x:\tau) \to \refT{y}{\top}{x \equiv y}$.

%% file: 05-experience.tex
\newcommand{\sepmargin}{2pt}
\newcommand{\sep}{\vspace{\sepmargin}\hrule\vspace{\sepmargin}}

\section{Checking Network Invariants}
\label{sec:experience}

We now show that dependent types are a good match for P4, by demonstrating that
\name's type system can be used to (i) check real network protocol invariants
and (ii) verify a variety of basic and advanced safety properties. 
We showcase properties %
that are also studied in the context of other P4 verification
tools~\cite{Liu2018,Stoenescu2018}.
All examples, in this section and the next, have been implemented in our \name prototype.

In most P4 programs, packet-forwarding behavior is specified using a
predefined record of type \texttt{standard\_metadata\_t}. In
particular, the \texttt{egress\_spec} field instructs the switch to
forward the packet out on a specific port. We assume that the field is
initialized to \texttt{0x00}, indicating that no forwarding decision
has been made, and that by setting the field to \texttt{0x1FF} (i.e.,
the largest unsigned 9-bit value), the switch can be instructed to
drop the packet.\footnote{In the examples that follow, we use
  bitvector literals assuming that they are implicitly cast to the
  appropriate widths (following $\PPPP{16}$'s casting semantics). The
  implementation, however, requires these to be explicit lengths, e.g.
  \texttt{0b111111111} instead of \texttt{0x1FF} for a 9 bit field.}
For simplicity, we treat all P4 metadata as an ordinary header
instance called \texttt{stdmeta}.

\begin{figure}[t]
\begin{minipage}[t]{0.49\textwidth}
\begin{p4code}
/* Unsafe */
if(ipv4.valid) {
  stdmeta.egress_spec := 0x1;
  ipv4.ttl := ipv4.ttl - 0x1
}
\end{p4code}
\end{minipage} \hfill
\begin{minipage}[t]{0.49\textwidth}
\begin{p4code}
/* Safe */
if(ipv4.valid) {
  if(ipv4.ttl == 0x00) {
    stdmeta.egress_spec := 0x1FF
  } else {
    stdmeta.egress_spec := 0x1;
    ipv4.ttl := ipv4.ttl - 0x1
  }
}
\end{p4code}
\end{minipage}
\sep
\begin{p4code}[numbers=none]
(x:{y:$\weak{\texttt{ipv4}}$ | y.meta.valid}) ->
    {y:$\weak{\texttt{ipv4}}$ | y.meta.valid && (x.ipv4.ttl==0x0 => y.meta.egress_spec==0x1FF)}
  \end{p4code}
\caption{IPv4 TTL example. Top left: doesn't typecheck; top right: typechecks; bottom: \name type encoding the TTL invariant.}
\label{fig:ipv4-ttl}
\end{figure}

\begin{figure}[t]
\begin{minipage}[t]{0.49\textwidth}
\begin{p4code}
/* Unsafe */
extract(ether);
if(ether.etherType == 0x0800) {
  extract(ipv4)
}
\end{p4code}
\end{minipage}\hfill
\begin{minipage}[t]{0.49\textwidth}
\begin{p4code}
/* Safe */
extract(ether);
if(ether.etherType == 0x0800) {
  extract(ipv4);
  if(ipv4.ihl != 0x5) {
    extract(ipv4opt)
  }
}
\end{p4code}
\end{minipage}
\sep
\begin{p4code}[numbers=none]
(x:{y:$\epsilon$|y.pkt_in.length > 280}) ->
    {y:$\top$|((y.ipv4.valid && y.ipv4.ihl != 0x5) => y.ipv4opt.valid) &&
        ((y.ipv4.valid && y.ipv4.ihl == 0x5) => >>!<<y.ipv4opt.valid)}
\end{p4code}
\caption{IPv4 Options example. Top left: doesn't typecheck; right: typechecks; bottom: \name type encoding the IPv4-Option specification.}
\label{fig:ipv4-opt}
\end{figure}

\subsection{Protocol Conformance}

We start with examples showing how \name's type system can be used to
ensure that a program conforms with standard network protocols.

\newcommand{\safe}{\textit{\textcolor{green}{Safe}}}
\newcommand{\unsafe}{\textit{\textcolor{red}{Unsafe}}}

\paragraph{IPv4 --- Time To Live.}
For Internet Protocol (IP) packets, the time to live (TTL) limits how
often a packet can be forwarded from one network switch to another.
Every time a packet is forwarded, TTL is decremented; when  TTL
is zero before the packet has reached its destination, forwarding
halts to eliminate the risk of infinite loops.\footnote{Strictly
  speaking, IPv4 requires a special ICMP message to be returned to the
  sender to indicate the error.}
The code snippet in the top left side of \cref{fig:ipv4-ttl} violates
the property because the packet is always forwarded
on the same port while  TTL is decremented. We can detect this
violation by checking the program with the type at the bottom of \cref{fig:ipv4-ttl},
which reads as:
Starting in a heap where at least IPv4 is valid, after executing the
ingress code, still at least IPv4 is valid and if the IPv4 TTL is
zero, the value of \texttt{egress\_spec} indicates that the packet
will be dropped. 
The program in the top right of \cref{fig:ipv4-ttl}  successfully typechecks with the type.

\paragraph{IPv4 Options.}
The standard IPv4 header consists of at least 160 bits, but it may
also carry additional data in optional fields. The \textit{Internet
  Header Length (IHL)} field specifies the length of the header as
multiples of 32 and indicates whether additional data is available.
The minimum IHL is 5 ($5*32=160$) and the maximum is 15. Due to their
flexibility, IP options are notoriously difficult to parse, and many
real-world network devices handle them incorrectly. We can use \name's
type system to ensure that we also extract the IPv4 options from the
input packet, whenever IPv4 is valid and IHL > 5. The type shown in
the bottom of \cref{fig:ipv4-opt} states that executing the parser in
the empty heap where enough bits are available to extract Ethernet,
IPv4 and IPv4 options, produces a heap satisfying the constraint that
when IPv4 is valid and IHL is 5, IPv4 options are not valid, and when
IPv4 is valid and IHL > 5, then IPv4 options are valid.
\cref{fig:ipv4-opt} shows one example where this property is violated
(top left) and one where it holds (top right).

\begin{figure}[t]
\begin{minipage}[t]{0.49\textwidth}
\begin{p4code}
/* Unsafe */
extract(ether);
extract(ipv4)
\end{p4code}
\end{minipage}\hfill
\begin{minipage}[t]{0.49\textwidth}
\begin{p4code}
/* Safe */
extract(ether);
if(ether.etherType == 0x0800) {
  extract(ipv4)
}
\end{p4code}
\end{minipage}
\sep
\begin{p4code}[numbers=none]
(x: {y:$\epsilon$ | y.pkt_in.length > 272}) ->
          {y:$\top$ | y.ipv4.valid => y.ether.etherType == 0x0800}
\end{p4code}
\caption{Header dependency example. Top left: doesn't typecheck; top right: typechecks; bottom: \name type encoding IPv4's dependency on Ethernet.}
\label{fig:header-dependency}
\end{figure}

\paragraph{Header Dependencies.}
Most protocols have some way of keeping track of what other protocols
are encapsulated in the payload of a packet---i.e., which header
follows next. The correspondence between field values and protocols is
typically defined as part of the protocol standard. For example, an
Ethernet frame uses the EtherType field (written \texttt{ether.etherType})
for this purpose: a value of \texttt{0x0800} indicates that the next
header is an IPv4 header, while a value of \texttt{0x86dd} indicates
that the next header is an IPv6 header. 
This is specified in the type at the bottom of 
\cref{fig:header-dependency}.
The code snippet on the top left of \cref{fig:header-dependency} violates the
dependency between the IPv4 header and the EtherType field of the Ethernet
header. Our type checker detects this by checking that executing the
parser on an empty heap with enough bits to extract both Ethernet and IPv4,
produces a heap with either an invalid or a valid IPv4 header and an
EtherType value of \texttt{0x0800}. The code on the top right of
\cref{fig:header-dependency} 
fixes the error by only
extracting \texttt{ipv4} when \texttt{ether.etherType} is \texttt{0x0800}.

\subsection{Basic Safety Properties}
\name's type system can be also used to ensure safety properties. We
discussed how it detects accesses to invalid header instances in
Section \ref{sec:motivation}. Here we present an example showing how
it can be used to enforce \emph{determined
  forwarding}~\cite{Liu2018,Stoenescu2018}. Typical P4 programs
contain thousands of paths on which a packet can be processed. To
avoid situations where packets are dropped unexpectedly, a desirable
invariant is that each program path contains an explicit forwarding
decision---i.e., packets are either forwarded on some switch port or
dropped. Our type checker is able to detect violations of this
property. The type in the bottom of \cref{fig:determined-forwarding}
shows one way of encoding this specification as a type. Under the
assumption that the egress specification is initialized with a dummy
value of \texttt{0x0}, the type asserts that it is, at some point,
modified during the execution of the pipeline, i.e., a forwarding
decision is made for every packet. The program on the top left of
\cref{fig:determined-forwarding} fails to typecheck with the type,
because the egress specification is unset for packets with
\texttt{ipv4.dst} equal to $\texttt{0x0a0a0a0a}$. The program on the
top right of \cref{fig:determined-forwarding} fixes this issue via an
else-case that assigns the egress specification to \texttt{0x1FF}.

\begin{figure}[t]
\begin{minipage}[t]{0.49\textwidth}
\begin{p4code}
/* Unsafe */
if(ipv4.valid) {
  if(ipv4.dst != 0x0a0a0a0a) {
    stdmeta.egress_spec := 0x1FF
  }
}
\end{p4code}
\end{minipage}\hfill
\begin{minipage}[t]{0.49\textwidth}
\begin{p4code}
/* Safe */
if(ipv4.valid) {
  if(ipv4.dst != 0x0A0A0A0A) {
    stdmeta.egress_spec := 0x1
  } else {
    stdmeta.egress_spec := 0x1FF
  }
}
\end{p4code}
\end{minipage}
\sep
\begin{p4code}[numbers=none]
(x:{y:$\weak{\texttt{ipv4}}$| y.stdmeta.valid}) -> {y:$\top$| y.stdmeta.egress_spec != 0x0}
\end{p4code}
\caption{Determined forwarding example. Top left: ill-typed; top right: well-typed; bottom: determined forwarding specification encoded as \name type.}
\label{fig:determined-forwarding}
\end{figure}

\subsection{Parser-Deparser Compatibility}
\label{sec:roundtrip}

A P4 program typically defines the parser, controls for ingress and
egress pipelines, and the deparser.\footnote{Why this four-phase
  structure? Having separate ingress and egress pipelines allows
  packet processing to occur both before and after packets are
  scheduled, typically using one or more queues.} In practice, parsing
and deparsing may also happen between the ingress and egress
stages---i.e., the deparser code is additionally executed at the end
of the ingress followed by the parser code, before the egress. In such
cases, it is important to ensure that data intended to be carried from
ingress to egress is serialized and deserialized correctly. Otherwise,
headers may be unexpectedly removed from the packet.

For example, assume that the parser shown in
\cref{fig:roundtrip} successfully parses the Ethernet and IPv4
headers from the input packet, but not a VLAN header. From the code we
can conclude that EtherType must be $0x0800$. Let's further assume
that the programmer intends the ingress control  in the middle
right of \cref{fig:roundtrip}. After parsing, the switch checks if
a VLAN header is present. If a VLAN header was already parsed from the
input packet, no changes are made. Otherwise a VLAN header is added
(\cref{lst:roundtrip-vlan-valid}) and the EtherType of the
Ethernet header is updated accordingly. If an IPv4 header is present,
the EtherType must be updated accordingly
(\cref{lst:roundtrip-ethertype-80}) to obtain a
protocol-conformant packet.
Now, assume that the programmer forgot the statement on
\cref{lst:roundtrip-ethertype-80}, i.e., didn't update
\texttt{ether.etherType} (this unsafe example is in the left of
\cref{fig:roundtrip}).
After running the deparser at the end of ingress, all three headers
are serialized: The first 112 bits correspond to the Ethernet header,
followed by 32 bits of the VLAN header, and another 160 bits of the
IPv4 header. Since the programmer forgot to update 
EtherType, bits 96 to 112 contain the value \texttt{0x0800}. If
the parser runs with this bitstream as the input, it will first
parse the Ethernet header, then look at the \texttt{etherType} and
given the value $0x0800$, it will continue to parse the IPv4 header.
Hence, the bits of the VLAN header are parsed as an IPv4 header,
leading to a corrupted packet.

To avoid such errors, we want to enforce the invariant that all instances valid
at the end of ingress are equivalent to those obtained after deparsing and
re-parsing. We instruct our type checker to verify this property via the type on
\cref{lst:roundtrip-ascription}, by checking the whole program with a type
that ensures there are enough bits to parse the headers, shown on \cref{lst:roundtrip-type}.

\subsection{Mutual Exclusion of Headers}

\begin{figure}[t]
\begin{minipage}[t]{0.5\textwidth}
\begin{p4code}
Parser $\triangleq$
  extract(ether);
  if(ether.etherType == 0x8100) {
    extract(vlan);
    if(vlan.etherType == 0x0800) {
      extract(ipv4)
    }
  } else {
    if(ether.etherType == 0x0800) {
      extract(ipv4)
    }
  }
\end{p4code}
\begin{p4code}[firstnumber=last]
Deparser $\triangleq$
  if (ether.valid) { remit(ether) };
  if (vlan.valid) { remit(vlan) };
  if (ipv4.valid) { remit(ipv4) }
\end{p4code}
\end{minipage}%
\begin{minipage}[t]{0.5\textwidth}
\begin{p4code}[firstnumber=last]
UnsafeIngress $\triangleq$
  if(>>!<<vlan.valid) {
    add(vlan);
    vlan.etherType := 0x0;
    if(ipv4.valid) {
      vlan.etherType := 0x0800
    }
  };
\end{p4code}
\begin{p4code}[escapeinside={@}{@},firstnumber=last]
SafeIngress $\triangleq$
  if(>>!<<vlan.valid) {
    @\label{lst:roundtrip-vlan-valid}@add(vlan);
    @\label{lst:roundtrip-vlan-ethertype-00}@vlan.etherType := 0x0;
    @\label{lst:roundtrip-ethertype-81}@ether.etherType := 0x8100;
    if(ipv4.valid) {
      @\label{lst:roundtrip-ethertype-80}@vlan.etherType := 0x0800;
    }
  }
\end{p4code}
\end{minipage}
\sep
\begin{p4code}[escapeinside={@}{@},firstnumber=last]
Parser; Ingress; /* Ingress is either UnsafeIngress or SafeIngress  */
(Deparser; remit; Parser) as
    @\label{lst:roundtrip-ascription}@(x:{z:$\weak{\texttt{ether}}$|z.ether.etherType == 0x8100 && z.vlan.valid &&
                (z.ipv4.valid <=> z.vlan.etherType == 0x0800) &&
                z.pkt_out.length == 0 &&
                z.pkt_in.length > 0}) -> {y:$\top$| x === y}
\end{p4code}
\sep
\begin{p4code}[escapeinside={@}{@},numbers=none]
@\label{lst:roundtrip-type}@(x:{y:$\epsilon$|y.pkt_out.length == 0 && y.pkt_in.length > 304}) -> $\top$
\end{p4code}

 \caption{Roundtripping Definitions. Top left: common parser and deparser; top right: unsafe and safe ingress code; middle: the pipeline, which typechecks with $\texttt{Ingress} \mapsto \texttt{UnsafeIngress}$, but not with $\texttt{Ingress} \mapsto \texttt{UnsafeIngress}$; bottom: the type at which to check the full pipeline.}
\label{fig:roundtrip}
\end{figure}

The parser shown on the top of \cref{fig:mutual-exclusion} conditionally
parses either IPv4 or IPv6. Because only one of the paths is taken at runtime,
it should never happen that both instances are valid at the same time. This
property might be exploited in an implementation, allowing the same memory to be
used to store both headers. In this small example, it is easy to see that this
invariant holds. But in larger programs it is difficult to track which header
instances are valid on which execution paths.
We can check that the property continues to hold in the ingress in the
middle of \cref{fig:mutual-exclusion}
using the type in the ascription on \cref{lst:mutex-ing-type}.
The ingress code in the middle left of \cref{fig:mutual-exclusion}
exemplifies a violation of the property: If a packet enters the control
block with a valid IPv4 header, it will leave with both a valid IPv4 and a valid
IPv6 header; a violation of our property.
The code on the middle right is safe because it includes a conditional that
explicitly checks the validity of IPv4 before adding IPv6. The combination of
union types and refinement types makes our type system capable of such precise
path-dependent reasoning.

\begin{figure}[t]
\begin{p4code}
(extract(ether);
if (ether.etherType == 0x86dd) { extract(ipv6) }
else { if(ether.etherType == 0x0800) { extract(ipv4) } })
  as (x:{y:$\epsilon$|y.pkt_in.length>432}) -> {y:$\weak{\texttt{ether}}$|>>!<<(y.ipv4.valid&&y.ipv6.valid)};
Ingress /* Can be SafeIngress or UnsafeIngress */
  as (x: {y:ether~|>>!<<(y.ipv4.valid && y.ipv6.valid)}) ->
         {y:ether~|>>!<<(y.ipv4.valid && y.ipv6.valid)});
if (ether.valid) { remit(ether) };
if (ipv4.valid) { remit(ipv4) };
if (ipv6.valid) { remit(ipv6) }
\end{p4code}
\sep
\begin{minipage}[t]{0.49\textwidth}
\begin{p4code}[firstnumber=last]
UnsafeIngress $\triangleq$
  add(ipv6)
  ether.etherType := 0x86DD
\end{p4code}
\end{minipage}\hfill
\begin{minipage}[t]{0.49\textwidth}
\begin{p4code}[firstnumber=last]
SafeIngress $\triangleq$
  if(ipv4.valid) {
    add(ipv6);
    ether.etherType := 0x86DD
  }
\end{p4code}
\end{minipage}
\sep
\begin{p4code}[escapeinside={@}{@},numbers=none]
@\label{lst:mutex-ing-type}@(x:{y:$\epsilon$|y.pkt_in.length>432}) -> $\top$
\end{p4code}
\caption{Mutual exclusion example: IPv4 and IPv6 should never be simultaneously valid. Top: Common pipeline; middle left: unsafe \texttt{Ingress} code; middle right: safe \texttt{Ingress} code; bottom: whole program type.}
\label{fig:mutual-exclusion}
\end{figure}

\subsection{Limitations}
There are a few P4 features that our \name prototype does not support,
mostly because they pose challenges to SMT-based approaches to
verification: hash functions, externs (a kind of foreign function
interface into the hardware), and registers. The unpredictability of
hash functions is difficult to verify, but we can either
over-approximate them as uninterpreted functions, or use a more
fine-grained approach such as concolic verification. Externs either
need to be annotated with specific types, or over-approximated as
uninterpreted functions. Registers are on-switch state that can be
modified by the packet or the controller and persists between packets.
This is tricky to represent in the semantics and has some distributed
computing concerns. We could over-approximate the behavior by
havoc-ing the values every time the register is read.

%% file: 06-case-study.tex
\section[Modularity Reasoning with Pi4]{Modularity Reasoning with \name}
\label{sec:case-study}

\begin{figure}[tp]
  \includegraphics[width=0.75\textwidth]{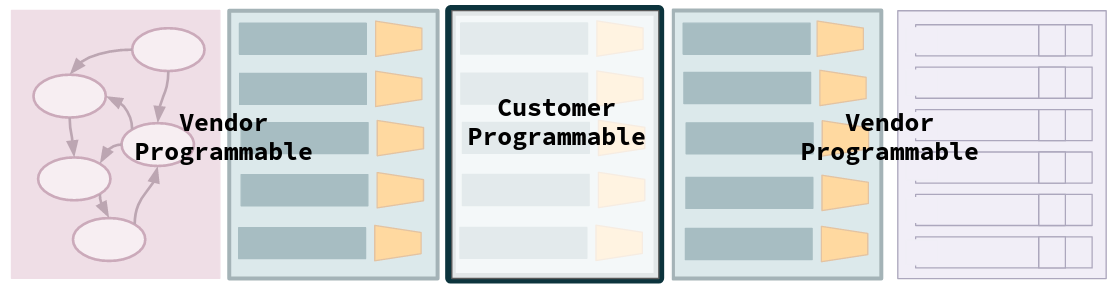}
  \vspace{-1em}
  \caption{Modular router design}
  \label{fig:cisco-arch}
\end{figure}

An emerging design pattern for data plane switches is partial
programmability, e.g., Cisco's daPIPE~\cite{daPIPE}, which is
designed for the Nexus 3400 switch~\cite{CiscoNexus3400}.
The idea is that a device vendor provides a partially-implemented pipeline
together with a set of program points where customers can inject custom code
as shown in \cref{fig:cisco-arch}. The designer requires that customer
programs satisfy certain properties, but in current architectures, they are not automatically checked.

To illustrate, consider a deployment of the customizable pipeline in a
campus network
where network engineers want to experiment with in-band network
telemetry (INT) without perturbing the VLAN tag, which is used to
enforce security policies. Let's say there are four classes of
traffic, \texttt{Visitor}, \texttt{Student}, \texttt{Faculty}, and
\texttt{Staff}, each with unique VLAN identifiers. We want to ensure
that no matter how \texttt{Ingress} is instantiated in the left of
\cref{fig:modular-router-ex}, it cannot cause students and visitors to
acquire privileges of faculty or staff---e.g., this might leak
confidential data.
With \name, we can design a modular system that statically checks invariants on customer
programs.
 Practically, we can
ensure that VLAN is not changed by checking that customer's code has a type
like:
$(x : \tau) \to \{y : \tau' \mid x.\texttt{vlan}.\texttt{vid} = y.\texttt{vlan}.\texttt{vid}\}$,
where the $\tau$ and $\tau'$ are appropriate for the specific
pipeline. We check, once-and-for-all, that the surrounding switch code
composes with this type, and incrementally check that the customer
code has this type (for an appropriate $\tau$).

\subsection{Specifying an Invariant}

\begin{figure}
\begin{p4code}[escapeinside={>!}{!<}]
extract(vlan);
Ingress /* Customer Specified. Default, Overwrite, Table, or UnsafeActions */
>!\label{lst:mod-type}!<  as (x:$\Sigma$x:stdmeta.ipv4) -> {y:$\Sigma$y:stdmeta~.ipv4~|y.vlan == x.vlan};
remit(vlan)
\end{p4code}
\caption{Instantiation of modular router design; the parser (\emph{Pre}) and deparser (\emph{Post}) are provided by the vendor, the \texttt{Ingress} code is provided by the customer.}
\label{fig:modular-router-ex}
\end{figure}

To further illustrate, consider the toy example shown the right of
\cref{fig:modular-router-ex}, which has a VLAN instance (32 bits)
and the standard metadata used in the P4 switch model (325 bits),
including a 9-bit egress specification \texttt{stdmeta.egress\_spec}, and a
12-bit vlan tag field, \texttt{vlan.vid}. The control flow simply
extracts the VLAN instance, executes the modular \texttt{Ingress}
control, and then emits the VLAN header. We want the program to
typecheck with type $(x : \refT x {\texttt{stdmeta}} {|x.\pIn| > 32}) \to
\sigmaT{y}{\weak{\texttt{stdmeta}}}{\weak{\texttt{vlan}}}$.

In this example, the type ascription provides compositional reasoning---i.e.,
we don't need to re-check that the whole pipeline is well-typed. Instead, 
we check once-and-for all that $\cExtract{\mathtt{vlan}}$ has type
$(x : \refT x {\texttt{stdmeta}} {|x.\pIn| > 32}) \to (x:(\Sigma x : \texttt{stdmeta}.\texttt{ipv4}))$,
and that $\cRemit{\mathtt{vlan}}$ has type
$(x:(\refT x {\Sigma x : \texttt{stdmeta}.\texttt{ipv4}} {x.\texttt{vlan} = y.\texttt{vlan}})) \to \sigmaT{y}{\weak{\texttt{stdmeta}}}{\weak{\texttt{vlan}}}$
in context $(y : \sigmaT x {\texttt{stdmeta}} {\texttt{ipv4}})$. Both are easy to check.

Now, when we swap in different implementations for \texttt{Ingress},
we only need to check that it has its ascribed type on
Line~\ref{lst:mod-type} of \cref{fig:modular-router-ex}, without
rechecking the surrounding code. With the infrastructure \name's type
system provides, network engineers can make changes to their
experimental module \texttt{Ingress} and check its compatibility with
the switch without re-checking the feasibility of the whole switch in a
modular fashion.

\subsection{Checking Customer Programs}

We now consider a collection of customer programs as shown in
\cref{fig:customer-impls} that an engineer may want to install into
the switch and how \name prevents security vulnerabilities by ensuring
the customer code has the type annotated on Line~\ref{lst:mod-type} of
\cref{fig:modular-router-ex}.

\begin{figure}
\newcommand{\modnum}{6}
\begin{minipage}{0.49\textwidth}
\begin{p4code}[firstnumber=last]
Default $\triangleq$ skip
\end{p4code}
\begin{p4code}[firstnumber=last]
Table $\triangleq$
  add(_vlan_table);
  if (_vlan_table.vid_key == 
      vlan.vid){
    if (_vlan_table.act == 0b0){
      stdmeta.egress_spec := 0x1FF
    } else {
      stdmeta.egress_spec := 0x1
    }
 }
\end{p4code}
\end{minipage}
\begin{minipage}{0.49\textwidth}
\begin{p4code}[firstnumber=last]
Overwrite $\triangleq$ vlan.vid := Faculty
\end{p4code}
\begin{p4code}[firstnumber=last]
UnsafeActions $\triangleq$
  add(_vlan_table);
  if (_vlan_table.vid_key == 
      vlan.vid) {
    if (_vlan_table.act == 0b0) {
      vlan.vid := Faculty
    } else { vlan.vid := Staff }
  } else {
    vlan.vid := Visitor
  }
\end{p4code}

\end{minipage}
\caption{A collection of safe and unsafe customer implementations for the \texttt{Ingress} module from Figure~\ref{fig:modular-router-ex}. Top Left: \emph{Default}; Top Right: \emph{Overwrite}; Bottom Left: \emph{Table}; Bottom Right: \emph{Unsafe Actions} }
\label{fig:customer-impls}
\end{figure}

\paragraph{Default.} Consider the empty program, shown in the top left of
\cref{fig:modular-router-ex}, which would surely be the default behavior when
the programmer hasn't written any code yet. To typecheck this no-op module, we
check that $\cSkip$ has the ascribed type, which it clearly does, since it does
not change the value of the program.

\paragraph{Overwrite.} Conversely, if the customer were to install a blatantly
incorrect program as the one in the top right of
\cref{fig:modular-router-ex}, which always overwrites the VLAN tag
with the identifier reserved for faculty members, the type system
complains that the following subtyping check fails---when $x.\texttt{vlan}.\texttt{vid}$ is, say, \texttt{Student},
the two types denote disjoint sets of heaps.
\[\begin{array}{l}
    (x:\sigmaT{y}{\texttt{stdmeta}}{\texttt{vlan}})\\
    \quad\vdash \{w : \sigmaT{y}{\weak{\texttt{stdmeta}}}{\weak{\texttt{vlan}}} \mid w.\texttt{vlan}.\texttt{vid} =  \texttt{Faculty}\} \\
    \qquad<:~
    \{w : \sigmaT{y}{\weak{\texttt{stdmeta}}}{\weak{\texttt{vlan}}} \mid w.\texttt{vlan}.\texttt{vid} = x.\texttt{vlan}.\texttt{vid}\},
\end{array}
\]

\paragraph{Table.} We model match-action tables using an encoding similar to the
one used in \texttt{p4v}~\cite{Liu2018}, where we create an extra
header that captures the keys and selected action. Consider the
\texttt{vlan} table on the bottom left of
\cref{fig:modular-router-ex}, which matches on 
$\texttt{vlan}.\texttt{vid}$ and selects one of two actions:
the first sets the \texttt{egress\_spec} to \texttt{0x1FF};
the second sets it to \texttt{0x001}. To encode this table, we create a new
header \texttt{\_vlan\_table} with a $12$-bit field
\texttt{vid\_key} and a $1$-bit field \texttt{act}, modelling
the table application via the code shown in the bottom left of \cref{fig:customer-impls}.
This will typecheck since no branch of the code modifies the
$\texttt{vlan}.\texttt{vid}$ field, and \texttt{\_vlan\_table} is
permitted to be valid.

\paragraph{Unsafe Actions.} 
Now consider a \texttt{vlan} table where each action
\emph{does} modify the VLAN id. For example, the table shown in the bottom right
of \cref{fig:modular-router-ex} can either set the VID to one of
$\{\mathtt{Faculty}, \mathtt{Staff}\}$, or, if the packet misses in the table,
to \texttt{Visitor}. Here, the VLAN id is clobbered whenever this table is applied,
triggering a violation of the subset check just as with
\emph{Overwrite}.

%% file: 07-related-work.tex
\section{Related Work}

\paragraph{Formal Reasoning for P4 Programs}
A number of verification approaches have been proposed 
for P4 programs. \texttt{p4v}~\cite{Liu2018} applies classical
techniques based on predicate transformer semantics to achieve
monolithic verification of P4 programs. Vera~\cite{Stoenescu2018} and
P4-Assert~\cite{p4-Assert} are symbolic execution engines for P4. The
\texttt{bf4} tool~\cite{Dumitrescu2020} follows the approach of
\texttt{p4v}, but also attempts to infer control-plane constraints
that are sufficiently strong to establish correctness, and offers
heuristics for repairing programs when verification fails.
SafeP4~\cite{Eichholz2019} uses a simple type system to track header
validity. Petr4~\cite{Doenges2021} develops a formal semantics for P4
but does not itself offer verification tools. In constrast to this
earlier work, \name uses dependent types and offers compositional
verification.

\paragraph{Dependent Types.}
Early work by 
\citet{Xi1999} showed
how dependent types could be used to eliminate run-time safety
checks---e.g., array bounds checks in imperative programs.
Xanadu~\cite{Xi2000} adds dependent typing to imperative programming,
but does not capture the effect of mutations in the type. 
\citet{Xi2001} 
later showed how dependent types could be applied to assembly
code. Deputy~\cite{Condit2007} used dependent types to
reason about complex, heap-allocated data structures. Similar to
Deputy, \name's typing rule for modification of header fields is also
inspired by the Hoare axiom for assignment. \name is different in that
typechecking has no effect on the run time and also supports
path-sensitive reasoning. Similar to \name, Hoare Type Theory
(HTT)~\cite{Nanevski2006} statically tracks how the heap evolves
during execution. Typing of computations in HTT is similar to the
dependent function types \name
-- the type captures
the state before and after execution, possibly relating the output
type with the input type. In our domain, this requires bit-by-bit
transformations on the input type, provided by $\chomp$. Other type
systems like Ynot~\cite{Nanevski2008}, FCSL~\cite{Nanevski2014}, and
F$^*$~\cite{Swamy2016} provide dependent types for low-level
imperative programming. 
They target general functional verification and often require manual
programs-as-proofs to do so. \name is designed with domain-specific
properties of network programming in mind and is fully automatic.

\paragraph{Solver-Aided Tools.}
Recent work on dependently-typed languages has focused on
automation, building on advances in SAT/SMT solvers 
to make dependent types usable by ordinary programs. A
prominent example
is Liquid Haskell~\cite{Rondon2008}, which extends Haskell with
decidable refinement types.
Under the hood, proof obligations generated
during type checking are transparently handled by an SMT solver.
Just as Liquid Haskell requires its refinements to be in the theory of
quantifier-free integer linear arithmetic in order to be decidable,
\name stipulates that types must denote finite sets---a restriction
justified by the domain. This assumption lets us encode types
into the effectively propositional fragment of first-order logic over bit
vectors. 

\paragraph{Formalizing Protocols}
Another line of work focuses on language-based specifications of protocols.
CMU's FoxNet project used SML to specify the behavior of an entire networking
stack~\cite{Biagioni1994}. 
\citet{McCann2000}
used a type-based approach to give
abstract specifications of protocols. Grammar-based tools such
as PADS~\cite{Fisher2005}, Narcissus~\cite{Delaware2019}, and Yakker
\cite{Jim2010}, enable specifying the syntax of complex, dependent formats
including network protocols, and provide tools for serializing and
deserializing data. 
They focus exclusively on deriving correct
parsers from a typed representation of data and may be
suitable to describe the header formats and the parser, but there is no equivalent
to our chomp operator that allows us to statically capture how the input packet
changes during parsing.

%% file: 08-conclusion.tex
\section{Conclusion and Future Work}
This paper presented \name---the first dependently-typed language for
data plane programming---a domain with difficult challenges where
programming language theory can have a big impact. In particular,
low-level data plane languages like P4 seem to be a sweet spot for
dependent types. On the one hand, precise types are necessary because
critical correctness properties often hinge on intricate, bit-level
packet formats, where the first few bits of a packet determine the
format, the length, and the processing of the following ones. On the
other hand, a high degree of automation is possible due to the
restricted nature of the language, which does not support pointers,
loops, or other features that often complicate very precise type
systems. Yet, thus far, dependent typing has not been explored for
data plane programs---the community has relied on verification tools
that lack compositional reasoning.

\name's type system is innovative in its combination of refinement
types, dependent function types, a limited form of regular types,
including unions, explicit substitutions, and a primitive ``chomp''
operation, reminiscent of regular expression
derivatives~\cite{brzozowski64}, which can be used to give a precise
type to P4's parsing constructs.
It is capable of statically checking advanced properties of data plane programs
that combine packet serialization and deserialization 
operations with imperative control-flow. 
Under the hood, an SMT solver automatically discharges the formulas
generated during type checking without requiring any manual proof. We
define \name formally and prove type soundness and decidability. Our
case studies demonstrate how \name supports modular reasoning in
scenarios ranging from basic safety properties to intricate
invariants.

There are a number of interesting directions for future work.
We plan to investigate connections that our verified approach to parsing
using derivatives may have to other domains, e.g., verified serializers and
deserializers like EverParse \cite{EverParse} and Narcissus \cite{Delaware2019}.
Another direction is to consider the effect of modularity on verification times;
if a tool incrementally caches verification results for ascribed code
blocks, it would only have to check the portions of the code that
change between runs of the typechecker. Some preliminary experiments
indicate that modular typechecking offers significant benefits, but an
empirical study to answer this question carefully is
left for future work, after we have optimized our prototype. %
We also plan to extend \name to handle more complicated features of P4
perhaps requiring concolic techniques~\cite{DART_Concolic}. Further,
understanding whether dependent types are the right interface for
modular verification of dataplane programs is important. In fact, user
studies investigating the appropriate typing interfaces, such as
gradual typing and type inference, would be important for guiding the
design of impactful systems for modular data plane verification.

%% file: appendix/definitions.tex
\section{Definitions}

\subsection{Chomp / HeapRef}

\begin{align*}
  \chomp_1                                         & : \tau \times B_n \rightarrow \tau                                                                     \\
  \chomp_1(\sigmaT{x}{\tau_1}{\tau_2}, \bitvar{n}) & \triangleq \sigmaT{x}{\chomp_1(\tau_1, \bitvar{n})}{\chompRef_1(\tau_2, x, \bitvar{n})}~+              \\
                                                   & \phantom{{}\triangleq{}} \sigmaT{x}{\refT{y}{\tau_1}{\length{y.\pIn}=0}}{\chomp_1(\tau_2, \bitvar{n})} \\
  \chomp_1(\tau_1 + \tau_2, \bitvar{n})            & \triangleq \chomp_1(\tau_1, \bitvar{n}) + \chomp_1(\tau_2, b_0)                                        \\
  \chomp_1(\rT{\tau}{e}, \bitvar{n})               & \triangleq \rT{\chomp_1(\tau, \bitvar{n})}{\chompForm_1(e, x, \bitvar{n})}                             \\
  \chomp_1(\tau_1[x\mapsto\tau_2], \bitvar{n})     & \triangleq \chomp_1(\tau_1, \bitvar{n})[x\mapsto\tau_2]                                                \\
  \chomp_1(\tau, \_)                               & \triangleq \tau
\end{align*}

\begin{align*}
  \chompRef_1                                            & : \tau \times \mathcal{X} \times B_n \rightarrow \tau                                         \\
  \chompRef_1(\sigmaT{x}{\tau_1}{\tau_2}, x, \bitvar{n}) & \triangleq \sigmaT{x}{\chompRef_1(\tau_1, x, \bitvar{n})}{\chompRef_1(\tau_2, x, \bitvar{n})} \\
  \chompRef_1(\tau_1 + \tau_2, x, \bitvar{n})            & \triangleq \chompRef_1(\tau_1, x, \bitvar{n}) + \chompRef_1(\tau_2, x, \bitvar{n})            \\
  \chompRef_1(\rT{\tau}{\varphi}, x, \bitvar{n})         & \triangleq \rT{\chompRef_1(\tau, x, \bitvar{n})}{\chompForm_1(\varphi, x, \bitvar{n})}        \\
  \chompRef_1(\tau_1[y\mapsto\tau_2], x, \bitvar{n})     & \triangleq \chompRef_1(\tau_1, x, \bitvar{n})[y\mapsto\chompRef(\tau_2, x, \bitvar{n})]       \\
  \chompRef_1(\tau, \_, \_)                              & \triangleq \tau
\end{align*}

\begin{align*}
  \chompForm_1                                           & : \varphi \times \mathcal{X}  \times B_n \rightarrow \varphi                                 \\
  \chompForm_1(t_1 = t_2, x, \bitvar{n})                 & \triangleq \chompExpr_1(t_1, x, \bitvar{n}) = \chompExpr_1(t_2, x, \bitvar{n})               \\
  \chompForm_1(t_1 > t_2, x, \bitvar{n})                 & \triangleq \chompExpr_1(t_1, x, \bitvar{n}) > \chompExpr_1(t_2, x, \bitvar{n})               \\
  \chompForm_1(\varphi_1 \land \varphi_2, x, \bitvar{n}) & \triangleq \chompForm_1(\varphi_1, x, \bitvar{n}) \land \chompForm(\varphi_2, x, \bitvar{n}) \\
  \chompForm_1(\neg \varphi, x, \bitvar{n})              & \triangleq \neg\chompForm_1(\varphi, x, \bitvar{n})                                          \\
  \chompForm_1(\varphi, \_, \_)                          & \triangleq \varphi
\end{align*}

\begin{align*}
  \chompExpr_1                                      & : e \times \mathcal{X} \times B_n \rightarrow e                                                             \\
  \chompExpr_1(\length{x.\pIn}, y, \_)              & \triangleq \begin{cases}
                                                                   \length{x.\pIn} + 1 & \text{if } x=y   \\
                                                                   \length{x.\pIn}     & \text{otherwise}
                                                                 \end{cases}                                                           \\
  \chompExpr_1(x.\pIn, y, \bitvar{n})               & \triangleq \begin{cases}
                                                                   \bvconcat{\bv{\bitvar{n}}}{x.\pIn} & \text{if } x=y   \\
                                                                   x.\pIn                             & \text{otherwise}
                                                                 \end{cases}                                     \\
  \chompExpr_1(\slice{x.\pIn}{l}{r}, y, \bitvar{n}) & \triangleq \begin{cases}
                                                                   \slice{x.\pIn}{l}{r}                               & \text{if } x \neq y             \\
                                                                   \bv{\bitvar{n}}                                    & \text{if } x = y \land r \leq 1 \\
                                                                   \bvconcat{\bv{\bitvar{n}}}{\slice{x.\pIn}{0}{r-1}} & \text{if } x = y \land l = 0    \\
                                                                   \slice{x.\pIn}{l-1}{r-1}                           & \text{if } x = y \land l \neq 0
                                                                 \end{cases} \\
  \chompExpr_1(n + m,y,\bitvar{n})                  & \triangleq \chompExpr_1(n,y,\bitvar{n}) + \chompExpr_1(m,y,\bitvar{n})                                      \\
  \chompExpr_1(\bvconcat{bv_1}{bv_2},y,\bitvar{n})  & \triangleq \bvconcat{\chompExpr_1(bv_1,y,\bitvar{n})}{\chompExpr_1(bv_2,y,\bitvar{n})}                      \\
  \chompExpr_1(e, \_, \_)                           & \triangleq e
\end{align*}

\begin{align*}
  \heapRef_1                                                  & : \tau \times \bitvar{n} \times \mathcal{X} \times \inst \times \tNat \rightarrow \tau \\
  \heapRef_1(\sigmaT{x}{\tau_1}{\tau_2},\bitvar{n},y,\inst,n) & \triangleq \sigmaT{x}{\heapRef_1(\tau_1,\bitvar{n},y,\inst,n)}{                        \\&\phantom{{}\triangleq\Sigma x:{}} \heapRef_1(\tau_2, \bitvar{n},y,\inst,n)} \\
  \heapRef_1(\tau_1 + \tau_2,\bitvar{n},y,\inst,n)            & \triangleq \heapRef_1(\tau_1,\bitvar{n},y,\inst,n) +                                   \\&\phantom{{}\triangleq{}}\heapRef_1(\tau_2,\bitvar{n},y,\inst,n) \\
  \heapRef_1(\rT{\tau}{\varphi},\bitvar{n},y,\inst,n)         & \triangleq \rT{\heapRef_1(\tau,\bitvar{n},y,\inst,n)}{                                 \\&\phantom{{}\triangleq\lbrace x:{}}\heapRef_1(\varphi,\bitvar{n},y,\inst,n)} \\
  \heapRef_1(\tau_1[x\mapsto\tau_2],\bitvar{n},y,\inst,n)     & \triangleq \heapRef_1(\tau_1,\bitvar{n},y,\inst,n)[x\mapsto                            \\&\hspace{2em}\heapRef_1(\tau_2,\bitvar{n},y,\inst,n)]\\
  \heapRef_1(\tau,\_,\_,\_,\_)                                & \triangleq \tau
\end{align*}

\begin{align*}
  \heapRef_1                                                & : \varphi \times \bitvar{n} \times \mathcal{X} \times \inst \times n \rightarrow \varphi               \\
  \heapRef_1(t_1=t_2,\bitvar{n},x,\inst,n)                  & \triangleq \heapRef_1(t_1,\bitvar{n},x,\inst,n) = \heapRef_1(t_2,\bitvar{n},x,\inst,n)                 \\
  \heapRef_1(t_1>t_2,\bitvar{n},x,\inst,n)                  & \triangleq \heapRef_1(t_1,\bitvar{n},x,\inst,n) > \heapRef_1(t_2,\bitvar{n},x,\inst,n)                 \\
  \heapRef_1(\varphi_1\land \varphi_2,\bitvar{n},x,\inst,n) & \triangleq \heapRef_1(\varphi_1,\bitvar{n},x,\inst,n) \land \heapRef_1(\varphi_2,\bitvar{n},x,\inst,n) \\
  \heapRef_1(\neg \varphi,\bitvar{n},x,\inst,n)             & \triangleq \neg \heapRef_1(\varphi,\bitvar{n},x,\inst,n)                                               \\
  \heapRef_1(\varphi,\_,\_,\_,\_)                           & \triangleq \varphi                                                                                     \\
\end{align*}

\begin{align*}
  \heapRef_1                              & : e \times \bitvar{n} \times \mathcal{X} \times \inst \times n \rightarrow e                                                                                                         \\
  \heapRef_1(b::bv,\bitvar{n},x, \inst,n) & \triangleq \begin{cases}
                                                         x.\inst[\bvconcat{\sizeof(\inst)-n:\sizeof(\inst)-n+1]}{                   \\\hspace{3em}\heapRef_1(bv, \bitvar{n}, x, \inst, n)} &\text{if } b = \bitvar{n} \\
                                                         \bvconcat{\bv{b}}{\heapRef_1(bv,\bitvar{n},x, \inst,n)} & \text{otherwise}
                                                       \end{cases} \\
  \heapRef_1(e,\_,\_,\_)                  & \triangleq e
\end{align*}

\subsection{Additional Rules for the Operational Semantics}

\begin{mathpar}
  \inferrule[E-If]{
    \langle I,O,H,\varphi \rangle \rightarrow \varphi'
  }{
    \langle I,O,H,\cIf{\varphi}{c_1}{c_2}\rangle \rightarrow \\ \langle I,O,H,\cIf{\varphi'}{c_1}{c_2}\rangle
  }
  \and
  \inferrule[E-IfTrue]{\ }{
    \langle I,O,H,\cIf{\true}{c_1}{c_2}\rangle \rightarrow \langle I,O,H,c_1\rangle
  }
  \and
  \inferrule[E-IfFalse]{\ }{
    \langle I,O,H,\cIf{\false}{c_1}{c_2}\rangle \rightarrow \langle I,O,H,c_2\rangle
  }
  \and
  \inferrule[E-Seq]{
    \
  }{
    \langle I,O,H,\cSkip ;c_2\rangle \rightarrow \langle I,O,H,c_2\rangle
  }
  \and
  \inferrule[E-Seq1]{
    \langle I,O,H,c_1\rangle \rightarrow \langle I',O',H',c_1'\rangle
  }{
    \langle I,O,H,c_1;c_2\rangle \rightarrow \langle I',O',H',c_1';c_2\rangle
  }
  \and
  \inferrule[E-Mod1]{
    \langle I,O,H,e\rangle \rightarrow e'
  }{
    \langle I,O,H,\inst.f=e\rangle \rightarrow \langle I,O,H,\inst.f:=e'\rangle
  }
\end{mathpar}

%% file: appendix/safety.tex
\section{Safety}
\label{sec:safety}

We assume that the environment $\env$ is empty whenever it is omitted from the entailment relation or the semantics of expressions and formulae.
In the context of the semantics of heap types, we use $(I,O,H)$ as a shorthand notation for the heap $H[\pIn\mapsto I, \pOut\mapsto O]$.
For example, we write $(I,O,H)\in\semantics{\tau}_\env$ instead of $H[\pIn\mapsto I,\pOut\mapsto O]\in\semantics{\tau}_\env$.

\input{appendix/lem-entailment.tex}

\begin{lemma}[Included Instances in Domain]
  \label{lem:included-instances-in-domain}
  If $(I,O,H)\models_\env\tau$ and $\includes{\tau}{\inst}$, then $\inst\in \dom(H)$.
\end{lemma}
\begin{proof}
  By Lemma \ref{lem:semantic-entailment}, $(I,O,H)\in\semantics{\tau}_\env$.
  By assumption $\includes{\tau}{\inst}$ and by definition of inclusion, $\forall h\in\semantics{\tau}_\env.\inst\in\dom(h)$, we can conclude that $\inst\in\dom(H)$.
\end{proof}

\begin{lemma}[Excluded Instances not in Domain]
  \label{lem:excluded-instances-not-in-domain}
  If $(I,O,H)\models_\env\tau$ and $\excludes{\tau}{\inst}$, then $\inst\not\in \dom(H)$.
\end{lemma}
\begin{proof}
  By Lemma \ref{lem:semantic-entailment}, $(I,O,H)\in\semantics{\tau}_\env$.
  By assumption $\excludes{\tau}{\inst}$ and by definition of exclusion, $\forall h\in\semantics{\tau}_\env.\inst\not\in\dom(h)$, we can conclude that $\inst\not\in\dom(H)$.
\end{proof}

\input{appendix/safety/progress.tex}
\input{appendix/safety/lem-input-type-strengthening.tex}
\input{appendix/safety/lem-chomp.tex}
\input{appendix/safety/preservation.tex}

%% file: appendix/lem-entailment.tex
\begin{lemma}[Semantic Entailment]
  \label{lem:semantic-entailment}
  If $(I,O,H)\models_\env\tau$, then $(I,O,H)\in\semantics{\tau}_\env$
\end{lemma}

\begin{proof}
  By induction on $\tau$.
  \begin{description}
    \case{$\tau = \varnothing$}
    Immediate, since $(I,O,H)\models_\env\varnothing$ is a contradiction.

    \case{$\tau = \top$}
    Immediate, since $\semantics{\top}_\env = \Heaps$.

    \case{$\tau = \sigmaT{x}{\tau_1}{\tau_2}$}
    By inversion of entailment, we get
    \begin{pf}{Sigma}
      \item \label{semantic-entailment-sigma-heap-split} $(I,O,H)=(\bvconcat{I_1}{I_2},\bvconcat{O_1}{O_2},H_1\cup H_2)$ and
      \item \label{semantic-entailment-sigma-heap-left} $(I_1,O_1,H_1)\models_\env\tau_1$ and
      \item \label{semantic-entailment-sigma-heap-right} $(I_2,O_2,H_2)\models_{\env[x\mapsto (I_1,O_1,H_1)]} \tau_2$
    \end{pf}

    By \ref{semantic-entailment-sigma-heap-left} respectively \ref{semantic-entailment-sigma-heap-right} and the induction hypothesis, we get
    \begin{pf*}{Sigma}
      \item \label{semantic-entailment-sigma-heap-left-contained} $(I_1,O_1,H_1)\in\semantics{\tau_1}_\env$ and
      \item \label{semantic-entailment-sigma-heap-right-contained} $(I_2,O_2,H_2)\in\semantics{\tau_2}_{\env[x\mapsto (I_1,O_1,H_1)]}$.
    \end{pf*}

    To show that $(I,O,H)\in\semantics{\sigmaT{x}{\tau_1}{\tau_2}}_\env = \{\concat{h_1}{h_2}\mid h_1\in\semantics{\tau_1}_\env \land h_2\in\semantics{\tau_2}_{\env[x\mapsto h_1]}\}$,
    we have to show that $(I,O,H)$ is the concatenation of two heaps $h_1$ and $h_2$, where $h_1\in\semantics{\tau_1}_\env$ and
    $h_2\in\semantics{\tau_2}_{\env[x\mapsto h_1]}$, which follows from \ref{semantic-entailment-sigma-heap-split}, \ref{semantic-entailment-sigma-heap-left-contained} and \ref{semantic-entailment-sigma-heap-right-contained}.

    \case{$\tau = \tau_1+\tau_2$}
    By inversion of entailment, either $(I,O,H)\models_\env\tau_1$ or $(I,O,H)\models_\env\tau_2$.
    To show that $(I,O,H)\in\semantics{\tau_1+\tau_2}_\env = \semantics{\tau_1}_\env \cup \semantics{\tau_2}_\env$,
    we have to show that $(I,O,H)\in\semantics{\tau_1}_\env$ or $(I,O,H)\in\semantics{\tau_2}_\env$.
    \begin{description}
      \subcase{$(I,O,H)\models_\env\tau_1$}
      By induction hypothesis, $(I,O,H)\in\semantics{\tau_1}_\env$.
      We can conclude $(I,O,H)\in\semantics{\tau_1+\tau_2}_\env$.

      \subcase{$(I,O,H)\models_\env\tau_2$}
      Symmetric to previous subcase.
    \end{description}

    \case{$\tau = \rT{\tau_1}{\varphi}$}
    By inversion of entailment, we get
    \begin{pf}{Refinement}
      \item \label{semantic-entailment-ref-heap-entails} $(I,O,H)\models_\env\tau$ and
      \item \label{semantic-entailment-ref-istrue} $\semantics{\varphi}_{\env[x\mapsto (I,O,H)]}=true$
    \end{pf}
    To show that $(I,O,H)\in\semantics{\rT{\tau}{\varphi}}_\env=\{h\mid h\in\semantics{\tau}_\env \land \semantics{\varphi}_{\env[x\mapsto h]} \}$, we have to show that $(I,O,H)\in\semantics{\tau}_\env$ and that $\semantics{\varphi}_{\env[x\mapsto (I,O,H)]}=true$.
    The first follows by induction hypothesis and \ref{semantic-entailment-ref-heap-entails} and the latter by \ref{semantic-entailment-ref-istrue}.

    \case{$\tau = \tau_1{[x\mapsto\tau_2]}$}
    By inversion of entailment, we get
    \begin{pf}{Substitution}
      \item \label{semantic-entailment-subst-entails-t2} $(I_2,O_2,H_2)\models_\env\tau_2$ for some $I_2,O_2,H_2$ and
      \item \label{semantic-entailment-subst-entails-t1} $(I,O,H)\models_{\env[x\mapsto (I_2,O_2,H_2)]} \tau_1$
    \end{pf}
    To show that $(I,O,H)\in\semantics{\tau_1[x\mapsto\tau_2]}_\env=\{h\mid h_2\in\semantics{\tau_2}_\env \land h\in\semantics{\tau_1}_{\env[x\mapsto h_2]} \}$, we have to show that $(I,O,H)\in\semantics{\tau_1}_{\env[x\mapsto h_2]}$ where $h_2\in\semantics{\tau_2}_\env$.
    By induction hypothesis and \ref{semantic-entailment-subst-entails-t1} follows that $(I,O,H)\in\semantics{\tau_1}_{\env[x\mapsto (I_2,O_2,H_2)]}$.
    $(I_2,O_2,H_2)\in\semantics{\tau_2}_\env$ follows by induction hypothesis and \ref{semantic-entailment-subst-entails-t2}, which concludes this case.
  \end{description}
\end{proof}

\begin{lemma}[Semantic Containment Entails]
  \label{lem:semantic-containment-entails}
  If $(I,O,H)\in\semantics{\tau}_\env$, then $(I,O,H)\models_\env\tau$.
\end{lemma}
\begin{proof}
  By induction on $\tau$.
  \begin{description}
    \case{$\tau=\varnothing$}
    Immediate, since there is no heap in $\semantics{\varnothing}_\env$.

    \case{$\tau=\top$}
    Result directly follows by \entTop.

    \case{$\tau=\sigmaT{x}{\tau_1}{\tau_2}$}
    By the semantics of heap types, all heaps $h\in\semantics{\sigmaT{x}{\tau_1}{\tau_2}}$ have the form $h=\concat{h_1}{h_2}$, where $h_1=(I_1,O_1,H_1)\in\semantics{\tau_1}_\env$ and $h_2=(I_2,O_2,H_2)\in\semantics{\tau_2}_{\env[x\mapsto h1]}$.
    By applying the induction hypothesis, we get $(I_1,O_1,H_1)\models_\env\tau_1$ and $(I_2,O_2,H_2)\models_{\env[x\mapsto h_1]}\tau_2$.
    The result directly follows by \entSigma.

    \case{$\tau=\tau_1+\tau_2$}
    By the semantics of heap types, for any $h\in\semantics{\tau_1+\tau_2}_\env$ holds that either $h\in\semantics{\tau_1}_\env$ or $h\in\semantics{\tau_2}_\env$.
    \begin{description}
      \subcase{$h\in\semantics{\tau_1}_\env$}
      By induction hypothesis, $h\models_\env\tau_1$. The result directly follows by \entChoiceL.

      \subcase{$h\in\semantics{\tau_1}_\env$}
      Symmetric to previous subcase.

    \end{description}

    \case{$\tau=\refT{y}{\tau_1}{\varphi}$}
    By the semantics of heap types, $h\in\semantics{\tau_1}_\env$ and $\semantics{\varphi}_{\env[x\mapsto h]}=\mathit{true}$.
    By induction hypothesis, $h\models_\env\tau$.
    The result directly follows by \entRefine.

    \case{$\tau=\tau_1{[x\mapsto\tau_2]}$}
    By the semantics of heap types, $h\in\semantics{\tau_1}_{\env[x\mapsto h_2]}$ where $h_2\in\semantics{\tau_2}_\env$.
    By induction hypothesis, $h_2\models_\env\tau_2$ and $h\models_{\env[x\mapsto h_2]}\tau_1$.
    The result directly follows by \entSubst.
  \end{description}
\end{proof}

\begin{lemma}[Subtype Entailment]
  \label{lem:subtype-entailment}
  If $(I,O,H) \models_\env \tau_1$ and $\env \models \Gamma$ and $\subtype{\tau_1}{\tau_2}$, then $(I,O,H) \models_\env \tau_2$.
\end{lemma}
\begin{proof}
  By \cref{lem:semantic-entailment}, $(I,O,H) \in \semantics{\tau_1}_\env$.
  With $\env \models \Gamma$ and by definition of subtyping,
  $(I,O,H)\in \semantics{\tau_2}_\env$.
  The result follows by \cref{lem:semantic-containment-entails}.
\end{proof}

\begin{lemma}[Extended Environment Entails]
  \label{lem:extended-env-entails}
  If $\env \models \Gamma$ and $(I,O,H)\models_\env \tau$ and $x\not\in\dom(\env)$, then $\env[x \mapsto (I,O,H)] \models \Gamma,x:\tau$.
\end{lemma}
\begin{proof}
  By definition of entailment between environments and typing contexts and by assumptions.
\end{proof}

%% file: appendix/safety/progress.tex
\begin{lemma}[Expression Progress]
  \label{lem:expr-progress}
  If $\Gamma;\tau \vdash e:\sigma$ and
  $\env \models \Gamma$ and
  $(I,O,H)\models_\env\tau$,
  then either $e$ is a value or
  $\exists e'.\langle I,O,H,e\rangle\rightarrow e'$.
\end{lemma}

\begin{lemma}[Formulae Progress]
  \label{lem:formulae-progress}
  If $\Gamma;\tau \vdash \varphi:\tBool$ and $\env \models \Gamma$ and $(I,O,H) \models_\env \tau$,
  then either $\varphi$ is a value or $\exists \varphi'.\langle I,O,H,\varphi\rangle\rightarrow \varphi'$
\end{lemma}

\begin{theorem}[Progress]
  \label{thm:progress}
  If $\cmdType{\Gamma}{c}{(x:\tau_1)\rightarrow\tau_2}$ and $\env\models\Gamma$ and $(I,O,H)\models_\env\tau_1$, then either $c=\cSkip$ or
  there exists $\langle I',O',H',c'\rangle$ such that $\langle I,O,H,c\rangle\rightarrow\langle I',O',H',c'\rangle$.
\end{theorem}
\begin{proof}
  By induction on typing derivations of $\cmdType{\Gamma}{c}{(x:\tau_1)\rightarrow\tau_2}$.
  \begin{description}
    \case{\tSkip}
    $c=\cSkip$\\
    The result is immediate.

    \case{\tExtract}
    $c=\cExtract{\inst}$ and $\Gamma \vdash \sizeof_{\pIn}(\tau_1)\geq\sizeof(\inst)$\\
    By inversion of $(I,O,H)\models_\env\tau_1$, we know that $I$ contains enough bits such that $\mathit{deserialize}_\eta(I)$ does not fail.
    Let $(v,I') = \mathit{deserialize_\eta(I)}$ and $O'=O$ and $H'=H[\inst \mapsto v]$ and $c'=\cSkip$.
    The result follows by \eExtract.

    \case{\tReset}
    $c=\cReset$\\
    Let $I'=\bvconcat{O}{I},O'=\bvNil,H'=[]$ and $c'=\cSkip$.
    The result follows by \eReset.

    \case{\tRemit}
    $c=\cRemit{\inst}$ and
    $\includes{\tau_1}{\inst}$ \\
    By \cref{lem:included-instances-in-domain} we know $\inst\in dom(H)$.
    Let $I'=I,O'=\bvconcat{O}{\mathit{serialize_\eta}(H(\inst))},H'=H$ and $c'=\cSkip$.
    The result follows by \eRemit

    \case{\tMod}
    $c=\inst.f:=e$ and
    $\includes{\tau_1}{\inst}$ and
    $\mathcal{F}(\inst, f) = \tBv$ and
    $\Gamma;\tau_1 \vdash e: \tBv$ and
    $\tau_2 = \refT{y}{\top}{\varphi_{pkt} \wedge \varphi_{\inst} \wedge \varphi_{f} \wedge y.\inst.f = e[x/\cmdVar]}$\\
    By \cref{lem:expr-progress}, either $e$ is a value
    or there is some $e'$ such that $\langle I,O,H,e\rangle\rightarrow e'$.
    \begin{description}
      \subcase{$e=v$}
      By \cref{lem:included-instances-in-domain}, $\inst\in dom(H)$.
      Let $r=H(\inst)$ and $r'=\{r \textit{ with } f=v \}$.
      Let $I'=I, O'=O, H'=H[\inst\mapsto r]$ and $c'=\cSkip$. The result follows by \eMod.
      \subcase{$\langle I,O,H,e\rangle\rightarrow e'$}
      Let $I'=I,O'=O,H'=H'$ and $c'=\inst.f:=t'$.
      The result follows by \eModOne.
    \end{description}

    \case{\tSeq}
    $c=c_1;c_2$ and
    $\cmdType{\Gamma}{c_1}{(x:\tau_1)\rightarrow\tau_1'}$ and
    $\cmdType{\Gamma}{c_2}{(x:\tau_1')\rightarrow\tau_2}$
    
    By IH, $c_1$ is either $\cSkip$ or there is some $\langle I',O',H',c_1'\rangle$, such that $\langle I,O,H,c_1\rangle\rightarrow\langle I',O',H',c_1'\rangle$.
    If $c_1=\cSkip$, let $I'=I,O'=O,H'=H$ and $c'=c_2$. 
    
    The result follows by \eSeq.
    Otherwise, the result follows by \eSeqOne.

    \case{\tIf}
    $c=\cIf{\varphi}{c_1}{c_2}$ and
    $\Gamma;\tau_1\vdash e:\tBool$\\
    By \cref{lem:formulae-progress}, we have that $\varphi$ is either $\true, \false$ or there is some $\varphi'$ such that $\langle I,O,H,\varphi\rangle\rightarrow \varphi'$.
    \begin{description}
      \subcase{$\varphi=\true$}
      Let $I'=I,O'=O,H'=H$ and $c'=c_1$. The result follows by \eIfTrue.
      \subcase{$\varphi=\false$}
      Symmetric to previous subcase.
      \subcase{$\langle I,O,H,\varphi\rangle\rightarrow \varphi'$}
      Let $I'=I$ and $O'=O$ and $H'=H$.

      Further, let $c'=\cIf{\varphi'}{c_1}{c_2}$.
      The result follows by \eIf.
    \end{description}

    \case{\tAdd}
    $c=\cAdd{\inst}$ and
    $\excludes{\tau}{\inst}$.\\
    By \cref{lem:excluded-instances-not-in-domain}, $\inst\not\in\dom(H)$.
    The result follows by \eAdd.

    \case{\tAscribe}
    $c=\cAscribe{c_a}{(x:\tau_{a_1}) \rightarrow \tau_{a_2}}$.
    Let $I'=I$ , $O'=O$, $H'=H$ and $c'=c_a$.
    The result follows by \eAscribe.

    \case{\tSub}
    $\subtypeCtx{\Gamma}{\tau_1}{\tau_3}$ and
    $\subtypeCtx{\Gamma,x:\tau_1}{\tau_4}{\tau_2}$ and
    $\cmdT{c}{(x:\tau_3) \rightarrow \tau_4}$.
    By \cref{lem:subtype-entailment}, $(I,O,H) \models_\env \tau_3$.
    By IH, $c=\cSkip$ or there exists $I',O',H',c'$ s.t.
    $\langle I,O,H,c \rangle \rightarrow \langle I',O',H',c' \rangle$.
    The result follows directly.
  \end{description}
\end{proof}

%% file: appendix/safety/lem-input-type-strengthening.tex
\begin{lemma}[Weakening]
  \label{lem:weakening}
  If $\cmdType{\Gamma}{c}{(x:\tau_1) \to \tau_2}$ and variable $z$ does not appear free in $\tau_1$ or $\tau_2$, then
  $\cmdType{\Gamma,z:\tau}{c}{(x:\tau_1) \to \tau_2}$ for any heap type $\tau$.
\end{lemma}
\begin{proof}
  By induction on the typing derivation.
\end{proof}

\begin{lemma}[Input Type Strengthening]
  \label{lem:input-type-strengthening}
  If $\Gamma \vdash c : (x:\tau_1) \rightarrow \tau_{2}$ and
  $\semantics{\tau_1'}_{\env'} \subseteq \semantics{\tau_1}_\env$ and
  $\env \models \Gamma$ and
  $\env' \models \Gamma'$ and
  $\Gamma \subseteq \Gamma'$ and
  $\env \subseteq \env'$, then
  $\exists \tau_2'.\Gamma' \vdash c : (x:\tau_1') \rightarrow \tau_2'$ and
  $\forall h'\in\semantics{\tau_1'}_{\env'}.\semantics{\tau_2'}_{\env'[x \mapsto h']} \subseteq \semantics{\tau_2}_{\env[x \mapsto h']}$
\end{lemma}

\begin{proof}
  By induction on a derivation of $\Gamma \vdash c :(x:\tau_1) \rightarrow \tau_2$ with case analysis on the last rule used.
  We refer to the proof goals as follows:
  \begin{enumerate}[label=(\arabic*)]
    \item \label{input-type-strengthening-g1} $\exists \tau_2'.\Gamma' \vdash c : (x:\tau_1') \rightarrow \tau_2'$
    \item \label{input-type-strengthening-g2} $\forall h'\in\semantics{\tau_1'}_{\env'}.\semantics{\tau_2'}_{\env'[x \mapsto h']} \subseteq \semantics{\tau_2}_{\env[x \mapsto h']}$
  \end{enumerate}
  We refer to the assumptions as follows:
  \begin{enumerate}[label=(\Alph*)]
    \item \label{input-type-strengthening-input-sub} $\semantics{\tau_1'}_{\env'} \subseteq \semantics{\tau_1}_\env$
    \item \label{input-type-strengthening-env-entails} $\env \models \Gamma$
    \item \label{input-type-strengthening-envprime-entails} $\env' \models \Gamma'$
    \item \label{input-type-strengthening-gamma-sub} $\Gamma \subseteq \Gamma'$
    \item \label{input-type-strengthening-env-sub} $\env \subseteq \env'$
  \end{enumerate}
  \begin{description}
    \case{\tAdd}~

    By inversion of rule \tAdd, we get
    \begin{pf}{T-Add}
      \item \label{input-type-strengthening-tadd-excludes} $\excludes{\tau}{\inst}$ and
      \item \label{input-type-strengthening-tadd-v} $init_{\HT(\inst)} = v$
      \item \label{input-type-strengthening-tadd-t2} $\tau_2 = \sigmaT{y}{\refT{z}{\tau_1}{z \equiv x}}{\refT{z}{\inst}{z.\pOut = z.\pIn = \bvNil \wedge z.\inst = v}}$.
    \end{pf}
    Let $\tau_2' = \sigmaT{y}{\refT{z}{\tau_1'}{z \equiv x}}{\refT{z}{\inst}{z.\pOut = z.\pIn = \bvNil \wedge z.\inst = v}}$.
    By assumptions \ref{input-type-strengthening-tadd-excludes}, \ref{input-type-strengthening-input-sub} and \ref{input-type-strengthening-envprime-entails} we can conclude that $\excludesCtx{\tau_1'}{\inst}{\Gamma'}$ must also hold.
    \ref{input-type-strengthening-g1} follows by \tAdd.

    Let $h'\in\semantics{\tau_1'}_{\env'}$ be arbitrary.
    \begin{align*}
      \phantom{{}\Leftrightarrow{}} & \semantics{\sigmaT{y}{\refT{z}{\tau_1'}{z \equiv x}}{\refT{z}{\inst}{z.\pOut = z.\pIn = \bvNil \wedge z.\inst = v}}}_{\env'[x \mapsto h']} \subseteq                \\
                                    & \semantics{\sigmaT{y}{\refT{z}{\tau_1}{z \equiv x}}{\refT{z}{\inst}{z.\pOut = z.\pIn = \bvNil \wedge z.\inst = v}}}_{\env[x \mapsto h']}                            \\
      \Leftrightarrow               & \lbrace \concat{h_1}{h_2} \mid h_1 \in \semantics{\refT{z}{\tau_1'}{z \equiv x}}_{\env'[x \mapsto h']}\ \wedge                                                      \\
                                    & \hspace{46pt}h_2\in\semantics{\refT{z}{\inst}{z.\pOut = z.\pIn = \bvNil \wedge z.\inst = v}}_{\env'[x \mapsto h', y \mapsto h_1]} \rbrace \subseteq                 \\
      \phantom{{}\Leftrightarrow{}} & \lbrace \concat{h_1}{h_2} \mid h_1 \in \semantics{\refT{z}{\tau_1}{z \equiv x}}_{\env[x \mapsto h']}\ \wedge                                                        \\
                                    & \hspace{46pt}h_2\in\semantics{\refT{z}{\inst}{z.\pOut = z.\pIn = \bvNil \wedge z.\inst = v}}_{\env[x \mapsto h', y \mapsto h_1]} \rbrace                            \\
      \Leftrightarrow               & \lbrace \concat{h'}{h_2} \mid h_2\in\semantics{\refT{z}{\inst}{z.\pOut = z.\pIn = \bvNil \wedge z.\inst = v}}_{\env'[x \mapsto h', y \mapsto h']} \rbrace \subseteq \\
                                    & \lbrace \concat{h'}{h_2} \mid h_2\in\semantics{\refT{z}{\inst}{z.\pOut = z.\pIn = \bvNil \wedge z.\inst = v}}_{\env[x \mapsto h', y \mapsto h']} \rbrace
    \end{align*}
    The type $\refT{z}{\inst}{z.\pOut = z.\pIn = \bvNil \wedge z.\inst = v}$ does not contain any free variables, so the semantics does not depend on the environment.
    In fact, the sets of heaps described by $\tau_2'$ and $\tau_2$ is actually equivalent, which shows \ref{input-type-strengthening-g2}.

    \case{\tAscribe}~

    By inversion of rule \tAscribe, we get
    \begin{pf}{T-Ascribe}
      \item $c = \cAscribe{c_a}{(x:\tau_1) \to \tau_2}$ and
      \item \label{input-type-strengthening-tascribe-ca} $\cmdType{\Gamma}{c_a}{(x:\tau_1) \to \tau_2}$
    \end{pf}
    From assumptions \ref{input-type-strengthening-tascribe-ca} and \ref{input-type-strengthening-gamma-sub} together with \cref{lem:weakening} follows that
    \begin{pf*}{T-Ascribe}
      \item \label{input-type-strengthening-tascribe-ca-gammaprime} $\cmdType{\Gamma'}{c_a}{(x:\tau_1) \to \tau_2}$
    \end{pf*}
    Since $\env'$ differs from $\env$ only in that it potentially contains additional bindings, we can conclude that
    \begin{pf*}{T-Ascribe}
      \item $\semantics{\tau_1}_\env = \semantics{\tau_1}_{\env'}$ and together with assumption \ref{input-type-strengthening-env-entails}
      \item $\Gamma' \vdash \tau_1' <: \tau_1$.
    \end{pf*}
    By assumption \ref{input-type-strengthening-tascribe-ca-gammaprime} and \tAscribe we get
    \begin{pf*}{T-Ascribe}
      \item $\cmdType{\Gamma'}{c_a\ as\ (x:\tau_1) \to \tau_2}{(x:\tau_1) \to \tau_2}$
    \end{pf*}
    Let $\tau_2' = \tau_2$.
    \ref{input-type-strengthening-g1} follows by \tSub.

    For \ref{input-type-strengthening-g2}, we have to show that $\forall h'\in\semantics{\tau_1'}_{\env'}.\semantics{\tau_2}_{\env'[x \mapsto h']} \subseteq \semantics{\tau_2}_{\env[x \mapsto h']}$.
    In fact $\tau_2$ describes the same set of heaps, both in $\env$ and $\env'$.
    Variable $x$ binds to the same heap and both environments provide the same bindings for any other free variable in $\tau_2$.

    \case{\tExtract}~

    By inversion of rule \tExtract, we get
    \begin{pf}{T-Extract}
      \item \label{input-type-strengthening-textract-size} $\Gamma \vdash \sizeof_\pIn(\tau_1) \ge \sizeof(\inst)$
      \item \label{input-type-strengthening-textract-t2} $\tau_2 = \sigmaT{y}{\refT{z}{\inst}{z.\pIn = z.\pOut = \bvNil}}{\refT{z}{\chomp(\tau_1,\inst,y)}{\\
            \bvconcat{y.\inst}{z.\pIn} = x.\pIn \wedge z.\pOut = x.\pOut \wedge z \equiv_\inst x}}$
    \end{pf}

    Let $\tau_2' = \sigmaT{y}{\refT{z}{\inst}{z.\pIn = z.\pOut = \bvNil}}{\refT{z}{\chomp(\tau_1',\inst,y)}{\\
          \bvconcat{y.\inst}{z.\pIn} = x.\pIn \wedge z.\pOut = x.\pOut \wedge z \equiv_\inst x}}$.
    By assumptions \ref{input-type-strengthening-input-sub} and \ref{input-type-strengthening-textract-size} follows that $\sizeof_\pIn(\tau_1') \ge \sizeof(\inst)$
    \ref{input-type-strengthening-g1} follows by \tExtract.

    Let $h'\in\semantics{\tau_1'}_{\env'}$ be arbitrary.
    \begin{align*}
      \phantom{{}\Leftrightarrow{}} & \semantics{\sigmaT{y}{\refT{z}{\inst}{z.\pIn = z.\pOut = \bvNil}}{                                                              \\
                                    & \hspace{23pt}\refT{z}{\chomp(\tau_1',\inst,y)}{\bvconcat{y.\inst}{z.\pIn} = x.\pIn\ \wedge                                      \\
                                    & \hspace{109pt} z.\pOut = x.\pOut \wedge z \equiv_\inst x}}}_{\env'[x \mapsto h']} \subseteq                                     \\
                                    & \semantics{\sigmaT{y}{\refT{z}{\inst}{z.\pIn = z.\pOut = \bvNil}}{                                                              \\
                                    & \hspace{23pt}\refT{z}{\chomp(\tau_1,\inst,y)}{\bvconcat{y.\inst}{z.\pIn} = x.\pIn\ \wedge                                       \\
                                    & \hspace{109pt}z.\pOut = x.\pOut \wedge z \equiv_\inst x}}}_{\env[x \mapsto h']}                                                 \\
      \Leftrightarrow               & \lbrace \concat{h_1}{h_2} \mid h_1 \in \semantics{\refT{z}{\inst}{z.\pIn = z.\pOut = \bvNil}}_{\env[x \mapsto h']}\ \wedge      \\
                                    & \hspace{46pt} h_2 \in \semantics{\refT{z}{\chomp(\tau_1',\inst,y)}{\bvconcat{y.\inst}{z.\pIn} = x.\pIn\ \wedge                  \\
                                    & \hspace{156pt} z.\pOut = x.\pOut\ \wedge                                                                                        \\
                                    & \hspace{156pt} z \equiv_\inst x}}_{\env[x \mapsto h', y \mapsto h_1]}\rbrace \subseteq                                          \\
                                    & \lbrace \concat{h_1}{h_2} \mid h_1 \in \semantics{\refT{z}{\inst}{z.\pIn = z.\pOut = \bvNil}}_{\env[x \mapsto h']}\ \wedge      \\
                                    & \hspace{46pt} h_2 \in \semantics{\refT{z}{\chomp(\tau_1,\inst,y)}{\bvconcat{y.\inst}{z.\pIn} = x.\pIn\ \wedge                   \\
                                    & \hspace{156pt} z.\pOut = x.\pOut\ \wedge                                                                                        \\
                                    & \hspace{156pt} z \equiv_\inst x}}_{\env[x \mapsto h', y \mapsto h_1]}\rbrace                                                    \\
      \Leftrightarrow               & \lbrace \concat{h_1}{h_2} \mid h_1 \in \semantics{\refT{z}{\inst}{z.\pIn = z.\pOut = \bvNil}}_{\env[x \mapsto h']}\ \wedge      \\
                                    & \hspace{46pt} h_2 \in \lbrace h_{22} | h_{22}\in\semantics{\chomp(\tau_1',\inst,y)}_{\env[x \mapsto h', y \mapsto h_1]}\ \wedge \\
                                    & \hspace{107pt} \semantics{\bvconcat{y.\inst}{z.\pIn} = x.\pIn\ \wedge z.\pOut = x.\pOut\ \wedge                                 \\
                                    & \hspace{112pt} z \equiv_\inst x}_{\env[x \mapsto h', y \mapsto h_1, z \mapsto h_{22}]}\rbrace \rbrace \subseteq                 \\
                                    & \lbrace \concat{h_1}{h_2} \mid h_1 \in \semantics{\refT{z}{\inst}{z.\pIn = z.\pOut = \bvNil}}_{\env[x \mapsto h']}\ \wedge      \\
                                    & \hspace{46pt} h_2 \in \lbrace h_{22} | h_{22}\in\semantics{\chomp(\tau_1,\inst,y)}_{\env[x \mapsto h', y \mapsto h_1]}\ \wedge  \\
                                    & \hspace{107pt} \semantics{\bvconcat{y.\inst}{z.\pIn} = x.\pIn\ \wedge z.\pOut = x.\pOut\ \wedge                                 \\
                                    & \hspace{112pt} z \equiv_\inst x}_{\env[x \mapsto h', y \mapsto h_1, z \mapsto h_{22}]}\rbrace \rbrace
    \end{align*}

    By \cref{lem:semantic-chomp}, we obtain all heaps contained in $\semantics{\chomp(\tau_1',\inst,y)}_{\env[x \mapsto h',y\mapsto h_1]}$ by taking all heaps from $\semantics{\tau_1'}_{\env'}$ and removing the first $\sizeof(\inst)$ bits from the input packet.
    From assumption \ref{input-type-strengthening-input-sub} we know that all heaps described by $\tau_1'$ are also contained in the set of heaps described by $\tau_1$ and when we remove the first $\sizeof(\inst)$ bits from the input packet, the relation still holds.
    Since the rest of the types are identical this also holds for the concatenated heaps.
    This shows \ref{input-type-strengthening-g2} and concludes the case.

    \case{\tIf}~

    By inversion of rule \tIf, we get
    \begin{pf}{T-If}
      \item $\Gamma;\tau_1 \vdash e : \tBool$
      \item $\Gamma \vdash c_1 : (x:\refT{y}{\tau_1}{\varphi[y/\cmdVar]}) \to \tau_{12}$
      \item $\Gamma \vdash c_2 : (x:\refT{y}{\tau_1}{\neg \varphi[y/\cmdVar]}) \to \tau_{22}$
      \item $\tau_2 = \refT{y}{\tau_{12}}{\varphi[x/\cmdVar]} + \refT{y}{\tau_{22}}{\varphi[x/\cmdVar]}$
      \item $c = \cIf{\varphi}{c_1}{c_2}$
    \end{pf}

    To be able to conclude (1) by \tIf, we must show that
    \begin{enumerate}[label=(1.\arabic*)]
      \item \label{input-type-strengthening-tif-g1} $\Gamma';\tau_1' \vdash \varphi : \tBool$
      \item \label{input-type-strengthening-tif-g2} $\Gamma' \vdash c_1 : (x:\refT{y}{\tau_1'}{\varphi[y/\cmdVar]}) \to \tau_{12}'$
      \item \label{input-type-strengthening-tif-g3} $\Gamma' \vdash c_2 : (x:\refT{y}{\tau_1'}{\neg \varphi[y/\cmdVar]}) \to \tau_{22}'$
    \end{enumerate}

    To apply the IH to $c_1$, we need some $\tau'_{IH_1}$ such that
    \[
      \semantics{\tau'_{IH_1}}_{\env'} \subseteq \semantics{\refT{y}{\tau_1}{\varphi[y/\cmdVar]}}_\env
    \]
    Let $\tau'_{IH_1} = \refT{y}{\tau_1'}{\varphi[y/\cmdVar]}$.

    By IH, there exists $\tau_{12}'$ such that
    \begin{pf*}{T-If}
      \item $\Gamma' \vdash c_1 : (x:\refT{y}{\tau_1'}{\varphi[y/\cmdVar]}) \to \tau_{12}'$
      \item \label{input-type-strengthening-tif-t12-sub} $\forall h_1'\in\semantics{\refT{y}{\tau_1'}{\varphi[y/\cmdVar]}}_{\env'}.\semantics{\tau_{12}'}_{\env'[y \mapsto h_1']} \subseteq \semantics{\tau_{12}}_{\env[y \mapsto h_1']}$
    \end{pf*}

    With a similar argument as before, also by IH, there exists $\tau_{22}'$ such that
    \begin{pf*}{T-If}
      \item $\Gamma' \vdash c_2 : (x:\refT{y}{\tau_1'}{\neg \varphi[y/\cmdVar]}) \to \tau_{22}'$
      \item \label{input-type-strengthening-tif-t22-sub} $\forall h_2'\in\semantics{\refT{y}{\tau_1'}{\neg \varphi[y/\cmdVar]}}_{\env'}.\semantics{\tau_{22}'}_{\env'[y \mapsto h_2']} = \semantics{\tau_{22}}_{\env[y \mapsto h_2']}$
    \end{pf*}

    $\Gamma';\tau_1' \vdash \varphi : \tBool$ also holds, because the subtyping relation between $\tau_1'$ and $\tau_1$ ensures that heaps described by $\tau_1'$ have the same shape (i.e., the same instances are valid) and thus we can typecheck formula $e$ in the context of type $\tau_1'$.

    \ref{input-type-strengthening-g1} follows by \tIf.

    Let $\tau_2' = \refT{y}{\tau_{12}'}{\varphi[x/\cmdVar]} + \refT{y}{\tau_{22}'}{\neg \varphi[x/\cmdVar]}$

    To show $\forall h_1'\in\semantics{\tau_1'}_{\env'}.\semantics{\refT{y}{\tau_{12}'}{\varphi[x/\cmdVar]} + \refT{y}{\tau_{22}}{\neg \varphi[x/\cmdVar]}}_{\env'[x \mapsto h_1']} \subseteq \semantics{\refT{y}{\tau_{12}}{\varphi[x/\cmdVar]} + \refT{y}{\tau_{22}}{\neg \varphi[x/\cmdVar]}}_{\env[x \mapsto h_1']}$

    Let $h_1'\in\semantics{\tau_1'}_{\env'}$ be arbitrary.
    Case distinction on wether the formula $\varphi$ in $h_1'$ evaluates to true or false.
    \begin{description}
      \subcase{$e$ evaluates to true}
      \begin{align*}
        \phantom{{}\Leftrightarrow{}} & \semantics{\refT{y}{\tau_{12}'}{\varphi[x/\cmdVar]} + \refT{y}{\tau_{22}}{\neg \varphi[x/\cmdVar]}}_{\env'[x \mapsto h_1']} \subseteq \\
                                      & \semantics{\refT{y}{\tau_{12}}{\varphi[x/\cmdVar]} + \refT{y}{\tau_{22}}{\neg \varphi[x/\cmdVar]}}_{\env[x \mapsto h_1']}             \\
        \Leftrightarrow               & \semantics{\refT{y}{\tau_{12}'}{\true} + \refT{y}{\tau_{22}}{\false}}_{\env'[x \mapsto h_1']} \subseteq                               \\
                                      & \semantics{\refT{y}{\tau_{12}}{\true} + \refT{y}{\tau_{22}}{\false}}_{\env[x \mapsto h_1']}                                           \\
        \Leftrightarrow               & \semantics{\tau_{12}'}_{\env'[x \mapsto h_1']} \subseteq \semantics{\tau_{12}}_{\env[x \mapsto h_1']}
      \end{align*}
      The result follows by \ref{input-type-strengthening-tif-t12-sub}.
      \subcase{$e$ evaluates to false}
      Symmetric to previous subcase. The result follows by \ref{input-type-strengthening-tif-t22-sub}.
    \end{description}

    \case{\tMod}~
    \begin{pf}{T-Mod}
      \item $\tau_2 = \refT{y}{\top}{\varphi_{pkt} \wedge \varphi_\inst \wedge \varphi_f \wedge y.\inst.f = e[x/\cmdVar]}$
      \item \label{input-type-strengthening-tmod-includes} $\includes{\tau_1}{\inst}$
      \item $\Gamma;\tau \vdash t:\tBv$
      \item $\mathcal{F}(\inst,f) = \tBv$
      \item $c = \inst.f := e$
    \end{pf}

    To show: There exists $\tau_2'$ such that
    \begin{enumerate}[label=(\arabic*)]
      \item $\Gamma' \vdash \inst.f := e : (x:\tau_1') \to \tau_2'$ and
      \item $\forall h_1' \in \semantics{\tau_1'}_{\env'}.\semantics{\tau_2'}_{\env[x \mapsto h_1']} \subseteq \semantics{\tau_2}_{\env[x \mapsto h_1']}$
    \end{enumerate}

    Let $\tau_2' = \tau_2$.

    $\includes{\tau_1'}{\inst}$ follows by assumptions \ref{input-type-strengthening-tmod-includes} and \ref{input-type-strengthening-input-sub} and set theory.
    By assumption \ref{input-type-strengthening-input-sub}, we know that $\tau_1'$ has the same shape (contains the same instances) as $\tau_1$,
    so we can typecheck expression $e$ in context $\tau_1'$ with a bit vector type,
    from which follows that $\Gamma';\tau_1' \vdash e:\tBv$.
    \ref{input-type-strengthening-g1} follows by \tMod.

    Let $h_1'\in\semantics{\tau_1'}_{\env'}$ be arbitrary.
    To show \ref{input-type-strengthening-g2}, we must show that
    \begin{align*}
      \phantom{{}\Leftrightarrow{}} & \semantics{\tau_2}_{\env'[x \mapsto h_1']} \subseteq \semantics{\tau_2}_{\env[x \mapsto h_1']}                                                    \\
      \Leftrightarrow               & \semantics{\refT{y}{\top}{\varphi_{pkt} \wedge \varphi_\inst \wedge \varphi_f \wedge y.\inst.f = t[x/\cmdVar]}}_{\env'[x \mapsto h_1']} \subseteq \\
                                    & \semantics{\refT{y}{\top}{\varphi_{pkt} \wedge \varphi_\inst \wedge \varphi_f \wedge y.\inst.f = t[x/\cmdVar]}}_{\env[x \mapsto h_1']}
      \intertext{Since the only free variable is $x$}
      \Leftrightarrow               & \semantics{\refT{y}{\top}{\varphi_{pkt} \wedge \varphi_\inst \wedge \varphi_f \wedge y.\inst.f = t[x/\cmdVar]}}_{[x \mapsto h_1']} \subseteq      \\
                                    & \semantics{\refT{y}{\top}{\varphi_{pkt} \wedge \varphi_\inst \wedge \varphi_f \wedge y.\inst.f = t[x/\cmdVar]}}_{[x \mapsto h_1']}
    \end{align*}
    The result is immediate.

    \case{\tRemit}~

    By inversion of rule \tRemit, we get
    \begin{pf}{T-Remit}
      \item \label{input-type-strengthening-tremit-includes} $\includes{\tau}{\inst}$
      \item $\tau_2 = \sigmaT{y}{\refT{z}{\tau_1}{z \equiv x}}{\refT{z}{\epsilon}{z.\pIn = \bvNil \wedge z.\pOut = x.\inst}}$
    \end{pf}
    Let $\tau_2' = \sigmaT{y}{\refT{z}{\tau_1'}{z \equiv x}}{\refT{z}{\epsilon}{z.\pIn = \bvNil \wedge z.\pOut = x.\inst}}$.

    $\includesCtx{\Gamma'}{\tau_1'}{\inst}$ follows by assumptions \ref{input-type-strengthening-tremit-includes} and \ref{input-type-strengthening-input-sub} and set theory.
    \ref{input-type-strengthening-g1} follows by \tRemit.
    Let $h\in\semantics{\tau_1'}_\env$ be arbitrary.
    \begin{align*}
      \phantom{{}={}} & \semantics{\sigmaT{y}{\refT{z}{\tau_1}{z \equiv x}}{\refT{z}{\epsilon}{z.\pIn = \bvNil \wedge z.\pOut = x.\inst}}}_{\env[x \mapsto h]}                  \\
      =               & \lbrace \concat{h_1}{h_2} \mid h_1 \in \semantics{\refT{z}{\tau_1}{z \equiv x}}_{\env[x \mapsto h]} \wedge                                              \\
                      & \hspace{45.5pt} h_2 \in \semantics{\refT{z}{\epsilon}{z.\pIn = \bvNil \wedge z.\pOut = x.\inst}}_{\env[x \mapsto h, y \mapsto h_1]} \rbrace             \\
      =               & \lbrace \concat{h}{h_2} \mid h_2 \in \semantics{\refT{z}{\epsilon}{z.\pIn = \bvNil \wedge z.\pOut = x.\inst}}_{\env[x \mapsto h, y \mapsto h]} \rbrace  \\
      \intertext{$x$ is the only free variable in $\refT{z}{\epsilon}{z.\pIn = \bvNil \wedge z.\pOut = x.\inst}$, which maps to the same heap $h$ in both environments $\env[x \mapsto h, y \mapsto h]$ and $\env'[x \mapsto h, y \mapsto h]$.}
      =               & \lbrace \concat{h}{h_2} \mid h_2 \in \semantics{\refT{z}{\epsilon}{z.\pIn = \bvNil \wedge z.\pOut = x.\inst}}_{\env'[x \mapsto h, y \mapsto h]} \rbrace \\
      =               & \lbrace \concat{h_1}{h_2} \mid h_1 \in \semantics{\refT{z}{\tau_1'}{z \equiv x}}_{\env'[x \mapsto h]} \wedge                                            \\
                      & \hspace{45.5pt} h_2 \in \semantics{\refT{z}{\epsilon}{z.\pIn = \bvNil \wedge z.\pOut = x.\inst}}_{\env'[x \mapsto h, y \mapsto h_1]} \rbrace            \\
      =               & \semantics{\sigmaT{y}{\refT{z}{\tau_1'}{z \equiv x}}{\refT{z}{\epsilon}{z.\pIn = \bvNil \wedge z.\pOut = x.\inst}}}_{\env'[x \mapsto h]}
    \end{align*}
    This concludes the case by showing \ref{input-type-strengthening-g2}.

    \case{\tReset}~

    By inversion of rule \tReset, we get
    \begin{pf}{T-Reset}
      \item $c = \cReset$
      \item $\tau_2 = \sigmaT{y}{\refT{z}{\epsilon}{z.\pOut = \bvNil \wedge z.\pIn = x.\pOut}}{\refT{z}{\epsilon}{z.\pOut = \bvNil \wedge z.\pIn = x.\pIn}}$
    \end{pf}
    Let $\tau_2' = \tau_2$. \ref{input-type-strengthening-g1} follows by \tReset.
    Let $h_1'\in\semantics{\tau_1'}_{\env'}$ be arbitrary.
    \begin{align*}
      \phantom{{}\Leftrightarrow{}} & \semantics{\sigmaT{y}{\refT{z}{\epsilon}{z.\pOut = \bvNil \wedge z.\pIn = x.\pOut}}{                          \\
                                    & \hspace{22.5pt}\refT{z}{\epsilon}{z.\pOut = \bvNil \wedge z.\pIn = x.\pIn}}}_{\env'[x \mapsto h_1']} \subseteq \\
                                    & \semantics{\sigmaT{y}{\refT{z}{\epsilon}{z.\pOut = \bvNil \wedge z.\pIn = x.\pOut}}{                          \\
                                    & \hspace{22.5pt}\refT{z}{\epsilon}{z.\pOut = \bvNil \wedge z.\pIn = x.\pIn}}}_{\env[x \mapsto h_1']}          \\
    \end{align*}
    Both sets are actually equal, because $x$ is the only free variable in $\tau_2$ and $\tau_2'$ respectively.
    Thus, all other bindings in the environments $\env$ and $\env'$ have no effect on the semantics of $\tau_2$ and $\tau_2'$ respectively.
    This shows \ref{input-type-strengthening-g2} and concludes the case.

    \case{\tSeq}~

    By inversion of rule \tSeq, we get
    \begin{pf}{T-Seq}
      \item $c = c_1;c_2$
      \item \label{input-type-strengthening-tseq-c1} $\Gamma \vdash c_1 : (x:\tau_1) \to \tau_{12}$
      \item \label{input-type-strengthening-tseq-c2} $\Gamma, x:\tau_1 \vdash c_2 : (y:\tau_{12}) \to \tau_{22}$
      \item $\tau_2 = \tau_{22}[y \mapsto \tau_{12}]$
    \end{pf}
    By IH with \ref{input-type-strengthening-tseq-c1}, \ref{input-type-strengthening-input-sub}, \ref{input-type-strengthening-env-entails} and \ref{input-type-strengthening-envprime-entails}, 
    there exists some $\tau_{12}'$ such that
    \begin{pf*}{T-Seq}
      \item $\Gamma' \vdash c_1: (x:\tau_1') \to \tau_{12}'$
      \item \label{input-type-strengthening-tseq-c1-out-sub} $\forall h_1'\in\semantics{\tau_1'}_{\env'}.\semantics{\tau_{12}'}_{\env'[x \mapsto h_1']}\subseteq \semantics{\tau_{12}}_{\env[x \mapsto h_1']}$
    \end{pf*}
    Apply the IH again to $c_2$:

    Let $h_1'\in\semantics{\tau_1'}_{\env'}$ be arbitrary.
    By \ref{input-type-strengthening-tseq-c2} $\Gamma, x:\tau_1 \vdash c_2 : (y:\tau_{12}) \to \tau_{22}$.
    By \ref{input-type-strengthening-tseq-c1-out-sub}, $\semantics{\tau_{12}'}_{\env'[x \mapsto h_1']} \subseteq \semantics{\tau_{12}}_{\env[x \mapsto h_1']}$.
    $\env[x \mapsto h_1'] \models \Gamma,x:\tau_1$ because by assumption $\env \models \Gamma$ the entailment holds for all $x_i \neq x$.
    For $x$ there exists a binding to heap $h_1'\in\semantics{\tau_1}_\env$ (with assumption \ref{input-type-strengthening-input-sub}) and the entailment between $h$ and $\tau_1$ trivially holds.

    To show $\env'[x \mapsto h_1'] \models \Gamma',x:\tau_1'$, we must show that
    $\forall x_i,\tau_i. \Gamma'(x_i) = \tau_i \Rightarrow \env'[x \mapsto h_1'](x_i) = h_i \wedge h_i \models_{\env[x \mapsto h_1']} \tau_i$.
    Case $x_i \neq x$: this holds by assumption \ref{input-type-strengthening-envprime-entails}.
    Case $x_i = x$. $\env'[x \mapsto h_1'](x) = h_1'$.
    To show that $h_1' \models_{\env'[x \mapsto h_1']} \tau_1' \Leftrightarrow h_1'\in\semantics{\tau_1'}_{\env'[x \mapsto h_1']}$.
    By assumption, $x$ is not free in $\tau_1'$, so we can equivalently show that $h_1'\in\semantics{\tau_1'}_{\env'}$, which holds by assumption.

    Again by IH, there exists some $\tau_{22}'$ such that
    \begin{pf*}{T-Seq}
      \item $\Gamma', x:\tau_1' \vdash c_2 (y:\tau_{12}') \to \tau_{22}'$
      \item \label{input-type-strengthening-tseq-c2-out-sub} $\forall h_{12}'\in\semantics{\tau_{12}'}_{\env'[x \mapsto h_1']}.\semantics{\tau_{22}'}_{\env'[x \mapsto h_1, y \mapsto h_{12}']} \subseteq \semantics{\tau_{22}}_{\env[x \mapsto h_1, y \mapsto h_{12}']}$
    \end{pf*}
    Let $\tau_2' = \tau_{22}'[y \mapsto \tau_{12}']$. \ref{input-type-strengthening-g1} follows by \tSeq.

    For \ref{input-type-strengthening-g2}, we must show that $\forall h_1'\in\semantics{\tau_1'}_{\env'}.\semantics{\tau_2'}_{\env'[x \mapsto h_1']} \subseteq \semantics{\tau_2}_{\env[x \mapsto h_1']}$

    Let $h_1'\in\semantics{\tau_1'}_{\env'}$ be arbitrary.
    \begin{align*}
      \phantom{{}\Leftrightarrow{}} & \semantics{\tau_{22}'[y \mapsto \tau_{12}']}_{\env'[x \mapsto h_1']} \subseteq \semantics{\tau_{22}[y \mapsto \tau_{12}]}_{\env[x \mapsto h_1']} \\
      \Leftrightarrow               & \bigcup_{h_{12}'\in\semantics{\tau_{12}'}_{\env'[x \mapsto h_1']}} \semantics{\tau_{22}'}_{\env'[x \mapsto h_1', y \mapsto h_{12}']} \subseteq
      \bigcup_{h_{12}\in\semantics{\tau_{12}}_{\env[x \mapsto h_1']}} \semantics{\tau_{22}}_{\env[x \mapsto h_1', y \mapsto h_{12}]}
    \end{align*}
    The result follows by \ref{input-type-strengthening-tseq-c1-out-sub}, \ref{input-type-strengthening-tseq-c2-out-sub} and set theory.

    \case{\tSkip}~
    
    By inversion of rule \tSkip, we get
    \begin{pf}{T-Skip}
      \item $c = \cSkip$
      \item $\tau_2 = \refT{y}{\tau_1}{y \equiv x}$
    \end{pf}
    Let $\tau_2' = \refT{y}{\tau_1'}{y \equiv x}$.
    \ref{input-type-strengthening-g1} follows by \tSkip.

    To show \ref{input-type-strengthening-g2}, let $h'\in\semantics{\tau_1'}_{\env'}$ be an arbitrary heap.
    \begin{align*}
      \phantom{{}\Leftrightarrow{}} & \semantics{\refT{y}{\tau_1'}{y \equiv x}}_{\env'[x \mapsto h']} \subseteq \semantics{\refT{y}{\tau_1}{y \equiv x}}_{\env[x \mapsto h']}                 \\
      \Leftrightarrow               & \lbrace h'\rbrace \subseteq \semantics{\refT{y}{\tau_1}{y \equiv x}}_{\env[x \mapsto h']}                                                               \\
      \Leftrightarrow               & \lbrace h'\rbrace \subseteq \lbrace h'\rbrace                                                                                           & \text{by \ref{input-type-strengthening-input-sub}}
    \end{align*}

    \case{\tSub}~

    By inversion of rule \tSub, we get
    \begin{pf}{T-Sub}
      \item $\cmdT{c}{(x:\tau_3) \to \tau_4}$
      \item \label{input-type-strengthening-tsub-t1-sub} $\subtypeCtx{\Gamma}{\tau_1}{\tau_3}$
      \item \label{input-type-strengthening-tsub-t4-sub} $\subtypeCtx{\Gamma, x\!:\!\tau_1}{\tau_4}{\tau_2}$
    \end{pf}

    By assumption
    $\semantics{\tau_1'}_{\env'} \subseteq \semantics{\tau_1}_\env$
    and from \ref{input-type-strengthening-tsub-t1-sub} follows that
    $\semantics{\tau_1}_\env \subseteq \semantics{\tau_3}_\env$
    and thus
    $\semantics{\tau_1'}_{\env'} \subseteq \semantics{\tau_3}_\env$.
    By IH, there exists $\tau_4'$ such that
    \begin{pf*}{T-Sub}
      \item \label{input-type-strengthening-tsub-c-prime} $\Gamma' \vdash c :(x:\tau_1') \to \tau_4'$
      \item \label{input-type-strengthening-tsub-t4-prime-sub} $\forall h'\in\semantics{\tau_1'}_{\env'}.\semantics{\tau_4'}_{\env'[x \mapsto h']} \subseteq \semantics{\tau_4}_{\env[x \mapsto h']}$
    \end{pf*}

    Let $\tau_2' = \tau_4'$.
    \ref{input-type-strengthening-g1} follows by \ref{input-type-strengthening-tsub-c-prime}.

    For \ref{input-type-strengthening-g1}, we have to show that $\forall h'\in\semantics{\tau_1'}_{\env'}.\semantics{\tau_4'}_{\env'[x \mapsto h']} \subseteq \semantics{\tau_2}_{\env[x \mapsto h']}$, which follows by \ref{input-type-strengthening-tsub-t4-sub} and \ref{input-type-strengthening-tsub-t4-prime-sub} and by set theory.
  \end{description}
\end{proof}

%% file: appendix/safety/lem-chomp.tex
\begin{lemma}[Semantic Chomp Expression]
  \label{lem:semantic-chompt}
  For all expressions $e$,
  heaps $h$ and $h'$,
  environments $\env$ and $\env'$ and variables $x$,
  if $h'=\chompS(h,1)$, and
  $\env'=\env[x\mapsto (\bvNil, \bvNil, [\inst\mapsto v])]$ and,
  if $x\in\dom(\env)$, $v=\env(x)(\inst)@\slice{h(\pIn)}{0}{1}$ and
  $\env(x)(\pIn)=\bvNil$ and $\env(x)(\pOut)=\bvNil$,
  and otherwise $v=\slice{h(\pIn)}{0}{1}$ and $x$ not free in $e$,
  then
  \[
    \semantics{e}_{\env[y\mapsto h]} = \semantics{\heapRef_1(\chompExpr_1(e, y, \bitvar{0}), \bitvar{0},x,\inst,1)}_{\env'[y \mapsto h']}
  \]
\end{lemma}
\begin{proof}
  Proof by induction on $e$.
  We only consider expressions referencing $\pIn$.
  All other expressions are not affected by chomping, and therefore the semantic is unchanged.
  \begin{description}
    \case{$e=\slice{z.\pIn}{l}{r}$}~
    Case distinction on $z=y$:
    \begin{description}
      \subcase{$z\neq y$}
      \begin{align*}
         & \phantom{{}={}}\semantics{\heapRef_1(\chompExpr_1(\slice{z.\pIn}{l}{r}, y, \bitvar{0}), \bitvar{0},x,\inst,1)}_{\env'[y\mapsto h']} & \\
         & = \semantics{\heapRef_1(\slice{z.\pIn}{l}{r},\bitvar{0},x,\inst,1)}_{\env'[y\mapsto h']}                                            & \\
         & = \semantics{\slice{z.\pIn}{l}{r}}_{\env'[y\mapsto h']}                                                                             & \\
        \intertext{If $z\neq x$, $z$ binds to some heap in $\env$, which must also be contained in $\env'$ unchanged.
          If $z=x$, by assumption, $x.\pIn$ maps to the empty bit vector, both in $\env$ and $\env'$.}
         & = \semantics{\slice{z.\pIn}{l}{r}}_{\env[y\mapsto h]}                                                                               & \\
      \end{align*}
      \subcase{$z=y, r \leq 1$}
      \begin{align*}
         & \phantom{{}={}}\semantics{\heapRef_1(\chompExpr_1(\slice{y.\pIn}{0}{1}, y, \bitvar{0}), \bitvar{0},x,\inst,1)}_{\env'[y\mapsto h']} & \\
         & = \semantics{\heapRef_1(\chompExpr_1(\slice{y.\pIn}{0}{1},y,\bitvar{0}),\bitvar{0},x,\inst,1)}_{\env'[y\mapsto h']}                 & \\
         & = \semantics{\heapRef_1(\bitvar{0}::\bvNil,\bitvar{0},x,\inst,1)}_{\env'[y\mapsto h']}                                              & \\
         & = \semantics{\bvconcat{\slice{x.\inst}{\sizeof(\inst)-1}{\sizeof(\inst)-1+1}}{\bvNil}}_{\env'[y\mapsto h']}                         & \\
         & = \semantics{\slice{x.\inst}{\length{v}-1}{\length{v}}}_{\env'[y\mapsto h']}                                                        & \\
         & = \env'(x)(\inst)[\length{v}-1]                                                                                                     & \\
         & = \slice{h(\pIn)}{0}{1}                                                                                                             & \\
         & = \semantics{\slice{y.\pIn}{0}{1}}_{\env[y\mapsto h]}
      \end{align*}
      \subcase{$z=y, l=0$}
      \begin{align*}
         & \phantom{{}={}}\semantics{\heapRef_1(\chompExpr_1(\slice{y.\pIn}{0}{r}, y, \bitvar{0}), \bitvar{0},x,\inst,1)}_{\env'[y\mapsto h']} & \\
         & = \semantics{\heapRef_1(\chompExpr_1(\slice{y.\pIn}{0}{r},y,\bitvar{0}),\bitvar{0},x,\inst,1)}_{\env'[y\mapsto h']}                 & \\
         & = \semantics{\heapRef_1(\bitvar{0}::\slice{y.\pIn}{0}{r-1},\bitvar{0},x,\inst,1)}_{\env'[y\mapsto h']}                              & \\
         & = \semantics{\bvconcat{\slice{x.\inst}{\sizeof(\inst)-1}{\sizeof(\inst)-1+1}}{\slice{y.\pIn}{0}{r-1}}}_{\env'[y\mapsto h']}         & \\
         & = \semantics{\bvconcat{\slice{x.\inst}{\length{v}-1}{\length{v}}}{\slice{y.\pIn}{0}{r-1}}}_{\env'[y\mapsto h']}                     & \\
         & = \bvconcat{\slice{\env'(x)(\inst)}{\length{v}-1}{\length{v}}}{\slice{h'(\pIn)}{0}{r-1}}                                            & \\
        \intertext{with $v=\env(x)(\inst)@h(\pIn)[0]$ follows}
         & = \bvconcat{\slice{h(\pIn)}{0}{1}}{\slice{h'(\pIn)}{0}{r-1}}                                                                        & \\
        \intertext{with $h'=\chompS(h,1)$ follows}
         & = \bvconcat{\slice{h(\pIn)}{0}{1}}{\slice{h(\pIn)}{1}{r}}                                                                           & \\
         & = \slice{h(\pIn)}{0}{r}                                                                                                             & \\
         & = \semantics{\slice{y.\pIn}{0}{r}}_{\env[y\mapsto h]}
      \end{align*}
      \subcase{$z=y, l \neq 0$}
      \begin{align*}
         & \phantom{{}={}}\semantics{\heapRef_1(\chompExpr_1(\slice{y.\pIn}{l}{r}, y, \bitvar{0}), \bitvar{0},x,\inst,1)}_{\env'[y\mapsto h']} & \\
         & = \semantics{\heapRef_1(\chompExpr_1(\slice{y.\pIn}{l}{r},y,\bitvar{0}),\bitvar{0},x,\inst,1)}_{\env'[y\mapsto h']}                 & \\
         & = \semantics{\heapRef_1(\slice{y.\pIn}{l-1}{r-1},\bitvar{0},x,\inst,1)}_{\env'[y\mapsto h']}                                        & \\
         & = \semantics{\slice{y.\pIn}{l-1}{r-1}}_{\env'[y\mapsto h']}                                                                         & \\
         & = \slice{h'(\pIn)}{l-1}{r-1}                                                                                                        & \\
        \intertext{with $h'=\chompS(h,1)$ follows}
         & = \slice{h(\pIn)}{l}{r}                                                                                                             & \\
         & = \semantics{\slice{y.\pIn}{l}{r}}_{\env[y\mapsto h]}
      \end{align*}
    \end{description}

    \case{$e=z.\pIn$}
    Case distinction on $z=y$:
    \begin{description}
      \subcase{$z\neq y$} Symmetric to first subcase of previous case.
      \subcase{$z=y$}
      \begin{align*}
         & \phantom{{}={}}\semantics{\heapRef_1(\chompExpr_1(y.\pIn, y, \bitvar{0}), \bitvar{0},x,\inst,1)}_{\env'[y\mapsto h']}   \\
         & = \semantics{\heapRef_1(\bitvar{0}::y.\pIn, \bitvar{0},x,\inst,1)}_{\env'[y\mapsto h']}                                 \\
         & = \semantics{\bvconcat{\slice{x.\inst}{\sizeof(\inst)-1}{\sizeof(\inst)-1+1}}{y.\pIn}}_{\env'[y\mapsto h']}             \\
         & = \semantics{\bvconcat{\slice{x.\inst}{\length{v}-1}{\length{v}}}{y.\pIn}}_{\env'[y\mapsto h']}                         \\
         & = \bvconcat{\slice{h(\pIn)}{0}{1}}{h'(\pIn)}                                                                            \\
        \intertext{with $h'=\chompS(h,1)$ follows}
         & = h(\pIn)                                                                                                             & \\
         & =\semantics{y.\pIn}_{\env[y\mapsto h]}
      \end{align*}
    \end{description}
    \case{$e=\length{z.\pIn}$}
    Case distinction on $z=y$:
    \begin{description}
      \subcase{$z\neq y$}
      \begin{align*}
         & \phantom{{}={}}\semantics{\heapRef_1(\chompExpr_1(\length{z.\pIn}, y, \bitvar{0}), \bitvar{0},x,\inst,1)}_{\env'[y\mapsto h']} \\
         & =\semantics{\length{z.\pIn}}_{\env'[y\mapsto h']}                                                                              \\
        \intertext{If $z=x$, the length of $x.\pIn=0$ in both environments and otherwise, $z.\pIn$ refers to the same heap in both $\env$ and $\env'$.}
         & =\semantics{\length{z.\pIn}}_{\env[y\mapsto h]}
      \end{align*}
      \subcase{$z=y$}
      \begin{align*}
         & \phantom{{}={}}\semantics{\heapRef_1(\chompExpr_1(\length{y.\pIn}, y, \bitvar{0}), \bitvar{0},x,\inst,1)}_{\env'[y\mapsto h']} \\
         & =\semantics{\length{y.\pIn}+1}_{\env'[y\mapsto h']}                                                                            \\
        \intertext{with $h'=\chompS(h,1)$ follows}
         & =\semantics{\length{y.\pIn}}_{\env[y\mapsto h]}                                                                                \\
      \end{align*}
    \end{description}
    \case{$e=b::bv$}
    \begin{align*}
       & \phantom{{}={}}\semantics{\heapRef_1(\chompExpr_1(b::bv, y, \bitvar{0}), \bitvar{0},x,\inst,1)}_{\env'[y\mapsto h']}                   \\
       & =\semantics{\heapRef_1(b::\chompExpr_1(bv, y, \bitvar{0}), \bitvar{0},x,\inst,1)}_{\env'[y\mapsto h']}                                 \\
       & =\semantics{b::\heapRef_1(\chompExpr_1(bv, y, \bitvar{0}), \bitvar{0},x,\inst,1)}_{\env'[y\mapsto h']}                                 \\
       & =\semantics{b}_{\env'[y\mapsto h']}::\semantics{\heapRef_1(\chompExpr_1(bv, y, \bitvar{0}),\bitvar{0},x,\inst,1)}_{\env'[y\mapsto h']} \\
       & =\semantics{b}_{\env'[y\mapsto h']}::\semantics{bv}_{\env[y\mapsto h]}                                                                 \\
      \intertext{by IH follows}
       & =\semantics{b}_{\env[y\mapsto h]}::\semantics{bv}_{\env[y\mapsto h]}                                                                   \\
      \intertext{since $b$ is either $0$ or $1$}
       & =\semantics{b::bv}_{\env[y\mapsto h]}
    \end{align*}
    \case{$e=\bvconcat{bv_1}{bv_2}$}
    \begin{align*}
       & \phantom{{}={}}\semantics{\heapRef_1(\chompExpr_1(\bvconcat{bv_1}{bv_2}, y, \bitvar{0}), \bitvar{0},x,\inst,1)}_{\env'[y\mapsto h']}               \\
       & =\semantics{\heapRef_1(\bvconcat{\chompExpr_1(bv_1, y, \bitvar{0})}{\chompExpr_1(bv_2, y, \bitvar{0})},\bitvar{0},x,\inst,1)}_{\env'[y\mapsto h']} \\
       & =\bvconcat{\semantics{\heapRef_1(\chompExpr_1(bv_1, y, \bitvar{0}), \bitvar{0},x,\inst,1)}_{\env'[y\mapsto h']}}{                                  \\
       & \phantom{{}={}}\semantics{\heapRef_1(\chompExpr_1(bv_1, y, \bitvar{0}), \bitvar{0},x,\inst,1)}_{\env'[y\mapsto h']}}                               \\
       & =\bvconcat{\semantics{bv_1}_{\env[y\mapsto h]}}{\semantics{bv_2}_{\env[y\mapsto h]}}                                                               \\
      \intertext{by IH}
       & =\semantics{\bvconcat{bv_1}{bv_2}}_{\env[y\mapsto h]}
    \end{align*}
    \case{$e=n+m$} Symmetric to previous case.
  \end{description}
\end{proof}

\begin{lemma}[Semantic Chomp Formulae]
  \label{lem:semantic-chompe}
  For all formulae $\varphi$, heaps $h$ and $h'$,
  environments $\env$ and $\env'$ and variables $x$,
  if $h'=\chompS(h,1)$, and
  $\env'=\env[x \mapsto (\bvNil, \bvNil, [\inst \mapsto v])]$,
  and if $x\in\dom$, $v=\env(x)(\inst)@\slice{h(\pIn)}{0}{1}$
  and $\env(x)(\pIn)=\bvNil$ and $\env(x)(\pOut)=\bvNil$ and
  otherwise $v=\slice{h(\pIn)}{0}{1}$ and $x$ not free in $\varphi$, then
  \[
    \semantics{\varphi}_{\env[y\mapsto h]} = \semantics{\heapRef_1(\chompForm_1(\varphi, y, \bitvar{0}), \bitvar{0},x,\inst,1)}_{\env'[y \mapsto h']}
  \]
\end{lemma}
\begin{proof}
  By induction on $\varphi$.
  \begin{description}
    \case{$\varphi=e_1=e_2$}
    \begin{align*}
       & \phantom{{}={}}\semantics{\heapRef_1(\chompForm_1(e_1=e_2,y,\bitvar{0}),\bitvar{0},x,\inst)}_{\env'[y\mapsto h']} \\
       & =\semantics{\heapRef_1(\chompForm_1(e_1,y,\bitvar{0}),\bitvar{0},x,\inst)\ =                                      \\
       & \hspace{14.5pt} \heapRef_1(\chompForm_1(e_1,y,\bitvar{0}),\bitvar{0},x,\inst)}_{\env'[y\mapsto h']}               \\
       & =\semantics{\heapRef_1(\chompForm_1(e_1,y,\bitvar{0}),\bitvar{0},x,\inst)}_{\env'[y\mapsto h']} =                 \\
       & \phantom{{}={}}\semantics{\heapRef_1(\chompForm_1(e_1,y,\bitvar{0}),\bitvar{0},x,\inst)}_{\env'[y\mapsto h']}     \\
      \intertext{by \cref{lem:semantic-chompt} follows}
       & =\semantics{e_1}_{\env[y\mapsto h]}=\semantics{e_2}_{\env[y\mapsto h]}                                            \\
       & =\semantics{e_1=e_2}_{\env[y\mapsto h]}
    \end{align*}

    \case{$\varphi=\varphi_1 \land \varphi_2$}
    \begin{align*}
       & \phantom{{}={}} \semantics{\heapRef_1(\chompForm_1(\varphi_1 \land \varphi_2, y, \bitvar{0}), \bitvar{0},x,\inst,1)}_{\env'[y \mapsto h']} \\
       & = \semantics{\heapRef_1(\chompForm_1(\varphi_1, y, \bitvar{0}), \bitvar{0},x,\inst,1) \land                                                \\
       & \phantom{{}={}\llbracket}\heapRef_1(\chompForm_1(\varphi_2, y, \bitvar{0}), \bitvar{0},x,\inst,1)}_{\env'[y \mapsto h']}                   \\
       & = \semantics{\heapRef_1(\chompForm_1(\varphi_1, y, \bitvar{0}), \bitvar{0},x,\inst,1)}_{\env'[y \mapsto h']} \land                         \\
       & \phantom{{}={}} \semantics{\heapRef_1(\chompForm_1(\varphi_2, y, \bitvar{0}), \bitvar{0},x,\inst,1)}_{\env'[y \mapsto h']}                 \\
      \intertext{by IH follows}
       & = \semantics{\varphi_1}_{\env[y \mapsto h]} \land \semantics{\varphi_2}_{\env[y \mapsto h]}                                                \\
       & = \semantics{\varphi_1 \land \varphi_2}_{\env[y \mapsto h]}
    \end{align*}

    \case{$\varphi=\neg \varphi_1$}
    \begin{align*}
       & \phantom{{}={}} \semantics{\heapRef_1(\chompForm_1(\neg \varphi_1, y, \bitvar{0}), \bitvar{0},x,\inst,1)}_{\env'[y \mapsto h]} \\
       & = \semantics{\neg \heapRef_1(\chompForm_1(\varphi_1, y, \bitvar{0}), \bitvar{0},x,\inst,1)}_{\env'[y \mapsto h]}               \\
       & = \neg \semantics{\heapRef_1(\chompForm_1(\varphi_1, y, \bitvar{0}), \bitvar{0},x,\inst,1)}_{\env'[y \mapsto h]}               \\
      \intertext{by IH follows}
       & = \neg \semantics{\varphi}_{\env[y \mapsto h]}                                                                                 \\
       & = \semantics{\neg \varphi}_{\env[y \mapsto h]}                                                                                 \\
    \end{align*}

    \case{$\varphi=z.\inst'.valid$}
    \begin{align*}
       & \phantom{{}={}} \semantics{\heapRef_1(\chompForm_1(z.\inst'.valid, y, \bitvar{0}), \bitvar{0},x,\inst,1)}_{\env'[y \mapsto h']} \\
       & = \semantics{z.\inst'.valid}_{\env'[y \mapsto h']}                                                                              \\
       & = \inst' \in dom(\env'[y \mapsto h'](z))                                                                                        \\
       & = \inst' \in dom(\env[y \mapsto h](z))                                                                                          \\
      \intertext{by definition of $\env'$ and $h'$ follows}
       & = \semantics{z.\inst'.valid}_{\env[y \mapsto h]}                                                                                \\
    \end{align*}

    \case{$\varphi=\true$}
    \begin{align*}
       & \phantom{{}={}} \semantics{\heapRef_1(\chompForm_1(\true, y, \bitvar{0}), \bitvar{0},x,\inst,1)}_{\env'[y \mapsto h]} \\
       & = \semantics{\true}_{\env'[y \mapsto h]}                                                                              \\
       & = \true                                                                                                               \\
       & = \semantics{\true}_{\env[y \mapsto h]}                                                                               \\
    \end{align*}

    \case{$\varphi=\false$} Symmetric to previous case.
  \end{description}
\end{proof}

\begin{lemma}[Semantic Chomp Refinement]
  \label{lem:semantic-chompref1}
  For all heap types $\tau$, heaps $h$ and $h'$, environments $\env$ and $\env'$ and variables $x$,
  if $h'=\chompS(h,1)$, and $\env'=\env[x \mapsto (\bvNil, \bvNil, [\inst \mapsto v])]$, and,
  if $x\in\dom$, $v=\env(x)(\inst)@\slice{h(\pIn)}{0}{1}$ and $\env(x)(\pIn)=\bvNil$ and $\env(x)(\pOut)=\bvNil$,
  and otherwise $v=\slice{h(\pIn)}{0}{1}$ and $x$ not free in $\tau$, then
  \[
    \semantics{\tau}_{\env[y\mapsto h]} = \semantics{\heapRef_1(\chompRef_1(\tau, y, \bitvar{0}), \bitvar{0},x,\inst,1)}_{\env'[y \mapsto h']}
  \]
\end{lemma}
\begin{proof}
  Proof by induction on $\tau$.
  \begin{description}
    \case{$\tau=\varnothing$}
    \begin{align*}
       & \phantom{{}={}}\semantics{\heapRef_1(\chompRef_1(\varnothing, y, \bitvar{0}), \bitvar{0},x,\inst,1)}_{\env'[y \mapsto h']} \\
       & =\semantics{\varnothing}_{\env'[y \mapsto h']}                                                                             \\
       & =\{\}                                                                                                                      \\
       & =\semantics{\tau}_{\env[y\mapsto h]}
    \end{align*}

    \case{$\tau=\top$}
    \begin{align*}
       & \phantom{{}={}}\semantics{\heapRef_1(\chompRef_1(\top, y, \bitvar{0}), \bitvar{0},x,\inst,1)}_{\env'[y \mapsto h']} \\
       & =\semantics{\top}_{\env'[y \mapsto h']}                                                                             \\
       & = H                                                                                                                 \\
       & =\semantics{\top}_{\env}
    \end{align*}

    \case{$\tau=\sigmaT{z}{\tau_1}{\tau_2}$}
    \begin{align*}
       & \phantom{{}={}}\semantics{\heapRef_1(\chompRef_1(\sigmaT{z}{\tau_1}{\tau_2}, y, \bitvar{0}),\bitvar{0},x,\inst,1)}_{\env'[y \mapsto h']}    \\
       & =\semantics{\sigmaT{z}{\heapRef_1(\chompRef_1(\tau_1,x,\bitvar{0}),\bitvar{0},x,\inst,1)}{                                                  \\
       & \phantom{{}=\llbracket\Sigma z:\!!}\heapRef_1(\chompRef_1(\tau_2,x,\bitvar{0}),\bitvar{0},x,\inst,1)}}_{\env'[y\mapsto h']}                 \\
       & =\{\concat{h_1'}{h_2'}\mid h_1'\in\semantics{\heapRef_1(\chompRef_1(\tau_1,x,\bitvar{0}), \bitvar{0},x,\inst,1)}_{\env'[y\mapsto h']}~\land \\
       & \hspace{58pt}h_2'\in\semantics{\heapRef_1( \chompRef_1(\tau_2,x,\bitvar{0}), \bitvar{0},x,\inst,1)}_{\env'[y\mapsto h',z\mapsto h_1']} \}   \\
       & =\{\concat{h_1'}{h_2'}\mid h_1'\in\semantics{\tau_1}_{\env[y\mapsto h]} \land h_2'\in\semantics{\tau_2}_{\env[y\mapsto h,z\mapsto h_1']}\}  \\
       & =\semantics{\sigmaT{z}{\tau_1}{\tau_2}}_{\env[y\mapsto h]}
    \end{align*}

    \case{$\tau=\tau_1+\tau_2$}
    \begin{align*}
       & \phantom{{}={}}\semantics{\heapRef_1(\chompRef_1(\tau_1+\tau_2, y, \bitvar{0}), \bitvar{0},x,\inst,1)}_{\env'[y \mapsto h']} \\
       & =\semantics{\heapRef_1(\chompRef_1(\tau_1,x,\bitvar{0}),\bitvar{0},x,\inst,1)~+                                              \\
       & \hspace{1.5em}\heapRef_1(\chompRef_1(\tau_2,x,\bitvar{0}),\bitvar{0},x,\inst,1)}_{\env'[y\mapsto h']}                        \\
       & =\semantics{\heapRef_1(\chompRef_1(\tau_1,x,\bitvar{0}),\bitvar{0},x,\inst,1)}_{\env'[y\mapsto h']}~\cup                     \\
       & \phantom{{}={}}\semantics{\heapRef_1(\chompRef_1(\tau_2,x,\bitvar{0}),\bitvar{0},x,\inst,1)}_{\env'[y\mapsto h']}            \\
       & =\semantics{\tau_1}_{\env[y\mapsto h]}\cup\semantics{\tau_1}_{\env[y\mapsto h]}                                              \\
       & =\semantics{\tau_1+\tau_2}_{\env[y\mapsto h]}
    \end{align*}

    \case{$\tau=\refT{z}{\tau_1}{\varphi}$}
    \begin{align*}
       & \phantom{{}={}}\semantics{\heapRef_1(\chompRef_1(\refT{z}{\tau_1}{\varphi}, y, \bitvar{0}),\bitvar{0},x,\inst,1)}_{\env'[y \mapsto h']} \\
       & =\semantics{\refT{z}{\heapRef_1(\chompRef_1(\tau_1,y,\bitvar{0}),\bitvar{0},x,\inst,1)}{                                                \\
       & \hspace{34pt}\heapRef_1(\chompForm_1(\varphi,y,\bitvar{0}),\bitvar{0},x,\inst,1)}}_{\env'[y \mapsto h']}                                \\
       & =\{h\mid h\in\semantics{\heapRef_1(\chompRef_1(\tau_1,y,\bitvar{0}), \bitvar{0},x,\inst,1)}_{\env'[y \mapsto h']}~\land                 \\
       & \hspace{46pt}\semantics{\heapRef_1(\chompForm_1(\varphi,y,\bitvar{0}), \bitvar{0},x,\inst,1)}_{\env'[y \mapsto h']}=\true \}            \\
       & =\{h\mid h\in\semantics{\tau_1}_{\env[y\mapsto h]} \land \semantics{\varphi}_{\env[y\mapsto h]}=\true\}                                 \\
       & =\semantics{\refT{z}{\tau_1}{\varphi}}_{\env[y\mapsto h]}
    \end{align*}

    \case{$\tau=\tau_1{[z\mapsto\tau_2]}$}
    \begin{align*}
       & \phantom{{}={}}\semantics{\heapRef_1(\chompRef_1(\tau_1[z\mapsto\tau_2], y, \bitvar{0}),\bitvar{0},x,\inst,1)}_{\env'[y \mapsto h']}   \\
       & =\semantics{\heapRef_1(\chompRef_1(\tau_1, y, \bitvar{0}), \bitvar{0},x,\inst,1)[z\mapsto                                              \\
       & \hspace{40pt}\heapRef_1(\chompRef_1(\tau_2, y, \bitvar{0}), \bitvar{0},x,\inst,1)]}_{\env'[y \mapsto h']}                              \\
       & =\{ h\mid h_2\in\semantics{\heapRef_1(\chompRef_1(\tau_2, y, \bitvar{0}),\bitvar{0},x,\inst,1)}_{\env'[y \mapsto h']} \land            \\
       & \hspace{30pt} h\in\semantics{\heapRef_1(\chompRef_1(\tau_1, y, \bitvar{0}),\bitvar{0},x,\inst,1}_{\env'[y \mapsto h',z\mapsto h_2]} \} \\
       & =\{h \mid h_2\in\semantics{\tau_2}_{\env[y\mapsto h]} \land h\in\semantics{\tau_1}_{\env[y\mapsto h,z\mapsto h_2]}\}                   \\
       & =\semantics{\tau_1[z\mapsto\tau_2]}_{\env[y\mapsto h]}
    \end{align*}
  \end{description}
\end{proof}

\begin{lemma}[Semantic Chomp\textsubscript{1}]
  \label{lem:semantic-chomp1}
  For all heap types $\tau$, environments $\env$ and $\env'$, and variables $x$,
  if $\env'=\env[x \mapsto (\bvNil, \bvNil, [\inst\mapsto v])]$, and
  if $x\in\dom(\env)$, $v=\bvconcat{\env(x)(\inst)}{\slice{h(\pIn)}{0}{1}}$ and $\env(x)(\pIn)=\env(x)(\pOut)=\bvNil$, otherwise $v=\slice{h(\pIn)}{0}{1}$, then
  $\forall h\in\semantics{\tau}_\env.\length{h(\pIn)}\geq 1\implies \exists h'\in\semantics{\chompRec(\tau,1,\inst,x)}_{\env'}.h'=\chompS(h,1)$
\end{lemma}
\begin{proof}
  \begin{align*}
     & \phantom{{}\Leftrightarrow{}}\forall h\in\semantics{\tau}_\env.\length{h(\pIn)}\geq 1\implies \exists h'\in\semantics{\chompRec(\tau,1,\inst,x)}_{\env'}.h'=\chompS(h,1) \\
     & \Leftrightarrow (\text{By definition of } \chompRec)                                                                                                                     \\
     & \phantom{{}\Leftrightarrow{}}\forall h\in\semantics{\tau}_\env.\length{h(\pIn)}\geq 1\implies                                                                            \\
     & \hspace{63pt}\exists h'\in\semantics{\heapRef_1(\chomp_1(\tau,\bitvar{0}),\bitvar{0},x,\inst,1)}_{\env'}.h'=\chompS(h,1)
  \end{align*}
  Proof by induction on $\tau$.
  \begin{description}
    \case{$\tau=\varnothing$}
    $\semantics{\varnothing}_\env = \{\}$. As there are no heaps in the semantics, the case holds.

    \case{$\tau=\top$}
    Let $h$ be some heap from $\semantics{\top}_\env = H$.
    Let $h'=h$ except that $h'(\pIn)=\slice{h(\pIn)}{1}{~}$.

    By definition of $\chomp_1$ and $\heapRef_1$, $\heapRef_1(\chomp_1(\top, \bitvar{0}), \bitvar{0}, x, \inst, 1) = \top$
    and $\semantics{\top}_{\env'} = H$.

    We can conclude that $h' \in \semantics{\heapRef_1(\chomp_1(\top, \bitvar{0}), \bitvar{0}, x, \inst, 1)}_{\env'}$
    and $h' = \chompS(h, 1)$ follows by construction of $h'(\pIn)=\slice{h(\pIn)}{1}{~}$.

    \case{$\tau=\sigmaT{y}{\tau_1}{\tau_2}$}
    Let $h$ be some heap from $\semantics{\sigmaT{y}{\tau_1}{\tau_2}}_\env$.
    We know $h = \concat{h1}{h2}$, for some $h_1 \in \semantics{\tau_1}_\env$ and some $h_2 \in \semantics{\tau_2}_{\env[y \mapsto h1]}$.

    We have to show that there exists some
    \[
      h'\in\semantics{\heapRef_1(\chomp_1(\sigmaT{y}{\tau_1}{\tau_2}, \bitvar{0}),\bitvar{0},x,\inst,1)}_{\env'}
    \] such that $h'=\chompS(h,1)$.

    We deconstruct $\semantics{\heapRef_1(\chomp_1(\tau, \bitvar{0}),\bitvar{0},x,\inst,1)}_{\env'}$:
    \begin{align*}
       & \phantom{{}={}} \semantics{\heapRef_1(\chomp_1(\sigmaT{y}{\tau_1}{\tau_2}, \bitvar{0}),\bitvar{0},x,\inst,1)}_{\env'}                              \\
       & = \semantics{\heapRef_1(\sigmaT{y}{\chomp_1(\tau_1,\bitvar{0})}{\chompRef_1(\tau_2, y, \bitvar{0})}\ +                                             \\
       & \phantom{{}=\llbracket{}}\sigmaT{y}{\refT{z}{\tau_1}{\length{z.\pIn}=0}}{\chomp_1(\tau_2,\bitvar{0})}, \bitvar{0},x,\inst,1)}_{\env'}              \\
       & = \semantics{\sigmaT{y}{\heapRef_1(\chomp_1(\tau_1, \bitvar{0}),\bitvar{0},x,\inst,1)}{                                                            \\
       & \hspace{32pt} \heapRef_1(\chompRef_1(\tau_2, x, \bitvar{0}),\bitvar{0},x,\inst,1)}\ +                                                              \\
       & \phantom{{}=\llbracket} \sigmaT{y}{\heapRef_1(\refT{z}{\tau_1}{\length{z.\pIn}=0}, \bitvar{0},x,\inst,1)}{                                         \\
       & \hspace{32pt} \heapRef_1(\chomp_1(\tau_2,\bitvar{0}), \bitvar{0},x,\inst,1)}}_{\env'}                                                              \\
       & = \semantics{\sigmaT{y}{\heapRef_1(\chomp_1(\tau_1, \bitvar{0}), \bitvar{0},x,\inst,1)}{                                                           \\
       & \hspace{32pt} \heapRef_1(\chompRef_1(\tau_2, x, \bitvar{0}),\bitvar{0},x,\inst,1)}}_{\env'}\ \cup                                                  \\
       & \phantom{{}={}} \semantics{\sigmaT{y}{\refT{z}{\tau_1}{\length{z.\pIn}=0}}{\heapRef_1(\chomp_1(\tau_2,\bitvar{0}), \bitvar{0},x,\inst,1)}}_{\env'} \\
       & = \{ \concat{h_1'}{h_2'} | h_1' \in \semantics{\heapRef_1(\chomp_1(\tau_1, \bitvar{0}), \bitvar{0},x,\inst,1)}_{\env'}~\land                       \\
       & \hspace{52pt}h_2' \in \semantics{\heapRef_1(\chompRef_1(\tau_2, x, \bitvar{0}))}_{\env'[y\mapsto h_1']}\}~\cup                                     \\
       & \phantom{{}={}} \{ \concat{h_1'}{h_2'} | h_1' \in \semantics{\refT{z}{\tau_1}{\length{z.\pIn}=0}}_{\env'}~\land                                    \\
       & \hspace{52pt}h_2' \in \semantics{\heapRef_1(\chomp_1(\tau_2,\bitvar{0}), \bitvar{0},x,\inst,1)}_{\env'[y\mapsto h_1']}\}
    \end{align*}

    By case distinction on the length of $\pIn$ in $h_1$.
    \begin{description}
      \subcase{$\length{h_1(\pIn)}=0$}~

      By definition of $\env'$,
      $\semantics{\tau_1}_\env=\semantics{\tau_1}_{\env'}$, because, $\tau_1$ can't contain a reference to the newly added bit in $\env'(x)(\inst)$, from which follows that $h_1\in\semantics{\tau_1}_{\env'}$.

      By semantics of heap types, $h_1 \in \semantics{\refT{z}{\tau_1}{\length{z.\pIn}=0}}_{\env'}$.

      By IH there exists
      \[
        h_2' \in \semantics{\heapRef_1(\chomp_1(\tau_2,\bitvar{0}), \bitvar{0},x,\inst,1)}_{\env'[y\mapsto h_1]}
      \] such that $h_2'=\chompS(h_2,1)$.

      Let $h' = \concat{h_1}{h_2'}$.

      We conclude that
      \begin{equation*}
        \begin{split}
          h' \in \{ \concat{h_1'}{h_2'} | &h_1' \in \semantics{\refT{z}{\tau_1}{\length{z.\pIn}=0}}_{\env'}~\land \\
          & h_2' \in \semantics{\heapRef_1(\chomp_1(\tau_2,\bitvar{0}), \bitvar{0},x,\inst,1)}_{\env'[y\mapsto h_1']}\}
        \end{split}
      \end{equation*}
      and thus
      $h' \in \semantics{\heapRef_1(\chomp_1(\sigmaT{y}{\tau_1}{\tau_2}, \bitvar{0}),\bitvar{0},x,\inst,1)}_{\env'}$.

      By assumption $\length{h_1(\pIn)}=0$, we can conclude that
      \begin{equation*}
        \concat{h_1}{\chompS(h_2,1)}=\chompS(\concat{h_1}{h_2},1)
      \end{equation*}

      thus
      \begin{equation*}
        \begin{split}
          h' &= \concat{h_1}{h_2'}\\
          &=\concat{h_1}{\chompS(h_2,1)} \\
          &=\chompS(\concat{h_1}{h_2},1) \\
          &=\chompS(h,1)
        \end{split}
      \end{equation*}

      \subcase{$\length{h_1(\pIn)}\neq 0$}~

      By IH there exists $h_1' \in \semantics{\heapRef_1(\chomp_1(\tau_1, \bitvar{0}), \bitvar{0},x,\inst,1)}_{\env'}$, such that 
      
      $h_1'=\chompS(h_1,1)$.

      By Lemma \ref{lem:semantic-chompref1} follows that
      \[
        h_2 \in \semantics{\heapRef_1(\chompRef_1(\tau_2, x, \bitvar{0}),\bitvar{0},x,\inst,1)}_{\env'[y\mapsto h_1']}
      \]

      Let $h' = \concat{h_1'}{h_2}$.

      We conclude that
      \begin{equation*}
        \begin{split}
          h' \in \{ \concat{h_1'}{h_2} | & h_1' \in \semantics{\heapRef_1(\chomp_1(\tau_1, \bitvar{0}), \bitvar{0},x,\inst,1)}_{\env'}\ \land \\
          & h_2 \in \semantics{\heapRef_1(\chompRef_1(\tau_2, x, \bitvar{0}))}_{\env'[y\mapsto h_1']}\}
        \end{split}
      \end{equation*}
      and thus $h' \in \semantics{\heapRef_1(\chomp_1(\sigmaT{y}{\tau_1}{\tau_2}, \bitvar{0}),\bitvar{0},x,\inst,1)}_{\env'}$.

      With
      \begin{equation*}
        \begin{split}
          h' &= \concat{h_1'}{h_2} \\
          &= \concat{\chompS(h_1,1)}{h_2} \\
          &= \chompS(\concat{h_1}{h_2},1) \\
          &= \chompS(h,1)
        \end{split}
      \end{equation*}
      we can conclude this case.
    \end{description}

    \case{$\tau=\tau_1+\tau_2$}~

    Let $h$ be some heap from $\semantics{\tau_1+\tau_2}_\env$.
    By the semantics of heap types, we know $h \in \semantics{\tau_1}_\env$ or $h \in \semantics{\tau_2}_\env$.
    \begin{description}
      \subcase{$h \in \semantics{\tau_1}_\env$}~

      By IH we know that there exists
      \[
        h' \in \semantics{\heapRef_1(\chomp_1(\tau_1, \bitvar{0}),\bitvar{0},x,\inst,1)}_{\env'}
      \]
      such that $h'=\chompS(h,1)$.

      By set theory and
      \begin{align*}
         & \phantom{{}={}}\semantics{\heapRef_1(\chomp_1(\tau_1+\tau_2, \bitvar{0}),\bitvar{0},x,\inst,1)}_{\env'}          \\
         & = \semantics{\heapRef_1(\chomp_1(\tau_1, \bitvar{0})+\chomp_1(\tau_2, \bitvar{0}),\bitvar{0},x,\inst,1)}_{\env'} \\
         & = \semantics{\heapRef_1(\chomp_1(\tau_1, \bitvar{0}),\bitvar{0},x,\inst,1)\ +                                    \\
         & \hspace{14.5pt} \heapRef_1(\chomp_1(\tau_2, \bitvar{0}),\bitvar{0},x,\inst,1)}_{\env'}                           \\
         & = \semantics{\heapRef_1(\chomp_1(\tau_1, \bitvar{0}),\bitvar{0},x,\inst,1)}_{\env'}~\cup                         \\
         & \phantom{{}={}}\semantics{\heapRef_1(\chomp_1(\tau_2, \bitvar{0}),\bitvar{0},x,\inst,1)}_{\env'}
      \end{align*}

      we conclude $h' \in \semantics{\heapRef_1(\chomp_1(\tau_1+\tau_2, \bitvar{0}),\bitvar{0},x,\inst,1)}_{\env'}$.
      \item[\textit{Subcase} $h \in \semantics{\tau_2}_\env$:] Symmetric to previous subcase.
    \end{description}

    \case{$\tau=\refT{y}{\tau_1}{\varphi}$}~

    Let $h$ be some heap from $\semantics{\refT{y}{\tau_1}{\varphi}}_\env$.

    By the semantics of heap types, we know that $h \in \semantics{\tau_1}_\env$ and $\semantics{\varphi}_{\env[y \mapsto h]} = \true$.
    By induction hypothesis there exists $h' \in \semantics{\heapRef_1(\chomp_1(\tau_1, \bitvar{0}), \bitvar{0},x,\inst,1)}_{\env'}$ such that $h'=\chompS(h,1)$.

    By \cref{lem:semantic-chompe} we know that
    \[
      \semantics{\varphi}_{\env[y\mapsto h]} = \semantics{\heapRef_1(\chompForm_1(\varphi, y, \bitvar{0}), \bitvar{0},x,\inst,1)}_{\env'[y \mapsto h']}
    \]
    To apply \cref{lem:semantic-chompe}, we must show that $x$ is not free in $\varphi$, if $x\not\in\dom(\env)$.
    If this does not hold, there is no $h\in\semantics{\tau}_\env$, which violates our initial assumption.
    With
    \begin{align*}
       & \phantom{{}={}} \semantics{\heapRef_1(\chomp_1(\refT{y}{\tau_1}{\varphi}, \bitvar{0}), \bitvar{0},x,\inst,1)}_{\env'}              \\
       & = \semantics{\heapRef_1(\refT{y}{\chomp_1(\tau_1, \bitvar{0})}{\chompForm_1(\varphi,y,\bitvar{0})}, \bitvar{0},x,\inst,1)}_{\env'} \\
       & = \semantics{\refT{y}{\heapRef_1(\chomp_1(\tau_1, \bitvar{0}), \bitvar{0},x,\inst,1)}{                                             \\
       & \hspace{32pt} \heapRef_1(\chompForm_1(\varphi, y, \bitvar{0}), \bitvar{0},x,\inst,1)}}_{\env'}                                     \\
       & = \{ h | h \in \semantics{\heapRef_1(\chomp_1(\tau_1, \bitvar{0}), \bitvar{0},x,\inst,1)}_{\env'}\ \land                           \\
       & \hspace{38.5pt}\semantics{\heapRef_1(\chompForm_1(\varphi, y, \bitvar{0}), \bitvar{0},x,\inst,1)}_{\env'[y \mapsto h]}\}
    \end{align*}

    and with our assumptions from Lemma \ref{lem:semantic-chompe} and the induction hypothesis, we conclude that
    $h' \in \semantics{\heapRef_1(\chomp_1(\refT{y}{\tau_1}{\varphi}, \bitvar{0}), \bitvar{0},x,\inst,1)}_{\env'}$ such that $h'=\chompS(h,1)$.

    \case{$\tau=\tau_1{[}y\mapsto\tau_2{]}$}~

    Let $h$ be some heap from $\semantics{\tau_1[y \mapsto \tau_2]}_\env$.
    We know that $h \in \semantics{\tau_1}_{\env[y \mapsto h_2]}$ for some $h_2 \in \semantics{\tau_2}_\env$.

    By IH there exists some $h' \in \semantics{\heapRef_1(\chomp_1(\tau_1, \bitvar{0}), \bitvar{0},x,\inst,1)}_{\env'[y \mapsto h_2]}$
    such that $h'=\chompS(h,1)$.

    To conclude this case, we must show that
    \[
      h'\in\semantics{\heapRef_1(\chomp_1(\tau_1[y\mapsto\tau_2], \bitvar{0}), \bitvar{0},x,\inst,1)}_{\env'}
    \]
    From $\heapRef_1$, $\chomp_1$ and the semantics of heap types, we get:
    \begin{align*}
       & \phantom{{}={}} \semantics{\heapRef_1(\chomp_1(\tau_1[y\mapsto\tau_2], \bitvar{0}), \bitvar{0},x,\inst,1)}_{\env'}                                                          \\
       & = \semantics{\heapRef_1(\chomp_1(\tau_1, \bitvar{0})[y\mapsto\tau_2], \bitvar{0},x,\inst,1)}_{\env'}                                                                        \\
       & = \semantics{\heapRef_1(\chomp_1(\tau_1, \bitvar{0}), \bitvar{0},x,\inst,1)[y\mapsto\tau_2]}_{\env'}                                                                        \\
       & = \{ h_{11} | h_{22} \in \semantics{\tau_2}_{\env'} \land h_{11} \in \semantics{\heapRef_1(\chomp_1(\tau_1, \bitvar{0}), \bitvar{0},x,\inst,1)}_{\env'[y\mapsto h_{22}]} \}
    \end{align*}

    With $h_{11}=h'$ and $h_{22}=h_2$, we can conclude that
    \[
      h'\in\semantics{\heapRef_1(\chomp_1(\tau_1[y\mapsto\tau_2], \bitvar{0}), \bitvar{0},x,\inst,1)}_{\env'}
    \]

    To argue that $h_2\in\semantics{\tau_2}_{\env'}$, we make a case distinction on $x\in\dom(\env)$.
    \begin{description}
      \subcase{$x\not\in\dom(\env)$}
      $x$ cannot appear free in $\tau$ and thereby also not in $\tau_2$, otherwise there would be no $h\in\semantics{\tau}_\env$, thus for all $h_2\in\semantics{\tau_2}_\env$, $h_2\in\semantics{\tau_2}_{\env'}$.

      \subcase{$x\in\dom(\env)$}
      By assumption, there is some $h_2\in\semantics{\tau_2}_\env$.
      By semantics of heap types, all formulae in $\tau_2$ referencing $x$, evaluate to true, i.e., they only refer to information contained in $\env(x)$.
      By assumption, $\env(x)(\pIn)=\env'(x)(\pIn)$, $\env(x)(\pOut)=\env'(x)(\pOut)$, and
      $\bvconcat{\env(x)(\inst)}{\slice{h(\pIn)}{0}{1}}=\env'(x)(\inst)$.
      Since all information of $\env(x)$ is preserved in $\env'$, $h_2 \in \semantics{\tau_2}_{\env'}$.
    \end{description}
  \end{description}
\end{proof}

\begin{lemma}[ChompRec Unroll]
  \label{lem:chomprec-unfold}
  For all instances $\inst$, if $m+1\leq\sizeof(\inst)$,
  then
  \[
    \chompRec(\chompRec(\tau,m,x,\inst),1,x,\inst)=\chompRec(\tau,m+1,x,\inst)
  \]
\end{lemma}
\begin{proof}
  By induction on $n$.
  \begin{description}
    \case{$n=0$}
    \begin{align*}
       & \phantom{{}={}}\chompRec(\chompRec(\tau,0,x,\inst),1,x,\inst) \\
       & = \chompRec(\tau,1,x,\inst)
    \end{align*}

    \case{$n=1$}
    \begin{align*}
       & \phantom{{}={}}\chompRec(\chompRec(\tau,1,x,\inst),1,x,\inst)                                       \\
       & =\chompRec(\chompRec(\heapRef_1(\chomp_1(\tau,\bitvar{0}),\bitvar{0},\inst,1),0,x,\inst),1,x,\inst) \\
       & =\chompRec(\heapRef_1(\chomp_1(\tau,\bitvar{0}),\bitvar{0},\inst,1),1,x,\inst)                      \\
       & =\chompRec(\tau,2,x,\inst)
    \end{align*}

    \case{$n=m$}~

    We assume that the lemma holds for $n=m$.
    We now have to show that the lemma also holds for $n=m+1$.
    \begin{align*}
       & \phantom{{}={}}\chompRec(\chompRec(\tau,m+1,x,\inst),1,x,\inst)                             \\
       & =\chompRec(\chompRec(\heapRef_1(\chomp_1(\tau,\bitvar{0}),\bitvar{0},\inst,m+1),m,x,\inst), \\
       & \hspace{58pt} 1,x,\inst)                                                                    \\
       & =\chompRec(\heapRef_1(\chomp_1(\tau,\bitvar{0}),\bitvar{0},\inst,m+1),m+1,x,\inst)          \\
       & =\chompRec(\tau,m+2,x,\inst)
    \end{align*}
  \end{description}
\end{proof}

\begin{lemma}[Semantic Chomp Unroll]
  \label{lem:semantic-chomp-unroll}
  For all heaps $h$ and all $n\in\mathbb{N}$, if $\length{h(\pIn)}\geq n+1$, then $\chompS(\chompS(h,n),1) = \chompS(h,n+1)$
\end{lemma}
\begin{proof}
  By definition of $\chompS$,
  \begin{align*}
     & \phantom{{}={}}\chompS(\chompS(h,n),1)                                                                         \\
     & =h[\pIn\mapsto \slice{h(\pIn)}{n}{\ }][\pIn\mapsto \slice{h[\pIn\mapsto \slice{h(\pIn)}{n}{\ }](\pIn)}{1}{\ }] \\
     & =h[\pIn\mapsto \slice{h[\pIn\mapsto \slice{h(\pIn)}{n}{\ }](\pIn)}{1}{\ }]                                     \\
     & =h[\pIn\mapsto \slice{h(\pIn)}{n+1}{\ }]                                                                       \\
     & =\chompS(h,n+1)
  \end{align*}
\end{proof}

\begin{lemma}[Chomp Slice]
  \label{lem:chomp-slice}
  For all heaps $h$ and all $n\in\tNat$, if $\length{h(\pIn)}\geq n+1$, then
  \[ \slice{\chompS(h,n)(\pIn)}{0}{1}=\slice{h(\pIn)}{n}{n+1} \]
\end{lemma}
\begin{proof}
  By definition of $\chompS$,
  \[
    \phantom{{}={}}\slice{\chompS(h,n)(\pIn)}{0}{1}=\slice{h[\pIn\mapsto\slice{h(\pIn)}{n}{\ }](\pIn)}{0}{1}
  \]

  Let $bv=h(\pIn)=\langle b_0,...,b_n,...,b_m\rangle$.

  Let $bv'$ be the bit vector we obtain after removing the first $n$ bits from $bv'$, $bv'=\langle b_n,...,b_m\rangle$.

  Accessing the first bit of $bv'$ gives us bit $b_n$, which is also the n-th bit in $bv$, i.e., $\slice{bv}{n}{n+1}$.
\end{proof}

\begin{lemma}[Semantic ChompRec]
  \label{lem:semantic-chomprec}
  \todo{Fix overflow}
  For all heap types $\tau$, environments $\env$ and $\env'$, variables $x$ and $n\in\mathbb{N}$,
  if $x$ does not appear free in $\tau$, and $\env'=\env[x \mapsto (\bvNil, \bvNil, [\inst \mapsto \slice{h(\pIn)}{0}{n}])]$,
  then $\forall h\in\semantics{\tau}_\env.\length{h(\pIn)}\geq n\implies \exists h'\in\semantics{\chompRec(\tau,n,\inst,x)}_{\env'}.h'=\chompS(h,n)$.
\end{lemma}
\begin{proof}
  Proof by induction on $n$.
  \begin{description}

    \case{$n=0$}
    \begin{align*}
       & \phantom{{}\Leftrightarrow{}}\forall h\in\semantics{\tau}_\env.\length{h(\pIn)}\geq 0\implies \exists h'\in\semantics{\chompRec(\tau,0,\inst,x)}_{\env'}.h'=\chompS(h,0) \\
       & \Leftrightarrow\forall h\in\semantics{\tau}_\env.\exists h'\in\semantics{\tau}_{\env'}.h'=h
    \end{align*}
    Let $h'=h$, i.e., we have to show that $h\in\semantics{\tau}_{\env'}$.
    By assumption, $x$ is not free in $\tau$, i.e., the binding of $x$ in $\env'$ has no effect on the semantics of $\tau$.
    Since $\env$ and $\env'$ are otherwise identical, $\tau$ evaluated in both environments is described by the same set of heaps, from which we can conclude that $h\in\semantics{\tau}_{\env'}$.

    \case{$n=1$}
    \begin{equation*}
      \phantom{{}\Leftrightarrow{}}\forall h\in\semantics{\tau}_\env.\length{h(\pIn)}\geq 1\implies \exists h'\in\semantics{\chompRec(\tau,1,\inst,x)}_{\env'}.h'=\chompS(h,1)
    \end{equation*}
    The result directly follows by \cref{lem:semantic-chomp1}.

    \case{$n=m+1$}~

    We assume that the lemma holds for $n=m$, i.e.,
    \begin{equation*}
      \forall h\in\semantics{\tau}_\env.\length{h(\pIn)}\geq m\implies
      \exists h'\in\semantics{\chompRec(\tau,m,\inst,x)}_{\env'}.h'=\chompS(h,m)
    \end{equation*}
    We have to show that the lemma also holds for $n=m+1$, i.e.,
    \begin{equation*}
      \forall h\in\semantics{\tau}_\env.\length{h(\pIn)}\geq m+1\implies \exists h'\in\semantics{\chompRec(\tau,m+1,\inst,x)}_{\env'}.h'=\chompS(h,m+1)
    \end{equation*}

    where $\env'=\env[x\mapsto[\inst\mapsto \slice{h(\pIn)}{0}{m+1},\pIn\mapsto\bvNil,\pOut\mapsto\bvNil]]$.

    Let $h$ be some heap $h\in\semantics{\tau}_{\env_0}$.

    By induction hypothesis, there exists some $h'\in\semantics{\chompRec(\tau,m,\inst,x)}_{\env_0'}$, such that $h'=\chompS(h,m)$.

    Let $\env_1 = \env_0' = \env_0[x \mapsto (\bvNil, \bvNil, [\inst\mapsto \slice{h(\pIn)}{0}{m}])]$.
    We use (A) to refer to this assumption.

    By \cref{lem:semantic-chomp1}, for all $h_1\in\semantics{\chompRec(\tau,m,\inst,x)}_{\env_1}$, there exists some heap $h_1'$ such that $h_1'\in\semantics{\chompRec(\chompRec(\tau,m,\inst,x),1,\inst,x)}_{\env_1'}$ and $\env_1' = \env_1[x \mapsto (\bvNil, \bvNil, [\inst\mapsto\bvconcat{\env_1(x)(\inst)}{\slice{h_1(\pIn)}{0}{1}}])]$, and $h_1'=\chompS(h_1,1)$.

    Since, by assumption, $h'\in\semantics{\chompRec(\tau,m,\inst,x)}_{\env_0'}$ and also $\env_1=\env_0'$, we can define $h_1$ to be equal to $h'$, i.e., $h_1'=\chompS(h',1)$.

    From $h'=\chompS(h,m)$ follows $h_1'=\chompS(\chompS(h,m),1)$.

    From \cref{lem:semantic-chomp-unroll} also follows that $h_1'=\chompS(h,m+1)$.

    We must show that $h_1'\in\semantics{\chompRec(\tau,m+1,\inst,x)}_{\env'}$.

    We know that
    \[
      h_1'\in\semantics{\chompRec(\chompRec(\tau,m,\inst,x),1,\inst,x)}_{\env_1'}
    \] and by \cref{lem:chomprec-unfold}, $h_1'\in\semantics{\chompRec(\tau,m+1,\inst,x)}_{\env_1'}$, so we must show that $\env'=\env_1'$.

    By assumption, $\env_1' = \env_1[x \mapsto (\bvNil, \bvNil, [\inst\mapsto\bvconcat{\env_1(x)(\inst)}{\slice{h_1(\pIn)}{0}{1}}])]$, where $h_1=\chompS(h,m)$ (by IH).

    Also by assumption, $\env(x)(\inst)=h(\pIn)[0:m]$, i.e., $\env_1' = \env[x \mapsto (\bvNil, \bvNil, [\inst\mapsto \bvconcat{\slice{h(\pIn)}{0}{m}}{\slice{h_1(\pIn)}{0}{1}}])]$.

    Again, substituting $h_1$ with $h'$, and by $h'=\chompS(h,m)$, we obtain $\env_1'=\env[x\mapsto(\bvNil, \bvNil, [\inst\mapsto \bvconcat{\slice{h(\pIn)}{0}{m}}{\slice{\chompS(h,m)(\pIn)}{0}{1}}])]$.

    By \cref{lem:chomp-slice} and by definition of bit vector concatenation, $\env_1' = \env[x \mapsto (\bvNil, \bvNil, [\inst\mapsto \slice{h(\pIn)}{0}{m+1}])] = \env'$.
  \end{description}
\end{proof}

\begin{lemma}[Semantic Chomp]
  \label{lem:semantic-chomp}
  If $x$ does not appear free in $\tau$, then forall heaps $h\in\semantics{\tau}_\env$ where $\length{h(\pIn)}\geq \sizeof(\inst)$,
  there exists $h'\in\semantics{\chomp(\tau,\inst,x)}_{\env'}$ such that $h'=\chompS(h,\sizeof(\inst))$ where $\env'=\env[x\mapsto(\bvNil, \bvNil, [\inst\mapsto h(\pIn)[0:\sizeof(\inst)]])]$.
\end{lemma}
\begin{proof}
  By definition of $\chomp$, we know that
  \[
    \chomp(\tau,\inst,x) = \chompRec(\tau,\sizeof(\inst),x,\inst)
  \]
  The result follows from \cref{lem:semantic-chomprec}.
\end{proof}

\begin{lemma}[Semantic Chomp\textsubscript{1} Inverse]
  \label{lem:semantic-chomp1-inverse}
  For all $x$, $v$, $\tau$, $\env'$ and $h'$ such that
  $\env'(x) = (\bvNil, \bvNil, [\inst \mapsto v])$ and $h'\in\semantics{\chompRec(\tau,1,x,\inst)}_{\env'}$ and $x$ not free in $\tau$ and $\sizeof(v) \ge 1$,
  there exists $h$ and $\env$ such that,
  \begin{enumerate}[label=(\arabic*)]
    \item \label{semantic-chomp1-inverse-g1} $h\in\semantics{\tau}_\env$ and
    \item \label{semantic-chomp1-inverse-g2} $h' = \chompS(h, 1)$ and
  \end{enumerate}
  and
  \begin{enumerate}[start=3,label=(\arabic*)]
    \item \label{semantic-chomp1-inverse-g31} $\env = \env'\setminus x$ and
    \item \label{semantic-chomp1-inverse-g41} $v = h(\pIn)[0:1]$
  \end{enumerate}
  or
  \begin{enumerate}[start=3,label=(\arabic*)]
    \item \label{semantic-chomp1-inverse-g32} $x\in\dom(\env)$ and
    \item \label{semantic-chomp1-inverse-g42} $v = \bvconcat{\env(x)(\inst)}{\slice{h(\pIn)}{0}{1}}$ and
    \item \label{semantic-chomp1-inverse-g52} $\env = \env'[x \mapsto (\bvNil, \bvNil,[\inst \mapsto v[0:\sizeof(v) - 1 ]])]$
  \end{enumerate}
\end{lemma}
\begin{proof}
  We refer to the general assumptions as follows:
  \begin{enumerate}[label=(\Alph*)]
    \item \label{semantic-chomp1-inverse-x} $\env'(x) = (\bvNil, \bvNil, [\inst \mapsto v])$
    \item \label{semantic-chomp1-inverse-hprime} $h'\in\semantics{\chompRec(\tau,1,x,\inst)}_{\env'}$ and
    \item \label{semantic-chomp1-inverse-x-notfree} $x$ not free in $\tau$ and
    \item \label{semantic-chomp1-inverse-sizeof-v} $\sizeof(v) \ge 1$
  \end{enumerate}

  Proof by induction on $\tau$.
  By definition of $\chompRec$ follows that
  \[
    \semantics{\chompRec(\tau,1,x,\inst)}_\env = \semantics{\heapRef_1(\chomp_1(\tau,\bvar{0}),\bvar{0}, x,\inst,1)}_\env
  \]

  \begin{description}
    \case{$\tau = \varnothing$}
    \begin{align*}
       & \phantom{{}={}}\semantics{\heapRef_1(\chomp_1(\varnothing,\bvar{0}),\bvar{0}, x,\inst,1)}_{\env'} \\
       & = \semantics{\heapRef_1(\varnothing,\bvar{0}, x,\inst,1)}_{\env'}                                 \\
       & = \semantics{\varnothing}_{\env'}                                                                 \\
       & = \{\}
    \end{align*}
    As there is no $h'\in\semantics{\chompRec(\tau,1,\inst,x)}_{\env'}$, this case is immediate.

    \case{$\tau=\top$}
    \begin{align*}
       & \phantom{{}={}}\semantics{\heapRef_1(\chomp_1(\top,\bvar{0}),\bvar{0}, x,\inst,1)}_{\env'} \\
       & = \semantics{\heapRef_1(\top,\bvar{0}, x,\inst,1)}_{\env'}                                 \\
       & = \semantics{\top}_{\env'}                                                                 \\
       & = \Heaps
    \end{align*}
    Let $\env'$ where $\env'(x) = (\bvNil, \bvNil,[\inst \mapsto v])$ be arbitrary.

    Let $h'\in\semantics{\chompRec(\top,1,\inst,x)}_{\env'} = \Heaps$ be arbitrary.
    We have to distinguish two cases.
    \begin{description}
      \subcase{$\sizeof(v) = 1$}~

      Let $\env = \env'\setminus x$ and let $h=h'[\pIn \mapsto \bvconcat{v}{h'(\pIn)}]$, i.e., $\slice{h(\pIn)}{0}{1} = v$.
      \ref{semantic-chomp1-inverse-g1} follows by the semantics of heap types.
      \ref{semantic-chomp1-inverse-g2} follows by the definition of $\chompS$.
      \ref{semantic-chomp1-inverse-g31} and \ref{semantic-chomp1-inverse-g41} immediately follow from the definition of $h$ and $\env$.
      \subcase{$\sizeof(v) > 1$}~

      Let $\env = \env'[x \mapsto (\bvNil, \bvNil, [\inst \mapsto v[0:\sizeof(v) - 1]])]$ and let $h=h'[\pIn \mapsto \bvconcat{\slice{v}{\sizeof(v)-1}{\sizeof(v)}}{h'(\pIn)}]$.
      \ref{semantic-chomp1-inverse-g1} follows by the semantics of heap types.
      \ref{semantic-chomp1-inverse-g2} follows by the definition of $\chompS$
      \ref{semantic-chomp1-inverse-g31} and \ref{semantic-chomp1-inverse-g41} immediately follow from the definition of $h$ and $\env$
    \end{description}

    \case{$\tau = \tau_1 + \tau_2$}

    Let $\env'$ where $\env'(x) = (\bvNil, \bvNil,[\inst \mapsto v])$ be arbitrary.

    Let $h'\in\semantics{\chompRec(\tau_1 + \tau_2, 1, \inst,x)}_{\env'}$ be arbitrary.
    \begin{align*}
       & \phantom{{}={}}\semantics{\chompRec(\tau_1 + \tau_2, 1, \inst,x)}_{\env'}                            \\
       & = \semantics{\heapRef_1(\chomp_1(\tau_1 + \tau_2, \bvar{0}), \bvar{0}, x, \inst, 1)}_{\env'}         \\
       & = \semantics{\heapRef_1(\chomp_1(\tau_1, \bvar{0}), \bvar{0}, x, \inst, 1)}_{\env'}\ \cup            \\
       & \hspace{10.5pt} \semantics{\heapRef_1(\chomp_1(\tau_1, \bvar{0}), \bvar{0}, x, \inst, 1)}_{\env'}    \\
       & = \semantics{\chompRec(\tau_1,1,x\inst)}_{\env'} \cup \semantics{\chompRec(\tau_2,1,x\inst)}_{\env'}
    \end{align*}

    We have to distinguish two cases,
    \begin{enumerate}
      \item $h'\in\semantics{\chompRec(\tau_1,1,x,\inst)}_{\env'}$ and
      \item $h'\in\semantics{\chompRec(\tau_2,1,x\inst)}_{\env'}$.
    \end{enumerate}
    \begin{description}
      \subcase{$h'\in\semantics{\chompRec(\tau_1,1,x\inst)}_{\env'}$}
      We further distinguish between the size of $v$.
      \begin{description}
        \subcase{$\sizeof(v) = 1$}~

        By IH, there exists $h_1$, $\env_1$, such that
        \begin{pf}{Inverse-Chomp1-Refinment-Size1}
          \item \label{semantic-chomp-inverse} $h_1\in\semantics{\tau_1}_{\env_1}$
          \item $h' = \chompS(h_1, 1)$
          \item $\env_1 = \env' \setminus x$
          \item $v = \slice{h_1(\pIn)}{0}{1}$
        \end{pf}
        Let $h = h_1$ and $\env = \env_1$.
        $h\in\semantics{\tau_1 + \tau_2}_\env$ follows from the semantics of heap types and by (A1).
        The rest is immediate.

        \subcase{$\sizeof(v) > 1$}
        By IH, there exists $h_1$, $\env_1$, such that
        \begin{pf}{Inverse-Chomp1-Refinment-Size2}
          \item $h_1\in\semantics{\tau_1}_{\env_1}$
          \item $h' = \chompS(h_1, 1)$
          \item $\env_1 = \env'[x \mapsto (\bvNil, \bvNil, [\inst \mapsto \slice{v}{0}{\sizeof(v) - 1}])]$
          \item $v = \bvconcat{\env(x)(\inst)}{\slice{h_1(\pIn)}{0}{1}}$
        \end{pf}
        Let $h=h_1$ and $\env=\env_1$.
        $h\in\semantics{\tau_1 + \tau_2}_\env$ follows from the semantics of heap types and by (A1).
        The rest is immediate.
      \end{description}
      \subcase{$h'\in\semantics{\chompRec(\tau_2,1,x,\inst)}_{\env'}$}~

      Symmetric to previous subcase.
    \end{description}

    \case{$\tau = \sigmaT{y}{\tau_1}{\tau_2}$}
    Let $\env'$ where $\env'(x) = (\bvNil, \bvNil,[\inst \mapsto v])$ be arbitrary.
    Let $h'\in\semantics{\chompRec(\sigmaT{y}{\tau_1}{\tau_2}, 1, \inst,x)}_{\env'}$ be arbitrary.
    \begin{align*}
       & \phantom{{}={}}\semantics{\chompRec(\sigmaT{y}{\tau_1}{\tau_2}, 1, \inst,x)}_{\env'}                            \\
       & = \{ \concat{h_1'}{h_2'} | h_1'\in\semantics{\chompRec(\tau_1,1,\inst,x)}_{\env'}\ \wedge                       \\
       & \hspace{53pt}h_2'\in\semantics{\heapRef_1(\chompRef_1(\tau_2,x,\bvar{0}))}_{\env'[y \mapsto h_1']}\} ~\cup      \\
       & \phantom{{}={}}\{ \concat{h_1'}{h_2'} | h_1'\in\semantics{\refT{z}{\tau_1}{\length{z.\pIn} = 0}}_{\env'} \wedge \\
       & \hspace{53pt}h_2'\in\semantics{\chompRec(\tau_2,1,\inst,x)}_{\env'[y \mapsto h_1']}\}
    \end{align*}
    Case distinction on the membership of $h'$.
    \begin{description}
      \subcase{$h'\ \mathit{contained\ in\ the\ first\ subset}$}~

      \begin{pf}{Inverse-Chomp1-Sigma}
        \item $h' = \concat{h_1'}{h_2'}$ and
        \item $h_1'\in \semantics{\chompRec(\tau_1,1,\inst,x)}_{\env'}$ and
        \item $h_2'\in\semantics{\heapRef_1(\chompRef_1(\tau_2,x,\bvar{0}))}_{\env'[y \mapsto h_1']}$.
      \end{pf}
      We distinguish two additional cases.
      \begin{description}
        \subcase{$\sizeof(v) = 1$}
        By IH, there exists $h_1$, $\env_1$, such that
        \begin{pf*}{Inverse-Chomp1-Sigma}
          \item \label{semantic-chomp1-inverse-sigma1-veq1-h1} $h_1\in\semantics{\tau_1}_{\env_1}$
          \item $h_1' = \chompS(h_1, 1)$
          \item \label{semantic-chomp1-inverse-sigma1-veq1-env1} $\env_1 = \env' \setminus x$
          \item \label{semantic-chomp1-inverse-sigma1-veq1-v} $v = \slice{h_1(\pIn)}{0}{1}$
        \end{pf*}
        By \cref{lem:semantic-chompref1}, $h_2'\in\semantics{\tau_2}_{\env'[y \mapsto h_1]}$.
        Since $\concat{h_1'}{h_2'}$ is defined, i.e., they have disjoint sets of headers and $\chompS$ does not affect the validity of headers, $\concat{h_1}{h_2'}$ is defined.

        Let $h=\concat{h_1}{h_2'}$ and $\env = \env_1$.

        \ref{semantic-chomp1-inverse-g1} follows by \ref{semantic-chomp1-inverse-sigma1-veq1-h1} and $h_2'\in\semantics{\tau_2}_{\env_1[y \mapsto h_1]}$. The latter holds, because $x$ is not free in $\tau_2$ by assumption.

        To show \ref{semantic-chomp1-inverse-g2}, we must show that $\concat{h_1'}{h_2'} = \chompS(\concat{h_1}{h_2'}, 1) \Leftrightarrow \concat{\chompS(h_1, 1)}{h_2'} = \chompS(\concat{h_1}{h_2'}, 1)$.
        This equality holds, because chomping of one bit from the input packet of $h_1$ and then concatenating $h_2'$ yields the same heap as concatenating both heaps and then removing the first bit of the input packet.

        \ref{semantic-chomp1-inverse-g31} follows by \ref{semantic-chomp1-inverse-sigma1-veq1-env1} and \ref{semantic-chomp1-inverse-g41} follows by \ref{semantic-chomp1-inverse-sigma1-veq1-v}.
        \subcase{$\sizeof(v) > 1$}
        By IH, there exists $h_1$, $\env_1$, such that
        \begin{pf*}{Inverse-Chomp1-Sigma}
          \item \label{semantic-chomp1-inverse-sigma1-vgt1-h1} $h_1\in\semantics{\tau_1}_{\env_1}$
          \item $h_1' = \chompS(h_1, 1)$
          \item \label{semantic-chomp1-inverse-sigma1-vgt1-env1} $\env_1 = \env'[x \mapsto (\bvNil, \bvNil, [\inst \mapsto \slice{v}{0}{\sizeof(v)-1}])]$
          \item \label{semantic-chomp1-inverse-sigma1-vgt1-v} $v = \bvconcat{\env_1(x)(\inst)}{\slice{h_1(\pIn)}{0}{1}}$
        \end{pf*}
        By \cref{lem:semantic-chompref1}, $h_2'\in\semantics{\tau_2}_{\env'[y \mapsto h_1]}$.
        Since $\concat{h_1'}{h_2'}$ is defined, i.e., they have disjoint sets of headers and $\chompS$ does not affect the validity of headers, $\concat{h_1}{h_2'}$ is defined.

        Let $h=\concat{h_1}{h_2'}$ and $\env = \env_1$.

        \ref{semantic-chomp1-inverse-g1} follows by \ref{semantic-chomp1-inverse-sigma1-vgt1-h1} and $h_2'\in\semantics{\tau_2}_{\env_1[y \mapsto h_1]}$. The latter holds, because $x$ is not free in $\tau_2$ by assumption.

        To show \ref{semantic-chomp1-inverse-g2}, we must show that $\concat{h_1'}{h_2'} = \chompS(\concat{h_1}{h_2'}, 1) \Leftrightarrow \concat{\chompS(h_1, 1)}{h_2'} = \chompS(\concat{h_1}{h_2'}, 1)$.
        This equality holds, because chomping of one bit from the input packet of $h_1$ and then concatenating $h_2'$ yields the same heap as concatenating both heaps and then removing the first bit of the input packet.

        \ref{semantic-chomp1-inverse-g32} follows by \ref{semantic-chomp1-inverse-sigma1-vgt1-env1} and \ref{semantic-chomp1-inverse-g42} follows by \ref{semantic-chomp1-inverse-sigma1-vgt1-v} and \ref{semantic-chomp1-inverse-g52} follows by \ref{semantic-chomp1-inverse-sigma1-vgt1-env1}.
      \end{description}

      \subcase{$h'\ \mathit{contained\ in\ the\ second\ subset}$}~

      \begin{pf}{Inverse-Chomp1-Sigma}
        \item $h' = \concat{h_1'}{h_2'}$ and
        \item \label{semantic-chomp1-inverse-sigma2-h1prime-empty} $h_1'\in \semantics{\refT{z}{\tau_1}{\length{z.\pIn} = 0}}_{\env'}$ and
        \item $h_2'\in\semantics{\chompRec(\tau_2,1,\inst,x)}_{\env'[y \mapsto h_1']}$.
      \end{pf}
      We distinguish two cases.
      \begin{description}
        \subcase{$\sizeof(v) = 1$}
        By IH, for every $h_1'\in\semantics{\refT{z}{\tau}{\length{z.\pIn} = 0}}_{\env'}$, there exists $h_2$, $\env_2$, such that
        \begin{pf*}{Inverse-Chomp1-Sigma}
          \item \label{semantic-chomp1-inverse-sigma2-veq1-h2} $h_2\in\semantics{\tau_2}_{\env_2}$
          \item $h_2' = \chompS(h_2, 1)$
          \item \label{semantic-chomp1-inverse-sigma2-veq1-env2} $\env_2 = \env'[y \mapsto h_1'] \setminus x$
          \item \label{semantic-chomp1-inverse-sigma2-veq1-v} $v = \slice{h_2(\pIn)}{0}{1}$
        \end{pf*}
        Let $h=\concat{h_1'}{h_2}$ and $\env = \env_2 \setminus y$.
        We have to show that $h\in\semantics{\sigmaT{x}{\tau_1}{\tau_2}}_\env$.
        By assumption, $x$ is not free in $\tau_1$ and $\tau_2$.
        By \ref{semantic-chomp1-inverse-sigma2-h1prime-empty} and by the fact that $\env = \env'\setminus x$, $h_1'\in\semantics{\refT{z}{\tau_1}{\length{z.\pIn} = 0}}_{\env}$ and by subtyping, $h_1'\in\semantics{\tau_1}_{\env}$.

        \ref{semantic-chomp1-inverse-g1} follows together with \ref{semantic-chomp1-inverse-sigma2-veq1-h2}.

        To show \ref{semantic-chomp1-inverse-g2}, we must show that $\concat{h_1'}{h_2'} = \chompS(\concat{h_1'}{h_2},1) \Leftrightarrow \concat{h_1'}{\chompS(h_2, 1)} = \chompS(\concat{h_1'}{h_2},1)$.
        Since by \ref{semantic-chomp1-inverse-sigma2-h1prime-empty}, the input packet of $h_1'$ is empty, the input packet of both heaps are equal.

        \ref{semantic-chomp1-inverse-g31} follows by \ref{semantic-chomp1-inverse-sigma2-veq1-env2} and \ref{semantic-chomp1-inverse-g41} follows by \ref{semantic-chomp1-inverse-sigma2-veq1-v}.

        \subcase{$\sizeof(v) > 1$}
        By IH, for every $h_1'\in\semantics{\refT{z}{\tau}{\length{z.\pIn} = 0}}_{\env'}$, there exists $h_2$, $\env_2$, such that
        \begin{pf*}{Inverse-Chomp1-Sigma}
          \item \label{semantic-chomp1-inverse-sigma2-vgt1-h2} $h_2\in\semantics{\tau_2}_{\env_2}$
          \item $h_2' = \chompS(h_2, 1)$
          \item \label{semantic-chomp1-inverse-sigma2-vgt1-env2} $\env_2 = \env'[y \mapsto h_1', x \mapsto (\bvNil, \bvNil, [\inst \mapsto \slice{v}{0}{\sizeof(v) - 1}])]$
          \item \label{semantic-chomp1-inverse-sigma2-vgt1-v} $v = \bvconcat{\env_2(x)(\inst)}{\slice{h_2(\pIn)}{0}{1}}$
        \end{pf*}
        Let $h=\concat{h_1'}{h_2}$ and $\env = \env_2 \setminus y$.
        To show that $h\in\semantics{\sigmaT{x}{\tau_1}{\tau_2}}_\env$.
        By assumption that $x$ is not free in $\tau_1$ and by \ref{semantic-chomp1-inverse-sigma2-h1prime-empty}, $h_1'\in\semantics{\refT{z}{\tau_1}{\length{z.\pIn} = 0}}_{\env}$ and by subtyping, $h_1'\in\semantics{\tau_1}_{\env}$.

        \ref{semantic-chomp1-inverse-g1} follows together with \ref{semantic-chomp1-inverse-sigma2-vgt1-h2}.

        To show \ref{semantic-chomp1-inverse-g2}, we must show that $\concat{h_1'}{h_2'} = \chompS(\concat{h_1'}{h_2},1) \Leftrightarrow \concat{h_1'}{\chompS(h_2, 1)} = \chompS(\concat{h_1'}{h_2},1)$.
        Since by \ref{semantic-chomp1-inverse-sigma2-h1prime-empty}, the input packet of $h_1'$ is empty, the input packet of both heaps are equal.

        \ref{semantic-chomp1-inverse-g32} follows by \ref{semantic-chomp1-inverse-sigma2-vgt1-env2} and \ref{semantic-chomp1-inverse-g42} follows by \ref{semantic-chomp1-inverse-sigma2-vgt1-v} and \ref{semantic-chomp1-inverse-g52} follows by \ref{semantic-chomp1-inverse-sigma2-vgt1-env2}.
      \end{description}
    \end{description}

    \case{$\tau = \refT{y}{\tau_1}{\varphi}$}~

    Let $\env'$ where $\env'(x) = (\bvNil, \bvNil,[\inst \mapsto v])$ be arbitrary. \\
    Let $h'\in\semantics{\chompRec(\refT{y}{\tau_1}{\varphi}, 1, \inst,x)}_{\env'}$ be arbitrary.
    \begin{align*}
       & \phantom{{}={}}\semantics{\chompRec(\refT{y}{\tau_1}{\varphi}, 1, \inst,x)}_{\env'}                                     \\
       & = \semantics{\heapRef_1(\chomp_1(\refT{y}{\tau_1}{\varphi}, \bvar{0}), \bvar{0},x,\inst, 1)}_{\env'}                    \\
       & = \{h' | h'\in\semantics{\chompRec(\tau_1, 1, \inst, x)}_{\env'}\ \wedge                                                \\
       & \hspace{30pt}\semantics{\heapRef_1(\chompForm_1(\varphi, y, \bvar{0}), \bvar{0}, x, \inst, 1)}_{\env'[y \mapsto h']} \}
    \end{align*}
    We distinguish two cases.
    \begin{description}
      \subcase{$\sizeof(v) = 1$}~

      By IH, there exists $h_1$, $\env_1$, such that
      \begin{pf}{Inverse-Chomp1-Refine}
        \item \label{semantic-chomp1-inverse-refine-veq1-h1} $h_1\in\semantics{\tau_1}_{\env_1}$
        \item \label{semantic-chomp1-inverse-refine-veq1-hprime} $h' = \chompS(h_1, 1)$
        \item \label{semantic-chomp1-inverse-refine-veq1-env1} $\env_1 = \env' \setminus x$
        \item \label{semantic-chomp1-inverse-refine-veq1-v} $v = \slice{h_1(\pIn)}{0}{1}$
      \end{pf}
      Let $\env = \env_1$ and $h=h_1$.
      To show \ref{semantic-chomp1-inverse-g1}, we must show that $h_1\in\semantics{\refT{y}{\tau_1}{\varphi}}_{\env_1}$, which follows by \ref{semantic-chomp1-inverse-refine-veq1-h1} and \cref{lem:semantic-chompe}.

      \ref{semantic-chomp1-inverse-g2} follows by \ref{semantic-chomp1-inverse-refine-veq1-hprime}, \ref{semantic-chomp1-inverse-g31} follows by \ref{semantic-chomp1-inverse-refine-veq1-env1} and \ref{semantic-chomp1-inverse-g41} follows by \ref{semantic-chomp1-inverse-refine-veq1-v}.

      \subcase{$\sizeof(v) > 1$}~

      By IH, there exists $h_1$, $\env_1$, such that
      \begin{pf}{Inverse-Chomp1-Refine}
        \item \label{semantic-chomp1-inverse-refine-vgt1-h1} $h_1\in\semantics{\tau_1}_{\env_1}$
        \item \label{semantic-chomp1-inverse-refine-vgt1-hprime} $h' = \chompS(h_1, 1)$
        \item \label{semantic-chomp1-inverse-refine-vgt1-env1} $\env_1 = \env'[x \mapsto (\bvNil, \bvNil, [\inst \mapsto \slice{v}{0}{\sizeof(v) - 1}])]$
        \item \label{semantic-chomp1-inverse-refine-vgt1-v} $v = \bvconcat{\env_1(x)(\inst)}{\slice{h_1(\pIn)}{0}{1}}$
      \end{pf}
      Let $\env = \env_1$ and $h=h_1$.
      To show \ref{semantic-chomp1-inverse-g1}, we must show $h_1\in\semantics{\refT{y}{\tau_1}{\varphi}}_{\env_1}$, which follows by \ref{semantic-chomp1-inverse-refine-vgt1-h1} and \cref{lem:semantic-chompe}.

      \ref{semantic-chomp1-inverse-g2} follows by \ref{semantic-chomp1-inverse-refine-vgt1-hprime}, \ref{semantic-chomp1-inverse-g32} follows by \ref{semantic-chomp1-inverse-refine-vgt1-env1}, \ref{semantic-chomp1-inverse-g42} follows by \ref{semantic-chomp1-inverse-refine-vgt1-v} and \ref{semantic-chomp1-inverse-g52} follows by \ref{semantic-chomp1-inverse-refine-vgt1-env1}.
    \end{description}

    \case{$\tau = \tau_1\lbrack y \mapsto \tau_2\rbrack$}~

    Let $\env'$ where $\env'(x) = (\bvNil, \bvNil,[\inst \mapsto v])$ be arbitrary. \\
    Let $h'\in\semantics{\chompRec(\tau_1[y \mapsto \tau_2], 1, \inst,x)}_{\env'}$ be arbitrary.
    \begin{align*}
       & \phantom{{}={}}\semantics{\chompRec(\tau_1[y \mapsto \tau_2], 1, \inst,x)}_{\env'}                                                  \\
       & = \semantics{\heapRef_1(\chomp_1(\tau_1[y \mapsto \tau_2], \bvar{0}), \bvar{0}, x,\inst,1)}_{\env'}                                 \\
       & = \semantics{\heapRef_1(\chomp_1(\tau_1, \bvar{0})[y \mapsto \tau_2], \bvar{0}, x,\inst,1)}_{\env'}                                 \\
       & = \semantics{\heapRef_1(\chomp_1(\tau_1, \bvar{0}), \bvar{0}, x,\inst,1)[y \mapsto \heapRef_1(\tau_2, \bvar{0},x,\inst,1)]}_{\env'} \\
       & = \semantics{\heapRef_1(\chomp_1(\tau_1, \bvar{0}), \bvar{0}, x,\inst,1)[y \mapsto \tau_2]}_{\env'}                                 \\
       & = \{ h_1' | h_2\in\semantics{\tau_2}_{\env'} \wedge h_1'\in\semantics{\chompRec(\tau_1, 1, x, \inst)}_{\env'[y \mapsto h_2]} \}
    \end{align*}
    We distinguish two cases.
    \begin{description}
      \subcase{$\sizeof(v) = 1$}~

      By IH, for every $h_2\in\semantics{\tau_2}_{\env'}$, there exists $h_1$, $\env_1$, such that
      \begin{pf}{Inverse-Chomp1-Subst}
        \item \label{semantic-chomp1-inverse-subst-veq1-h1} $h_1\in\semantics{\tau_1}_{\env_1[y \mapsto h_2]}$
        \item \label{semantic-chomp1-inverse-subst-veq1-h1chomp} $h_1' = \chompS(h_1, 1)$
        \item \label{semantic-chomp1-inverse-subst-veq1-env1} $\env_1 = \env'[y \mapsto h_2] \setminus x$
        \item \label{semantic-chomp1-inverse-subst-veq1-v} $v = \slice{h_1(\pIn)}{0}{1}$
      \end{pf}
      Let $\env = \env_1$ and $h=h_1$.
      Since $x$ and $y$ not free in $\tau_2$, for every heap $h_2\in\semantics{\tau_2}_{\env'}$ also holds that $h_2\in\semantics{\tau_2}_{\env_1}$,
      from which we can conclude \ref{semantic-chomp1-inverse-g1}, i.e., $h\in\semantics{\tau_1[y \mapsto \tau_2]}_\env$.

      To show \ref{semantic-chomp1-inverse-g2}, we must show that $h' = \chompS(h, 1) = \chompS(h_1, 1)$, which follows from \ref{semantic-chomp1-inverse-subst-veq1-h1chomp} and the fact that $h'=h_1'$.

      \ref{semantic-chomp1-inverse-g31} follows by choice of $\env$ and \ref{semantic-chomp1-inverse-subst-veq1-env1}.
      \ref{semantic-chomp1-inverse-g41} follows by \ref{semantic-chomp1-inverse-subst-veq1-v}.

      \subcase{$\sizeof(v) > 1$}~

      By IH, for every $h_2\in\semantics{\tau_2}_{\env'}$, there exists $h_1$, $\env_1$, such that
      \begin{pf}{Inverse-Chomp1-Subst}
        \item \label{semantic-chomp1-inverse-subst-vgt1-h1} $h_1\in\semantics{\tau_1}_{\env_1[y \mapsto h_2]}$
        \item \label{semantic-chomp1-inverse-subst-vgt1-h1chomp} $h_1' = \chompS(h_1, 1)$
        \item \label{semantic-chomp1-inverse-subst-vgt1-env1} $\env_1 = \env'[y \mapsto h_2, x \mapsto (\bvNil, \bvNil, [\inst \mapsto \slice{v}{0}{\sizeof(v) - 1}])]$
        \item \label{semantic-chomp1-inverse-subst-vgt1-v} $v = \bvconcat{\env_1(x)(\inst)}{\slice{h_1(\pIn)}{0}{1}}$
      \end{pf}
      Let $\env = \env_1$ and $h = h_1$.

      To show \ref{semantic-chomp1-inverse-g1}, we must show that $h\in\semantics{\tau_1[y \mapsto \tau_2]}_{\env_1}$.
      Since $x$ and $y$ not free in $\tau_2$, for every heap $h_2\in\semantics{\tau_2}_{\env'}$ also holds that $h_2\in\semantics{\tau_2}_{\env_1}$.
      The result follows by the semantics of heap types.

      To show \ref{semantic-chomp1-inverse-g2}, we must show that $h' = \chompS(h_1,1)$, which follows from \ref{semantic-chomp1-inverse-subst-vgt1-h1chomp} and $h'=h_1'$.

      \ref{semantic-chomp1-inverse-g32} follows by \ref{semantic-chomp1-inverse-subst-vgt1-env1}, \ref{semantic-chomp1-inverse-g42} follows by \ref{semantic-chomp1-inverse-subst-vgt1-v} and \ref{semantic-chomp1-inverse-g52} follows also by \ref{semantic-chomp1-inverse-subst-vgt1-env1}.
    \end{description}
  \end{description}
\end{proof}

\begin{lemma}[Semantic ChompRec Inverse]
  \label{lem:semantic-chomprec-inverse}
  For all variables $x$, values $v$, $n\in\mathbb{N}$, heap types $\tau$, environments $\env'$ and heaps $h'$ such that
  $\env'(x) = (\bvNil, \bvNil, [\inst \mapsto v])$ and $h'\in\semantics{\chompRec(\tau,n,\inst,x)}_{\env'}$ and $x$ not free in $\tau$ and $\sizeof(v) = n$,
  there exists $h$ and $\env$ such that
  $h\in\semantics{\tau}_\env$ and $\env = \env' \setminus x$ and $\chompS(h,n) = h'$.
\end{lemma}
\begin{proof}
  Proof by induction on $n$.
  \begin{description}
    \case{$n=0$}~

    By definition of $\chompRec$, $\chompRec(\tau,0,\inst,x) = \tau$.
    
    Together with assumption $h'\in\semantics{\chompRec(\tau,n,\inst,x)}_{\env'}$, we know that $h'\in\semantics{\tau}_{\env'}$.

    Let $h=h'$.
    By assumption, $x$ is not free in $\tau$, thus the binding of $x$ in $\env'$ does not affect the semantics of $\tau$.
    We can therefore remove the binding altogether, so $\tau$ describes the same set of heaps both in $\env'$ and in $\env$.
    We can conclude that $h\in\semantics{\tau}_\env$.
    $\chompS(h,0) = h'$ follows from the definition of semantic chomp.

    \case{$n=1$}~

    The result directly follows by \cref{lem:semantic-chomp1-inverse}.

    \case{$n=m+1$}~

    We assume that the lemma holds for $n=m$.
    We have to show that the lemma also holds for $n=m+1$.
    Let $h_0'$ be some heap such that $h_0'\in\semantics{\chompRec(\tau,m+1,\inst,x)}_{\env_0'}$.
    Together with \cref{lem:chomprec-unfold}, we can conclude that
    \[
      h_0'\in\semantics{\chompRec(\chompRec(\tau,m,\inst,x), 1,\inst,x)}_{\env_0'}
    \]
    By \cref{lem:semantic-chomp1-inverse}, there is some $h_1'$, $\env_1'$ such that $h_1'\in\semantics{\chompRec(\tau,m,\inst,x)}_{\env_1'}$ and $h_0'=\chompS(h_1', 1)$
    where $\env_1' = \env_0'[x \mapsto (\bvNil, \bvNil, [\inst \mapsto \slice{v}{0}{m}])]$.
    By IH,
    there exists a $h_1$ and $\env_1$, such that $h_1\in\semantics{\tau}_{\env_1}$ where $\env_1 = \env_1'\setminus x$ and $h_1' = \chompS(h_1, m)$.
    From $h_0' = \chompS(\chompS(h_1,m),1)$ and \cref{lem:semantic-chomp-unroll}, follows
    $h_0' = \chompS(h_1, m+1)$.
  \end{description}
\end{proof}

\begin{lemma}[Semantic Chomp Inverse]
  \label{lem:semantic-chomp-inverse}
  For all variables $x$, values $v$, instances $\inst$, heap types $\tau$, environments $\env'$ and heaps $h'$ such that
  $\env'(x) = (\bvNil, \bvNil, [\inst \mapsto v])$ and $h'\in\semantics{\chomp(\tau,\inst,x)}_{\env'}$ and $x$ not free in $\tau$,
  there exists $h$ and $\env$ such that
  $h\in\semantics{\tau}_\env$ and $\env = \env' \setminus x$ and $\chompS(h,\sizeof(\inst)) = h'$.
\end{lemma}
\begin{proof}
  By definition of $\chomp$, we know that
  \[
    \chomp(\tau,\inst,x)=\chompRec(\tau,\sizeof(\inst),x,\inst)
  \]
  The result follows by \cref{lem:semantic-chomprec-inverse}.
\end{proof}

%% file: appendix/safety/preservation.tex
\begin{lemma}[Formulae Preservation]
  \label{lem:expr-preservation}
  If $\Gamma;\tau \vdash \varphi:\tBool$ and
  $\env \models \Gamma$ and
  $(I,O,H) \models_\env \tau$ and
  $\langle I,O,H,\varphi\rangle\rightarrow \varphi'$
  then $\Gamma;\tau \vdash \varphi':\tBool$.
\end{lemma}
\begin{proof}
  By induction on a derivation of $\Gamma;\tau \vdash \varphi:\tBool$.
\end{proof}

\begin{lemma}[Semantic Formulae Preservation]
  \label{lem:semantic-expr-preservation}
  If $\Gamma;\tau \vdash \varphi:\tBool$ and
  $\env \models \Gamma$ and
  $(I,O,H) \models_\env \tau$ and
  $\langle I,O,H,\varphi\rangle\rightarrow \varphi'$
  then
  \[
    \semantics{\varphi[x/\cmdVar]}_{\env[x \mapsto (I,O,H)]} = \semantics{\varphi'[x/\cmdVar]}_{\env[x \mapsto (I,O,H)]}
  \]
\end{lemma}
\begin{proof}
  By induction on a derivation of $\Gamma;\tau \vdash \varphi:\tBool$.
\end{proof}

\begin{lemma}[Expression Preservation]
  \label{lem:term-preservation}
  If $\Gamma;\tau \vdash e:\sigma$ and
  $\env \models \Gamma$ and
  $(I,O,H) \models_\env \tau$ and
  $\langle I,O,H,e\rangle\rightarrow e'$
  then $\Gamma;\tau \vdash e':\sigma$.
\end{lemma}
\begin{proof}
  By induction on a typing derivation of $\Gamma;\tau \vdash e:\sigma$.
\end{proof}

\begin{lemma}[Semantic Expression Preservation]
  \label{lem:semantic-term-preservation}
  If $\Gamma;\tau \vdash e:\sigma$ and
  $\env \models \Gamma$ and
  $(I,O,H) \models_\env \tau$ and
  $\langle I,O,H,e\rangle\rightarrow e'$
  then
  \[
    \semantics{e[x/\ cmdVar]}_{\env[x \mapsto (I,O,H)]} = \semantics{e'[x/\cmdVar]}_{\env[x \mapsto (I,O,H)]}
  \]
\end{lemma}
\begin{proof}
  By induction on a typing derivation of $\Gamma;\tau \vdash e:\sigma$.
\end{proof}

\begin{theorem}[Preservation]
  \label{thm:preservation}
  If $\Gamma \vdash c:(x:\tau_1) \rightarrow \tau_2$,
  $\langle I,O,H,c\rangle \rightarrow \langle I',O',H',c'\rangle$,
  $\env \models \Gamma$ and $(I,O,H)\models_\env \tau_1$, then
  $\exists \Gamma',\env',x',\tau_1',\tau_2'.\Gamma' \vdash c':(x':\tau_1') \rightarrow \tau_2'$ and
  $\env' \models \Gamma'$ and
  $(I',O',H')\models_{\env'}\tau_1'$ and
  $\semantics{\tau_2'}_{\env'[x' \mapsto (I',O',H')]} \subseteq \semantics{\tau_2}_{\env[x \mapsto (I,O,H)]} $
\end{theorem}
\begin{proof}
  Directly follows by the slightly stronger \cref{lem:preservation-aux}.
\end{proof}

\begin{lemma}[Preservation AUX]
  \label{lem:preservation-aux}
  If $\Gamma \vdash c:(x:\tau_1) \rightarrow \tau_2$,
  $\langle I,O,H,c\rangle \rightarrow \langle I',O',H',c'\rangle$,
  $\env \models \Gamma$ and $(I,O,H)\models_\env \tau_1$, then
  $\exists \Gamma',\env',x',\tau_1',\tau_2'.\Gamma' \vdash c':(x':\tau_1') \rightarrow \tau_2'$ and
  $\env' \models \Gamma'$ and
  $\Gamma \subseteq \Gamma'$ and
  $\env \subseteq \env'$ and
  $(I',O',H')\models_{\env'}\tau_1'$ and
  $\semantics{\tau_2'}_{\env'[x' \mapsto (I',O',H')]} \subseteq \semantics{\tau_2}_{\env[x \mapsto (I,O,H)]} $
\end{lemma}

\begin{proof}
  By induction on a derivation of $\Gamma \vdash c :(x:\tau_1) \rightarrow \tau_2$ with case analysis on the last rule used.
  We refer to the proof goals as follows:
  \begin{enumerate}[label=(\arabic*)]
    \item \label{preservation-g1} $\Gamma' \vdash c' : (x':\tau_1') \rightarrow  \tau_2'$
    \item \label{preservation-g2} $\env' \models \Gamma'$
    \item \label{preservation-g3} $\Gamma \subseteq \Gamma'$
    \item \label{preservation-g4} $\env \subseteq \env'$
    \item \label{preservation-g5} $(I',O',H') \models_{\env'} \tau_1'$
    \item \label{preservation-g6} $\semantics{\tau_2'}_{\env'[x' \mapsto (I',O',H')]} \subseteq \semantics{\tau_2}_{\env[x \mapsto (I,O,H)]}$
  \end{enumerate}

  General assumptions:
  \begin{enumerate}[label=(\Alph*)]
    \item \label{preservation-c} $\Gamma \vdash c : (x:\tau_1) \rightarrow \tau_2$
    \item \label{preservation-step} $\langle I,O,H,c\rangle \rightarrow \langle I',O',H',c'\rangle$
    \item \label{preservation-env} $\env \models \Gamma$
    \item \label{preservation-heap} $(I,O,H) \models_\env \tau_1$
  \end{enumerate}
  \begin{description}
    \case{\tAdd}~

    By inversion of rule \tAdd, we get
    \begin{pf}{T-Add}
      \item $c = \cAdd{\inst}$
      \item $\excludes{\tau_1}{\inst}$
      \item $\tau_2 = \sigmaT{y}{\refT{z}{\tau_1}{z \equiv x}}{\refT{z}{\inst}{z.\pIn = z.\pOut = \bvNil \wedge z.\inst=v}}$
    \end{pf}
    Only evaluation rule \eAdd applies to $c$:
    \begin{pf*}{T-Add}
      \item $\inst\not\in\dom(H)$
      \item $\HT(\inst) = \eta$
      \item $\mathit{init} = v$.
      \item $I'=I$ and $O'=O$ and $H' = H[\inst \mapsto v]$ and $c'=\cSkip$
    \end{pf*}
    Let $\Gamma' = \Gamma,x:\tau_1$ and $\env' = \env[x \mapsto (I,O,H)]$.

    Let $\tau_1' = \sigmaT{y}{\refT{z}{\tau_1}{z \equiv x}}{\refT{z}{\inst}{z.\pIn = z.\pOut = \bvNil \wedge z.\inst=v}}$ and

    $\tau_2' = \refT{w}{\sigmaT{y}{\refT{z}{\tau_1}{z \equiv x}}{\refT{z}{\inst}{z.\pIn = z.\pOut = \bvNil \wedge z.\inst=v}}}{w \equiv x'}$.

    \ref{preservation-g1} follows by \tSkip and \ref{preservation-g2} follows by assumptions \ref{preservation-env} and \ref{preservation-heap} and \cref{lem:extended-env-entails}.
    \ref{preservation-g3} and \ref{preservation-g4} are immediate.

    To show \ref{preservation-g5}, we must show that $(I,O,H[\inst \mapsto v]) \models_{\env[x \mapsto (I,O,H)]} \sigmaT{y}{\refT{z}{\tau_1}{z \equiv x}}{\refT{z}{\inst}{z.\pIn = z.\pOut = \bvNil \wedge z.\inst=v}}$.
    By \entSigma, we must show that
    \begin{enumerate}[label=(5.\arabic*)]
      \item \label{preservation-tadd-g51} $(I,O,H) \models_{\env[x \mapsto (I,O,H)]} \refT{z}{\tau_1}{z \equiv x}$ and
      \item \label{preservation-tadd-g52} $(\bvNil,\bvNil,[\inst \mapsto v]) \models_{\env[x \mapsto (I,O,H) y \mapsto (I,O,H)]} \refT{z}{\inst}{z.\pIn = \bvNil \wedge z.\pOut = \bvNil \wedge z.\inst=v}$.
    \end{enumerate}
    \ref{preservation-tadd-g51} follows by \entRefine and \ref{preservation-heap}.
    To show \ref{preservation-tadd-g52}, by \entRefine, we must show that
    \begin{enumerate}[label=(5.2.\arabic*)]
      \item \label{preservation-tadd-g521} $(\bvNil, \bvNil, [\inst \mapsto v]) \models_{\env[y \mapsto (I, O, H)]} \refT{z}{\top}{z.\inst.\valid \wedge \bigwedge_{\kappa \in \dom(\HT)}\neg \kappa.\valid}$ and
      \item \label{preservation-tadd-g522} $\semantics{z.\pIn = \bvNil \wedge z.\pOut = \bvNil \wedge z.\inst=v}_{\env[y \mapsto (I, O, H), z \mapsto (\bvNil,\bvNil,[\inst \mapsto v])]} = \mathit{true}$
    \end{enumerate}
    \ref{preservation-tadd-g521} follows by \entRefine, \entTop and the semantics of formulae.
    \ref{preservation-tadd-g522} follows from the semantics of formulae.

    \ref{preservation-g6} follows by
    \begin{align*}
      \phantom{{}={}} & \semantics{\sigmaT{y}{\refT{z}{\tau_1}{z \equiv x}}{\refT{z}{\inst}{z.\pIn = z.\pOut = \bvNil \wedge z.\inst=v}}}_{\env[x \mapsto (I,O,H)]} \\
      =               & \{ (I,O,H[\inst \mapsto v]) \}                                                                                                              \\
      =               & \{ h \mid h\in\semantics{\sigmaT{y}{\refT{z}{\tau_1}{z \equiv x}}{\refT{z}{\inst}{z.\pIn = z.\pOut = \bvNil\ \wedge                         \\
                      & \hspace{141.5pt} z.\inst=v}}}_{\env[x \mapsto (I,O,H), x' \mapsto (I,O,H[\inst \mapsto v])]}\ \wedge                                        \\
                      & \hspace{33pt}\semantics{w \equiv x'}_{\env[x \mapsto (I,O,H), x' \mapsto (I,O,H[\inst \mapsto v]), w \mapsto h]}\}                          \\
      =               & \semantics{\refT{w}{\sigmaT{y}{\refT{z}{\tau_1}{z \equiv x}}{\refT{z}{\inst}{z.\pIn = z.\pOut = \bvNil \wedge z.\inst=v}}}{                 \\
                      & \hspace{23pt} w \equiv x'}}_{\env[x \mapsto (I,O,H), x' \mapsto (I,O,H[\inst \mapsto v])]}
    \end{align*}

    \case{\tAscribe}~

    By inversion of rule \tAscribe, we get
    \begin{pf}{T-Ascribe}
      \item $c = \cAscribe{c_a}{(x:\tau_1) \rightarrow \tau_2}$
      \item \label{preservation-tascribe-ca} $\Gamma \vdash c_a : (x:\tau_1) \rightarrow \tau_2$
    \end{pf}
    There is one evaluation rule that applies to $c$, \eAscribe, so
    $I'=I$ and $O'=O$ and $H'=H$ and $c' = c_a$.
    Let $\Gamma'=\Gamma$, $\env'=\env$, $\tau_1' = \tau_1$ and $\tau_2' = \tau_2$.
    \ref{preservation-g1} follows by \ref{preservation-tascribe-ca}, \ref{preservation-g2} follows by assumption \ref{preservation-env}.
    \ref{preservation-g3} and \ref{preservation-g4} are immediate.
    \ref{preservation-g5} follows by assumption \ref{preservation-heap} and \ref{preservation-g6} follows from the equality of $\tau_2$ and $\tau_2'$, which itself follows by reflexivity.

    \case{\tExtract}~

    By inversion of rule \tExtract, we get
    \begin{pf}{T-Extract}
      \item $c = \cExtract{\inst}$
      \item $\Gamma \vdash \sizeof_\pIn(\tau_1) = \sizeof(\inst)$
      \item $\tau_2 = \sigmaT{y}{\refT{z}{\inst}{z.\pIn = z.\pOut = \bvNil}}{\refT{z}{\chomp(\tau_1, \inst, y)}{ \\
            \bvconcat{y.\inst}{z.\pIn} = x.\pIn \wedge z.\pOut = x.\pOut \wedge z\equiv_\inst x}}$
    \end{pf}
    Only evaluation rule \eExtract applies to $c$:
    \begin{pf*}{T-Extract}
      \item $\mathit{deserialize_{\HT(\inst)}}(I) = (v, I')$
      \item $O' = O$
      \item $H' = H[\inst \mapsto v]$
      \item $c' = \cSkip$
    \end{pf*}
    Let $\Gamma' = \Gamma,x:\tau_1, \env' = \env[x \mapsto (I,O,H)]$.

    Let $\tau_1' = \sigmaT{y}{\refT{z}{\inst}{z.\pIn = z.\pOut = \bvNil}}{\refT{z}{\chomp(\tau_1, \inst, y)}{\bvconcat{y.\inst}{z.\pIn} = x.\pIn \wedge z.\pOut = x.\pOut \wedge z\equiv_\inst x}}$.

    Let $\tau_2' = \refT{v}{\sigmaT{y}{\refT{z}{\inst}{z.\pIn = z.\pOut = \bvNil}}{\refT{z}{\chomp(\tau_1, \inst, y)}{\bvconcat{y.\inst}{z.\pIn} = x.\pIn \wedge z.\pOut = x.\pOut \wedge z\equiv_\inst x}}}{v \equiv x}$.

    \ref{preservation-g1} follows by \tSkip and \ref{preservation-g2} follows by assumptions \ref{preservation-env} and \ref{preservation-heap} and \cref{lem:extended-env-entails}.
    \ref{preservation-g3} and \ref{preservation-g4} are immediate.

    To show \ref{preservation-g5}, we must show that $(I',O,H[\inst \mapsto v]) \models_{\env[x \mapsto (I,O,H)]} \sigmaT{y}{\refT{z}{\inst}{z.\pIn = z.\pOut = \bvNil}}{\refT{z}{\chomp(\tau_1, \inst, y)}{\bvconcat{y.\inst}{z.\pIn} = x.\pIn \wedge z.\pOut = x.\pOut \wedge z\equiv_\inst x}}$.

    By \entSigma, we must show that
    \begin{enumerate}[label=(5.\arabic*)]
      \item $(\bvNil, \bvNil,[\inst \mapsto v]) \models_{\env[x \mapsto (I,O,H)]} \refT{z}{\inst}{z.\pIn = z.\pOut = \bvNil}$, which follows by \entRefine and the semantics of types and
      \item $(I',O,H) \models_{\env[x \mapsto (I,O,H), y \mapsto (\bvNil, \bvNil, [\inst \mapsto v])]} \refT{z}{\chomp(\tau_1, \inst, y)}{\bvconcat{y.\inst}{z.\pIn} = x.\pIn \wedge z.\pOut = x.\pOut \wedge z \equiv_\inst x}$

            By \entRefine, we must show that
            \begin{enumerate}[label=(5.2.\arabic*)]
              \item $(I',O,H) \models_{\env[x \mapsto (I,O,H), y \mapsto (\bvNil, \bvNil, [\inst \mapsto v])]} \chomp(\tau_1, \inst, y)$

                    By \cref{lem:semantic-containment-entails}, it is sufficient to show that
                    \[
                      (I',O,H)\in\semantics{\chomp(\tau_1, \inst, y)}_{\env[x \mapsto (I,O,H), y \mapsto (\bvNil, \bvNil, [\inst \mapsto v])]}
                    \]

                    By assumption (D) and by \cref{lem:semantic-containment-entails} follows $(I,O,H)\in\semantics{\tau_1}_\env$.

                    By \cref{lem:semantic-chomp}, there exists some heap
                    \[
                      h\in\semantics{\chomp(\tau_1, \inst, y)}_{\env[x \mapsto (I,O,H), y \mapsto (\bvNil, \bvNil, [\inst \mapsto v])]}
                    \] such that $h = \chompS((I,O,H),\sizeof(\inst))$.

                    From the definition of $\chompS$ follows that
                    \[
                      \chompS((I,O,H), \sizeof(\inst)) = (I'',O,H)
                    \] where $I'' = \slice{I}{\sizeof(\inst)}{\ } = I'$.

              \item $\semantics{\bvconcat{y.\inst}{z.\pIn} = x.\pIn\ \wedge\ z.\pOut = x.\pOut\ \wedge\ \\
                        z \equiv_\inst x}_{\env[x \mapsto (I,O,H), y \mapsto (\bvNil, \bvNil,[\inst \mapsto v]), z \mapsto (I',O,H)]}$, which follows by the definition of \textit{deserialize} and the semantics of formulae and expressions.
            \end{enumerate}
    \end{enumerate}

    Finally, we must show that $\semantics{\tau_2}_{\env[x \mapsto (I,O,H)]} \subseteq \semantics{\tau_2'}_{\env[x \mapsto (I',O,H[\inst \mapsto v])]}$.

    \begin{align*}
      \phantom{{}={}} & \semantics{\tau_2}_{\env[x \mapsto (I,O,H)]}                                                                                       \\
      =               & \semantics{\sigmaT{y}{\refT{z}{\inst}{z.\pIn = z.\pOut = \bvNil}}{                                                                 \\
                      & \hspace{23pt}\refT{z}{\chomp(\tau_1, \inst, y)}{\bvconcat{y.\inst}{z.\pIn} = x.\pIn \wedge z.\pOut = x.\pOut\ \wedge               \\
                      & \hspace{109pt} z\equiv_\inst x}}}_{\env[x \mapsto (I,O,H)]}                                                                        \\
      =               & \{ \concat{h_1}{h_2} \mid h_1\in\semantics{\refT{z}{\inst}{z.\pIn = z.\pOut = \bvNil}}_{\env[x \mapsto (I,O,H)]}\ \wedge           \\
                      & \hspace{45pt} h_2\in\semantics{\refT{z}{\chomp(\tau_1, \inst, y)}{\bvconcat{y.\inst}{z.\pIn} = x.\pIn\ \wedge                      \\
                      & \hspace{109pt} z.\pOut = x.\pOut \wedge z\equiv_\inst x}}_{\env[x \mapsto (I,O,H), y \mapsto h_1]} \}                              \\
      =               & \{ \concat{h_1}{h_2} \mid h_1\in\semantics{\refT{z}{\inst}{z.\pIn = z.\pOut = \bvNil}}_{\env[x \mapsto (I,O,H)]}\ \wedge           \\
                      & \hspace{45pt} h_2\in\{ h_2' \in\semantics{\chomp(\tau_1, \inst, y)}_{\env[x \mapsto (I,O,H), y \mapsto h_1]} \wedge                \\
                      & \hspace{90pt} \semantics{\bvconcat{y.\inst}{z.\pIn} = x.\pIn \wedge z.\pOut = x.\pOut\ \wedge                                      \\
                      & \hspace{95pt} z\equiv_\inst x}_{\env[x \mapsto (I,O,H), y \mapsto h_1, z \mapsto h_2']} \} \}                                      \\
      \intertext{By \cref{lem:semantic-chomp}}
      =               & \{ \concat{h_1}{h_2} \mid h_1\in\semantics{\refT{z}{\inst}{z.\pIn = z.\pOut = \bvNil}}_{\env[x \mapsto (I,O,H)]}\ \wedge           \\
                      & \hspace{45pt} h_2\in\{ h\in\semantics{\tau_1}_{\env[x \mapsto (I,O,H)]} \wedge h_2'=\chompS(h,\sizeof(\inst))\ \wedge              \\
                      & \hspace{86pt} \semantics{\bvconcat{y.\inst}{z.\pIn} = x.\pIn \wedge z.\pOut = x.\pOut\ \wedge                                      \\
                      & \hspace{91pt} z\equiv_\inst x}_{\env[x \mapsto (I,O,H), y \mapsto h_1, z \mapsto h_2']} \} \}
      \intertext{By definition of $\chompS$ and semantics of types}
      =               & \{ (I',O,H[\iota \mapsto v]) \}                                                                                                    \\
      =               & \{ h \mid h\in\semantics{\sigmaT{y}{\refT{z}{\inst}{z.\pIn = z.\pOut = \bvNil}}{                                                   \\
                      & \hspace{55pt} \refT{z}{\chomp(\tau_1, \inst, y)}{\bvconcat{y.\inst}{z.\pIn} = x.\pIn\ \wedge                                       \\
                      & \hspace{60pt} z.\pOut = x.\pOut\ \wedge z\equiv_\inst x}}}_{\env[x \mapsto (I,O,H), x' \mapsto (I',O,H[\iota \mapsto v])]}\ \wedge \\
                      & \hspace{35pt}\semantics{v \equiv x'}_{\env[x \mapsto (I,O,H), x' \mapsto (I',O,H[\iota \mapsto v]), v \mapsto h]} \}               \\
      =               & \semantics{\refT{v}{\sigmaT{y}{\refT{z}{\inst}{z.\pIn = z.\pOut = \bvNil}}{                                                        \\
                      & \hspace{38.5pt}\refT{z}{\chomp(\tau_1, \inst, y)}{\bvconcat{y.\inst}{z.\pIn} = x.\pIn\ \wedge                                      \\
                      & \hspace{125pt} z.\pOut = x.\pOut\ \wedge z\equiv_\inst x}}}{                                                                       \\
                      & \hspace{23pt} v \equiv x}}_{\env[x \mapsto (I,O,H), x' \mapsto (I',O,H[\inst \mapsto v])]}
    \end{align*}
    This shows \ref{preservation-g6} and concludes the case.

    \case{\tIf}~

    By inversion of rule \tIf, we get
    \begin{pf}{T-If}
      \item $\Gamma;\tau_1 \vdash \varphi : \tBool$
      \item \label{preservation-tif-then} $\Gamma \vdash c_1: (x:\refT{y}{\tau_1}{\varphi[y/\cmdVar]}) \rightarrow \tau_{12}$
      \item \label{preservation-tif-else} $\Gamma \vdash c_2: (x:\refT{y}{\tau_1}{\neg \varphi[y/\cmdVar]}) \rightarrow \tau_{22}$
      \item $\tau_2 = \refT{y}{\tau_{12}}{\varphi[x/\cmdVar]} + \refT{y}{\tau_{22}}{\neg \varphi[x/\cmdVar]}$
    \end{pf}{T-If}
    There are three evaluation rules that apply to $c$.
    \begin{description}
      \subcase{\eIf}~

      \begin{pf*}{T-If}
        \item $c' = \cIf{\varphi'}{c_1}{c_2}$
        \item $I'=I, O'=O, H'=H$
      \end{pf*}
      Let $\Gamma' = \Gamma$, $\env' = \env$, $x'=x$,

      $\tau_1' = \tau_1$ and

      $\tau_2' = \refT{y}{\tau_{12}}{\varphi'[x/\cmdVar]} + \refT{y}{\tau_{22}}{\neg \varphi'[x/\cmdVar]}$.

      By \cref{lem:expr-preservation}, $\Gamma;\tau_1 \vdash \varphi' : \tBool$.

      By \cref{lem:semantic-expr-preservation}, $\semantics{\varphi[x/\cmdVar]}_{\env[x \mapsto (I,O,H)]} = \semantics{\varphi'[x/\cmdVar]}_{\env[x \mapsto (I,O,H)]}$.

      From \ref{preservation-tif-then} follows that $\Gamma \vdash c_1: (x:\refT{y}{\tau_1}{\varphi'[y/\cmdVar]}) \rightarrow \tau_{12}$ and from \ref{preservation-tif-else} follows that
      $\Gamma \vdash c_2: (x:\refT{y}{\tau_1}{\neg \varphi'[y/\cmdVar]}) \rightarrow \tau_{22}$
      \ref{preservation-g1} follows by \tIf.

      \ref{preservation-g2} follows by assumption \ref{preservation-env}, \ref{preservation-g3} and \ref{preservation-g4} are immediate.

      \ref{preservation-g5} follows by assumption \ref{preservation-heap}.

      \ref{preservation-g6} follows together with the assumption
      \[
        \semantics{\varphi[x/\cmdVar]}_{\env[x \mapsto (I,O,H)]} = \semantics{\varphi'[x/\cmdVar]}_{\env[x \mapsto (I,O,H)]}
      \] from the equality $\semantics{\tau_2'}_{\env'[x \mapsto (I',O',H')]} = \semantics{\tau_2}_{\env[x \mapsto (I,O,H)]}$.
      \begin{align*}
        \phantom{{}={}} & \semantics{\tau_2}_{\env[x \mapsto (I,O,H)]}                                                                       \\
        =               & \semantics{\refT{y}{\tau_{12}}{\varphi[x/\cmdVar]} + \refT{y}{\tau_{22}}{\neg \varphi[x/\cmdVar]}}_{\env[x \mapsto (I,O,H)]}   \\
        =               & \semantics{\refT{y}{\tau_{12}}{\varphi'[x/\cmdVar]} + \refT{y}{\tau_{22}}{\neg \varphi'[x/\cmdVar]}}_{\env[x \mapsto (I,O,H)]} \\
        =               & \semantics{\tau_2'}_{\env'[x \mapsto (I',O',H')]}
      \end{align*}

      \subcase{\eIfTrue}~

      \begin{pf*}{T-If}
        \item $c' = c_1$
        \item $I'=I, O'=O, H'=H$
        \item $\varphi = \true$
      \end{pf*}
      Let $\Gamma' = \Gamma, \env' = \env, x'=x$,

      $\tau_1' = \refT{y}{\tau_1}{\varphi[y/\cmdVar]} = \refT{y}{\tau_1}{\true} = \tau_1$ and

      $\tau_2' = \tau_{12}$.

      \ref{preservation-g1} holds by assumption \ref{preservation-tif-then}, \ref{preservation-g2} holds by assumption \ref{preservation-env}.
      \ref{preservation-g3} and \ref{preservation-g4} are immediate and \ref{preservation-g5}
      $(I',O',H') \models_{\env'} \tau_1' \Leftrightarrow (I,O,H) \models_{\env} \tau_1$ holds by assumption \ref{preservation-heap}.

      \ref{preservation-g6} follows from
      \begin{align*}
        \phantom{{}={}} & \semantics{\refT{y}{\tau_{12}}{\varphi[x/\cmdVar]} + \refT{y}{\tau_{22}}{\neg \varphi[x/\cmdVar]}}_{\env[x \mapsto (I,O,H)]} \\
        =               & \semantics{\refT{y}{\tau_{12}}{\true} + \refT{y}{\tau_{22}}{\false}}_{\env[x \mapsto (I,O,H)]}                   \\
        =               & \semantics{\tau_{12}}_{\env[x \mapsto (I,O,H)]}                                                                  \\
        =               & \semantics{\tau_2'}_{\env'[x \mapsto (I',O',H')]}
      \end{align*}

      \subcase{\eIfFalse}~

      Symmetric to previous subcase.
    \end{description}

    \case{\tMod}~

    By inversion of rule \tMod, we get
    \begin{pf}{T-Mod}
      \item $c = \inst.f := e$
      \item $\includes{\tau_1}{\inst}$
      \item $\mathcal{F}(\inst, f) = \tBv$
      \item $\Gamma;\tau_1 \vdash e : \tBv$
      \item $\tau_2 = \refT{y}{\top}{\varphi_{pkt} \wedge \varphi_\inst \wedge \varphi_f \wedge y.\inst.f = e[x/\cmdVar]}$
      \item $\varphi_{pkt} \triangleq y.\pIn=x.\pIn \wedge y.\pOut=x.\pOut$
      \item $\varphi_\inst \triangleq \forall \kappa\in\dom(\HT).\inst\neq\kappa \rightarrow y.\kappa = x.\kappa$
      \item $\varphi_f \triangleq \forall g\in\dom(\HT(\inst)).f\neq g \rightarrow y.\inst.g = x.\inst.g$
    \end{pf}
    There are two evaluation rule that apply to $c$.

    \begin{description}
      \subcase{\eMod}~
      \begin{pf*}{T-Mod}
        \item $H(\inst) = r$
        \item $r' \triangleq \lbrace \text{r with } f = v\rbrace$
        \item $(I',O',H') = (I,O,H[\inst \mapsto r'])$
        \item $c' = \cSkip$
        \item $e = v$
        \item $\tau_2 = \refT{y}{\top}{\varphi_{pkt} \wedge \varphi_\inst \wedge \varphi_f \wedge y.\inst.f = v[x/\cmdVar]}$
      \end{pf*}
      Let $\Gamma' = \Gamma,x:\tau_1$ and $\env' = \env[x \mapsto (I,O,H)]$ and $\tau_1' = \tau_2$ and $\tau_2' = \refT{z}{\tau_2}{z \equiv x}$.

      \ref{preservation-g1} follows by \tSkip and \tSub.
      \ref{preservation-g2} follows by assumptions \ref{preservation-env} and \ref{preservation-heap} and \cref{lem:extended-env-entails}.
      \ref{preservation-g3} and \ref{preservation-g4} are immediate.

      To show \ref{preservation-g5}, we must show that $(I,O,H[\inst \mapsto r']) \models_{\env[x \mapsto (I,O,H)]} \top$ and
      $\semantics{\varphi_{pkt} \wedge \varphi_\inst \wedge \varphi_f \wedge y.\inst.f = v[x/\cmdVar]}_{\env[x \mapsto (I,O,H), y \mapsto (I,O,H[\inst\mapsto r'])]}=\true$.
      Since $\env(x)(I) = \env(y)(I)$ and $\env(x)(O) = \env(y)(O)$, $\varphi_{pkt}$ holds.
      Similarly, $\env(x)(H) = \env(y)(H)$ in every aspect, except for field $f$ of instance $\inst$, so $\varphi_{f}$ and $\varphi_{inst}$ also hold.
      $y.\inst.f = v[x/\cmdVar]$ also holds, because
      \begin{align*}
        \phantom{{}\Leftrightarrow{}} & \semantics{y.\inst.f}_{\env[x \mapsto (I,O,H), y \mapsto (I,O,H[\inst\mapsto r'])]} =  \\
                                      & \semantics{v[x/\cmdVar]}_{\env[x \mapsto (I,O,H), y \mapsto (I,O,H[\inst\mapsto r'])]} \\
        {}\Leftrightarrow{}           & v = \semantics{v[x/\cmdVar]}_{[x \mapsto (I,O,H)]}                                     \\
        {}\Leftrightarrow{}           & v = v
      \end{align*}

      To show
      \begin{align*}
        \phantom{{}\Leftrightarrow{}} & \semantics{\tau_2'}_{\env'[x' \mapsto (I',O',H')]} \subseteq \semantics{\tau_2}_{\env[x \mapsto (I,O,H)]}                                        \\
        {}\Leftrightarrow{}           & \semantics{\refT{z}{\tau_2}{z \equiv x}}_{\env[x \mapsto (I,O,H), x' \mapsto (I',O',H')]} \subseteq \semantics{\tau_2}_{\env[x \mapsto (I,O,H)]} \\
        {}\Leftrightarrow{}           & \{ (I',O',H') \} \subseteq \semantics{\tau_2}_{\env[x \mapsto (I,O,H)]}                                                                          \\
        {}\Leftrightarrow{}           & \{ (I,O,H[\inst \mapsto r']) \} \subseteq \semantics{\tau_2}_{\env[x \mapsto (I,O,H)]}
      \end{align*}

      To show \ref{preservation-g6},
      let $h = (I,O,H[\inst \mapsto r'])$.
      Therefore,
      \begin{pf*}{T-Mod}
        \item $h(\pIn) = I$ and
        \item $h(\pOut) = O$ and
        \item for all $\kappa\in\dom(\HT)$ such that $\kappa \neq \inst$, $h(\kappa) = H(\kappa)$
        \item for all $g\in\dom(\HT(\inst))$ such that $g\neq f$, $h(\inst)(g) = H(\inst)(g)$
        \item $h(\inst)(f) = v[x/\cmdVar] = v$
      \end{pf*}
      From the semantics of types follows that $h\in\semantics{\tau_2}_{\env[x \mapsto (I,O,H)]}$.

      \subcase{\eModOne}~

      \begin{pf*}{T-Mod}
        \item $\langle I,O,H,t\rangle \rightarrow e'$
        \item $c' = \inst.f := e'$
        \item $I'=I, O'=O$ and $H'=H$
      \end{pf*}
      Let $\Gamma' = \Gamma$ and $\env' = \env$ and $x' = x$.

      Let $\tau_1' = \tau_1$ and

      $\tau_2' = \refT{y}{\top}{\varphi_{pkt} \wedge \varphi_\inst \wedge \varphi_f \wedge y.\inst.f = e'[x/\cmdVar]}$.

      By semantic expression preservation (\autoref{lem:semantic-term-preservation}), we know that \\
      $\semantics{e[x/\cmdVar]}_{\env[x\mapsto(I,O,H)]} = \semantics{e'[x/\cmdVar]}_{\env[x \mapsto (I,O,H)]}$.

      If $\Gamma;\tau_1 \vdash e : \tBv$ and $\semantics{e[x/\cmdVar]}_{\env[x\mapsto(I,O,H)]} = \semantics{e'[x/\cmdVar]}_{\env[x \mapsto (I,O,H)]}$ holds,
      then it must hold that $\Gamma;\tau_1 \vdash  e':\tBv$.

      \ref{preservation-g1} follows by \tMod, \ref{preservation-g2} follows from assumption \ref{preservation-env}.
      \ref{preservation-g3} and \ref{preservation-g4} are immediate.
      \ref{preservation-g5} follows from assumption \ref{preservation-heap}.

      \ref{preservation-g6} follows from
      \begin{align*}
        \phantom{{}\Leftrightarrow{}} & \semantics{\tau_2'}_{\env'[x' \mapsto (I',O',H')]} \subseteq \semantics{\tau_2}_{\env[x \mapsto (I,O,H)]}                                            \\
        \Leftrightarrow               & \semantics{\tau_2'}_{\env[x \mapsto (I,O,H)]} \subseteq \semantics{\tau_2}_{\env[x \mapsto (I,O,H)]}                                                 \\
        \Leftrightarrow               & \semantics{\refT{y}{\top}{\varphi_{pkt} \wedge \varphi_\inst \wedge \varphi_f \wedge y.\inst.f = e'[x/\cmdVar]}}_{\env[x \mapsto (I,O,H)]} \subseteq \\
                                      & \semantics{\refT{y}{\top}{\varphi_{pkt} \wedge \varphi_\inst \wedge \varphi_f \wedge y.\inst.f = e[x/\cmdVar]}}_{\env[x \mapsto (I,O,H)]}            \\
      \end{align*}

      together with assumption
      \[
        \semantics{e[x/\cmdVar]}_{\env[x\mapsto(I,O,H)]} = \semantics{e'[x/\cmdVar]}_{\env[x \mapsto (I,O,H)]}
      \] and the semantics of types.

    \end{description}

    \case{\tRemit}~

    By inversion of rule \tRemit, we get
    \begin{pf}{T-Remit}
      \item $c = remit$
      \item $\tau_2 = \sigmaT{y}{\refT{z}{\tau_1}{z \equiv x}}{\refT{z}{\epsilon}{z.\pIn = \bvNil \wedge z.\pOut = x.\inst}}$
      \item $\includes{\tau_1}{\inst}$
    \end{pf}
    There is only evaluation rule \eRemit that applies to $c$.
    \begin{pf*}{T-Remit}
      \item $\inst\in\dom(H)$
      \item $\mathcal{HT}(\inst) = \eta$
      \item \label{preservation-tremit-serialize} $\mathit{serialize}_\eta(H(\inst)) = bv$
      \item $I'=I$, $O'=\bvconcat{O}{bv}$, $H'=H$, $c'=\cSkip$
    \end{pf*}
    Let $\Gamma' = \Gamma,x:\tau_1$ and $\env' = \env[x \mapsto (I,O,H)]$.

    Let $\tau_1' = \sigmaT{y}{\refT{z}{\tau_1}{z \equiv x}}{\refT{z}{\epsilon}{z.\pIn = \bvNil \wedge z.\pOut = x.\inst}}$ and

    $\tau_2' = \refT{v}{\sigmaT{y}{\refT{z}{\tau_1}{z \equiv x}}{\refT{z}{\epsilon}{z.\pIn = \bvNil \wedge z.\pOut = x.\inst}}}{v \equiv x'}$.

    \ref{preservation-g1} follows by \tSkip and \ref{preservation-g2} follows by assumptions \ref{preservation-env} and \ref{preservation-heap} and \cref{lem:extended-env-entails}.

    \ref{preservation-g3} and \ref{preservation-g4} are immediate.

    For \ref{preservation-g5} we have to show that $(I,\bvconcat{O}{bv},H) \models_{\env[x \mapsto (I,O,H)]} \sigmaT{y}{\refT{z}{\tau_1}{z \equiv x}}{\refT{z}{\epsilon}{z.\pIn = \bvNil \wedge z.\pOut = x.\inst}}$.
    By \entSigma, we must show that
    \begin{enumerate}[label=(5.\arabic*)]
      \item \label{preservation-tremit-g51} $(I,O,H) \models_{\env[x \mapsto (I,O,H)]} \refT{z}{\tau_1}{z \equiv x}$ and
      \item \label{preservation-tremit-g52} $(\bvNil,bv,[]) \models_{\env[x \mapsto (I,O,H), y \mapsto (I,O,H)]} \refT{z}{\epsilon}{z.\pIn = \bvNil \wedge z.\pOut = x.\inst}$
    \end{enumerate}
    For \ref{preservation-tremit-g51} we must show by \entRefine that $(I,O,H) \models_{\env[x \mapsto (I,O,H)]} \tau_1$, which follows by assumption \ref{preservation-heap} and the fact that x is not free in $\tau_1$.
    We must also show that $\semantics{z \equiv x}_{\env[x \mapsto (I,O,H), z \mapsto (I,O,H)]}$, which follows by the semantics of formulae and by reflexivity.

    To show \ref{preservation-tremit-g52}, by \entRefine, we must show that
    \[
      (\bvNil, bv, []) \models_{\env[x \mapsto (I,O,H), y \mapsto (I,O,H)]} \epsilon
    \] and that
    \[
      \semantics{z.\pIn = \bvNil \wedge z.\pOut = x.\inst}_{\env[x \mapsto (I,O,H), y \mapsto (I,O,H), z \mapsto (\bvNil, bv, [])]}=\true
    \]
    The first follows after unfolding the definition of $\epsilon$ by \entTop, \entRefine and the semantics of formulae.
    The second follows by the semantics of formulae and \ref{preservation-tremit-serialize}.

    \ref{preservation-g6} follows by
    \begin{align*}
      \phantom{{}={}} & \semantics{\tau_2}_{\env[x \mapsto (I,O,H)]}                                                                                                 \\
      =               & \semantics{\sigmaT{y}{\refT{z}{\tau_1}{z \equiv x}}{\refT{z}{\epsilon}{z.\pIn = \bvNil \wedge z.\pOut = x.\inst}}}_{\env[x \mapsto (I,O,H)]} \\
      =               & \lbrace \concat{h_1}{h_2} \mid h_1 \in \semantics{\refT{z}{\tau_1}{z \equiv x}}_{\env[x \mapsto (I,O,H)]} \wedge                             \\
                      & \hspace{4em} h_2\in\semantics{\refT{z}{\epsilon}{z.\pIn = \bvNil \wedge z.\pOut = x.\inst}}_{\env[x \mapsto (I,O,H), y \mapsto h_1]}\rbrace  \\
      =               & \lbrace \concat{(I,O,H)}{(\bvNil, bv, [])}\rbrace                                                                                            \\
      =               & \lbrace (I,\bvconcat{O}{bv},H) \rbrace                                                                                                       \\
      =               & \{ h \mid h\in\semantics{\sigmaT{y}{\refT{z}{\tau_1}{z \equiv x}}{                                                                           \\
                      & \hspace{54pt}\refT{z}{\epsilon}{z.\pIn = \bvNil\ \wedge                                                                                      \\
                      & \hspace{82pt} z.\pOut = x.\inst}}}_{\env[x \mapsto (I,O,H), x' \mapsto (I,\bvconcat{O}{bv},H)]}\ \wedge                                      \\
                      & \hspace{35pt} \semantics{v \equiv x'}_{\env[x \mapsto (I,O,H), x' \mapsto (I,\bvconcat{O}{bv},H), v \mapsto h]} \}                           \\
      =               & \semantics{\refT{v}{\sigmaT{y}{\refT{z}{\tau_1}{z \equiv x}}{                                                                                \\
                      & \hspace{40pt}\refT{z}{\epsilon}{z.\pIn = \bvNil \wedge z.\pOut = x.\inst}}}{                                                                 \\
                      & \hspace{21pt} v \equiv x'}}_{\env[x \mapsto (I,O,H), x' \mapsto (I,\bvconcat{O}{bv},H)]}                                                     \\
      =               & \semantics{\tau_2'}_{\env'[x' \mapsto (I',O',H')]}
    \end{align*}

    \case{\tReset}~

    By inversion of rule \tReset, we get
    \begin{pf}{T-Reset}
      \item $c = \cReset$
      \item $\tau_2 = \sigmaT{y}{\refT{z}{\epsilon}{z.\pOut = \bvNil \wedge z.\pIn = x.\pOut}}{\refT{z}{\epsilon}{z.\pOut = \bvNil \wedge z.\pIn = x.\pIn}}$
    \end{pf}
    There is only one evaluation rule that applies to $c$, \eReset.
    \begin{pf*}{T-Reset}
      \item $c' = skip$
      \item $I' = \bvconcat{O}{I}$, $O' = \bvNil$ and $H' = []$
    \end{pf*}

    Let $\env' = \env[x \mapsto (I,O,H)]$ and $\Gamma' = \Gamma,x:\tau_1$.

    Let $\tau_1' = \sigmaT{y}{\refT{z}{\epsilon}{z.\pOut = \bvNil \wedge z.\pIn = x.\pOut}}{\refT{z}{\epsilon}{z.\pOut = \bvNil \wedge z.\pIn = x.\pIn}}$

    Let $\tau_2' = \refT{v}{\sigmaT{y}{\refT{z}{\epsilon}{z.\pOut = \bvNil \wedge z.\pIn = x.\pOut}}{\refT{z}{\epsilon}{z.\pOut = \bvNil \wedge z.\pIn = x.\pIn}}}{v \equiv x'}$

    \ref{preservation-g1} follows by \tSkip and \ref{preservation-g2} follows by assumptions \ref{preservation-env} and \ref{preservation-heap} and \cref{lem:extended-env-entails}.

    \ref{preservation-g3} and \ref{preservation-g4} are immediate.

    To show \ref{preservation-g5}, we must show that
    \begin{equation*}
      \begin{split}
        (\bvconcat{O}{I},\bvNil,[]) \models_{\env[x \mapsto (I,O,H)]} \sigmaT{y}{& \refT{z}{\epsilon}{z.\pOut = \bvNil \wedge z.\pIn = x.\pOut}}{ \\
          & \refT{z}{\epsilon}{z.\pOut = \bvNil \wedge z.\pIn = x.\pIn}}
      \end{split}
    \end{equation*}

    By \entSigma, we must show that
    \[
      (O,\bvNil,[]) \models_{\env[x \mapsto (I,O,H)]} \refT{z}{\epsilon}{z.\pOut = \bvNil \wedge z.\pIn = x.\pOut}
    \]
    and
    \[
      (I,\bvNil,[]) \models_{\env[x \mapsto (I,O,H), y \mapsto (O,\bvNil,[])]} \refT{z}{\epsilon}{z.\pOut = \bvNil \wedge z.\pIn = x.\pIn}
    \]

    Both follow after unfolding the definition of $\epsilon$ by \entRefine, \entTop and the semantics of formulae.

    \ref{preservation-g6} follows by
    \begin{align*}
      \phantom{{}={}} & \semantics{\sigmaT{y}{\refT{z}{\epsilon}{z.\pOut = \bvNil \wedge z.\pIn = x.\pOut}}{                                                          \\
                      & \hspace{22pt}\refT{z}{\epsilon}{z.\pOut = \bvNil \wedge z.\pIn = x.\pIn}}}_{\env[x\mapsto (I,O,H)]}                                           \\
      =               & \lbrace \concat{h_1}{h_2} \mid h_1\in\semantics{\refT{z}{\epsilon}{z.\pOut = \bvNil \wedge z.\pIn = x.\pOut}_{\env[x\mapsto (I,O,H)]}} \wedge \\
                      & \hspace{46pt} h_2\in\semantics{\refT{z}{\epsilon}{z.\pOut = \bvNil \wedge z.\pIn = x.\pIn}}_{\env[x\mapsto (I,O,H), y \mapsto h_1]} \rbrace   \\
      =               & \lbrace \concat{(O,\bvNil,[])}{(I,\bvNil, [])}\rbrace                                                                                         \\
      =               & \lbrace (\bvconcat{O}{I}, \bvNil, []) \rbrace                                                                                                 \\
      =               & \semantics{\refT{v}{\sigmaT{y}{\refT{z}{\epsilon}{z.\pOut = \bvNil \wedge z.\pIn = x.\pOut}}{                                                 \\
                      & \hspace{39pt}\refT{z}{\epsilon}{z.\pOut = \bvNil \wedge z.\pIn = x.\pIn}}}{                                                                   \\
                      & \hspace{22pt}v \equiv x'}}_{\env[x\mapsto (\bvconcat{O}{I},\bvNil,[]), x' \mapsto (\bvconcat{O}{I},\bvNil,[])]}
    \end{align*}

    \case{\tSeq}~

    By inversion of rule \tSeq, we get
    \begin{pf}{T-Seq}
      \item $c = c1;c2$
      \item \label{preservation-tseq-c1} $\Gamma \vdash c_1 : (x:\tau_1) \rightarrow \tau_{12}$
      \item \label{preservation-tseq-c2} $\Gamma,x:\tau_1 \vdash c_2 : (y:\tau_{12}) \rightarrow \tau_{22}$
      \item $\tau_2 = \tau_{22}[y \mapsto \tau_{12}]$
    \end{pf}

    \begin{description}
      \subcase{\eSeq}~
      \begin{pf*}{T-Seq}
        \item $c_1 = \cSkip$
        \item $c' = c_2$
      \end{pf*}
      By \eSeq, $I'=I$, $O'=O$, $H'=H$.

      Let $\Gamma' = \Gamma,x:\tau_1$, $\env' = \env[x \mapsto (I,O,H)]$ and $x'=y$.

      Let $\tau_1' = \tau_{12}$ and $\tau_2' = \tau_{22}$.

      \ref{preservation-g1} follows by \ref{preservation-tseq-c2},
      \ref{preservation-g2} follows by assumptions \ref{preservation-env} and \ref{preservation-heap} and \cref{lem:extended-env-entails}.
      \ref{preservation-g3} and \ref{preservation-g4} are immediate.

      To show \ref{preservation-g5}, we must show that $(I,O,H) \models_{\env[x \mapsto (I,O,H)]} \refT{z}{\tau_1}{z \equiv x}$ holds.
      By \entRefine, we must show that $(I,O,H) \models_{\env[x \mapsto (I,O,H)]} \tau_1$ and $\semantics{z \equiv x}_{\env[x \mapsto (I,O,H), z \mapsto (I,O,H)]}$, which follows by assumption (D) and the semantics of formulae.

      For \ref{preservation-g6}, we must show that
      \begin{align*}
        \phantom{{}\Leftrightarrow{}} & \semantics{\tau_2'}_{\env'[x \mapsto (I',O',H')]} \subseteq \semantics{\tau_2}_{\env[x \mapsto (I,O,H)]}                                                             \\
        \Leftrightarrow               & \semantics{\tau_{22}}_{\env[x \mapsto (I,O,H), y \mapsto (I,O,H)]} \subseteq \semantics{\tau_{22}[y \mapsto \refT{z}{\tau_1}{z \equiv x}]}_{\env[x \mapsto (I,O,H)]}
      \end{align*}

      \begin{align*}
        \phantom{{}={}} & \semantics{\tau_{22}[y \mapsto \tau_{12}]}_{\env[x \mapsto (I,O.H)]}                                                                                                                    \\
        =               & \lbrace h_{22}| h_{12} \in \semantics{\refT{z}{\tau_1}{z \equiv x}}_{\env[x \mapsto (I,O,H)]} \wedge h_{22}\in\semantics{\tau_{22}}_{\env[x \mapsto (I,O,H), y \mapsto h_{12}]} \rbrace \\
        =               & \lbrace h_{22}| h_{12} = (I,O,H) \wedge h_{22}\in\semantics{\tau_{22}}_{\env[x \mapsto (I,O,H), y \mapsto h_{12}]} \rbrace                                                              \\
        =               & \lbrace h_{22}| h_{22}\in\semantics{\tau_{22}}_{\env[x \mapsto (I,O,H), y \mapsto (I,O,H)]} \rbrace                                                                                     \\
        =               & \semantics{\tau_{22}}_{\env[x \mapsto (I,O,H), y \mapsto (I,O,H)]}
      \end{align*}
      This shows \ref{preservation-g6} and concludes this subcase.
      \subcase{\eSeqOne}~

      \begin{pf*}{T-Seq}
        \item \label{preservation-tseq-esq1-cprime} $c'=c_1';c_2$
        \item $\langle I,O,H,c_1\rangle \rightarrow \langle I',O',H', c_1'\rangle$
      \end{pf*}

      By IH with \ref{preservation-tseq-c1}, \ref{preservation-tseq-esq1-cprime}, \ref{preservation-env} and \ref{preservation-heap}, there exists $\Gamma', \env', \tau_1', \tau_{12}',x'$, such that,
      \begin{pf*}{T-Seq}
        \item \label{preservation-tseq-seq1-c1prime-type} $\Gamma' \vdash c_1' :(x':\tau_1') \rightarrow \tau_{12}'$ where
        \item \label{preservation-tseq-seq1-envprime} $\env' \models \Gamma'$
        \item \label{preservation-tseq-seq1-gammaprime} $\Gamma \subseteq \Gamma'$
        \item \label{preservation-tseq-seq1-env-entails} $\env \subseteq \env'$
        \item \label{preservation-tseq-seq1-heapprime-entails} $(I',O',H') \models_{\env'} \tau_1'$
        \item \label{preservation-tseq-seq1-t12-sub} $\semantics{\tau_{12}'}_{\env'[x' \mapsto (I',O',H')]} \subseteq \semantics{\tau_{12}}_{\env[x \mapsto (I,O,H)]}$
      \end{pf*}

      \ref{preservation-g1} follows by \tSeq, if we can show that there exists some $\tau_{22}'$, such that
      $\Gamma',x':\tau_1' \vdash c_2 : (y:\tau_{12}') \rightarrow \tau_{22}'$ where
      $\tau_2' = \tau_{22}'[y \mapsto \tau_{12}']$:

      \begin{mathpar}
        \inferrule[\tSeq]{
        \Gamma' \vdash c_1':(x:\tau_1') \rightarrow \tau_{12}'\\
        \Gamma',x:\tau_1' \vdash c_2 : (y:\tau_{12}') \rightarrow \tau_{22}'
        }{
        \Gamma' \vdash c_1';c_2 : (x:\tau_1') \rightarrow \tau_{22}'[y \mapsto \tau_{12}']
        }
      \end{mathpar}

      By \autoref{lem:input-type-strengthening} with \ref{preservation-tseq-c2} and \ref{preservation-tseq-seq1-heapprime-entails}, there exists some $\tau_{22}'$ such that
      \begin{pf*}{T-Seq}
        \item $\Gamma',x':\tau_1' \vdash c_2 : (y:\tau_1') \rightarrow \tau_{22}'$
        \item \label{preservation-tseq-seq1-t22-sub} $\forall h'\in\semantics{\tau_{12}'}_{\env'[x' \mapsto (I',O',H')]}.\semantics{\tau_{22}'}_{\env'[x' \mapsto (I',O',H'), y \mapsto h']} \subseteq \\
                \semantics{\tau_{22}}_{\env[x \mapsto (I,O,H), y \mapsto h']}$
      \end{pf*}

      \ref{preservation-g2} follows by \ref{preservation-tseq-seq1-envprime} and \ref{preservation-g3} follows by \ref{preservation-tseq-seq1-gammaprime}, \ref{preservation-g4} follows by \ref{preservation-tseq-seq1-env-entails} and \ref{preservation-g5} follows by \ref{preservation-tseq-seq1-heapprime-entails}.

      \ref{preservation-g6} follows by

      \begin{align*}
        \phantom{{}\Leftrightarrow{}} & \semantics{\tau_2'}_{\env'[x' \mapsto (I',O',H')]} \subseteq \semantics{\tau_2}_{\env[x \mapsto (I,O,H)]}                                                 \\
        \Leftrightarrow               & \semantics{\tau_{22}'[y \mapsto \tau_{12}']}_{\env'[x \mapsto (I',O',H')]} \subseteq \semantics{\tau_{22}[y \mapsto \tau_{12}]}_{\env[x \mapsto (I,O,H)]} \\
        \Leftrightarrow               & \bigcup_{h'\in\semantics{\tau_{12}'}_{\env'[x \mapsto (I',O',H')]}} \semantics{\tau_{22}'}_{\env'[x \mapsto (I',O',H'), y \mapsto h']}\ \subseteq         \\
                                      & \bigcup_{h\in\semantics{\tau_{12}}_{\env[x \mapsto (I,O,H)]}} \semantics{\tau_{22}}_{\env[x \mapsto (I,O,H), y \mapsto h]}
      \end{align*}
      and by \ref{preservation-tseq-seq1-t12-sub}, \ref{preservation-tseq-seq1-t22-sub} and the semantics of heap types.
    \end{description}

    \case{\tSkip}~

    Immediately holds as there is no $c'$ such that $\langle I,O,H,c\rangle \rightarrow \langle I',O',H',c'\rangle$.

    \case{\tSub}~

    \begin{pf}{T-Sub}
      \item \label{preservation-tsub-c-t3-t4} $\cmdT{c}{(x:\tau_3) \rightarrow \tau_4}$
      \item \label{preservation-tsub-t1-sub-t3} $\subtypeCtx{\Gamma}{\tau_1}{\tau_3}$
      \item \label{preservation-tsub-t4-sub-t2} $\subtypeCtx{\Gamma,x\!:\!\tau_1}{\tau_4}{\tau_2}$
    \end{pf}

    By \cref{lem:subtype-entailment} with assumptions \ref{preservation-env}, \ref{preservation-heap} and \ref{preservation-tsub-t1-sub-t3},
    \begin{pf*}{T-Sub}
      \item \label{preservation-tsub-entails-t3} $(I,O,H) \models_\env \tau_3$.
    \end{pf*}
    By IH with \ref{preservation-tsub-c-t3-t4}, \ref{preservation-step}, \ref{preservation-env} and \ref{preservation-tsub-entails-t3}, there exists $\Gamma',\env',\tau_3', \tau_4',x'$ such that
    \begin{pf*}{T-Sub}
      \item \label{preservation-tsub-ih1} $\Gamma' \vdash c: (x':\tau_3') \rightarrow \tau_4'$
      \item \label{preservation-tsub-ih2} $\env' \models \Gamma'$
      \item \label{preservation-tsub-ih3} $\Gamma \subseteq \Gamma'$
      \item \label{preservation-tsub-ih4} $\env \subseteq \env'$
      \item \label{preservation-tsub-ih5} $(I',O',H') \models_{\env'} \tau_3'$
      \item \label{preservation-tsub-ih6} $\semantics{\tau_4'}_{\env'[x' \mapsto (I',O',H')]} \subseteq \semantics{\tau_4}_{\env[x \mapsto (I,O,H)]}$
    \end{pf*}

    Let $\tau_1' = \tau_3'$ and $\tau_2' = \tau_4'$.
    \ref{preservation-g1} follows by \ref{preservation-tsub-ih1},
    \ref{preservation-g2} follows by assumption \ref{preservation-tsub-ih2},
    \ref{preservation-g3} follows by assumption \ref{preservation-tsub-ih3},
    \ref{preservation-g4} follows by assumption \ref{preservation-tsub-ih4},
    \ref{preservation-g5} follows by assumption \ref{preservation-tsub-ih5} and
    \ref{preservation-g6} follows by \ref{preservation-tsub-t4-sub-t2} and \ref{preservation-tsub-ih6}.

  \end{description}
\end{proof}

%% file: appendix/algorithmic_types.tex
\section{Algorithmic Typing Rules}

The command typing rules are shown in
Figure~\ref{fig:algorithmic-command-typing-rules}.

\begin{figure}[t]
  \begin{mathpar}
    \inferrule[\tExtractAlgo]{
      \subtype{\tau_1}{\refT{x}{\top}{|x.\pIn| \geq \sizeof(\inst)}} \\
      \varphi_1 \triangleq z.\pIn = z.\pOut = \bvNil\\\\
      \varphi_2 \triangleq \bvconcat{y.\inst}{z.\pIn} = x.\pIn \land z.\pOut = x.\pOut \land z \equiv_\inst x
    }{
      \cmdTypeAlgo{\Gamma}{\mathit{extract(\inst)}}{x:\tau_1}{\sigmaT{y}{\refT{z}{\inst}{\varphi_1}}{\refT{z}{\chomp(\tau_1, \inst, y)}{\varphi_2}}}
    }
    \and
    \inferrule[\tAddAlgo]{
      \subtype{\tau_1}{\refT{x}{\top}{\neg x.\inst.\valid}} \\
      \mathit{init}_{\HT(\inst)} = v
    }{
      \cmdTypeAlgo{\Gamma}{\cAdd{\inst}}{x:\tau_1}{\sigmaT{y}{\refT{z}{\tau_1}{z \equiv x}}{\refT{z}{\inst}{z.\pIn=z.\pOut=\bvNil \wedge z.\inst = v}}}
    }
    \and
    \inferrule[\tModAlgo]{
      \subtype{\tau}{\instWeak} \\
      \mathcal{F}(\inst, f) = \tBv \\
      \Gamma;\tau_1 \vdash e : \tBv \\\\
      \varphi_{pkt} \triangleq y.\pIn=x.\pIn \land y.\pOut=x.\pOut \\\\
      \varphi_{\inst} \triangleq \forall \kappa \in \dom(\HT).~\inst\neq\kappa \Rightarrow y.\kappa=x.\kappa\\\\
      \varphi_{f} \triangleq \forall g \in \dom(\HT(\inst)).~f \neq g \Rightarrow y.\inst.g=x.\inst.g
    }{
      \cmdTypeAlgo{\Gamma}{\inst.f := e}{x:\tau_1}{\refT{y}{\top}{\varphi_{pkt} \land \varphi_{\inst} \land \varphi_{f} \land y.\inst.f = e[x/\cmdVar]}}
    }
    \and
    \inferrule[\tRemitAlgo]{
      \subtype{\tau_1}{\instWeak} \\
      \varphi \triangleq z.\pIn = \bvNil \land z.\pOut=x.\inst %
    }
    {
      \cmdTypeAlgo{\Gamma}{\mathit{remit(\inst)}}{x:\tau_1}{\sigmaT{y}{\refT{z}{\tau_1}{z\equiv x}}{\refT{z}{\epsilon}{\varphi}}}
    }
    \and
    \inferrule[\tResetAlgo]{
      \varphi_1 \triangleq z.\pOut = \bvNil \land z.\pIn = x.\pOut\\\\
      \varphi_2 \triangleq z.\pOut = \bvNil \land z.\pIn = x.\pIn
    }{
      \cmdTypeAlgo{\Gamma}{\mathit{reset}}{x: \tau_1}{
        \sigmaT{y}
        {\refT{z}{\epsilon}{\varphi_1}}
        {\refT{z}{\epsilon}{\varphi_2}}}
    }
    \and
    \inferrule[\tSeqAlgo]{
      \cmdTypeAlgo{\Gamma}{c_1}{x:\tau_1}{\tau_{12}}\\\\
      \cmdTypeAlgo{\Gamma, (x:\tau_1)}{c_2}{y:\tau_{12}}{\tau_{22}}\\
    }{
      \cmdTypeAlgo{\Gamma}{c_1;c_2}{x:\tau_1}{\tau_{22}[y\mapsto \tau_{12}]}
    }
    \and
    \inferrule[\tSkipAlgo]{
      \tau_2 \triangleq \refT{y}{\tau_1}{y \equiv x}
    }{
      \cmdTypeAlgo{\Gamma}{\cSkip}{x:\tau_1}{\tau_2}
    }
    \and
    \inferrule[\tAscribeAlgo]{
      \cmdTypeAlgo{\Gamma}{c}{x:\hat{\tau_1}}{\tau_c}\\
      \subtype{\tau_1}{\hat{\tau_1}} \\\\
      \subtypeCtx{\Gamma,x:\hat{\tau_1}}{\tau_c}{\hat{\tau_2}} \\
    }{
      \cmdTypeAlgo{\Gamma}{\cAscribe{c}{(x:\hat{\tau_1}) \to \hat{\tau_2}}}{x:\tau_1}{\hat{\tau_2}}
    }
    \and
    \inferrule[\tIfAlgo]{
      \Gamma;\tau_1 \vdash \varphi : \tBool \\
      \cmdTypeAlgo{\Gamma}{c_1}{x:\refT{y}{\tau_1}{\varphi[y/\cmdVar]}}{\tau_{12}} \\
      \cmdTypeAlgo{\Gamma}{c_2}{x:\refT{y}{\tau_1}{\neg \varphi[y/\cmdVar]}}{\tau_{22}}
    }{
      \cmdTypeAlgo{\Gamma}{\mathit{if(\varphi)\ c_1\ else\ c_2}}{x:\tau_1}{\refT{y}{\tau_{12}}{\varphi[x/\cmdVar]} + \refT{y}{\tau_{22}}{\neg \varphi[x/\cmdVar]}}
    }
  \end{mathpar}
  \caption{Algorithmic typing rules for \name. Assume binders are fresh in all rules}
  \label{fig:algorithmic-command-typing-rules}
\end{figure}

\begin{lemma}[Subtype Reflexivity]
  \label{lem:subtype-reflexivity}
  For all subtyping contexts $\Gamma$ and heap types $\tau$, $\subtype{\tau}{\tau}$.
\end{lemma}
\begin{proof}
  Immediate.
\end{proof}

\begin{lemma}[Subtype Transitivity]
  \label{lem:subtype-transitivity}
  If $\subtype{\tau_{1}}{\tau_{2}}$, and $\subtype{\tau_{2}}{\tau_{3}}$,
  then $\subtype{\tau_{1}}{\tau_{3}}$.
\end{lemma}
\begin{proof}
  Assume $\subtype{\tau_{1}}{\tau_{2}}$ and also assume $\subtype{\tau_{2}}{\tau_{3}}$.
  Let $\env \models \Gamma$ and $h \in \semantics{\tau_{1}}_{\env}$ be arbitrary.
  By the first assumption $h \in \semantics{\tau_{2}}_{\env}$.
  By the second assumption $h \in \semantics{\tau_{3}}_{\env}$.
\end{proof}

\begin{lemma}[Environment Entails Subtype]
  \label{lem:context-subtyping}
  If $\subtype{\tau_{1}}{\tau_{3}}$, and
  $\env \models \Gamma,x:\tau_{3}$, then
  $\env \models \Gamma,x:\tau_{1}$.
\end{lemma}
\begin{proof}
  Let $\subtype{\tau_{1}}{\tau_{3}}$ and $\env \models \Gamma,(x:\tau_{3})$.
  Let $\env \models \Gamma,(x:\tau_{1})$.
  We can write $\env = \env'[x \mapsto h_{1}]$, such that $\env' \models \Gamma$, and
  $h_{1} \in \semantics{\tau_{1}}_{\env'}$.
  The definition of subtyping gives $h_{1} \in \semantics{\tau_{3}}_{\env'}$.
  The result follows by definition of entailment.
\end{proof}

\begin{lemma}[Context Strengthening]
  \label{lem:context-strengthening}
  If $\subtype{\tau_{1}}{\tau_{3}}$ and
  $\subtypeCtx{\Gamma,x:\tau_{3}}{\tau_{2}}{\tau_{4}}$
  then
  $\subtypeCtx{\Gamma,x:\tau_{1}}{\tau_{2}}{\tau_{4}}$.
\end{lemma}
\begin{proof}
  Assume $\subtype{\tau_{1}}{\tau_{3}}$ and further assume $\subtypeCtx{\Gamma,x:\tau_{3}}{\tau_{2}}{\tau_{4}}$.
  Let $\env \models \Gamma,x:\tau_{1}$.
  By Lemma~\ref{lem:context-subtyping} and
  the first assumption, $\env \models \Gamma,(x:\tau_{3})$. Let
  $h \in \semantics{\tau_{2}}_{\env}$, the second assumption gives that
  $h \in \semantics{\tau_{2}}_{\env}$, and we're done.
\end{proof}

\begin{lemma}[Packet Bound Subtype]
  \label{lem:bound-subtype}
  $\Gamma \vdash \sizeof_{\pIn}(\tau) \geq N$ iff $\subtype{\tau}{\refT{x}{\top}{\length{x.\pIn} \geq \sizeof(\inst)}}$.
\end{lemma}
\begin{proof}
  We show each direction separately.
  \begin{enumerate}[align=left]
    \item[($\Rightarrow$)]
      Assume $\Gamma \vdash \sizeof_{\pIn}(\tau) \geq N$.
      Let $\env \models \Gamma$ and $h \in \semantics{\tau}_{\env}$ be arbitrary.
      By definition, $\length{h(\pIn)} \geq N$.
      By the definition of subtyping, it suffices to show $h \in \semantics{\refT{x}{\top}{\length{x.\pIn} \geq N}}_{\env}$.
      By definition, $\semantics{\refT{x}{\top}{\length{x.\pIn} \geq N}}_{\env} = \{h \mid h\in\Heaps \wedge h(\pIn) \geq N\}$, which concludes this case.
    \item[($\Leftarrow$)]
      Assume $\subtype{\tau}{\refT{x}{\top}{\length{x.\pIn} \geq N}}$.
      We have to show that $\Gamma \vdash \sizeof_{\pIn}(\tau) \geq N$.
      Let $\env \models \Gamma$ and $h \in \semantics{\tau}_{\env}$ be arbitrary.
      By the definition of subtyping, $h \in \semantics{\refT{x}{\top}{\length{x.\pIn} \geq N}}_{\env}$.
      By definition of the semantics, we can conclude $h(\pIn) \geq N$.
  \end{enumerate}
\end{proof}

\begin{lemma}[Chomp Subtype]
  \label{lem:chomp-subtype}
  If $x$ not free in $\tau$ and $\tau'$,
  and $\Gamma \vdash \sizeof_{\pIn}(\tau) \geq \sizeof(\inst)$
  and  $\subtype{\tau}{\tau'}$,
  then $\subtypeCtx{\Gamma,x:\refT{y}{\inst}{y.\pIn = y.\pOut = \bvNil}}{\chomp(\tau, \inst, x)}{\chomp(\tau', \inst, x)}$.
\end{lemma}
\begin{proof}
  Given some heap $h'\in\semantics{\chomp(\tau',\inst,x)}_{\env'}$.
  By \cref{lem:semantic-chomp-inverse} there exists some $\env$ and $h''\in\semantics{\tau'}_\env$,
  such that $\env' = \env[x \mapsto (\bvNil, \bvNil, [\inst \mapsto \slice{h''(\pIn)}{0}{\sizeof(\inst)}])]$ and \\
  $h'=\chompS(h'',\sizeof(\inst))$.
  By assumption $\subtype{\tau}{\tau'}$, we also know that $h''\in\semantics{\tau}_\env$.

  By \cref{lem:semantic-chomp}, we know that there exists $h\in\semantics{\chomp(\tau,\inst,x)}_{\env''}$ such that
  $h=\chompS(h'',\sizeof(\inst))$ and $\env''=\env[x \mapsto (\bvNil, \bvNil, [\inst \mapsto \slice{h''(\pIn)}{0}{\sizeof(\inst)}])]$.
  From $h'=\chompS(h'',\sizeof(\inst))$ and $h=\chompS(h'',\sizeof(\inst))$ follows by the transitivity of equality that $h'=h$.
  By the fact that $\env'=\env''$ follows that for every heap $h'\in\semantics{\chomp(\tau',\inst,x)}_{\env'}$ also holds that
  $h'\in\semantics{\chomp(\tau,\inst,y)}_{\env''}$.
\end{proof}

\begin{lemma}[Refinement Subtype]
  \label{lem:refinement-subtype}
  If $\subtype{\tau'}{\tau}$,
  then
  $\subtype{\refT{x}{\tau'}{\varphi}}{\refT{x}{\tau}{\varphi}}$.
\end{lemma}
\begin{proof}
  Let $\env \vdash \Gamma$ and $h \in \semantics{\refT{x}{\tau'}{\varphi}}_\env$.
  Then $h \in \semantics{\tau'}_{\env}$, and $\semantics{\varphi}_{\env[x \mapsto h]} = \mathit{true}$.
  By assumption $\subtype{\tau'}{\tau}$, $h \in \semantics{\tau}_{\env}$.
  Conclude $h \in \semantics{\refT{x}{\tau}{\varphi}}_{\env}$ by definition.
  The result follows.
\end{proof}

\begin{lemma}[Sigma Left-Subtype]
  \label{lem:sigma-left-subtype}
  If $\subtype{\tau_{1}'}{\tau_{1}}$
  then
  $\subtype{\sigmaT{x}{\tau_{1}'}{\tau_{2}}}{\sigmaT{x}{\tau_{1}}{\tau_{2}}}$.
\end{lemma}
\begin{proof}
  Let $\env \models \Gamma$, and $h \in \semantics{\sigmaT x {\tau_{1}'} {\tau_{2}}}_{\env}$.
  By definition of the semantics, $h = \concat{h_{1}}{h_{2}}$,where
  $h_{1} \in \semantics{\tau_{1}'}_{\env}$ and $h_{2} \in \semantics{\tau_{2}}_{\env[x \mapsto h]}$.
  By assumption $\subtype{\tau_{1}'}{\tau_{1}}$ follows, $h_{1} \in \semantics{\tau_{1}}_{\env}$.
  By the definition of the semantics,
  $h \in \semantics{\sigmaT x {\tau_{1}} {\tau_{2}}}_{\env}$.
  The result follows.
\end{proof}

\begin{lemma}[Sigma Right-Subtype]
  \label{lem:sigma-right-subtype}
  If $\subtypeCtx{\Gamma,x:\tau_{1}}{\tau_{2}'}{\tau_{2}}$
  then
  $\subtype{\sigmaT{x}{\tau_{1}} {\tau_{2}'}}{\sigmaT{x}{\tau_{1}} {\tau_{2}}}$.
\end{lemma}
\begin{proof}
  Let $\env \models \Gamma$ and $h \in \semantics{\sigmaT{x}{\tau_{1}}{\tau_{2}'}}_{\env}$.
  By definition,
  $h = \concat{h_{1}}{h_{2}}$ such that $h_{1} \in \semantics{\tau_{1}}_{\env}$,
  and $h_{2} \in \semantics{\tau_{2}'}_{\env[x \mapsto h_{1}]}$.
  Notice that $\env[x \mapsto h_{1}] \models \Gamma,(x:\tau_{1})$, so by assumption $\subtypeCtx{\Gamma,x:\tau_{1}}{\tau_{2}'}{\tau_{2}}$ follows $h_{2} \in \semantics{\tau_{2}}_{\env[x \mapsto h_{1}]}$.
  By the definition of the semantics
  $h \in \semantics{\sigmaT x {\tau_{1}}{\tau_{2}}}_{\env}$. The result follows.
\end{proof}

\begin{lemma}[Substitution Subtype]
  \label{lem:substitution-subtype}
  If $\subtype{\tau_{1}'}{\tau_{1}}$ and
  $\subtypeCtx{\Gamma,(x:\tau_{1}')}{\tau_{2}'}{\tau_{2}}$, then
  $\subtype{\tau_{2}'[x \mapsto \tau_{1}']}{\tau_{2}[x \mapsto \tau_{1}]}$.
\end{lemma}
\begin{proof}
  Let $\env \models \Gamma$ and $h_{2} \in \semantics{\tau_{2}'[x \mapsto \tau_{1}']}_{\env}$.
  Then we know $h_{1} \in \semantics{\tau_{1}'}_{\env}$ and $h_{2} \in \semantics{\tau_{2}'}_{\env[x \mapsto h_{1}]}$.
  Assumption $\subtype{\tau_{1}'}{\tau_{1}}$ tells us that $h_{1} \in \semantics{\tau_{1}}_{\env}$.
  Notice that $\env[x\mapsto h_{1}] \models \Gamma,(x:\tau_{1}')$.
  Assumption $\subtypeCtx{\Gamma,(x:\tau_{1})}{\tau_{2}'}{\tau_{2}}$ gives $h_{2} \in \semantics{\tau_{2}}_{\env[x\mapsto h_{1}]}$.
  By the definition of the semantics of heap types,
  $h_{2} \in \semantics{\tau_{2}[x \mapsto \tau_{1}]}_{\env}$.
\end{proof}

\begin{lemma}[Choice Subtype]
  \label{lem:choice-subtype}
  If $\subtype{\tau_{1}'}{\tau_{1}}$,
  and $\subtype{\tau_{2}'}{\tau_{2}}$,
  then
  $\subtype{\tau_{1}' + \tau_{2}'}{\tau_{1} + \tau_{2}}$
\end{lemma}
\begin{proof}
  Let $\env \models \Gamma$. Let $h \in \semantics{\tau_{1}' + \tau_{2}'}_{\env}$.
  By semantics of heap types, either $h \in \semantics{\tau_{1}'}_{\env}$ or $h \in \semantics{\tau_{2}'}_{\env}$.
  \begin{description}
    \subcase{$h \in \semantics{\tau_{1}'}_{\env}$}
    By assumption $\subtype{\tau_{1}'}{\tau_{1}}$ it also holds that $h \in \semantics{\tau_{1}}_{\env}$ and we can conclude that $h \in \semantics{\tau_{1} + \tau_{2}}_{\env}$.
    \subcase{$h \in \semantics{\tau_{2}'}_{\env}$}
    By assumption $\subtype{\tau_{2}'}{\tau_{2}}$ it also holds that $h \in \semantics{\tau_{2}}_{\env}$ and we can conclude that $h \in \semantics{\tau_{1} + \tau_{2}}_{\env}$.
  \end{description}
\end{proof}

\begin{lemma}[Context-Bound Refinement Subtype]
  \label{lem:context-bound-refinement-subtype}
  If $\cmdVar$ is the only free binder in $\varphi$,
  and $\subtypeCtx{\Gamma,x:\refT{y}{\tau_{1}}{\varphi[y/\cmdVar]}}{\tau_{2}'}{\tau_{2}}$
  then
  $\subtypeCtx{\Gamma,x:\tau_{1}}{\refT{y}{\tau_{2}'}{\varphi[x/\cmdVar]}}{\refT{y}{\tau_{2}}{\varphi[x/\cmdVar]}}$.
\end{lemma}

\begin{proof}
  Let $\env \models \Gamma,x:\tau_{1}$.
  We can write this as $\env = \env'[x \mapsto h_{1}]$, where $h_{1} \in \semantics{\tau_{1}}_{\env}$.
  Let $h_{2} \in \semantics{\refT{y}{\tau_{2}'}{\varphi[x/\cmdVar]}}_{\env}$.
  Then $h_{2} \in \semantics{\tau_{2}'}_{\env}$ and $\semantics{\varphi[x/\cmdVar]}_{\env[y \mapsto h_{2}]} = \mathit{true}$.
  Compute as follows, recalling that $\cmdVar$ is the only free binder in $\varphi$:
  \begin{align*}
    \phantom{{}={}} & \semantics{\varphi[x/\cmdVar]}_{\env[y \mapsto h_{2}]}  \\
    =               & \semantics{\varphi[x/\cmdVar]}_{\env}                   \\
    =               & \semantics{\varphi[x/\cmdVar]}_{\env'[x \mapsto h_{1}]} \\
    =               & \semantics{\varphi[y/\cmdVar]}_{\env'[y \mapsto h_{1}]}
  \end{align*}
  Together with assumption $h_{1} \in \semantics{\tau_{1}}_{\env}$, we get $h_{1}\in\semantics{\refT{y}{\tau_{1}}{\varphi[y/\cmdVar]}}_{\env'}$, and thus
  $\env'[x\mapsto h_{1}] \models \Gamma,x:\refT{y}{\tau_{1}} {\varphi[y/\cmdVar]}$.

  With assumption $\subtypeCtx{\Gamma,x:\refT{y}{\tau_{1}}{\varphi[y/\cmdVar]}}{\tau_{2}'}{\tau_{2}}$, we can conclude that $h_{2} \in \semantics{\tau_{2}}_{\env}$.
  Since we already have that $\semantics{\varphi[x/\cmdVar]}_{\env[y\mapsto h_{2}]} = \mathit{true}$, it follows that $h_{2} \in \semantics{\refT{y}{\tau_{2}}{\varphi[x/\cmdVar]}}_{\env}$, which is what we wanted to show.
\end{proof}

\begin{lemma}[If Choice Subtype]
  \label{lem:if-choice-subtype}
  If \cmdVar is the only free binder in $\varphi$ and
  $\subtypeCtx{\Gamma,x:\refT{y}{\tau_{1}'}{\varphi[y/\cmdVar]}}{\tau_{12}'}{\tau_{12}}$, and
  $\subtypeCtx{\Gamma,x:\refT{y}{\tau_{1}'}{\neg \varphi[y/\cmdVar]}}{\tau_{22}'}{\tau_{22}}$,
  then
  $\subtypeCtx{\Gamma,x:\tau_{1}'}
    {\refT{y}{\tau_{12}'}{\varphi[x/\cmdVar]} + \refT{y}{\tau_{22}'}{\varphi[x/\cmdVar]}}
    {\refT{y}{\tau_{12}}{\varphi[x/\cmdVar]} + \refT{y}{\tau_{22}}{\neg \varphi[x/\cmdVar]}}$
\end{lemma}
\begin{proof}
  By Lemmas~\ref{lem:context-bound-refinement-subtype} and~\ref{lem:choice-subtype}.
\end{proof}

\begin{lemma}[Algorithmic Weakening]
  \label{lem:algorithmic-weakening}
  If $\cmdTypeAlgo{\Gamma}{c}{x:\tau_1}{\tau_2}$ and variable $y$ does not appear free in $\tau_1$ or $\tau_2$,
  then $\cmdTypeAlgo{\Gamma,y:\tau}{c}{x:\tau_1}{\tau_2}$ for any heap type $\tau$.
\end{lemma}
\begin{proof}
  By induction on the typing derivation.
\end{proof}

\begin{lemma}[Typing Context Subtype]
  \label{lem:typing-context-subtype}
  If $\cmdTypeAlgo{\Gamma,x:\tau_1}{c}{y:\tau_{12}}{\tau_{22}}$ and
  $\subtype{\tau_1'}{\tau_1}$,
  then $\cmdTypeAlgo{\Gamma,x:\tau_1'}{c}{y:\tau_{12}}{\tau_{22}}$.
\end{lemma}
\begin{proof}
  If $x$ is not free in $\tau_{12}$ or $\tau_{22}$, the result follows from \cref{lem:algorithmic-weakening}.
  Otherwise we proceed by induction on the typing derivation.
  We refer to the general assumptions as follows:
  \begin{enumerate}[label=(\Alph*)]
    \item \label{typing-context-subtype-a1} $\cmdTypeAlgo{\Gamma,x:\tau_1}{c}{y:\tau_{12}}{\tau_{22}}$
    \item \label{typing-context-subtype-a2} $\subtype{\tau_1'}{\tau_1}$
  \end{enumerate}
  \begin{description}
    \case{\tExtractAlgo}~
    By inversion of \tExtractAlgo, we know
    \begin{pf}{T-Extract-Algo}
      \item $\cmdTypeAlgo{\Gamma,x:\tau_1}{\cExtract{\inst}}{y:\tau_{12}}{\tau_{22}}$
      \item $\tau_{22} = \sigmaT{z}{\refT{v}{\inst}{\varphi_1}}{\refT{v}{\chomp(\tau_{12},\inst,z)}{\varphi_2}}$
      \item \label{typing-context-subtype-textract-sizecheck} $\subtypeCtx{\Gamma,x:\tau_1}{\tau_{12}}{\refT{z}{\top}{\length{z.\pIn} \ge \sizeof(\inst)}}$
    \end{pf}
    By \cref{lem:context-strengthening} applied to \ref{typing-context-subtype-textract-sizecheck} and \ref{typing-context-subtype-a2} follows
    \begin{pf*}{T-Extract-Algo}
      \item $\subtypeCtx{\Gamma,x:\tau_1'}{\tau_{12}}{\refT{z}{\top}{\length{z.\pIn} \ge \sizeof(\inst)}}$
    \end{pf*}
    The result follows by \tExtractAlgo.

    \case{\tSeqAlgo}~
    By inversion of \tSeqAlgo, we know
    \begin{pf}{T-Seq-Algo}
      \item $\cmdTypeAlgo{\Gamma,x:\tau_1}{c_1;c_2}{y:\tau_{12}}{\tau_{22}}$
      \item \label{typing-context-subtype-tseq-c1} $\cmdTypeAlgo{\Gamma,x:\tau_1}{c_1}{y:\tau_{12}}{\tau_{12}'}$
      \item \label{typing-context-subtype-tseq-c2} $\cmdTypeAlgo{\Gamma,x:\tau_1,y:\tau_{12}}{c_2}{z:\tau_{12}'}{\tau_{22}'}$
      \item $\tau_{22} = \tau_{22}'[z \mapsto \tau_{12}']$
    \end{pf}
    By IH applied to \ref{typing-context-subtype-tseq-c1} and \ref{typing-context-subtype-a2} follows
    \begin{pf*}{T-Seq-Algo}
      \item \label{typing-context-subtype-tseq-ih1} $\cmdTypeAlgo{\Gamma,x:\tau_1'}{c_1}{y:\tau_{12}}{\tau_{12}'}$
    \end{pf*}
    By IH applied to \ref{typing-context-subtype-tseq-c2} and \ref{typing-context-subtype-a2} follows
    \begin{pf*}{T-Seq-Algo}
      \item \label{typing-context-subtype-tseq-ih2} $\cmdTypeAlgo{\Gamma,x:\tau_1',y:\tau_{12}}{c_2}{z:\tau_{12}'}{\tau_{22}'}$
    \end{pf*}
    The result follows by \tSeqAlgo with \ref{typing-context-subtype-tseq-ih1} and \ref{typing-context-subtype-tseq-ih2}.

    \case{\tSkipAlgo}~
    The result immediately follows by \tSkipAlgo.

    \case{\tRemitAlgo}~
    By inversion of \tRemitAlgo, we know
    \begin{pf}{T-Remit-Algo}
      \item \label{typing-context-subtype-tremit-includes} $\subtypeCtx{\Gamma,x:\tau_1}{\tau_{12}}{\inst_\approx}$
    \end{pf}
    By \cref{lem:context-strengthening} applied to \ref{typing-context-subtype-tremit-includes} and \ref{typing-context-subtype-a2} follows
    \begin{pf*}{T-Remit-Algo}
      \item $\subtypeCtx{\Gamma,x:\tau_1'}{\tau_{12}}{\inst_\approx}$
    \end{pf*}
    The result follows by \tRemitAlgo.

    \case{\tResetAlgo}~
    The result immediately follows by \tResetAlgo.

    \case{\tAscribeAlgo}~
    By inversion of \tAscribeAlgo, we know
    \begin{pf}{T-Ascribe-Algo}
      \item $\cmdTypeAlgo{\Gamma,x:\tau_1}{c_0\ as\ (y:\hat\tau_{12}) \to \tau_{22}}{y:\tau_{12}}{\tau_{22}}$
      \item \label{typing-context-subtype-tascribe-c0}$\cmdTypeAlgo{\Gamma,x:\tau_1}{c_0}{y:\hat\tau_{12}}{\tau_{22}'}$
      \item \label{typing-context-subtype-tascribe-subin} $\subtypeCtx{\Gamma,x:\tau_1}{\tau_{12}}{\hat\tau_{12}}$
      \item \label{typing-context-subtype-tascribe-subout} $\subtypeCtx{\Gamma,x:\tau_1,y:\hat\tau_{12}}{\tau_{22}'}{\tau_{22}}$
    \end{pf}
    By IH applied to \ref{typing-context-subtype-tascribe-c0} and \ref{typing-context-subtype-a2} follows
    \begin{pf*}{T-Ascribe-Algo}
      \item \label{typing-context-subtype-tascribe-concl1} $\cmdTypeAlgo{\Gamma,x:\tau_1'}{c_0}{y:\hat\tau_{12}}{\tau_{22}'}$
    \end{pf*}
    By \cref{lem:context-strengthening} applied to \ref{typing-context-subtype-tascribe-subin} and \ref{typing-context-subtype-a2} follows
    \begin{pf*}{T-Ascribe-Algo}
      \item \label{typing-context-subtype-tascribe-concl2} $\subtypeCtx{\Gamma,x:\tau_1'}{\tau_{12}}{\hat\tau_{12}}$
    \end{pf*}
    By \cref{lem:context-strengthening} applied to \ref{typing-context-subtype-tascribe-subout} and \ref{typing-context-subtype-a2} follows
    \begin{pf*}{T-Ascribe-Algo}
      \item \label{typing-context-subtype-tascribe-concl3} $\subtypeCtx{\Gamma,x:\tau_1',y:\hat\tau_{12}}{\tau_{22}'}{\tau_{22}}$
    \end{pf*}
    The result follows by \tAscribeAlgo with \ref{typing-context-subtype-tascribe-concl1}, \ref{typing-context-subtype-tascribe-concl2} and \ref{typing-context-subtype-tascribe-concl3}.

    \case{\tIfAlgo}~

    By inversion of \tIfAlgo, we know
    \begin{pf}{T-If-Algo}
      \item \label{typing-context-subtype-tif-then} $\cmdTypeAlgo{\Gamma,x:\tau_1}{c_1}{y:\refT{z}{\tau_{12}}{\varphi[z/\cmdVar]}}{\tau_{12}'}$
      \item \label{typing-context-subtype-tif-else} $\cmdTypeAlgo{\Gamma,x:\tau_1}{c_1}{y:\refT{z}{\tau_{12}}{\neg\varphi[z/\cmdVar]}}{\tau_{12}''}$
      \item $\tau_{22} = \refT{z}{\tau_{12}'}{\varphi[x/\cmdVar]} + \refT{z}{\tau_{12}''}{\neg \varphi[x/\cmdVar]}$
      \item $\Gamma,x:\tau_1;\tau_{12} \vdash \varphi : \mathbb{B}$
    \end{pf}
    By IH applied to \ref{typing-context-subtype-tif-then} and \ref{typing-context-subtype-a2} follows
    \begin{pf*}{T-If-Algo}
      \item $\cmdTypeAlgo{\Gamma,x:\tau_1'}{c_1}{y:\refT{z}{\tau_{12}}{\varphi[z/\cmdVar]}}{\tau_{12}'}$
    \end{pf*}
    By IH applied to \ref{typing-context-subtype-tif-else} and \ref{typing-context-subtype-a2} follows
    \begin{pf*}{T-If-Algo}
      \item $\cmdTypeAlgo{\Gamma,x:\tau_1'}{c_1}{y:\refT{z}{\tau_{12}}{\neg\varphi[z/\cmdVar]}}{\tau_{12}''}$
    \end{pf*}
    Since $\Gamma,x:\tau_1;\tau_{12} \vdash \varphi : \mathbb{B}$, it also holds that $\Gamma,x:\tau_1';\tau_{12} \vdash \varphi : \mathbb{B}$.
    The result then follows by \tIfAlgo.

    \case{\tModAlgo}~
    By inversion of \tModAlgo, we know
    \begin{pf}{T-Mod-Algo}
      \item \label{typing-context-subtype-tmod-includes} $\subtypeCtx{\Gamma,x:\tau_1}{\tau_{12}}{\inst_\approx}$
      \item $\Gamma,x:\tau_1;\tau_{12} \vdash e : \mathsf{BV}$
    \end{pf}
    By \cref{lem:context-strengthening} applied to \ref{typing-context-subtype-tmod-includes} and \ref{typing-context-subtype-a2} follows
    \begin{pf*}{T-Mod-Algo}
      \item $\subtypeCtx{\Gamma,x:\tau_1'}{\tau_{12}}{\inst_\approx}$
    \end{pf*}
    Since $\Gamma,x:\tau_1;\tau_{12} \vdash e : \mathsf{BV}$, it also holds that $\Gamma,x:\tau_1';\tau_{12} \vdash e : \mathsf{BV}$.
    The result follows by \tModAlgo.

    \case{\tAddAlgo}~
    By inversion of \tAddAlgo, we know
    \begin{pf}{T-Add-Algo}
      \item \label{typing-context-subtype-tadd-excludes}$\subtypeCtx{\Gamma,x:\tau_1}{\tau_{12}}{\refT{x}{\top}{\neg x.\inst.\valid}}$
    \end{pf}
    By \cref{lem:context-strengthening} applied to \ref{typing-context-subtype-tadd-excludes} and \ref{typing-context-subtype-a2} follows
    \begin{pf*}{T-Add-Algo}
      \item $\subtypeCtx{\Gamma,x:\tau_1'}{\tau_{12}}{\refT{x}{\top}{\neg x.\inst.\valid}}$
    \end{pf*}
    The result follows by \tAddAlgo.
  \end{description}
\end{proof}

\begin{lemma}[Algorithmic Input Subtype]
  \label{lem:algorithmic-input-subtype}
  If $\subtype{\tau_{1}'}{\tau_{1}}$ and $\cmdTypeAlgo{\Gamma}{c}{x:\tau_{1}}{\tau_{2}}$ such that $x$ is not free in $\tau_1$ or $\tau_1'$,
  then there exists $\tau_{2}'$ such that $\cmdTypeAlgo{\Gamma}{c}{x:\tau_{1}'}{\tau_{2}'}$ and $\subtypeCtx{\Gamma,x:\tau_{1}'}{\tau_{2}'}{\tau_{2}}$.
\end{lemma}

\begin{proof}
  By induction on the typing derivation.
  We refer to the general assumptions as follows:
  \begin{enumerate}[label=(\Alph*)]
    \item \label{algoin-subtype-in} $\subtype{\tau_{1}'}{\tau_{1}}$
    \item $\cmdTypeAlgo{\Gamma}{c}{x:\tau_{1}}{\tau_{2}}$
    \item \label{algoin-free-in}$x$ not free in $\tau_1$ or $\tau_1'$
  \end{enumerate}
  We refer to the proof goals as follows:
  \begin{enumerate}[label=(\arabic*)]
    \item $\cmdTypeAlgo{\Gamma}{c}{x:\tau_{1}'}{\tau_{2}'}$
    \item $\subtypeCtx{\Gamma,x:\tau_{1}'}{\tau_{2}'}{\tau_{2}}$
  \end{enumerate}
  \begin{description}
    \case{\tExtractAlgo}~
    By inversion of \tExtractAlgo, we know
    \begin{pf}{T-Extract-Algo-Input}
      \item $c = \cExtract\inst$
      \item \label{textract-algoin-bound} $\subtype{\tau_{1}}{\refT{x}{\top}{\length{x.\pIn} \geq \sizeof(\inst)}}$
      \item $\varphi_{1} \triangleq z.\pIn = z.\pOut = \bv{}$
      \item $\varphi_{2} \triangleq \bvconcat{y.\inst}{z.\pIn} \wedge z.\pOut = x.\pOut \wedge z \equiv_{\inst} x$
      \item $\tau_{21} \triangleq \refT z \inst {\varphi_{1}}$
      \item $\tau_{22} \triangleq \refT z {\chomp(\tau_{1},\inst, y)} {\varphi_{2}}$
      \item $\tau_{2} = \sigmaT y {\tau_{21}} {\tau_{22}}$
    \end{pf}

    By \cref{lem:subtype-transitivity} with \ref{algoin-subtype-in} and \ref{textract-algoin-bound},
    \begin{pf*}{T-Extract-Algo-Input}
      \item $\subtype{\tau_{1}'}{\refT{x}{\top}{\length{x.\pIn} \geq \sizeof(\inst)}}$
    \end{pf*}

    By \cref{lem:chomp-subtype} with \ref{textract-algoin-bound} and \ref{algoin-subtype-in},
    \begin{pf*}{T-Extract-Algo-Input}
      \item \label{textract-algoin-chomp}$\subtypeCtx{\Gamma,y:\tau_{21}}{\chomp(\tau_{1}', \inst, y)}{\chomp(\tau_{1}, \inst, y)}$
    \end{pf*}
    (1) follows by \tExtractAlgo.
    (2) follows by Lemmas~\ref{lem:refinement-subtype} and~\ref{lem:sigma-right-subtype} and \ref{textract-algoin-chomp}.

    \case{\tSeqAlgo}~
    By inversion of \tSeqAlgo, we know
    \begin{pf}{T-Seq-Algo-Input}
      \item $c = c_{1};c_{2}$
      \item \label{tseq-algoin-c1} $\cmdTypeAlgo{\Gamma}{c_{1}}{x:\tau_{1}}{\tau_{12}}$
      \item \label{tseq-algoin-c2} $\cmdTypeAlgo{\Gamma,x:\tau_{1}}{c_{2}}{y:\tau_{12}}{\tau_{22}}$
      \item $\tau_{2} = \tau_{22}[y \mapsto \tau_{12}]$
    \end{pf}
    By IH applied to \ref{tseq-algoin-c1} and \ref{algoin-subtype-in}, there is some $\tau_{12}'$ such that
    \begin{pf*}{T-Seq-Algo-Input}
      \item \label{tseq-algoin-c1-computed} $\cmdTypeAlgo{\Gamma}{c_{1}}{x:\tau_{1}'}{\tau_{12}'}$, and
      \item \label{tseq-algoin-subtype-join} $\subtypeCtx{\Gamma,x:\tau_{1}'}{\tau_{12}'}{\tau_{12}}$
    \end{pf*}
    By \cref{lem:typing-context-subtype} with \ref{algoin-subtype-in} and \ref{tseq-algoin-c2},
    \begin{pf*}{T-Seq-Algo-Input}
      \item \label{tseq-algoin-c2-subtype} $\cmdTypeAlgo{\Gamma,x:\tau_{1}'}{c_{2}}{y:\tau_{12}}{\tau_{22}}$
    \end{pf*}
    By IH applied to \ref{tseq-algoin-c2-subtype} and \ref{tseq-algoin-subtype-join}, there is some $\tau_{22}'$ such that
    \begin{pf*}{T-Seq-Algo-Input}
      \item \label{tseq-algoin-c2-computed} $\cmdTypeAlgo{\Gamma,x:\tau_{1}'}{c_{2}}{y:\tau_{12}'}{\tau_{22}'}$, and
      \item \label{tseq-algoin-subtype-outer} $\subtypeCtx{\Gamma,x:\tau_{1}',y:\tau_{12}'}{\tau_{22}'}{\tau_{22}}$
    \end{pf*}
    (1) follows by \tSeqAlgo with~\ref{tseq-algoin-c1-computed} and~\ref{tseq-algoin-c2-computed}.
    To show (2), we just need to show that $\subtypeCtx{\Gamma,x:\tau_{1}'}{\tau_{22}'[y\mapsto \tau_{12}']}{\tau_{22}[y \mapsto \tau_{12}]}$.
    This follows by \cref{lem:substitution-subtype} applied to \ref{tseq-algoin-subtype-join} and \ref{tseq-algoin-subtype-outer}.

    \case{\tSkipAlgo}~
    Immediate by \tSkipAlgo.

    \case{\tRemitAlgo}~
    By inversion of \tRemitAlgo, we know
    \begin{pf}{T-Remit-Algo-Input}
      \item \label{algoin-tremit-includes} $\subtype{\tau_1}{\inst_\approx}$
      \item $c = \cRemit{\inst}$
      \item $\tau_{2} = \sigmaT{y}{\refT{z}{\tau_{1}}{z \equiv x}}{\refT{z}{\epsilon}{z.\pIn = \bvNil \wedge z.\pOut = x.\inst}}$
    \end{pf}
    By \cref{lem:subtype-transitivity} with \ref{algoin-tremit-includes} and \ref{algoin-subtype-in} follows
    \begin{pf*}{T-Remit-Algo-Input}
      \item $\subtype{\tau_1'}{\inst_\approx}$
    \end{pf*}
    Let $\tau_2' = \sigmaT{y}{\refT{z}{\tau_{1}'}{z \equiv x}}{\refT{z}{\epsilon}{z.\pIn = \bvNil \wedge z.\pOut = x.\inst}}$.
    (1) follows by \tRemitAlgo.
    By Lemma~\ref{lem:context-strengthening}, and since $x$ does not occur free in $\tau_{1}$ or $\tau_{1}'$, we know
    \begin{pf*}{T-Remit-Algo-Input}
      \item $\subtypeCtx{\Gamma,x:\tau_{1}'}{\tau_{1}'}{\tau_{1}}$
    \end{pf*}
    By Lemma~\ref{lem:refinement-subtype} we know
    \begin{pf*}{T-Remit-Algo-Input}
      \item $\subtypeCtx{\Gamma,x:\tau_{1}'}{\refT{z}{\tau_{1}'}{z \equiv x}}{\refT{z}{\tau_{1}}{z \equiv x}}$
    \end{pf*}
    By \cref{lem:sigma-left-subtype} with $\tau_{22} = \refT{z}{\epsilon}{z.\pIn = \bvNil \wedge z.\pOut = x.\inst}$ follows
    \begin{pf*}{T-Remit-Algo-Input}
      \item $\subtypeCtx{\Gamma,x:\tau_{1}'}
        {\sigmaT{y}{\refT z {\tau_{1}'} {z \equiv x}} {\tau_{22}}}
        {\sigmaT{y}{\refT z {\tau_{1}} {z \equiv x}} {\tau_{22}}}$
    \end{pf*}
    This shows (2) and concludes this case.

    \case{\tResetAlgo}~
    By inversion of \tResetAlgo, we know
    \begin{pf}{T-Reset-Algo-Input}
      \item $\tau_{2} = \sigmaT{y}
        {\refT{z}{\epsilon}{z.\pOut = \bv{} \wedge z.\pIn = x.\pOut}}
        {\refT{z}{\epsilon}{z.\pOut = \bv{} \wedge z.\pIn = z.\pIn}}$
    \end{pf}
    Let $\tau_2' = \tau_2$.
    (1) follows by \tResetAlgo and (2) follows by \cref{lem:subtype-reflexivity}.

    \case{\tAscribeAlgo}~
    By inversion of \tAscribeAlgo, we know
    \begin{pf}{T-Ascribe-Algo-Input}
      \item $c = \cAscribe{c_{0}}{(x : \hat\tau_{1}) \to \tau_{2}}$
      \item \label{t-ascribe-algoin-c0} $\cmdTypeAlgo{\Gamma}{c_{0}}{x:\hat\tau_{1}}{\tau_{c}}$
      \item \label{t-ascribe-algoin-t1} $\subtype{\tau_{1}}{\hat\tau_{1}}$
      \item \label{t-ascribe-algoin-t2} $\subtypeCtx{\Gamma,x:\hat\tau_{1}}{\tau_{c}}{\tau_{2}}$
    \end{pf}
    By \cref{lem:subtype-transitivity} applied to \ref{algoin-subtype-in} and \ref{t-ascribe-algoin-t1} follows that
    \begin{pf*}{T-Ascribe-Algo-Input}
      \item \label{t-ascribe-algoin-t1-prime-hat} $\subtype{\tau_{1}'}{\hat\tau_{1}}$
    \end{pf*}
    Let $\tau_2' = \tau_2$.
    (1) follows by \tAscribeAlgo with \ref{t-ascribe-algoin-c0}, \ref{t-ascribe-algoin-t2} and \ref{t-ascribe-algoin-t1-prime-hat}.
    (2) follows by \cref{lem:subtype-reflexivity}.

    \case{\tIfAlgo}~
    By inversion of \tIfAlgo, we know
    \begin{pf}{T-If-Algo-Input}
      \item $c = \cIf \varphi {c_{1}} {c_{2}}$
      \item \label{t-if-algoin-e} $\Gamma;\tau_1\vdash \varphi : \tBool$
      \item \label{t-if-algoin-c1-type} $\cmdTypeAlgo{\Gamma}{c_{1}}{x:\refT{y}{\tau_{1}}{\varphi[y/\cmdVar]}} {\tau_{12}}$
      \item \label{t-if-algoin-c2-type} $\cmdTypeAlgo{\Gamma}{c_{2}}{x:\refT{y}{\tau_{1}}{\neg \varphi[y/\cmdVar]}} {\tau_{22}}$
      \item $\tau_{2} = \refT{y}{\tau_{12}}{\varphi[x/\cmdVar]} + \refT{y}{\tau_{22}}{\neg \varphi[x/\cmdVar]}$
    \end{pf}
    By \ref{algoin-subtype-in} and \cref{lem:refinement-subtype}, we know
    \begin{pf*}{T-If-Algo-Input}
      \item \label{t-if-algo-tru-refsub} $\subtype{\refT{y}{\tau_{1}'}{\varphi[y/\cmdVar]}}{\refT{y}{\tau_{1}}{\varphi[y/\cmdVar]}}$, and
      \item \label{t-if-algo-fls-refsub} $\subtype{\refT{y}{\tau_{1}'}{\neg \varphi[y/\cmdVar]}}{\refT{y}{\tau_{1}}{\neg \varphi[y/\cmdVar]}}$
    \end{pf*}
    By applying the IH to \ref{t-if-algo-tru-refsub} and \ref{t-if-algoin-c1-type} we get $\tau_{12}'$ such that
    \begin{pf*}{T-If-Algo-Input}
      \item \label{t-if-algo-c1-typecheck} $\cmdTypeAlgo{\Gamma}{c_{1}}{x:\refT{y}{\tau_{1}'}{\varphi[y/\cmdVar]}}{\tau_{12}'}$, and
      \item \label{t-if-algo-true-ctxsub} $\subtypeCtx{\Gamma,x:\refT{y}{\tau_{1}'}{\varphi[y/\cmdVar]}}{\tau_{12}'}{\tau_{12}}$
    \end{pf*}
    By applying the IH to \ref{t-if-algo-fls-refsub} and \ref{t-if-algoin-c2-type} we get $\tau_{22}'$ such that
    \begin{pf*}{T-If-Algo-Input}
      \item \label{t-if-algo-c2-typecheck} $\cmdTypeAlgo{\Gamma}{c_{2}}{x:\refT{y}{\tau_{1}'}{\neg \varphi[y/\cmdVar]}}{\tau_{22}'}$, and
      \item \label{t-if-algo-fls-ctxsub} $\subtypeCtx{\Gamma,x:\refT{y}{\tau_{1}'}{\neg \varphi[y/\cmdVar]}}{\tau_{22}'}{\tau_{22}}$
    \end{pf*}
    From \ref{t-if-algoin-e} and \ref{algoin-subtype-in}, we can conclude that
    \begin{pf*}{T-If-Algo-Input}
      \item \label{t-if-algoin-e-sub} $\Gamma;\tau_1' \vdash \varphi : \mathbb{B}$
    \end{pf*}

    Let $\tau_2' = \refT{y}{\tau_{12}'}{\varphi[x/\cmdVar]} + \refT{y}{\tau_{22}'}{\neg \varphi[x/\cmdVar]}$.
    (2) follows by \cref{lem:if-choice-subtype}.
    (1) follows by \tIfAlgo with \ref{t-if-algo-c1-typecheck}, \ref{t-if-algo-c2-typecheck} and \ref{t-if-algoin-e-sub}.

    \case{\tModAlgo}~
    By inversion of \tModAlgo, we know
    \begin{pf}{T-Mod-Algo-Input}
      \item \label{tmod-algoin-includes}$\subtype{\tau_1}{\inst_\approx}$
      \item \label{tmod-algoin-e-type} $\Gamma;\tau_1 \vdash e:\tBv$
      \item $\tau_2 = \refT{y}{\top}{\varphi \wedge \varphi_\inst \wedge \varphi_f \wedge y.\inst.f = e[x/\cmdVar]}$
    \end{pf}
    By \cref{lem:subtype-transitivity} with \ref{algoin-subtype-in} and \ref{tmod-algoin-includes} follows
    \begin{pf*}{T-Mod-Algo-Input}
      \item \label{tmod-algoin-includes-sub} $\subtype{\tau_1'}{\inst_\approx}$
    \end{pf*}
    From \ref{tmod-algoin-e-type} and \ref{algoin-subtype-in}, we can conclude
    \begin{pf*}{T-Mod-Algo-Input}
      \item \label{tmod-algoin-e-type-sub} $\Gamma;\tau_1' \vdash e:\tBv$
    \end{pf*}
    Let $\tau_2' = \tau_2$.
    (1) follows by \tModAlgo with \ref{tmod-algoin-includes-sub} and \ref{tmod-algoin-e-type-sub}.
    (2) follows by \cref{lem:subtype-reflexivity}.

    \case{\tAddAlgo}~
    By inversion of \tAddAlgo, we know
    \begin{pf}{T-Add-Algo-Input}
      \item \label{tadd-algoin-excludes} $\subtype{\tau_1}{\refT{x}{\top}{\neg x.\inst.\valid}}$
      \item $\tau_2 = \sigmaT{y}{\refT{z}{\tau_1}{z\equiv x}}{\refT{z}{\inst}{z.\pIn = z.\pOut = \bvNil \wedge z.\inst = v}}$
    \end{pf}
    By \cref{lem:subtype-transitivity} with \ref{tadd-algoin-excludes} and \ref{algoin-subtype-in} follows
    \begin{pf*}{T-Add-Algo-Input}
      \item \label{tadd-algoin-excludes-sub} $\subtype{\tau_1'}{\refT{x}{\top}{\neg x.\inst.\valid}}$
    \end{pf*}
    Let $\tau_2' = \sigmaT{y}{\refT{z}{\tau_1'}{z\equiv x}}{\refT{z}{\inst}{z.\pIn = z.\pOut = \bvNil \wedge z.\inst = v}}$.
    (1) follows by \tAddAlgo.
    By \cref{lem:context-strengthening} with \ref{algoin-subtype-in} and \ref{algoin-free-in},
    \begin{pf*}{T-Add-Algo-Input}
      \item $\subtypeCtx{\Gamma,x:\tau_1'}{\tau_1'}{\tau_1}$
    \end{pf*}
    By \cref{lem:refinement-subtype} follows
    \begin{pf*}{T-Add-Algo-Input}
      \item \label{tadd-algoin-subtype-left} $\subtypeCtx{\Gamma,x:\tau_1'}{\refT{z}{\tau_1'}{z\equiv x}}{\refT{z}{\tau_1}{z\equiv x}}$
    \end{pf*}
    (2) follows by \cref{lem:sigma-left-subtype} with \ref{tadd-algoin-subtype-left}.

  \end{description}
\end{proof}

\begin{lemma}[Includes Subtype]
  \label{lem:includes-subtype}
  $\includes \tau \inst \iff \subtype{\tau}{\instWeak}$
\end{lemma}
\begin{proof}
  Prove each direction separately
  \begin{enumerate}[align=left]
    \item[($\Rightarrow$)] Assume $\includes \tau \inst$. Let
      $\env \models \Gamma$ and $h \in \semantics{\tau}_{\env}$. By
      definition of the inclusion relation, $\inst \in \dom(h)$. By the
      definition of subtyping, it suffices to show
      $h \in \semantics{\instWeak}_{\env}$. By definition,
      $\semantics{\instWeak}_{\env} = \{h \mid \inst \in \dom(h)\}$, and
      we're done.

    \item[($\Leftarrow$)] Assume $\subtype{\tau}{\instWeak}$. Show
      $\includes \tau \inst$. To that end, let $\env \models \Gamma$ and
      $h \in \semantics{\tau}_{\env}$ be arbitrary.
      By the definition of subtyping, $h \in \semantics{\instWeak}_{\env}$.
      By definition of the semantics, conclude $\inst \in \dom(h)$.
  \end{enumerate}
\end{proof}

\begin{lemma}[Excludes Subtype]
  \label{lem:excludes-subtype}
  $\excludes \tau \inst$ iff $\subtype{\tau}{\refT{x}{\top}{\neg x.\inst.\valid}}$
\end{lemma}
\begin{proof}
  Prove each direction separately
  \begin{enumerate}[align=left]
    \item[($\Rightarrow$)]
      Assume $\excludes \tau \inst$.
      Let $\env \models \Gamma$ and $h \in \semantics{\tau}_{\env}$.
      By definition of the exclusion relation, $\inst \not\in \dom(h)$.
      By the definition of subtyping, it suffices to show $h \in \semantics{\refT x \top {\neg x.\inst.\valid}}_{\env}$.
      By definition, $\semantics{\refT x \top {\neg x.\inst.\valid}}_{\env} = \{h \mid \inst \not\in \dom(h)\}$, and we're done.

    \item[($\Leftarrow$)]
      Assume $\subtype{\tau}{\refT{x}{\top}{\neg x.\inst.\valid}}$.
      Show $\excludes \tau \inst$.
      To that end, let $\env \models \Gamma$ and $h \in \semantics{\tau}_{\env}$ be arbitrary.
      By the definition of subtyping, $h \in \semantics{\refT x \top {\neg x.\inst.\valid}}_{\env}$.
      By definition of the semantics, conclude $\inst \not\in \dom(h)$.
  \end{enumerate}
\end{proof}

\begin{theorem}[Algorithmic Typing Correctness]
  \label{thm:algorithmic-typing-correctness}
  For all subtyping contexts $\Gamma$, commands $c$, variables $x$, heap types $\tau_{1}$, and $\tau_{2}$,
  where $x$ is not free in $\tau_{1}$,
  $\cmdType{\Gamma}{c}{(x:\tau_{1}) \to \tau_{2}}$
  if and only if
  there is some $\tau_{2}'$ such that
  $\cmdTypeAlgo{\Gamma}{c}{x:\tau_{1}}{\tau_{2}'}$,
  and
  $\subtypeCtx{\Gamma,(x: \tau_{1})}{\tau_{2}'}{\tau_{2}}$.
\end{theorem}
\begin{proof}

  \begin{enumerate}[align=left]
    \item[($\Rightarrow$)]
      Assume $\cmdType{\Gamma}{c}{(x:\tau_{1}) \to \tau_{2}}$.
      Proceed by induction on the typing derivation, leaving $\Gamma$ general.
      We refer to the proof goals as follows:
      \begin{enumerate}[label=(\arabic*)]
        \item \label{algotypcorrect-forward-goal1}$\cmdTypeAlgo{\Gamma}{c}{x:\tau_{1}}{\tau_{2}'}$
        \item \label{algotypcorrect-forward-goal2}$\subtypeCtx{\Gamma,(x: \tau_{1})}{\tau_{2}'}{\tau_{2}}$
      \end{enumerate}

      \begin{description}
        \case{\tExtract}~
        \begin{pf}{T-Extract}
          \item $c = \cExtract\inst$
          \item \label{textract-sizeof} $\Gamma \vdash \sizeof_{\pIn}(\tau_{1}) \geq \sizeof(i)$
          \item $\varphi_{1} \triangleq z.\pIn = z.\pOut = \bv{}$
          \item $\varphi_{2} \triangleq y.\inst@z.\pIn = x.\pIn \wedge z.\pOut = x.\pOut \wedge z \equiv_{\inst} x$
          \item $\tau_2 = \sigmaT{y}{\refT{z}{\inst}{\varphi_{1}}}{\refT{z}{\chomp(\tau_{1},\inst,y)}{\varphi_{2}}}$
        \end{pf}
        The only algorithmic rule that applies to $\cExtract\inst$ is \tExtractAlgo.
        Since $\subtype{\tau_1}{\refT{x}{\top}{\length{x.\pIn} \geq \sizeof(\inst)}}$ by \ref{textract-sizeof} and \cref{lem:bound-subtype}, \tExtractAlgo produces $\tau_2'$ such that
        \begin{pf*}{T-Extract}
          \item $\varphi'_{1} \triangleq z.\pIn = z.\pOut = \bv{}$
          \item $\varphi'_{2} \triangleq y.\inst@z.\pIn = x.\pIn \wedge z.\pOut = x.\pOut \wedge z \equiv_{\inst} x$
          \item $\tau_2' = \sigmaT{y}{\refT{z}{\inst}{\varphi_{1}'}}{\refT{z}{\chomp(\tau_{1},\inst,y)}{\varphi_{2}'}}$
        \end{pf*}
        which shows \ref{algotypcorrect-forward-goal1}.
        \ref{algotypcorrect-forward-goal2} follows by \cref{lem:subtype-reflexivity}.

        \case{\tSeq}~
        \begin{pf}{T-Seq}
          \item $c = c_{1};c_{2}$
          \item \label{tseq-type-c1} $\cmdType{\Gamma}{c_{1}}{(x:\tau_{1}) \to \tau_{12}}$
          \item \label{tseq-type-c2} $\cmdType{\Gamma,(x:\tau_{1})}{c_{2}}{(y:\tau_{12}) \to \tau_{22}}$
          \item $\tau_{2} = \tau_{22}[x \mapsto \tau_{12}]$
        \end{pf}
        By applying the IH to \ref{tseq-type-c1}, we get $\tau_{12}'$ such that
        \begin{pf*}{T-Seq}
          \item \label{tseq-algotype-c1} $\cmdTypeAlgo{\Gamma}{c_{1}}{x:\tau_{1}}{\tau_{12}'}$, and
          \item \label{tseq-algosubtype-c1} $\subtypeCtx{\Gamma,x:\tau_{1}}{\tau_{12}'}{\tau_{12}}$
        \end{pf*}
        By applying the IH to \ref{tseq-type-c2}, we get $\tau_{22}'$ such that
        \begin{pf*}{T-Seq}
          \item \label{tseq-algotype-c2} $\cmdTypeAlgo{\Gamma,x:\tau_{1}}{c_{2}}{y:\tau_{12}}{\tau_{22}'}$, and
          \item \label{tseq-algosubtype-c2} $\subtypeCtx{\Gamma,x:\tau_{1},y:\tau_{12}}{\tau_{22}'}{\tau_{22}}$
        \end{pf*}
        By \cref{lem:algorithmic-input-subtype} with \ref{tseq-algosubtype-c1} and \ref{tseq-algotype-c2} there exists $\tau_{22}''$ such that
        \begin{pf*}{T-Seq}
          \item \label{tseq-algotype-subtype-c2} $\cmdTypeAlgo{\Gamma,x:\tau_{1}}{c_{2}}{y:\tau_{12}'}{\tau_{22}''}$, and
          \item \label{tseq-algosubtype-subtype-c2} $\subtypeCtx{\Gamma,x:\tau_{1},y:\tau_{12}'}{\tau_{22}''}{\tau_{22}'}$
        \end{pf*}
        By \tSeqAlgo with \ref{tseq-algotype-c1} and \ref{tseq-algotype-subtype-c2} follows
        \begin{pf*}{T-Seq}
          \item \label{tseq-algotype} $\cmdTypeAlgo{\Gamma}{c_{1};c_{2}}{x:\tau_{1}}{\tau_{22}''[y \mapsto \tau_{12}']}$
        \end{pf*}
        which shows \ref{algotypcorrect-forward-goal1}.
        By \cref{lem:context-strengthening} with \ref{tseq-algosubtype-c1} and \ref{tseq-algosubtype-c2} follows
        \begin{pf*}{T-Seq}
          \item \label{tseq-algosubtype-subtype-ctx-c2} $\subtypeCtx{\Gamma,x:\tau_{1},y:\tau_{12}'}{\tau_{22}'}{\tau_{22}}$
        \end{pf*}
        By \cref{lem:subtype-transitivity} with \ref{tseq-algosubtype-subtype-c2} and \ref{tseq-algosubtype-subtype-ctx-c2} follows
        \begin{pf*}{T-Seq}
          \item $\subtypeCtx{\Gamma,x:\tau_{1},y:\tau_{12}'}{\tau_{22}''}{\tau_{22}}$
        \end{pf*}
        By Lemma~\ref{lem:substitution-subtype} follows
        \begin{pf*}{T-Seq}
          \item \label{tseq-algosubtype-subst}
          $\subtypeCtx{\Gamma,x:\tau_{1}}{\tau_{22}''[y\mapsto \tau_{12}']}{\tau_{22}[y \mapsto \tau_{12}]}$
        \end{pf*}
        which shows \ref{algotypcorrect-forward-goal2} and concludes this case.

        \case{\tSkip}~
        \ref{algotypcorrect-forward-goal1} follows by \tSkipAlgo and \ref{algotypcorrect-forward-goal2} follows by \cref{lem:subtype-reflexivity}.

        \case{\tRemit}~
        By inversion of \tRemit, we know
        \begin{pf}{T-Remit}
          \item $c=\cRemit{\inst}$
          \item \label{tremit-inclusion} $\includes{\tau_1}{\inst}$
          \item $\varphi \triangleq z.\pIn = \bvNil \land z.\pOut=x.\inst$
          \item $\tau_{2} = \sigmaT y {\refT z {\tau_{1}} {z \equiv x}} {\refT z \epsilon \varphi}$
        \end{pf}
        By \cref{lem:includes-subtype} and \ref{tremit-inclusion}, \tRemitAlgo computes $\tau_2'$ such that
        \begin{pf*}{T-Remit}
          \item $\cmdTypeAlgo{\Gamma}{\cRemit{\inst_{i}}}{(x:\tau_{1})}{\tau_{2}'}$, and
          \item $\tau_{2}' = \sigmaT{y}{\refT z {\tau_{1}} {z \equiv x}} {\refT z \epsilon {z.\pIn = \bv{} \wedge z.\pIn = x.\inst}}$
        \end{pf*}
        which shows \ref{algotypcorrect-forward-goal1}.
        Since $\tau_2'=\tau_2$, \ref{algotypcorrect-forward-goal2} follows by \cref{lem:subtype-reflexivity}.

        \case{\tReset}~
        \ref{algotypcorrect-forward-goal1} follows by \tResetAlgo and \ref{algotypcorrect-forward-goal2} follows by \cref{lem:subtype-reflexivity}.

        \case{\tAscribe}~
        By inversion of \tAscribe, we know
        \begin{pf}{T-Ascribe}
          \item $c = \cAscribe{c_{0}}{(x:\tau_{1}) \to \tau_{2}} $
          \item \label{t-ascr-c0} $\cmdType{\Gamma}{c_{0}}{(x:\tau_{1}) \to \tau_{2}}$
        \end{pf}

        By IH applied to \ref{t-ascr-c0}, there exists $\hat\tau_2$ such that
        \begin{pf*}{T-Ascribe}
          \item \label{t-ascr-algo-c0} $\cmdTypeAlgo{\Gamma}{c_{0}}{x:\tau_{1}}{\hat\tau_{2}}$
          \item \label{t-ascr-algosub} $\subtypeCtx{\Gamma,x:\tau_{1}}{\hat\tau_{2}}{\tau_{2}}$
        \end{pf*}
        By \tAscribeAlgo with \ref{t-ascr-algo-c0}, \ref{t-ascr-algosub} and \cref{lem:subtype-reflexivity},
        \begin{pf*}{T-Ascribe}
          \item $\cmdTypeAlgo{\Gamma}{\cAscribe{c_{0}}{(x:\tau_{1}) \to \tau_{2}}}{x:\tau_{1}}{\tau_{2}}$
        \end{pf*}
        showing \ref{algotypcorrect-forward-goal1}.
        \ref{algotypcorrect-forward-goal2} follows by \cref{lem:subtype-reflexivity}.

        \case{\tIf}~
        By inversion of \tIf, we know
        \begin{pf}{T-If}
          \item $c = \cIf{\varphi}{c_{1}}{c_{2}}$
          \item \label{tif-etype}$\Gamma;\tau_1 \vdash \varphi : \tBool$
          \item \label{tif-type-c1} $\cmdType{\Gamma}{c_{1}}{(x : \refT y {\tau_{1}} {\varphi[y/\cmdVar]}) \to \tau_{12}}$
          \item \label{tif-type-c2} $\cmdType{\Gamma}{c_{2}}{(x : \refT y {\tau_{1}} {\neg \varphi[y/\cmdVar]}) \to \tau_{22}}$
          \item $\tau_{2} = \refT y {\tau_{12}} {\varphi[x/\cmdVar]} + \refT y {\tau_{22}} {\neg \varphi[x/\cmdVar]}$
        \end{pf}
        By the IH applied to \ref{tif-type-c1} there exists $\tau_{12}'$ such that
        \begin{pf*}{T-If}
          \item \label{tif-true-branch-algotype}$\cmdTypeAlgo{\Gamma}{c_{1}}{x : \refT y {\tau_{1}} {\varphi[y/\cmdVar]}}{\tau_{12}'}$
          \item \label{tif-true-branch-subtype} $\subtypeCtx{\Gamma,x:\refT{y}{\tau_{1}}{\varphi[y/\cmdVar]}}{\tau_{12}'}{\tau_{12}}$
        \end{pf*}
        By the IH applied to \ref{tif-type-c2} there exists $\tau_{22}'$ such that
        \begin{pf*}{T-If}
          \item \label{tif-false-branch-algotype} $\cmdTypeAlgo{\Gamma}{c_{2}}{x : \refT y {\tau_{1}} {\neg \varphi[y/\cmdVar]}}{\tau_{22}'}$
          \item \label{tif-false-branch-subtype} $\subtypeCtx{\Gamma,x:\refT{y}{\tau_{1}}{\neg \varphi[y/\cmdVar]}}{\tau_{22}'}{\tau_{22}}$
        \end{pf*}
        \ref{algotypcorrect-forward-goal1} follows by \tIfAlgo with \ref{tif-etype}, \ref{tif-true-branch-algotype} and \ref{tif-false-branch-algotype}.

        \ref{algotypcorrect-forward-goal2} follows by \cref{lem:if-choice-subtype} with \ref{tif-true-branch-subtype} and \ref{tif-false-branch-subtype}.

        \case{\tMod}~
        By inversion of \tMod, we know
        \begin{pf}{T-Mod}
          \item $c = \inst.f := e$
          \item \label{tmod-includes} $\includes {\tau_{1}}{\inst}$
          \item \label{tmod-F} $\mathcal{F}(\inst, f) = \tBv$
          \item \label{tmod-t} $\Gamma;\tau_{1}\vdash e : \tBv$
          \item \label{tmod-tau2} $\tau_{2} = \refT{y}{\top}{\varphi_{\mathit{pkt}} \wedge \varphi_{\inst} \wedge \varphi_{f} \wedge y.\inst.f = e[x/\cmdVar]}$
        \end{pf}

        By Lemma~\ref{lem:includes-subtype} and \ref{tmod-includes},
        \begin{pf*}{T-Mod}
          \item \label{tmod-includes-subtype} $\subtype{\tau_1}{\inst_{i}}$
        \end{pf*}
        \ref{algotypcorrect-forward-goal1} follows by \tModAlgo with \ref{tmod-F},\ref{tmod-t}, \ref{tmod-tau2}, and \ref{tmod-includes-subtype}.
        \ref{algotypcorrect-forward-goal2} follows by \cref{lem:subtype-reflexivity}.

        \case{\tAdd}~
        By inversion of \tAdd, we know
        \begin{pf}{T-Add}
          \item \label{algotypecorrect-tadd-excludes}$\excludes{\tau_1}{\inst}$
          \item \label{algotypecorrect-tadd-init}$\mathsf{init}_{\HT(\inst)} = v$
          \item $\tau_2 = \sigmaT{y}{\refT{z}{\tau_1}{z \equiv x}}{\refT{z}{\inst}{z.\pIn = z.\pOut = \bvNil \wedge z.\inst = v}}$
        \end{pf}
        By \cref{lem:excludes-subtype} and \ref{algotypecorrect-tadd-excludes},
        \begin{pf*}{T-Add}
          \item $\subtype{\tau}{\refT{x}{\top}{\neg x.\inst.\valid}}$
        \end{pf*}
        \ref{algotypcorrect-forward-goal1} follows by \tAddAlgo with \ref{algotypecorrect-tadd-excludes} and \ref{algotypecorrect-tadd-init}.
        \ref{algotypcorrect-forward-goal2} follows by \cref{lem:subtype-reflexivity}.

        \case{\tSub}~
        By inversion of \tSub, there exists some $\tau_3$ and $\tau_4$ such that
        \begin{pf}{T-Sub}
          \item \label{algotypecorrect-tsub-in} $\subtype{\tau_1}{\tau_3}$
          \item \label{algotypecorrect-tsub-out} $\subtypeCtx{\Gamma,x:\tau_1}{\tau_4}{\tau_2}$
          \item \label{algotypecorrect-tsub-c} $\cmdType{\Gamma}{c}{(x:\tau_3) \to \tau_4}$
        \end{pf}
        By applying the IH to \ref{algotypecorrect-tsub-c} there is some $\tau_4'$ such that
        \begin{pf*}{T-Sub}
          \item \label{algotypecorrect-tsub-algo-c} $\cmdTypeAlgo{\Gamma}{c}{x:\tau_{3}}{\tau_{4}'}$, and
          \item \label{algotypecorrect-tsub-algo-subtype} $\subtypeCtx{\Gamma,x:\tau_{3}}{\tau_{4}'}{\tau_{4}}$
        \end{pf*}
        By \cref{lem:context-strengthening} together with \ref{algotypecorrect-tsub-in} and \ref{algotypecorrect-tsub-algo-subtype}, follows
        \begin{pf*}{T-Sub}
          \item \label{algotypecorrect-tsub-algo-subtype-t1} $\subtypeCtx{\Gamma,x:\tau_{1}}{\tau_{4}'}{\tau_{4}}$
        \end{pf*}
        By applying \cref{lem:algorithmic-input-subtype} to \ref{algotypecorrect-tsub-in} and \ref{algotypecorrect-tsub-algo-c} we get $\tau_4''$ such that
        \begin{pf*}{T-Sub}
          \item \label{algotypecorrect-tsub-algo-t1-in} $\cmdTypeAlgo{\Gamma}{c}{x:\tau_{1}}{\tau_{4}''}$, and
          \item \label{algotypecorrect-tsub-algo-subtype2} $\subtypeCtx{\Gamma,x:\tau_{1}}{\tau_{4}''}{\tau_{4}'}$
        \end{pf*}
        \ref{algotypcorrect-forward-goal1} follows by \ref{algotypecorrect-tsub-algo-t1-in}.
        \ref{algotypcorrect-forward-goal2} follows by repeated application of \cref{lem:subtype-transitivity} with \ref{algotypecorrect-tsub-out}, \ref{algotypecorrect-tsub-algo-subtype-t1} and \ref{algotypecorrect-tsub-algo-subtype2}.

      \end{description}
    \item[($\Leftarrow)$]
      Proceed by induction on the typing derivation.
      We refer to the general assumptions as follows:
      \begin{enumerate}[label=(\Alph*)]
        \item \label{algotypcorrect-back-a1} $\cmdTypeAlgo{\Gamma}{c}{x:\tau_{1}}{\tau_{2}'}$
        \item \label{algotypcorrect-back-a2} $\subtypeCtx{\Gamma,(x: \tau_{1})}{\tau_{2}'}{\tau_{2}}$
      \end{enumerate}
      \begin{description}
        \case{\tExtractAlgo}~
        By inversion of \tExtractAlgo, we know
        \begin{pf}{T-Extract-Algo}
          \item $\subtype{\tau_1}{\refT{x}{\top}{\length{x.\pIn} \ge \sizeof(\inst)}}$
          \item $\tau_2' = \sigmaT{y}{\refT{z}{\inst}{\varphi_1}}{\refT{z}{\chomp(\tau_1,\inst,y)}{\varphi_2}}$
        \end{pf}
        By \cref{lem:bound-subtype} follows
        \begin{pf*}{T-Extract-Algo}
          \item \label{algotypcorrect-textract-algo-size}$\Gamma \vdash \sizeof_{\pIn}(\tau_1) \ge \sizeof(\inst)$
        \end{pf*}
        The result follows by \tExtract with \ref{algotypcorrect-textract-algo-size} and \cref{lem:subtype-reflexivity}.

        \case{\tSeqAlgo}~
        By inversion of \tSeqAlgo, we know
        \begin{pf}{T-Seq-Algo}
          \item $c = c_{1};c_{2}$
          \item \label{algotypcorrect-tseq-algo-c1} $\cmdTypeAlgo{\Gamma}{c_{1}}{(x:\tau_{1})}{\tau_{12}}$
          \item \label{algotypcorrect-tseq-algo-c2} $\cmdTypeAlgo{\Gamma,(x:\tau_{1})}{c_{2}}{(y:\tau_{12})}{\tau_{22}}$
          \item \label{algotypcorrect-tseq-algo-t2} $\tau_{2}' = \tau_{22}[y \mapsto \tau_{12}]$
          \item \label{algotypcorrect-tseq-algo-subtype} $\subtypeCtx{\Gamma,x:\tau_{1}}{\tau_{2}'}{\tau_{2}}$
        \end{pf}

        With \cref{lem:subtype-reflexivity} follows
        \begin{pf*}{T-Seq-Algo}
          \item \label{algotypcorrect-tseq-algo-c1-sub} $\subtypeCtx{\Gamma,x:\tau_1}{\tau_{12}}{\tau_{12}}$
        \end{pf*}
        By IH with \ref{algotypcorrect-tseq-algo-c1} and \ref{algotypcorrect-tseq-algo-c1-sub} follows
        \begin{pf*}{T-Seq-Algo}
          \item \label{algotypcorrect-tseq-algo-type-c1} $\cmdType{\Gamma}{c_{1}}{(x:\tau_{1}) \to \tau_{12}}$
        \end{pf*}
        Similarly, applying the IH to \ref{algotypcorrect-tseq-algo-c2} gives
        \begin{pf*}{T-Seq-Algo}
          \item \label{algotypcorrect-tseq-algo-type-c2} $\cmdType{\Gamma,(x:\tau_{1})}{c_{2}}{(x:\tau_{12}) \to \tau_{22}}$
        \end{pf*}
        By \tSeq with \ref{algotypcorrect-tseq-algo-type-c1} and \ref{algotypcorrect-tseq-algo-type-c2} follows
        \begin{pf*}{T-Seq-Algo}
          \item \label{algotypcorrect-tseq-algo-type-c1c2}
          $\cmdType{\Gamma}{c_{1};c_{2}}{(x:\tau_{1}) \to \tau_{22}[y \mapsto \tau_{12}]}$
        \end{pf*}
        By \ref{algotypcorrect-tseq-algo-t2}, \ref{algotypcorrect-tseq-algo-subtype}, \cref{lem:subtype-reflexivity} and \tSub follows
        \begin{pf*}{T-Seq-Algo}
          \item $\cmdType{\Gamma}{c_{1};c_{2}}{(x:\tau_{1}) \to \tau_{2}}$
        \end{pf*}
        which concludes this case.

        \case{\tSkipAlgo}~

        The result follows by \tSkip, \cref{lem:subtype-reflexivity}, and \tSub.

        \case{\tRemitAlgo}~
        The result follows by \tRemit, \cref{lem:includes-subtype}, \cref{lem:subtype-reflexivity}, and \tSub.

        \case{\tResetAlgo}~
        The result follows by \tReset, \cref{lem:subtype-reflexivity}, and \tSub.

        \case{\tAscribeAlgo}~
        By inversion of \tAscribeAlgo, we know
        \begin{pf}{T-Ascribe-Algo}
          \item $c = \cAscribe{c_{0}}{(x:\hat\tau_{1}) \to \tau_{2}'}$
          \item \label{algotypcorrect-tascr-algo-c0} $\cmdTypeAlgo{\Gamma}{c_{0}}{x:\hat\tau_{1}}{\tau_{2}''}$
          \item \label{algotypcorrect-tascr-algo-inv-subtype-in} $\subtype{\tau_{1}}{\hat\tau_{1}}$
          \item \label{algotypcorrect-tascr-algo-inv-subtype} $\subtypeCtx{\Gamma,x:\hat\tau_{1}}{\tau_{2}''}{\tau_{2}'}$
        \end{pf}
        By IH applied to \ref{algotypcorrect-tascr-algo-c0} and \ref{algotypcorrect-tascr-algo-inv-subtype}, we get
        \begin{pf*}{T-Ascribe-Algo}
          \item \label{algotypcorrect-tascr-algo-c0-typing} $\cmdType{\Gamma}{c_0}{(x:\hat\tau_1)\to\tau_2'}$
        \end{pf*}
        By \tAscribe follows from \ref{algotypcorrect-tascr-algo-c0-typing} that
        \begin{pf*}{T-Ascribe-Algo}
          \item \label{algotypcorrect-tascr-algo-c0-typing-ascribed} $\cmdType{\Gamma}{\cAscribe{c_0}{(x:\hat\tau_1)\to\tau_2'}}{(x:\hat\tau_1)\to\tau_2'}$
        \end{pf*}
        The result follows by \tSub with assumptions \ref{algotypcorrect-back-a2}, \ref{algotypcorrect-tascr-algo-inv-subtype-in} and \ref{algotypcorrect-tascr-algo-c0-typing-ascribed}.

        \case{\tIfAlgo}~
        By inversion of \tIfAlgo, we know
        \begin{pf}{T-If-Algo}
          \item $c = \cIf{\varphi}{c_{1}}{c_{2}}$
          \item \label{algotypcorrect-tif-algo-e}$\Gamma;\tau_1 \vdash \varphi : \tBool$
          \item \label{algotypcorrect-tif-algo-c1} $\cmdTypeAlgo{\Gamma}{c_{1}}{x:{\refT{y}{\tau_{1}}{\varphi[y/\cmdVar]}}}{\tau_{12}}$
          \item \label{algotypcorrect-tif-algo-c2} $\cmdTypeAlgo{\Gamma}{c_{2}}{x:{\refT{y}{\tau_{1}}{\neg\varphi[y/\cmdVar]}}}{\tau_{22}}$
          \item \label{algotypcorrect-tif-algo-t2prime} $\tau_{2}' = \refT y {\tau_{12}} {\varphi[x/\cmdVar]} + \refT y {\tau_{22}} {\neg \varphi [x/\cmdVar]}$
        \end{pf}
        By \cref{lem:subtype-reflexivity}, we can conclude
        \begin{pf*}{T-If-Algo}
          \item \label{algotypcorrect-tif-algo-sub-in} $\subtypeCtx{\Gamma,x:\refT{y}{\tau_{1}}{\varphi[y/\cmdVar]}}{\tau_{12}}{\tau_{12}}$
        \end{pf*}
        By IH with \ref{algotypcorrect-tif-algo-c1} and \ref{algotypcorrect-tif-algo-sub-in} follows
        \begin{pf*}{T-If-Algo}
          \item \label{algotypcorrect-tif-algo-c1-type} $\cmdType{\Gamma}{c_1}{(x:\refT{y}{\tau_{1}}{\varphi[y/\cmdVar]}) \to \tau_{12}}$
        \end{pf*}
        We can reason similarly as before to conclude
        \begin{pf*}{T-If-Algo}
          \item \label{algotypcorrect-tif-algo-c2-type} $\cmdType{\Gamma}{c_2}{(x:\refT{y}{\tau_{1}}{\neg\varphi[y/\cmdVar]}) \to \tau_{22}}$
        \end{pf*}
        The result follows by \tIf with \ref{algotypcorrect-tif-algo-e}, \ref{algotypcorrect-tif-algo-c1-type}, \ref{algotypcorrect-tif-algo-c2-type} and \ref{algotypcorrect-tif-algo-t2prime} and by \tSub with assumption \ref{algotypcorrect-back-a2}.

        \case{\tModAlgo}~
        The result follows by \tMod with \cref{lem:includes-subtype}, \cref{lem:subtype-reflexivity} and \tSub.

        \case{\tAddAlgo}~
        The result follows by \tAdd with \cref{lem:excludes-subtype}, \cref{lem:subtype-reflexivity} and \tSub.
      \end{description}
  \end{enumerate}
\end{proof}

%% file: appendix/mtu.tex
\section{MTU Bound}
\label{sec:mtu}

We want to prove a theorem that says that if we have an maximum transmission unit (MTU) $N$ for a packet,
then we can bound the number of bits in the types.

We want to prove that if a program typechecks with a bound on its input type,
then we can compute that maximum number of bits that we need to encode the input
type. Ideally, this would be the same bound; however, it is possible for the
text of the program to emit more bits from the incoming packet than is
allowed by the MTU. So we define a helper function $\emit c \in \mathbb N$ that
over-approximates the maximum number of bits that could be emitted  along
any program path in $c$. This is defined in Figure~\ref{fig:max-emit}

\begin{figure}[H]
  \[\begin{array}{l>{\triangleq}cl}
      \emit {\cExtract \inst}        &  & 0                                 \\
      \emit {\cIf b {c_{1}} {c_{2}}} &  & \max(\emit {c_{1}}, \emit{c_{2}}) \\
      \emit {c_{1};c_{2}}            &  & \emit{c_{1}} + \emit{c_{2}}       \\
      \emit {\inst.f := t}           &  & 0                                 \\
      \emit {\cRemit{\inst}}         &  & \sizeof(\inst)                    \\
      \emit {\cSkip}                 &  & 0                                 \\
      \emit {\cReset}                &  & 0                                 \\
      \emit {\cAdd(\inst)}           &  & 0                                 \\
      \emit {\cAscribe c \sigma}     &  & \emit c
    \end{array}\]

  \caption{$\emit c \in \mathbb N$ computes the maximum number of bits that can be
    emitted along any path in c.}
  \label{fig:max-emit}
\end{figure}

We also need a way to state that a type $\tau$ satisfies a given MTU $n$; that is
that for every denoted heap $h$, $\pIn$ and $\pOut$ use fewer than $n$ bits
combined. This is defined formally in Figure~\ref{fig:type-bound}.

\begin{figure}[H]
  \[ \Gamma \vdash \tau \leq n \triangleq \forall \env \models \Gamma, \forall h \in \semantics\tau_{\env}, |h(\pIn)| + |h(\pOut)| \leq n\]
  \caption{Bound the size of types}
  \label{fig:type-bound}
\end{figure}

We also need a few lemmas about how this interacts with various types

\begin{lemma}[Refinement Bound]
  \label{lem:refinement-bound}
  For every $\Gamma$, $x$, $\tau$, $\varphi$, $N$, such that $\bound{\Gamma}{\tau}{N}$, $\bound{\Gamma}{\refT{x}{\tau}{\varphi}}{N}$.
\end{lemma}
\begin{proof}
  Let $\Gamma$, $x$, $\tau$, $\varphi$, $N$, be given such that $\bound{\Gamma}{\tau}{N}$.
  Let $\env \models \Gamma$.
  Further, let $h \in \semantics{\refT{x}{\tau}{\varphi}}_{\env}$.
  By the semantics of heap types, we also know that $h \in \semantics{\tau}_\env$.
  Assumption $\bound{\Gamma}{\tau}{N}$ gives us that $|h(\pIn)| + |h(\pOut)| \leq N$, which is what we want to show.
\end{proof}

\begin{lemma}[Bound Constraints]
  \label{lem:bound-constraints}
  For every $\Gamma$, $x$, $y$, $\tau_{1}$, $\tau_{2}$, and $\varphi$, such that $\cmdVar$ is the only free variable in $\varphi$,
  $\bound{\Gamma,x:\tau_1}{\refT{y}{\tau_2}{\varphi[x/\cmdVar]}}{N}$, if and only if
  $\bound{\Gamma,x:\refT{y}{\tau_1}{\varphi[y/\cmdVar]}}{\tau_2}{N}$.
\end{lemma}
\begin{proof}
  Let $\Gamma$, $x$, $y$, $\tau_{1}$, $\tau_{2}$, and $\varphi$ be given.
  Prove each direction separately.

  \begin{enumerate}
    \item[$(\Rightarrow)$] Assume
      $\bound {\Gamma,x:\tau_1} {\refT{y}{\tau_2}{\varphi[x/\cmdVar]}}{N}$.

      Let $\env \models \Gamma,x:{\refT{y}{\tau_1}{\varphi[y/\cmdVar]}}$.
      We can write $\env = \env'[x \mapsto h]$ for some $\env' \models \Gamma$, and some $h_1 \in \semantics{\tau_{1}}_{\env'}$, such that $\semantics{\varphi[y/\cmdVar]}_{\env'[y \mapsto h_1]} = \semantics{\varphi}_{\env'[\cmdVar \mapsto h_1]} = \mathit{true}$.
      Since $y$ does not occur in $\varphi$, then we also have
      $\semantics{\varphi}_{\env[\cmdVar \mapsto h, y\mapsto h_{2}]} = \mathit{true}$.

      Now, consider $h_{2} \in \semantics{\tau_{2}}_{\env'[x \mapsto h_1]}$.
      To show that $|h_{2}(\pIn)| + |h_{2}(\pOut)| \leq N$.

      Since $\semantics{\varphi}_{\env[\cmdVar \mapsto h,y \mapsto h_{2}]} = \mathit{true} = \semantics{\varphi[x/\cmdVar]}_{\env[x \mapsto h,y \mapsto h_{2}]}$, we can conclude that $h_{2} \in \semantics{{\refT y {\tau_{2}} {\varphi[x/\cmdVar]}}}_{\env'[x\mapsto h]}$.
      Now, since $\env'[x\mapsto h] \models \Gamma,(x:\tau_{1})$, the result follows by our initial assumption
      assumption $\bound{\Gamma,(x:\tau_{1})}{\refT{y}{\tau_{2}}{\varphi[x/\cmdVar]}}{N}$.

    \item[$(\Leftarrow)$]
      Assume $\bound {\Gamma,(x:\refT{y}{\tau_{1}}{\varphi[y/\cmdVar]})} {\tau_{2}} {N}$.

      Let $\env \models \Gamma,(x:\tau_{1})$.
      We can write $\env = \env'[x \mapsto h_{1}]$ where $h_{1} \in \semantics{\tau_{1}}_{\env'}$ and $\env' \models \Gamma$.

      Now consider $h_{2} \in \semantics{\refT y {\tau_{2}} {\varphi[x/\cmdVar]}}_{\env'[x \mapsto h_{1}]}$.
      To show $\length{h_{2}(\pIn)} + \length{h_{2}(\pOut)} < N$.

      By the semantics of heap types, we have $\semantics{\varphi}_{\env'[x \mapsto h_1, y\mapsto h_2]}=\semantics{\varphi}_{\env'[\cmdVar \mapsto h_{1}, y \mapsto h_{2}]} = \mathit{true}$.
      Since $y$ is not free in $\varphi$, we also have $\semantics{\varphi}_{\env'[\cmdVar \mapsto h_{1}]}=\semantics{\varphi[y/\cmdVar]}_{\env'[y \mapsto h_1]}=\mathit{true}$, so we can conclude that $h_{1} \in \semantics{\refT {y} {\tau_{1}}{e[y/\cmdVar]}}_{\env'}$.
      By our initial assumption, every heap in $\tau_2$ is bounded and as such also heap $h_{2} \in \semantics{\refT{y}{\tau_2}{\varphi[x/\cmdVar]}}_{\env'[x \mapsto h_1]}$.
  \end{enumerate}
\end{proof}

\begin{lemma}[Bound Choice]
  \label{lem:bound-choice}
  If $\bound{\Gamma}{\tau_1}{N}$ and $\bound{\Gamma}{\tau_2}{M}$, then $\bound{\Gamma}{\tau_1 + \tau_2}{\max(M,N)}$.
\end{lemma}
\begin{proof}
  Let $\env\models\Gamma$ and $h\in\semantics{\tau_1 + \tau_2}_\env$.
  We have to show that $\length{h(\pIn)} + \length{h(\pOut)} \le \max(M,N)$.
  \begin{description}
    \case{$M = N$}~

    Assume $M = N$, so $\max(M,N) = M = N$.
    By the semantics of heap types $h\in\semantics{\tau_1}_\env$ or $h\in\semantics{\tau_2}_\env$.
    \begin{description}
      \subcase{$h\in\semantics{\tau_1}_\env$}~
      The result immediately follows by assumption $\bound{\Gamma}{\tau_1}{N}$.
      \subcase{$h\in\semantics{\tau_2}_\env$}~
      The result immediately follows by assumption $\bound{\Gamma}{\tau_2}{M}$.
    \end{description}
    \case{$M > N$}~
    Without loss of generality, we assume that $M>N$, so $\max(M,N)=M$.
    By the semantics of heap types $h\in\semantics{\tau_1}_\env$ or $h\in\semantics{\tau_2}_\env$.
    \begin{description}
      \subcase{$h\in\semantics{\tau_1}_\env$}~
      By assumption $\bound{\Gamma}{\tau_1}{N}$ and since by assumption $N < M$, it follows $\bound{\Gamma}{\tau_1}{M}$.
      \subcase{$h\in\semantics{\tau_2}_\env$}~
      The result immediately follows by assumption $\bound{\Gamma}{\tau_2}{M}$.

    \end{description}

  \end{description}
\end{proof}

\begin{lemma}[Bound Substitution]
  \label{lem:bound-subst}
  For all $\Gamma$, $y$, $\tau_{1}$, $\tau_{2}$, $N$,
  $\bound {\Gamma} {\tau_{2}[y \mapsto \tau_1]} N$ if and only if
  $\bound {\Gamma, (y:\tau_{1})} {\tau_{2}} N$.
\end{lemma}
\begin{proof}
  Let $\Gamma$, $y$, $\tau_{1}$, $\tau_{2}$, and $N$ be given.
  Prove each direction separately:
  \begin{enumerate}
    \item[($\Rightarrow$)]
      Assume $\bound\Gamma{\tau_{2}[y \mapsto \tau_{2}]} N$.
      Let $\env \models \Gamma, y:\tau_1$ such that $h_2\in\semantics{\tau_{2}}_{\env}$.
      This means there is some  $\env' \models \Gamma$ and $h_{1}\in\semantics{\tau_{1}}_{\env'}$
      such that $h_{2} \in \semantics{\tau_{2}}_{\env'[y\mapsto h_{1}]}$.
      By the semantics of heap types follows that $h_{2} \in \semantics{\tau_{2}[y \mapsto \tau_{1}]}_{\env'}$.
      With the initial assumption $\bound {\Gamma} {\tau_{2}[y \mapsto \tau_1]} N$, we can conclude that $\length{h_{2}(\pIn)} + \length{h_{2}(\pOut)} \leq N$.

    \item[($\Leftarrow$)]
      Assume $\Gamma, y : \tau_{1} \models \tau_{2} \leq N$.
      Let $\env \models \Gamma$ and let $h_2 \in \semantics{\tau_{2}[y \mapsto \tau_{1}]}_{\env}$.
      By the semantics of heap types, there is some $h_{1} \in \semantics{\tau_{1}}_{\env}$ such that $h_2 \in \semantics{\tau_{2}}_{\env[y \mapsto h_{1}]}$.
      Notice that $\env[y\mapsto h_{1}] \models \Gamma,y:\tau_1$.
      The initial assumption proves that $|h(\pIn)| + |h(\pOut)| \leq N$.

  \end{enumerate}
\end{proof}

We can then formulate the following two theorems.

\begin{theorem}[Forwards MTU Bound]
  \label{thm:forwards-mtu-bound}
  For every $\Gamma$, $c$, $x$, $\tau_{1}$, $\tau_{2}$, and $N\in\mathbb{N}$, if $\Gamma \vdash \tau_{1} \leq N$ and $\cmdTypeAlgo{\Gamma}{c}{x:\tau_1}{\tau_2}$ and every ascribed type in $c$ is also bounded by $N$, then
  $\bound{\Gamma,x:\tau_1}{\tau_2}{N + \emit{c}}$
\end{theorem}
\begin{proof}
  Proceed by induction on $c$, leaving $\Gamma$ and $N$ general.
  We refer to the general assumptions as follows:
  \begin{enumerate}[label=(\Alph*)]
    \item \label{forwardsmtubound-a1} $\bound{\Gamma}{\tau_1}{N}$
    \item \label{forwardsmtubound-a2} $\cmdTypeAlgo{\Gamma}{c}{x:\tau_1}{\tau_2}$
  \end{enumerate}
  \begin{description}
    \case{\cExtract{\inst}}~

    The only algorithmic typing rule that applies to \cExtract{\inst} is \tExtractAlgo.
    By inversion, we know
    \begin{pf}{T-Extract-Algo}
      \item $\tau_{2} = \sigmaT{y}{\refT{z}{\inst}{\varphi_{1}}}{\refT{z}{\chomp(\tau_{1},\inst,y)} {\varphi_{2}}}$
      \item $\varphi_{1} = z.\pIn = z.\pOut = \bv{}$
      \item $\varphi_{2} = y.\inst@z.\pIn = x.\pIn \wedge z.\pOut = x.\pOut \wedge z \equiv_{\inst} x$.
    \end{pf}
    Since $\emit{\cExtract{\inst}} = 0$, it suffices to show $\bound {\Gamma,x:\tau_{1}}{\tau_{2}}{N}$.

    Let $\env \models \Gamma,x:\tau_1$ and let $h_{2} \in \semantics{\tau_2}_{\env}$.
    We can write $\env$ as $\env'[x \mapsto h_1]$ where $h_1 \in \semantics{\tau_1}_{\env'}$.

    By definition of the semantics of heap types we know there are some $h_{21}$ and $h_{22}$ such that
    \begin{pf*}{T-Extract-Algo}
      \item $h_{2} = \concat{h_{21}}{h_{22}}$,
      \item $h_{21}(\pIn) = h_{21}(\pOut) = \bv{}$
      \item $h_{1}(\pIn) = h_{21}(\inst)@h_{22}(\pIn)$
      \item $h_{1}(\pOut) = h_{22}(\pOut)$
    \end{pf*}
    We can further conclude that
    \begin{pf*}{T-Extract-Algo}
      \item \label{forwardsmtubound-textract-algo-h2-pout} $h_{2}(\pOut) = h_{1}(\pOut)$,
      \item \label{forwardsmtubound-textract-algo-h2-pin} $h_{2}(\pIn) = h_{22}(\pIn) = \slice{h_{1}(\pIn)}{\length{\inst}}{\ }$
    \end{pf*}
    From assumption \ref{forwardsmtubound-a1} follows
    \begin{pf*}{T-Extract-Algo}
      \item $\length{h_1(\pIn)} + \length{h_1(\pOut)} \le N$
    \end{pf*}
    Together with \ref{forwardsmtubound-textract-algo-h2-pout} and \ref{forwardsmtubound-textract-algo-h2-pin}, we can conclude that $\length{h_2(\pIn)} + \length{h_2(\pOut)} < \length{h_1(\pIn)} + \length{h_1(\pOut)} \le N$.

    \case{\cAdd{\inst}}~

    The only algorithmic typing rule that applies to \cAdd{\inst} is \tAddAlgo.
    By inversion, we know
    \begin{pf}{T-Add-Algo}
      \item $\tau_{2} = \sigmaT{y}{\refT{z}{\tau_1}{z \equiv x}{\refT{z}{\inst}{z.\pIn = z.\pOut = \bv{} \wedge z.\inst = v}}}$
    \end{pf}
    Since $\emit{\cAdd{\inst}} = 0$, it suffices to show that $\bound{\Gamma,x : \tau_{1}}{\tau_{2}}{N}$.

    Let $\env \models \Gamma, x : \tau_{1}$, and $h_2 \in \semantics{\tau_{2}}_{\env}$.
    We can write $\env$ as $\env'[x \mapsto h_{1}]$ where $h_1 \in \semantics{\tau_{1}}_{\env'}$.

    By definition of the semantics of heap types we know there are some $h_{21}$ and $h_{22}$ such that
    \begin{pf*}{T-Add-Algo}
      \item $h_{2} = \concat{h_{21}}{h_{22}}$
      \item $h_{21} = h_1$
      \item $h_{22}(\pIn) = h_{22}(\pOut) = \bv{}$
    \end{pf*}
    From these three equations we can conclude that
    \begin{pf*}{T-Add-Algo}
      \item $h_2(\pIn) = h_1(\pIn)$
      \item $h_2(\pOut) = h_1(\pOut)$
    \end{pf*}
    The result follows by assumption \ref{forwardsmtubound-a1}.

    \case{\cMod{\inst.f}{e}}~

    The only algorithmic typing rule that applies to \cMod{\inst.f}{e} is \tModAlgo.
    By inversion, we know
    \begin{pf}{T-Mod-Algo}
      \item \label{forwardsmtubound-tmod-algo-out} $\tau_2 = \refT{y}{\top}{\varphi_{pkt} \wedge \varphi_\inst \wedge \varphi_{f} \wedge y.\inst.f=e[x/\cmdVar]}$
      \item $\varphi_{pkt} = y.\pIn = x.\pIn \wedge y.\pOut = x.\pOut$
    \end{pf}
    Since $\emit{\cMod{\inst.f}{e}} = 0$, it suffices to show that $\bound{\Gamma,x:\tau_1}{\tau_2}{N}$.

    Let $\env \models \Gamma,x:\tau_1$ and $h_2\in\semantics{\tau_2}_\env$.
    We can write $\env$ as $\env'[x \mapsto h_1]$ where $h_1\in\semantics{\tau_1}_{\env'}$.
    From assumption \ref{forwardsmtubound-tmod-algo-out} and by the semantics of heap types follows
    \begin{pf*}{T-Mod-Algo}
      \item $h_2(\pIn) = h_1(\pIn)$
      \item $h_2(\pOut) = h_1(\pOut)$
    \end{pf*}
    The result follows by assumption \ref{forwardsmtubound-a1}.

    \case{\cRemit{\inst}}~

    The only algorithmic typing rule that applies to \cRemit{\inst} is \tRemitAlgo.
    By inversion, we know
    \begin{pf}{T-Remit-Algo}
      \item $\tau_2 = \sigmaT{y}{\refT{z}{\tau_1}{z \equiv x}}{\refT{z}{\epsilon}{\varphi}}$
      \item $\varphi = z.\pIn = \bvNil \wedge z.\pOut = x.\inst$
    \end{pf}
    Since $\emit{\cRemit{\inst}} = \sizeof(\inst)$, we have to show that $\bound{\Gamma,x:\tau_1}{\tau_2}{N + \sizeof(\inst)}$.

    Let $\env \models \Gamma,x:\tau_1$ and $h_2\in\semantics{\tau_2}_\env$.
    We can write $\env$ as $\env'[x \mapsto h_1]$ where $h_1\in\semantics{\tau_1}_{\env'}$.
    By the semantics of heap types, there exists $h_{21}$ and $h_{22}$ such that
    \begin{pf*}{T-Remit-Algo}
      \item $h_2 = \concat{h_{21}}{h_{22}}$
      \item \label{forwardsmtubound-tremit-h21} $h_{21} = h_1$
      \item \label{forwardsmtubound-tremit-h22-pin} $h_{22}(\pIn) = \bvNil$
      \item \label{forwardsmtubound-tremit-h22-pout} $h_{22}(\pOut) = h_1(\inst)$
    \end{pf*}
    From \ref{forwardsmtubound-tremit-h21} and \ref{forwardsmtubound-tremit-h22-pin}, we can conclude that
    \begin{pf*}{T-Remit-Algo}
      \item \label{forwardsmtubound-tremit-h2-pin} $h_2(\pIn) = h_1(\pIn)$
    \end{pf*}
    From \ref{forwardsmtubound-tremit-h21} and \ref{forwardsmtubound-tremit-h22-pout}, we can further conclude that
    \begin{pf*}{T-Remit-Algo}
      \item \label{forwardsmtubound-tremit-h2-pout} $h_2(\pOut) = \bvconcat{h_1(\pOut)}{h_1(\inst)}$
    \end{pf*}
    From \ref{forwardsmtubound-tremit-h2-pin} and \ref{forwardsmtubound-tremit-h2-pout} then follows
    \begin{pf*}{T-Remit-Algo}
      \item $\length{h_2(\pIn)} + \length{h_2(\pOut)} = \length{h_1(\pIn)} + \length{h_1(\pOut)} + \sizeof(\inst)$
    \end{pf*}
    Together with assumption \ref{forwardsmtubound-a1}, we can conclude that $\length{h_2(\pIn)} + \length{h_2(\pOut)} \le N + \sizeof(\inst)$.

    \case{\cReset}~

    The only algorithmic typing rule that applies to \cReset is \tResetAlgo.
    By inversion, we know
    \begin{pf}{T-Reset-Algo}
      \item $\tau_{2} = \sigmaT{y}{\refT{z}{\epsilon}{\varphi_1}} {\refT{z}{\epsilon}{\varphi_2}}$
      \item $\varphi_1 = z.\pOut = \bvNil \wedge z.\pIn = x.\pOut$
      \item $\varphi_2 = z.\pOut = \bvNil \wedge z.\pIn = x.\pIn$
    \end{pf}

    Since $\emit{\cReset} = 0$, we have to show that $\bound{\Gamma,x:\tau_1}{\tau_2}{N}$.

    Let $\env \models \Gamma,x:\tau_1$ and $h_2\in\semantics{\tau_2}_\env$.
    We can write $\env$ as $\env'[x \mapsto h_1]$ where $h_1\in\semantics{\tau_1}_{\env'}$.
    By the semantics of heap types, there exists $h_{21}$ and $h_{22}$ such that
    \begin{pf*}{T-Reset-Algo}
      \item \label{forwardsmtubound-treset-h2} $h_2 = \concat{h_{21}}{h_{22}}$
      \item \label{forwardsmtubound-treset-h21-pout} $h_{21}(\pOut) = \bvNil$
      \item \label{forwardsmtubound-treset-h21-pin} $h_{21}(\pIn) = h_1(\pOut)$
      \item \label{forwardsmtubound-treset-h22-pout} $h_{22}(\pOut) = \bvNil$
      \item \label{forwardsmtubound-treset-h22-pin} $h_{22}(\pIn) = h_1(\pIn)$
    \end{pf*}
    By \ref{forwardsmtubound-treset-h2}, \ref{forwardsmtubound-treset-h21-pout} and \ref{forwardsmtubound-treset-h22-pout} follows
    \begin{pf*}{T-Reset-Algo}
      \item \label{forwardsmtubound-treset-h2-pout} $h_2(\pOut) = \bvNil$
    \end{pf*}
    and by \ref{forwardsmtubound-treset-h2}, \ref{forwardsmtubound-treset-h21-pin} and \ref{forwardsmtubound-treset-h22-pin} follows
    \begin{pf*}{T-Reset-Algo}
      \item \label{forwardsmtubound-treset-h2-pin} $h_2(\pIn) = \bvconcat{h_1(\pOut)}{h_1(\pIn)}$
    \end{pf*}
    Since by assumption \ref{forwardsmtubound-a1}, $|h_1(\pIn)| + |h_1(\pOut)| \leq N$, by \ref{forwardsmtubound-treset-h2-pout} and \ref{forwardsmtubound-treset-h2-pin}, $\length{h_2(\pIn)} + \length{h_2(\pOut)} \leq N$.

    \case{\cIf{\varphi}{c_1}{c_2}}~
    The only algorithmic typing rule that applies is \tIfAlgo.
    By inversion, we know
    \begin{pf}{T-If-Algo}
      \item $\tau_{2} = \refT{y}{\tau_{12}}{\varphi[x/\cmdVar]} + \refT{y}{\tau_{22}}{\neg \varphi[x/\cmdVar]}$
      \item \label{forwardsmtubound-tif-algo-c1-type} $\cmdTypeAlgo{\Gamma}{c_1}{x : \refT{y}{\tau_1}{\varphi[y/\cmdVar]}}{\tau_{12}}$
      \item \label{forwardsmtubound-tif-algo-c2-type} $\cmdTypeAlgo{\Gamma}{c_1}{x : \refT{y}{\tau_1}{\neg \varphi[y/\cmdVar]}}{\tau_{22}}$
    \end{pf}

    Since $\emit{\cIf{\varphi}{c_1}{c_2}} = \max(\emit{c_1},\emit{c_2})$, we have to show that $\bound{\Gamma,x:\tau_1}{\tau_2}{N + \max(\emit{c_1},\emit{c_2})}$.

    By \cref{lem:refinement-bound} and assumption \ref{forwardsmtubound-a1} follows
    \begin{pf*}{T-If-Algo}
      \item \label{forwardsmtubound-tif-algo-c1-type-bound} $\bound{\Gamma}{\refT{y}{\tau_1}{\varphi[y/\cmdVar]}}{N}$
      \item \label{forwardsmtubound-tif-algo-c2-type-bound} $\bound{\Gamma}{\refT{y}{\tau_1}{\neg \varphi[y/\cmdVar]}}{N}$
    \end{pf*}
    Applying the IH to \ref{forwardsmtubound-tif-algo-c1-type} and \ref{forwardsmtubound-tif-algo-c1-type} with \ref{forwardsmtubound-tif-algo-c1-type-bound} and \ref{forwardsmtubound-tif-algo-c2-type-bound} respectively, gives
    \begin{pf*}{T-If-Algo}
      \item \label{forwardsmtubound-tif-algo-t12-bound} $\bound{\Gamma,x:\refT{y}{\tau_1}{\varphi[y/\cmdVar]}}{\tau_{12}}{N + \emit{c_1}}$
      \item \label{forwardsmtubound-tif-algo-t22-bound} $\bound{\Gamma,x:\refT{y}{\tau_1}{\neg\varphi[y/\cmdVar]}}{\tau_{22}}{N + \emit{c_2}}$
    \end{pf*}
    By \cref{lem:bound-constraints} with \ref{forwardsmtubound-tif-algo-t12-bound} and \ref{forwardsmtubound-tif-algo-t22-bound} respectively follows
    \begin{pf*}{T-If-Algo}
      \item \label{forwardsmtubound-tif-algo-t12-ref-bound} $\bound{\Gamma,x:\tau_1}{\refT{y}{\tau_{12}}{\varphi[x/\cmdVar]}}{N + \emit{c_1}}$
      \item \label{forwardsmtubound-tif-algo-t22-ref-bound} $\bound{\Gamma,x:\tau_1}{\refT{y}{\tau_{22}}{\neg\varphi[x/\cmdVar]}}{N + \emit{c_2}}$
    \end{pf*}
    By \cref{lem:bound-choice} with \ref{forwardsmtubound-tif-algo-t12-ref-bound} and \ref{forwardsmtubound-tif-algo-t22-ref-bound} follows
    \begin{pf*}{T-If-Algo}
      \item $\bound{\Gamma,x:\tau_1}{\refT{y}{\tau_{12}}{\varphi[x/\cmdVar]}+\refT{y}{\tau_{22}}{\neg\varphi[x/\cmdVar]}}{\max(N+\emit{c_1},N+\emit{c_2})}$
    \end{pf*}
    The result follows together with the fact that $\max(A + B, A + C) = A + \max(B,C)$.

    \case{$c_1;c_2$}~

    The only algorithmic typing rule that applies to $c_1;c_2$ is \tSeqAlgo.
    By inversion, we know
    \begin{pf}{T-Seq-Algo}
      \item \label{forwardsmtubound-tseq-algo-c1} $\cmdTypeAlgo{\Gamma}{c_1}{x:\tau_1}{\tau_{12}}$
      \item \label{forwardsmtubound-tseq-algo-c2} $\cmdTypeAlgo{\Gamma,x:\tau_1}{c_2}{y:\tau_{12}}{\tau_{22}}$
      \item \label{forwardsmtubound-tseq-algo-t2} $\tau_2 = \tau_{22}[y \mapsto \tau_{12}]$
    \end{pf}
    Since $\emit{c_1;c_2} = \emit{c_1} + \emit{c_2}$, we have to show that $\bound{\Gamma,x:\tau_1}{\tau_2}{N + \emit{c_1} + \emit{c_2}}$.
    Let $\env \models \Gamma,x:\tau_1$ and let $h_2\in\semantics{\tau_2}_\env$.

    By applying the IH to \ref{forwardsmtubound-tseq-algo-c1}, we get
    \begin{pf*}{T-Seq-Algo}
      \item $\bound{\Gamma,x:\tau_1}{\tau_{12}}{N + \emit{c_1}}$
    \end{pf*}
    Since we left $\Gamma$ and $N$ general, we can apply the IH again to \ref{forwardsmtubound-tseq-algo-c2} and get
    \begin{pf*}{T-Seq-Algo}
      \item \label{forwardsmtubound-tseq-algo-bound-t22} $\bound{\Gamma,x:\tau_1,y:\tau_{12}}{\tau_{22}}{N + \emit{c_1} + \emit{c_2}}$
    \end{pf*}
    The result follows by \cref{lem:bound-subst} with \ref{forwardsmtubound-tseq-algo-bound-t22}.

    \case{\cSkip}~

    The only algorithmic typing rule that applies to \cSkip is \tSkipAlgo.
    By inversion, we know
    \begin{pf}{T-Skip-Algo}
      \item $\tau_2 = \refT{y}{\tau_1}{y \equiv x}$
    \end{pf}
    Since $\emit{\cSkip} = 0$, we have to show that $\bound{\Gamma,x:\tau_1}{\tau_2}{N}$.

    Let $\env \models \Gamma,x:\tau_1$ and $h_2\in\semantics{\tau_2}_\env$.
    We can write $\env$ as $\env'[x \mapsto h_1]$ where $h_1\in\semantics{\tau_1}_{\env'}$.

    By the semantics of heap types, follows
    \begin{pf*}{T-Skip-Algo}
      \item $h_2(\pIn) = h_1(\pIn)$
      \item $h_2(\pOut) = h_1(\pOut)$
    \end{pf*}
    The result follows by assumption \ref{forwardsmtubound-a1}.

    \case{\cAscribe{c_0}{(x:\hat\tau_1) \to \tau_2}}~

    The only algorithmic typing rule that applies is \tAscribeAlgo.
    By inversion, we know
    \begin{pf}{T-Ascribe-Algo}
      \item \label{forwardsmtubound-tascribe-algo-c0} $\cmdTypeAlgo{\Gamma}{c_0}{x:\hat\tau_1}{\tau_c}$
      \item \label{forwardsmtubound-tascribe-algo-t1} $\subtype{\tau_1}{\hat\tau_1}$
      \item \label{forwardsmtubound-tascribe-algo-t2} $\subtypeCtx{\Gamma,x:\hat\tau_1}{\tau_c}{\tau_2}$
    \end{pf}
    Since $\emit{\cAscribe{c_0}{\sigma}} = \emit{c_0}$, we have to show that $\bound{\Gamma,x:\tau_1}{\tau_2}{N + \emit{c_0}}$.
    By our initial assumption, every ascribed type is also bounded by $N$.
    We therefore have $\bound{\Gamma,x:\tau_1}{\tau_2}{N}$ from which the result immediately follows.
  \end{description}
\end{proof}

\begin{theorem}[Decidability]
  \label{thm:decidability}
  If $\Gamma$, $\tau_1$, $\tau_2$ and every ascribed type in $c$ are bounded by the MTU $N$,
  then $\cmdType{\Gamma}{c}{(x:\tau_1) \to \tau_2}$ is decidable.
\end{theorem}
\begin{proof}
  By \cref{thm:algorithmic-typing-correctness} (Algorithmic Typing Correctness), we can equivalently show that $\cmdTypeAlgo{\Gamma}{c}{x:\tau_1}{\tau_2'}$ and $\subtypeCtx{\Gamma,x:\tau_1}{\tau_2'}{\tau_2}$ are decidable.
  By \cref{thm:forwards-mtu-bound}, $\tau_2'$ is bounded.
  $\subtypeCtx{\Gamma,x:\tau_1}{\tau_2'}{\tau_2}$ is therefore decidable by finite enumeration.

  To show that $\cmdTypeAlgo{\Gamma}{c}{x:\tau_1}{\tau_2'}$ is decidable, we proceed by induction on the algorithmic typing derivation.
  \begin{description}
    \case{\tSkipAlgo}~

    Immediate, because \tSkipAlgo does not perform any subtyping checks.

    \case{\tResetAlgo}~ 

    Also immediate, because \tResetAlgo does not perform any subtyping checks.

    \case{\tSeqAlgo}~

    By inversion of \tSeqAlgo,
    \begin{pf}{T-Seq-Algo}
      \item \label{decidability-tseq-algo-c1} $\cmdTypeAlgo{\Gamma}{c_1}{x:\tau_1}{\tau_{12}}$
      \item \label{decidability-tseq-algo-c2} $\cmdTypeAlgo{\Gamma,x:\tau_1}{c_2}{y:\tau_{12}}{\tau_{22}}$
    \end{pf}
    By \ref{decidability-tseq-algo-c1} and \cref{thm:forwards-mtu-bound},
    \begin{pf*}{T-Seq-Algo}
      \item \label{decidability-tseq-algo-bound-t12} $\bound{\Gamma,x:\tau_1}{\tau_{12}}{N + \emit{c_1}}$
    \end{pf*}
    Applying the IH to \ref{decidability-tseq-algo-c1} with assumption $\bound{\Gamma}{\tau_1}{N}$ and \ref{decidability-tseq-algo-bound-t12} gives us that $\cmdTypeAlgo{\Gamma}{c_1}{x:\tau_1}{\tau_{12}}$ is decidable.

    Again, by \cref{thm:forwards-mtu-bound} with \ref{decidability-tseq-algo-c2} and \ref{decidability-tseq-algo-bound-t12}, follows
    \begin{pf*}{T-Seq-Algo}
      \item $\bound{\Gamma,x:\tau_1,y:\tau_{12}}{\tau_{22}}{N + \emit{c_1} + \emit{c_2}}$
    \end{pf*}
    By IH follows that $\cmdTypeAlgo{\Gamma,x:\tau_1}{c_2}{y:\tau_{12}}{\tau_{22}}$ is decidable and thus typechecking the sequence of both commands is decidable.

    \case{\tAddAlgo}~

    By inversion, we know that \tAddAlgo performs the subtyping check $\subtype{\tau_1}{\refT{x}{\top}{\neg x.\inst.\valid}}$.
    To show that typechecking is decidable in this case, we must show that $\subtype{\tau_1}{\refT{x}{\top}{\neg x.\inst.\valid}}$ is decidable.
    This is the case because we can finitely enumerate the heaps $h$ described by $\tau_1$ and check wether every $h$ is a member of $\refT{x}{\top}{\neg y.\inst.\valid}$.

    \case{\tExtractAlgo}~

    By inversion, we know that \tExtractAlgo performs the subtyping check $\subtype{\tau_1}{\refT{x}{\top}{\length{x.\pIn} \ge \sizeof(\inst)}}$.
    To show that typechecking is decidable in this case, we must show that $\subtype{\tau_1}{\refT{x}{\top}{\length{x.\pIn} \ge \sizeof(\inst)}}$ is decidable.
    This is the case because we can finitely enumerate the heaps $h$ described by $\tau_1$ and check wether every $h$ is a member of $\refT{x}{\top}{\length{x.\pIn} \ge \sizeof(\inst)}$.

    \case{\tRemitAlgo}~
    
    Identical to the previous subcase.

    \case{\tModAlgo}~

    Identical to the previous subcase.

    \case{\tIfAlgo}~

    Since $\tau_1$ is bounded by assumption and refining the input type does not increase the size, $\refT{y}{\tau_1}{\varphi[y/\cmdVar]}$ and $\refT{y}{\tau_1}{\neg \varphi[y/\cmdVar]}$ are still bounded.
    By \cref{thm:forwards-mtu-bound} then follows that the output types of $c_1$ and $c_2$ are also bounded.
    By IH applied to $c_1$ and $c_2$, we get that the algorithmic type checking applied to $c_1$ and $c_2$ respectively is decidable and thus checking the conditional is decidable.

    \case{\tAscribeAlgo}~

    By assumption, $\Gamma, \tau_1$ and $\hat\tau_1$ are bounded, so $\subtype{\tau_1}{\hat\tau_1}$ is decidable by finite enumeration.
    Since by assumption $\bound{\Gamma}{\tau_1}{N}$, by \cref{thm:forwards-mtu-bound} follows that $\bound{\Gamma,x:\hat\tau_1}{\tau_c}{N + \emit{c_0}}$.
    By IH then follows that $\cmdTypeAlgo{\Gamma}{c}{x:\tau_1}{\tau_2'}$ is decidable.
    Since $\tau_c$ is bounded and by assumption also $\tau_2'$ is bounded, we can finitely enumerate, so $\subtypeCtx{\Gamma,x:\hat\tau_1}{\tau_c}{\tau_2'}$ is also decidable and thus typechecking an ascribed command is decidable.

  \end{description}
\end{proof}

%% file: appendix/rewrites.tex
\section{Rewrite Optimization Correctness}

\begin{definition}[Type Equivalence]
  We write $\typeEquiv{\Gamma}{\tau_1}{\tau_2}$ for the equivalence of types $\tau_1$ and $\tau_2$ in context $\Gamma$, i.e., $\typeEquiv{\Gamma}{\tau_1}{\tau_2} \triangleq \forall \env\models \Gamma.\semantics{\tau_1}_\env = \semantics{\tau_2}_\env$
\end{definition}

\begin{lemma}[Rewriting Sigma Types]
  \label{lem:rewriting-sigma-types}
  In any context $\Gamma$,
  \[
    \begin{array}{c}
      \Gamma \vdash \sigmaT{x}{\tau_1}{\tau_2} \\
      \doteq                     \\
      \begin{array}{ll}
        \left\{ y:\top\;\middle|\;
        \begin{array}{l}
          \left( \begin{array}{l}
                     y.\pIn = \bvconcat{x.\pIn}{r.\pIn}\ \wedge \\
                     y.\pOut = \bvconcat{x.\pOut}{r.\pOut}
                   \end{array} \right)\ \wedge \\
          \bigwedge_{\inst\in\dom(\HT)}\left(
          \begin{array}{l}
              y.\inst.\valid = x.\inst.\valid \oplus r.\inst.\valid\ \wedge \\
              x.\inst.\valid \implies y.\inst = x.\inst\ \wedge             \\
              r.\inst.\valid \implies y.\inst = r.\inst
            \end{array}\right)
        \end{array} \right\}\left[r \mapsto \tau_2\right][x \mapsto \tau_1]
      \end{array}
    \end{array}
  \]
\end{lemma}
\begin{proof}
  Proof each direction separately.
  \begin{enumerate}
    \item[($\Rightarrow)$]
      Let $\env \models \Gamma$ and let $h \in \semantics{\sigmaT{x}{\tau_{1}}{\tau_{2}}}_{\env}$.
      By the semantics of heap types, we know there exists $h_{1}$ and $h_{2}$ such that $h = \concat{h_{1}}{h_{2}}$ and $h_{1} \in \semantics{\tau_{1}}_{\env}$ and $h_{2} \in \semantics{\tau_{2}}_{\env[x \mapsto h_{1}]}$.

      By definition of heap concatenation, $h(\pIn) = \bvconcat{h_{1}(\pIn)}{h_{2}(\pIn)}$ and also $h(\pOut) = \bvconcat{h_{1}(\pOut)}{h_{2}(\pOut)}$.
      Further, $\dom(h)$ is the disjoint union of $\dom(h_{1})$ and $\dom(h_{2})$ such that if $\inst \in \dom(h_{i})$, $h(\inst) = h_{i}(\inst)$ for each $i = 1,2$ and each $\inst \in \dom(\HT)$.
      The result follows by definition of the semantics.

    \item[($\Leftarrow)$]
      Let $\env \models \Gamma$.
      By the definition of the semantics, it suffices to show, for  $h_{1}\in\semantics{\tau_{1}}_{\env}$, and $h_{2}\in\semantics{\tau_{2}}_{\env[x \mapsto h_{1}]}$, and $h \in \semantics{\top}_{\env[x \mapsto h_{1}, r \mapsto h_{2}]}$ such that the above refinement holds for $h$, that $h \in \semantics{\sigmaT{x}{\tau_{1}}{\tau_{2}}}_\env$.
      By the semantics of heap types, it suffices to show that $h=\concat{h_1}{h_2}$.
      The refinement tells us that
      \begin{itemize}
        \item $h(\pIn) = \bvconcat{h_{1}(\pIn)}{h_{2}(\pIn)}$ and
        \item $h(\pOut) = \bvconcat{h_{1}(\pOut)}{h_{2}(\pOut)}$.
      \end{itemize}
      Further, $\dom(h)$ is the disjoint union of $\dom(h_{1})$ and $\dom(h_{2})$ such that if $\inst \in \dom(h_{i})$, $h(\inst) = h_{i}(\inst)$ for each $i = 1,2$ and each $\inst \in \dom(\HT)$.
  \end{enumerate}
\end{proof}

\begin{lemma}[Rewriting Refinement Types]
  For $\Gamma$, $\tau$, $\inst$, $x$, $y$, such that $x$ and $y$
  do not occur free in $\tau$,
  \[\typeEquiv{\Gamma,(x:\tau)}{\refT{y}{\tau}{x \equiv y}}{\refT{y}{\top}{x \equiv y}}\]
\end{lemma}
\begin{proof}
  Prove each direction separately.
  \begin{enumerate}[align=left]
    \item[($\Rightarrow$)]
      Let $\env \models \Gamma,(x:\tau)$.
      We know $\env = \env'[x\mapsto h_{1}]$ such that $h_{1} \in \semantics{\tau}_{\env'}$.
      Let $h_{2}\in\semantics{\refT{y}{\tau}{x \equiv y}}_{\env'[x \mapsto h_{1}]}$.
      Then $h_{2} \in \semantics{\tau}_{\env'[x\mapsto h_{1}]}$ and $\semantics{x \equiv y}_{\env'[x\mapsto h_{1},y\mapsto h_{2}]}=\mathit{true}$.
      From the latter, we can conclude that $h_2=h_1$.
      To show $h_2\in\semantics{\refT{y}{\top}{y\equiv x}}_{\env'[x \mapsto h_1]}$, we have to show that $h_2\in\semantics{\top}_{\env'[x \mapsto h_1]}$, which is immediate, and $\semantics{x\equiv y}_{\env'[x \mapsto h_1, y\mapsto h_2]}=\mathit{true}$, which immediately follows by the fact that $h_2=h_1$.

    \item[($\Leftarrow$)]
      Let $\env \models \Gamma,(x:\tau)$.
      We know $\env = \env'[x\mapsto h_{1}]$ such that $h_{1} \in \semantics{\tau}_{\env'}$.
      Let $h_{2} \in \semantics{\refT{y}{\top}{x \equiv y}}_{\env'[x \mapsto h_{1}]}$.
      Then $h_{2} \in \semantics{\top}_{\env'[x\mapsto h_{1}]}$,
      and $\semantics{x \equiv y}_{\env[x\mapsto h_{1},y\mapsto h_{2}]}=\mathit{true}$.
      Observe that $h_{1} = h_{2}$.
      To show that $h_2\in\semantics{\refT{y}{\tau}{y \equiv x}}_{\env'[x \mapsto h_1]}$, we must show that
      $h_2\in\semantics{\tau}_{\env'[x \mapsto h_1]}$ and $\semantics{x \equiv y}_{\env[x\mapsto h_{1},y\mapsto h_{2}]}=\mathit{true}$.
      The first follows by assumption that $h_1\in\semantics{\tau}_{\env'}$ and the fact that $h_2=h_1$.
      The second immediately follows from $h_2=h_1$.

  \end{enumerate}
\end{proof}

\begin{lemma}
  For all $\Gamma$, $x$, $\tau$ and $\inst$, if $\Gamma \vdash \sizeof_{\pIn}(\tau) \ge \sizeof(\inst)$ and  $x$ does not occur free in $\tau$, then
  \[
    \begin{array}{c}
      \Gamma,x:\tau \vdash            \\
      \text{\Large $\Sigma$} y : \left\{z : \inst\;\middle|\;
      \begin{array}[center]{l}
        z.\pIn = \bv{}\ \wedge \\
        z.\pOut = \bv{}
      \end{array}\right\} \mathrel{.}
      \left\{ z : \chomp(\tau, \inst, y)\;\middle|\;
      \begin{array}[center]{l}
        y.\inst@z.\pIn = x.\pIn\ \wedge \\
        z.\pOut = x.\pOut\ \wedge       \\
        z \equiv_{\inst} x
      \end{array} \right\} \\
      \doteq                          \\
      \left\{ y:\top \;\middle|\; \begin{array}{l}
                                    y.\inst.\valid \wedge \bigwedge_{\kappa\in\dom(\HT) \wedge \kappa\neq\inst} y.\kappa = x.\kappa\ \wedge \\
                                    \bvconcat{y.\inst}{y.\pIn} = x.\pIn \wedge y.\pOut\ = x.\pOut \wedge
                                  \end{array}\right\}
    \end{array}
  \]
\end{lemma}[Rewrite Sigma Extract]
\begin{proof}
  Proof each direction separately.
  \begin{enumerate}
    \item[($\Rightarrow$)]
      Let $\env \models \Gamma,x:\tau$.
      We know $\env = \env'[x \mapsto h]$ such that $h\in\semantics{\tau}_{\env'}$.
      Let $h_\Sigma\in\semantics{\sigmaT{y}{\refT{z}{\inst}{z.\pIn=z.\pOut=\bvNil}}{\refT{z}{\chomp(\tau,\inst,y)}{\bvconcat{y.\inst}{z.\pIn} = x.\pIn \wedge z.\pOut = x.\pOut \wedge z \equiv_\inst x}}}_{\env'[x \mapsto h]}$ be arbitrary.
      By the semantics of heap types follows
      \begin{pf}{Forwards}
        \item \label{rewrite-sigma-extract-forwards-hsigma-concat} $h_\Sigma=\concat{h_1}{h_2}$
        \item $h_1\in\semantics{\refT{z}{\inst}{z.\pIn=z.\pOut=\bvNil}}_{\env'[x \mapsto h]}$
        \item $h_2\in\semantics{\refT{z}{\chomp(\tau,\inst,y)}{\bvconcat{y.\inst}{z.\pIn} = x.\pIn \wedge z.\pOut = x.\pOut \wedge z \equiv_\inst x}}_{\env'[x \mapsto h, y \mapsto h_1]}$
        \item $h_2\in\semantics{\chomp(\tau,\inst,y)}_{\env'[x \mapsto h, y \mapsto h_1]}$
        \item \label{rewrite-sigma-extract-forwards-refine-h2} $\semantics{\bvconcat{y.\inst}{z.\pIn} = x.\pIn \wedge z.\pOut = x.\pOut \wedge z \equiv_\inst x}_{\env'[x \mapsto h, y \mapsto h_1, z \mapsto h_2]} = \mathit{true}$
      \end{pf}
      By \cref{lem:semantic-chomp-inverse}, there exists $\hat{h}_2\in\semantics{\tau}_{\env'[x \mapsto h]}$ such that
      \begin{pf*}{Forwards}
        \item \label{rewrite-sigma-extract-forwards-h2-chomp} $h_2 = \chompS(\hat{h}_2, \sizeof(\inst))$
      \end{pf*}
      Together with \ref{rewrite-sigma-extract-forwards-refine-h2}, we can conclude that
      \begin{pf*}{Forwards}
        \item \label{rewrite-sigma-extract-forwards-hath2-pout} $\hat{h}_2(\pOut) = h(\pOut)$
        \item \label{rewrite-sigma-extract-forwards-hath2-pin} $\hat{h}_2(\pIn) = \slice{h(\pIn)}{\sizeof(\inst)}{\ }$
        \item \label{rewrite-sigma-extract-forwards-hath2-kappa} $\forall \kappa\ne\inst.\hat{h}_2(\kappa) = h(\kappa)$
        \item \label{rewrite-sigma-extract-forwards-hath2-inst} $\inst\not\in\dom(\hat{h}_2)$
      \end{pf*}
      $h_\Sigma\in\semantics{\refT{y}{\top}{y.\inst.\valid \wedge \bigwedge_{\kappa\in\dom(\HT) \wedge \kappa\neq\inst} y.\kappa = x.\kappa\ \wedge \bvconcat{y.\inst}{y.\pIn} = x.\pIn \wedge y.\pOut\ = x.\pOut}}_\env$ follows by the semantics of heap types with \ref{rewrite-sigma-extract-forwards-hsigma-concat}, \ref{rewrite-sigma-extract-forwards-h2-chomp}, \ref{rewrite-sigma-extract-forwards-hath2-pout}, \ref{rewrite-sigma-extract-forwards-hath2-pin}, \ref{rewrite-sigma-extract-forwards-hath2-kappa} and \ref{rewrite-sigma-extract-forwards-hath2-inst}.

    \item[($\Leftarrow$)]
      Let $\env \models \Gamma,x:\tau$.
      We know $\env = \env'[x \mapsto h]$ such that $h\in\semantics{\tau}_{\env'}$.
      Let $\hat{h}\in\semantics{\refT{y}{\top}{y.\inst.\valid \wedge \bigwedge_{\kappa\in\dom(\HT) \wedge \kappa\neq\inst} y.\kappa = x.\kappa \wedge \bvconcat{y.\inst}{y.\pIn} = x.\pIn \wedge y.\pOut\ = x.\pOut}}_{\env'[x \mapsto h]}$

      By the semantics of heap types,
      \begin{pf}{Backwards}
        \item $\hat{h}(\inst) = \slice{h(\pIn)}{0}{\sizeof(\inst)}$
        \item $\forall\kappa\ne\inst.\hat{h}(\kappa) = h(\kappa)$
        \item $\hat{h}(\pOut) = h(\pOut)$
        \item $\bvconcat{\hat{h}(\inst)}{\hat{h}(\pIn)} = h(\pIn) \Leftrightarrow \hat{h}(\pIn) = \slice{h(\pIn)}{\sizeof(\inst)}{\ }$
      \end{pf}
      To show that $\hat{h}\in\semantics{\sigmaT{y}{\refT{z}{\inst}{z.\pIn=z.\pOut=\bvNil}}{\refT{z}{\chomp(\tau,\inst,y)}{\bvconcat{y.\inst}{z.\pIn} = x.\pIn \wedge z.\pOut = x.\pOut \wedge z \equiv_\inst x}}}_{\env'[x \mapsto h]}$, we have to show that there exists $h_1$ and $h_2$ such that $\hat{h}=\concat{h_1}{h_2}$ and $h_1\in\semantics{\refT{z}{\inst}{z.\pIn=z.\pOut=\bvNil}}_{\env'[x \mapsto h]}$ and $h_2\in\semantics{\refT{z}{\chomp(\tau,\inst,y)}{\bvconcat{y.\inst}{z.\pIn} = x.\pIn \wedge z.\pOut = x.\pOut \wedge z \equiv_\inst x}}_{\env'[x \mapsto h, y \mapsto h_1]}$.

      Let $h_1(\pIn) = h_1(\pOut) = \bvNil$ and $h_1(\inst) = \slice{h(\pIn)}{0}{\sizeof(\inst)}$ and no other instances be valid in heap $h_1$.

      $h_1\in\semantics{\refT{z}{\inst}{z.\pIn=z.\pOut=\bvNil}}_{\env'[x \mapsto h]}$ then follows by the semantics of heap types.
      By \cref{lem:semantic-chomp}, there exists $h_2\in\semantics{\chomp(\tau,\inst,y)}_{\env'[y \mapsto h_1]}$ such that $h_2 = \chompS(h,\sizeof(\inst))$.
      Since $x$ not free in $\tau$, it also holds that $h_2\in\semantics{\chomp(\tau,\inst,y)}_{\env'[x \mapsto h, y \mapsto h_1]}$.

      Since $\bvconcat{h_1(\inst)}{h_2(\pIn)} = %
        h(\pIn)$, $h_2(\pOut) = h(\pOut)$ and since $\chomp$ does not change already valid header instances also for all $\kappa\ne\inst$, $h_2(\kappa) = h(\kappa)$, we can conclude that
      $\semantics{\bvconcat{y.\inst}{z.\pIn} = x.\pIn \wedge z.\pOut = x.\pOut \wedge z \equiv_\inst x}_{\env'[x \mapsto h, y \mapsto h_1, z \mapsto h_2]} = \mathit{true}$ and thus
      $h_2\in\semantics{\refT{z}{\chomp(\tau,\inst,y)}{\bvconcat{y.\inst}{z.\pIn} = x.\pIn \wedge z.\pOut = x.\pOut \wedge z \equiv_\inst x}}_{\env'[x \mapsto h, y \mapsto h_1]}$.

      By the semantics of heap types, we can further conclude that
      \begin{pf*}{Backwards}
        \item $(\concat{h_1}{h_2})(\pIn) = h_2(\pIn) = h'(\pIn)$
        \item $(\concat{h_1}{h_2})(\pOut) = h(\pOut) = h'(\pOut)$
        \item $(\concat{h_1}{h_2})(\inst) = \slice{h(\pIn)}{0}{\sizeof(\inst)} = h'(\inst)$
        \item $\forall\kappa\ne\inst.(\concat{h_1}{h_2})(\kappa) = h_2(\kappa) = h(\kappa) = h'(\kappa)$
      \end{pf*}
      This show that actually $h'=\concat{h_1}{h_2}$ and concludes this case.
  \end{enumerate}
\end{proof}